\documentclass[useAMS,usenatbib]{mn2e}

\usepackage{graphicx}
\usepackage{amsmath}
\usepackage{amssymb}
\usepackage{mathrsfs}
\usepackage{boxedminipage}
\usepackage{aas_macros} 
\usepackage{natbib}
\usepackage{threeparttable}
\usepackage{appendix}
\usepackage{color,latexsym}
\usepackage{rotating}
\usepackage{longtable,arcs}
\usepackage{subfigure}
\usepackage{setspace}
\usepackage{lscape}
\usepackage{Times}
\pdfoutput=1
\newcommand{\myr}{${\rm M_{\sun}\,yr^{-1}}$}
\newcommand{\Msol}{${\rm M_{\sun}}$}
\newcommand{\Lsol}{${\rm L_{\sun}}$}
\newcommand{\um}{$\umu$m}
\newcommand{\uJy}{$\umu$Jy}

\begin{document}

\title[SMG source counts in high-z galaxy clusters]{Submillimetre Source Counts in the Fields of High-Redshift Galaxy Clusters}
\author[A.G. Noble et al.]{\parbox[t]{\textwidth}{A.G. Noble,$^{1}$\thanks{E-mail: nobleal@physics.mcgill.ca}
T.M.A. Webb,$^{1}$
E. Ellingson,$^{2}$
A.J. Faloon,$^{1}$
R.R. Gal,$^{3}$
M.D. Gladders,$^{4,5}$
A.K. Hicks,$^{6}$
H. Hoekstra,$^{7}$
B.C. Hsieh,$^{8}$
R.J. Ivison,$^{9,10}$
B.C. Lemaux,$^{11}$
L.M. Lubin,$^{11}$
D.V. O'Donnell,$^{1}$
H.K.C. Yee$^{12}$}\\\\
$^{1}$Department of Physics, McGill University, 3600 rue University, Montr\'{e}al, Qu\'{e}bec, H3A 2T8, Canada\\
$^{2}$Center for Astrophysics and Space Astronomy, University of Colorado at Boulder, Boulder, CO, 80309, USA\\
$^{3}$Institute for Astronomy, University of Hawaii, 2680 Woodlawn Dr., Honolulu, HI 96822, USA\\
$^{4}$Department of Astronomy \& Astrophysics, University of Chicago, 5640 South Ellis Avenue, Chicago, IL 60637, USA\\
$^{5}$Kavli Institute for Cosmological Physics, University of Chicago, 5640 South Ellis Avenue, Chicago, IL 60637, USA\\
$^{6}$Department of Physics \& Astronomy, Michigan State University, 3255 Biomedical and Physical Sciences Bldg., East Lansing, MI, 48824, USA\\
$^{7}$Leiden Observatory, Leiden University, Niels Bohrweg 2, 2333 CA, Leiden, The Netherlands\\
$^{8}$Institute of Astronomy and Astrophysics, Academia Sinica, P.O. Box 23-141, Taipei 10617, Taiwan, R.O.C.\\
$^{9}$UK Astronomy Technology Centre, Royal Observatory, Blackford Hill, Edinburgh EH9 3HJ\\
$^{10}$Institute for Astronomy, University of Edinburgh, Blackford Hill, Edinburgh EH9 3HJ, UK\\
$^{11}$Department of Physics, University of California, Davis, 1 Shields Avenue, Davis, CA 95616, USA\\
$^{12}$Department of Astronomy and Astrophysics, University of Toronto, 50 St George Street, Toronto, Ontario M5S 3H4, Canada\\}

\date{\today}

\pagerange{\pageref{firstpage}--\pageref{lastpage}} \pubyear{2011}

\maketitle

\label{firstpage}

\begin{abstract}
We present a submillimetre survey of seven high-redshift galaxy clusters ($0.64<z<1.0$) using the Submillimetre Common-User Bolometer Array (SCUBA) at 850 and 450\,\um.  The targets, of similar richness and redshift, are selected from the Red-sequence Cluster Survey (RCS). We use this sample to investigate the apparent excess of submillimetre source counts in the direction of cluster fields compared to blank fields, as seen in the literature.  The sample consists of three galaxy clusters that exhibit multiple optical arcs due to strong gravitational lensing, and a control group of four clusters with no apparent strong lensing.  A tentative excess of 2.7\,$\sigma$ is seen in the number density of submillimetre sources within the lensing cluster fields compared to that in the control group.  Ancillary observations at radio, mid-infrared, optical, and X-ray wavelengths allow for the identification of counterparts to many of the submillimetre luminous galaxies (SMGs), and provide improved astrometry and redshift constraints.  Utilizing photometric redshifts, we conclude that at least three of the galaxies within the lensing fields have redshifts consistent with the clusters and implied infrared luminosities of $\sim10^{12}$\,\Lsol.  The existence of submillimetre cluster members may therefore be boosting source counts in the lensing cluster fields, which might be an effect of the dynamical state of those clusters.  However, we find that the removal of potential cluster members from the counts analysis does not entirely eliminate the difference between the cluster samples.  We also investigate possible occurrences of lensing between background submillimetre sources and lower-redshift optical galaxies, though further observations are required to make any conclusive claims.  Although the excess counts between the two cluster samples have not been unambiguously accounted for, these results warrant caution for interpreting submillimetre source counts in cluster fields and point source contamination for Sunyaev-Zel'dovich surveys. 

\end{abstract}


\section{Introduction}
\label{sec:intro}

\subsection{Infrared-luminous Galaxies within Galaxy Cluster Fields}
\label{sec:IRintro}
The initial discovery of a population of dusty star-forming field galaxies with extreme luminosities in the high-redshift universe with the Submillimetre Common-User Bolometer Array (SCUBA; e.g.\ \citealp{Smail97,Barger98,Hughes98}) has since spawned numerous submillimetre studies (e.g.\ \citealp{Blain99ApJ, Dunne00, Eales00, Best02, Cowie02, Chapman02counts, Scott02, Borys03, Webb03VI, Webb05, Coppin06, Knudsen06}) aimed at determining the cosmic star formation history and constraining galaxy evolution models.  Deep blank-field surveys with SCUBA (e.g.\ \citealp{Scott02, Borys03, Webb03VI, Coppin06}) of high-redshift submillimetre luminous galaxies (SMGs) have provided robust statistics to estimate source counts down to the 2\,mJy level at 850\,\um.  The lower end of the submillimetre flux density distribution has been probed by exploiting massive cluster lenses at low redshift (typically at $z<0.4$) to magnify the population of distant SMGs behind them (e.g.\ \citealp{Smail97, Smail98, Smail02, Cowie02, Chapman02counts, Knudsen06}), thereby reducing the source density on the sky and illuminating faint sources.  These cluster surveys have led the way in understanding the properties of SMGs through numerous contributions.  The first identification of radio counterparts \citep{Ivison98, Ivison00, Smail00} secured the astrometry of individual SMGs, and thus facilitated the first spectroscopic redshifts \citep{Ivison98, Ivison00, Barger99}.  Moreover, follow-up observations of SMGs from cluster surveys yielded the first CO detections \citep{Frayer98, Frayer99, Ivison01} and the discovery of extremely red object (ERO) counterparts \citep{Smail99}. 

An elucidating comparison can be made between source counts in the field and in cluster environs.  Consistently, the cluster counts lie above the blank-field counts, which could be the result of unaccounted-for lensing or infrared galaxies associated with the clusters themselves \citep{Borys03,Webb05}.  This holds despite the fact that lower redshift cluster surveys typically remove central dominant (cD) galaxies from counts analyses \citep{Cowie02,Smail02}, or expect few cluster member detections \citep{Chapman02counts,Knudsen06}.  Cluster counts in single pointings or small maps can carry a potential bias due to human intervention in two ways.  For instance, the tendency to not publish blank maps could bias the available sample of clusters towards more crowded fields.  Secondly, real-time observers are more inclined to increase the integration time on areas that look more dense in small maps, and therefore, introduce additional sources that would otherwise go undetected.  However, it is impossible to quantify these possible biases in the literature (but see \S\ref{sec:cumul} for a discussion with our own work); we will therefore assume they are not issues in the above surveys and instead focus on the two aforementioned physical reasons for enhanced cluster counts.

With regards to the unaccounted lensing scenario, higher redshift clusters are generally less efficient lenses (i.e.\ have a lower lensing magnification) than their low-redshift counterparts \citep{Bartelmann98}.  This implies that unaccounted-for lensing is less likely to produce higher counts in high-redshift clusters fields.  However, high-redshift cluster surveys that have calculated negligible lensing corrections for the SMGs within the fields still observe a tentative excess of source counts compared to the field (\citealp{Best02, Webb05}; this work), though this relies on small number statistics.  
 
If the excess counts are instead due to submillimetre bright cluster members, we would expect the effect to become more pronounced in higher redshift clusters (i.e.\ the surveys of \citealp{Best02,Webb05}; this work) given the following trends with increasing redshift: a rise in the fraction of star-forming galaxies (e.g.\ \citealp{Butcher78,Saintonge08,Haines09}); an increase in field infall (e.g.\ \citealp{Bower91,Ellingson01,Loh08}) and younger infalling galaxies; and the less relaxed state of younger clusters.  Clarifying the source of excess counts will not only improve the submillimetre source counts at the faint end, but also provide information on the extreme phases of galaxy evolution in dense cluster environments, and aid cluster cosmology studies, which require constraints on the point source contamination within cluster fields, specifically Sunyaev-Zel'dovich surveys.

\subsection{Strong Lensing in Clusters}
\label{sec:strong_lensing}
The efficacy of a cluster to produce multiple strong arcs primarily depends on the cluster redshift, the central surface mass density, the distribution of background sources, and the symmetry of the structure; these requirements generally imply that high-redshift clusters are poor gravitational lenses compared to their low-redshift counterparts.  The critical surface mass density required depends on redshift, with low-redshift clusters being the most efficient lenses for background sources at $z\sim1$ \citep{Bartelmann98}.  The fourth efficiency factor dictates that asymmetric clusters with enhanced substructure typically produce larger arcs compared to those from spherically symmetric cluster models \citep{Bartelmann95}.

In contradiction to this redshift trend, a subset of high-redshift clusters with multiple giant arcs has been discovered with the first Red-sequence Cluster Survey (RCS-1; \citealp{Gladders05}) at $z>0.64$, while at lower redshifts, the survey failed to detect any lensing clusters, implying the property may be linked to the process of cluster formation \citep{Gladders03}.  While the excess of high redshift super-lenses can be explained with a higher value of  $\sigma_8$ \citep{Li06}, the same effect can be produced through a population of massive, concentrated clusters \citep{Dalal04} or clusters undergoing a major merger, with substantial substructure.  In each scenario, the central lensing cross-section can be temporarily enhanced, thereby boosting lensing efficiency (e.g.\ \citealp{Torri04}).  Thus, these super-lenses may represent clusters which are dynamically unrelaxed or merging, both of which have been shown to increase the activity within their cluster galaxies \citep{BlainJames99, Owen99, Miller03a}.

\subsection{The Study}
To investigate the possible excess of SMGs in cluster environments and the presence of giant optical arcs from strong lensing effects, we analyse seven galaxy clusters of similar richness and redshift (0.640 $< z <$ 1.045) drawn from the RCS-1 sample (see \S\ref{sec:RCS1}).  The clusters are divided into two subsamples: the strong lensing sample and the control sample.  The former consists of three galaxy clusters that exhibit a high incidence of optical arcs due to strong gravitational lensing within the central $\sim30$\,arcsec compared to theoretical predictions within a $\Lambda$CDM cosmology framework \citep{Gladders03}.  As a means of assessing the importance of the the cluster lensing efficiency, we also include four clusters (the control group) that cover the same range in richness and redshift, but lack any evidence of strong lensing effects. The original goal of the survey was to obtain five clusters for each sample, but SCUBA ceased working before the remaining observations were completed.  The entire sample is listed in Table \ref{tab:clusters}, along with relevant cluster properties.

\begin{table*}
\begin{minipage}{168mm}
\begin{center}
\caption{A summary of the seven galaxy clusters of the submillimetre survey.}
\label{tab:clusters}
\begin{tabular}{ccccccccl}
\hline 
\multicolumn{1}{c}{} &
\multicolumn{1}{c}{} &
\multicolumn{1}{c}{} &
\multicolumn{1}{c}{$B_{\rm{gc}}$$^{b}$} &
\multicolumn{1}{c}{M$_{200}$$^{c}$} &
\multicolumn{1}{c}{T$_{X}$$^{d}$} &
\multicolumn{1}{c}{$\sim$3$\,\sigma$ Depth$^{e}$} &
\multicolumn{1}{c}{Map Area$^{f}$}&
\multicolumn{1}{c}{Ancillary Data$^{g}$}\\
\multicolumn{1}{c}{Cluster Name} &
\multicolumn{1}{c}{Subsample} &
\multicolumn{1}{c}{Redshift$^{a}$}&
\multicolumn{1}{c}{(Mpc$^{1.77}$)} &
\multicolumn{1}{c}{($10^{14}$\,\Msol)} &
\multicolumn{1}{c}{(keV)} &
\multicolumn{1}{c}{(mJy/beam)} &
\multicolumn{1}{c}{(arcmin$^2$)}&
\multicolumn{1}{c}{}\\

\hline
\vspace{+1.5mm}	
RCS 022434$-$0002.5 & Lensing & 0.773 & 782 $\pm$ 201 &  $3.3\pm1.3$  & $5.0^{+1.2}_{-0.8}$ & 3.1 & 7.02 & SMIRX$z'R_cVB$\\		
\vspace{+1.5mm}
RCS 141910$+$5326.2 & Lensing & 0.640 & 1400 $\pm$ 256 & $9.5\pm3.3$ & $4.5^{+0.4}_{-0.3}$ & 4.3 & 6.64 & SMIRX$z'R_c$\\
\vspace{+1.5mm}
RCS 231953$+$0038.0 & Lensing & 0.9024 & 2107 $\pm$ 433 & $18.7\pm8.3$ & $6.2^{+0.9}_{-0.8}$ & 3.1 & 6.75 & SMIRX$z'R_cVB$\\		
\vspace{+1.5mm}
RCS 112225$+$2422.9 & Control & 0.940* & 1312 $\pm$ 350 & $7.1\pm3.5$ & --- & 2.8 & 6.65 & SMI$z'R_cVB$\\
\vspace{+1.5mm}
RCS 132629$+$2903.1 & Control & 1.045* & 1901 $\pm$ 451 & $13.8\pm6.9$ & $1.6^{+0.7}_{-0.3}$ & 2.5& 6.81 & SMIX$z'R_c$\\
\vspace{+1.5mm}
RCS 215248$-$0609.4 & Control & 0.650 & 935 $\pm$ 241& $4.7\pm1.9$ & --- & 3.0 & 6.62 & SMI$z'R_cVB$\\			
\vspace{+1.5mm}
RCS 231831$+$0034.2 & Control & 0.7518 & 1153 $\pm$ 252 & $6.2\pm2.4$ & $6.1^{+1.3}_{-0.9}$ & 2.7 & 6.60 & SRX$z'R_c$\\		

\hline
\end{tabular}
\begin{tablenotes}[normal]
$^{a}$Redshifts are all spectroscopic except for those denoted with an asterisk (*), which are based on the location of the red sequence and accurate to 10 per cent. \\
$^{b}$The amplitude of the cluster-centre galaxy correlation function.\\
$^{c}$The mass contained within the radius at which the cluster mass density is 200 times larger than the critical mass density of the universe.  The value of M$_{200}$ is calculated from the velocity dispersion, which is correlated with $B_{\rm{gc}}$ \citep{Yee03}.\\
$^{d}$The X-ray temperature, from \cite{Hicks08}, for the five clusters with X-ray coverage.\\
$^{e}$The average 3$\,\sigma$ depth over the 850\,\um\  maps.\\
$^{f}$The usable area of the 850\,\um\  maps over which the noise the uniform.\\
$^{g}$The ancillary data coverage for each field (see \S\ref{sec:ancillary}).  The symbols are as follows: SCUBA 850- and 450-\um\  imaging (S); MIPS 24-\um\  imaging (M); four-channel IRAC coverage (I); 1.4-GHz imaging with the VLA (R); Chandra X-ray data (X); original RCS filters ($z', R_c$); optical filters ($V, B$).  Photometric redshifts are only available for clusters with all four optical filters. 
\end{tablenotes}
\end{center}
\end{minipage}
\end{table*}

If excess source counts in cluster fields derive from lensing effects due to the cluster potential, we should not observe this effect in our seven cluster fields, assuming our sample contains inherently bad lenses compared to local clusters.  Although seemingly contradictory to our claim above, the efficiency of our lensing sample to produce arcs arises within the central $\sim30$\,arcsec of the cluster, not over the entire field.  The count excess in this study, however, is seen over a larger area ($\sim7$\,arcmin$^2$), where the strong lensing boosting factor is expected to be quite small.  Certainly, we should not see an excess in the control fields, which show no signs of strong lensing.   

Alternatively, if a trend towards higher counts exists in all seven clusters, then it is likely an evolutionary effect with cluster galaxy populations.  Moreover, if the excess is greater or limited to the super-lenses, then perhaps the physical mechanism driving this effect is the dynamical state of the cluster or excess lensing due to the cluster potential (but only in the cluster centre).  We can therefore attempt to disentangle lensing and cluster contamination, and further isolate the driver of contamination in the latter case.

We describe our observations and data reduction for the submillimetre and the ancillary data sets in \S\ref{sec:data}.  In \S\ref{sec:analysis}, we outline the submillimetre data analysis, and present 850\,\um\  source counts in \S\ref{sec:counts}.  The counterpart identification process with the multi-wavelength follow-up data is presented in \S\ref{sec:counters}.  We discuss both the candidate submillimetre cluster members and potential cases of lensed SMGs, and their respective effects on the submillimetre source counts in \S\ref{sec:disc}.  We conclude in \S\ref{sec:conclusions}, and include a discussion of each SMG in Appendix \ref{app:cases}.  We assume a flat, $\Omega_{\textup{M}}=0.3$, $\Omega_{\Lambda}=0.7$ cosmology, with $H_0=70$\,km\,s$^{-1}$\,Mpc$^{-1}$ throughout the paper.

\section{Observations and Data Reduction}
\label{sec:data}

\subsection{The Red-sequence Cluster Survey Sample}
\label{sec:RCS1}
To study the above issues, we have observed, using SCUBA at 850 and 450\,\um, a unique sample of seven galaxy clusters, drawn from the RCS-1 \citep{Gladders05}.   The three lensing clusters (RCS 0224-0002, RCS 1419+5326, and RCS 2319+0038) all lie within $0.640<z<0.9024$ and contain at least two optical arcs from strong-lensing.  The remaining four clusters (RCS 1122+2422, RCS 1326+2903, RCS 2152-0609, and RCS 2318+0034) have redshifts within $0.650<z<1.045$ and are used as a control sample.

The RCS utilizes the well-defined relation between colour and magnitude of early-type galaxies, known as the red-sequence, to find clusters. The method is robust with a contamination rate of $<5$ per cent and provides a redshift estimate to $\sim10$ per cent \citep{Gladders00, Gilbank07}.  Cluster significance relies on detecting simultaneous overdensities in colour, magnitude, and position.

Launching the largest systematic search for galaxy clusters up to $z \sim 1$ at the time, the RCS-1 imaged $\sim92$\,deg$^{2}$ and discovered $\sim1000$ galaxy clusters using only two optical filters ($R_C$ and $z'$).  The bands span the $4000\, \textup\AA$ break and provide superior colour discrimination at high redshift (up to $z\sim1$) compared to the $V$ and $I$ filter pair typically used in optical surveys.  Moreover, the survey includes an optical richness estimate based on the $B_{\textup{gc}}$ parameter \citep{Longair79}, the amplitude of the cluster-centre galaxy correlation function, which has been shown to be a good proxy for mass \citep{Yee99, Yee03}.  

\subsection{Submillimetre Data}
\subsubsection{SCUBA Observations}
The seven clusters were observed on the James Clerk Maxwell Telescope (JCMT) using SCUBA, at 850 and 450\,\um, over the course of several observing runs in 2001 and 2002.  The data were reduced following the standard SCUBA User Reduction Facility (SURF) routine \citep{Jenness98}.  A 30-arcsec chop throw in right ascension was used for each cluster observation.  Each map consists of either one or two overlapping pointings, resulting in areas all greater than $\sim$6.5 arcmin$^2$. The edges of each map, however, are unacceptably noisy due to lower integration times and are not used in further analysis (see \S\ref{sec:detection}).  Pointing stability was maintained and checked after each major telescope slew.  The sky opacity was monitored regularly through ``skydips" every 1--2 hours, allowing for a determination of signal attenuation due to a lack of atmospheric transmission.  The opacity at 850\,\um\  ranged from roughly 0.15--0.29, with a typical value below 0.2.

The presence of a 16-sample noise spike complicated the data reduction for the 2002 observations.  This strongly correlated noise signal was removed from corrupted bolometers as described in \cite{Webb05}.  The final corrected and uncorrected data were compared to the noncorrupted 2001 data and it was concluded that no robust point sources were either lost or introduced into the map, though the overall noise was reduced by $\sim20$ per cent.  The SHADES survey was similarly affected and also found the spike to have a negligible effect on the data \citep{Coppin06}.

\subsubsection{850\,$\mu$m Source Detection: Noise Analysis and Cleaning}
\label{sec:detection}  
Robust source detection and minimization of spurious sources rely on accurate noise analysis.  Due to the inconsistent noise properties of SCUBA data and the generally low signal-to-noise (S/N) at which one is working, standard chop and nod techniques are not sufficient to remove all the atmospheric noise.  Maps characterizing the noise are created from a set of Monte Carlo simulations for each bolometer time stream and a bootstrap algorithm (outlined in \citealp{Eales00, Webb03VI, Webb05}).

The jiggle-map scan pattern of SCUBA pointings produces uneven noise characteristics across the map; specifically, the edges of the map are much noisier compared to the centre due to the shorter integration time where the pointings do not overlap.  To prevent falsely classifying spurious noise spikes as real sources, a mask is created to eliminate map regions from subsequent analysis where the noise exceeds 2\,mJy, primarily in the outer 15\,arcsec of the map.  Using maps with more uniform noise and even coverage helps maintain consistency throughout the data analysis process.  The resulting map areas of the seven clusters range between 6.60 and 7.02 arcmin$^2$ (see Table \ref{tab:clusters}).  It is within these areas that both source extraction and Monte Carlo simulations are performed.

We create an initial source list of S/N $>3$ detections from the uniform noise map area after convolving the raw signal map with a beam template, which consists of a 14-arcsec point source kernel and the negative off-source chops on either side.  In low signal-to-noise images, a template-convolved map improves the accuracy of photometric parameters \citep{Phillipps91}, helps discriminate between real and spurious sources, and increases the signal-to-noise due to the inclusion of signal from the negative beams.

We use an iterative cleaning process (see \citealp{Eales00,Webb05} for more details) to isolate each potential source on the raw map, and separately convolve it with the beam template to reduce source confusion.  Our final catalogue, shown in Table \ref{tab:850cat}, consists of 26 sources above 3\,$\sigma$, with $2-6$ SMGs per cluster field.

\begin{table*}
\begin{minipage}{162mm}
\begin{center}
\caption{Positions and flux densities of the 850\,\um\  submillimetre sources.}
\label{tab:850cat}
\begin{tabular}{cccccccc}
\hline 
\multicolumn{3}{c}{} &
\multicolumn{1}{c}{$S_{850 \mu m}$$^{b}$} &
\multicolumn{1}{c}{$S_{850 \mu m}$$^{c}$} &
\multicolumn{1}{c}{} &
\multicolumn{1}{c}{$S_{450 \mu m}$$^{d}$} &
\multicolumn{1}{c}{} \\
\multicolumn{1}{c}{} &
\multicolumn{1}{c}{R.A} &
\multicolumn{1}{c}{Dec.} &
\multicolumn{1}{c}{(measured)} &
\multicolumn{1}{c}{(deboosted)} &
\multicolumn{1}{c}{850\,\um} &
\multicolumn{1}{c}{(measured)} &
\multicolumn{1}{c}{450\,\um}\\
\multicolumn{1}{c}{Object Name$^{a}$} &
\multicolumn{1}{c}{(J2000.0)} &
\multicolumn{1}{c}{(J2000.0)} &
\multicolumn{1}{c}{(mJy)} &
\multicolumn{1}{c}{(mJy)} &
\multicolumn{1}{c}{S/N} &
\multicolumn{1}{c}{(mJy)} &
\multicolumn{1}{c}{S/N} \\
\hline
SMM 0224.1 ........................ & 02 24 34.19 & $-$00 03 25 & 6.2 & 6.2 & 6.1 & 14.4 & 3.7\\
SMM 0224.2 ........................ & 02 24 33.46 & $-$00 03 56 & 4.9 & 3.8 & 4.3 & $<$54 & ... \\
SMM 0224.3 ........................ & 02 24 28.06 & $-$00 03 18 & 4.3 & 3.4 & 3.5 & $<$115 & ... \\
SMM 0224.4 ........................ & 02 24 29.79 & $-$00 02 59 & 4.2 & 3.2 & 3.9 & 15.8 & 3.3\\   
{\it SMM 0224.5} ........................ & {\it 02 24 32.66} & {\it $-$00 03 42} & {\it 3.3} & {\it 2.4} & {\it 3.4} & {\it 18.7} & {\it 2.8}\\ 
SMM 0224.6 ........................ & 02 24 32.93 & $-$00 02 32 & 3.1 & 2.2 & 3.7 & 13.2 & 3.1\\  
\hline
\textit{SMM 1122.1} ........................ & \textit{11 22 29.42} & \textit{$+$24 21 17} & \textit{4.0} & \textit{3.1} & \textit{3.2} & \textit{---} & \textit{---} \\
SMM 1122.2 ........................ & 11 22 31.69 & $+$24 22 09 & 3.6 & 2.6 & 4.3 & --- & --- \\
\textit{SMM 1122.3} ........................ & \textit{11 22 27.08} & \textit{$+$24 22 41} & \textit{2.5} & \textit{1.5} & \textit{3.4} & \textit{16.0} & \textit{5.6}\\
\hline
SMM 1326.1 ........................ & 13 26 28.18 & $+$29 03 13 & 4.6 & 3.5 & 6.5 & 22.4 & 6.2\\
SMM 1326.2 ........................ & 13 26 33.06 & $+$29 02 20 & 3.5 & 2.5 & 3.6 & $<$57 & ... \\
\textit{SMM 1326.3} ........................ & \textit{13 26 33.60} & \textit{$+$29 03 23} & \textit{2.9} & \textit{1.7} & \textit{3.4} & \textit{$<$31} & \textit{...} \\
\textit{SMM 1326.4} ........................ & \textit{13 26 25.28} &\textit{ $+$29 02 41} & \textit{2.5} & \textit{1.5} & \textit{3.2} & \textit{$<$10} & \textit{...} \\
\hline
SMM 1419.1 ........................ & 14 19 15.47 & $+$53 24 55 & 8.3 & 8.3 & 5.2 & --- & --- \\
\textit{SMM 1419.2} ........................ & \textit{14 19 08.76} & \textit{$+$53 24 56} & \textit{7.3} & \textit{7.3} & \textit{3.0} & --- & --- \\
SMM 1419.3 ........................ & 14 19 20.17 & $+$53 26 29 & 5.0 & 4.5 & 4.5 & 19.3 & 2.7\\
SMM 1419.4 ........................ & 14 19 05.51 & $+$53 26 20 & 4.7 & 3.6 & 3.8 & $<$16 & ... \\
SMM 1419.5 ........................ & 14 19 14.69 & $+$53 25 21 & 4.4 & 3.4 & 5.9 & $<$13 & ... \\
\hline
SMM 2152.1 ........................ & 21 52 46.62 & $-$06 08 49 & 3.4 & 2.4 & 3.5 & $<$26 & ... \\
\textit{SMM 2152.2} ........................ & \textit{21 52 43.74} & \textit{$-$06 10 20} & \textit{3.0} & \textit{2.1} & \textit{3.0} & \textit{$<$54} & ... \\
\hline
SMM 2318.1 ........................ & 23 18 27.89 & $+$00 34 55 & 14.9 & 14.9 & 17.0 & 36.5 & 7.1\\
SMM 2318.2 ........................ & 23 18 28.03 & $+$00 34 18 & 5.0 & 4.5 & 5.6 & 14.5 & 3.0\\
SMM 2318.3 ........................ & 23 18 32.63 & $+$00 34 22 & 3.6 & 2.6 & 4.0 & 14.2 & 3.3\\
\hline
SMM 2319.1 ........................ & 23 19 49.73 & $+$00 37 57 & 7.6 & 7.6 & 5.5 & 26.1 & 4.4\\
SMM 2319.2 ........................ & 23 19 56.86 & $+$00 37 21 & 7.4 & 7.4 & 7.5 & 44.8 & 4.4\\
SMM 2319.3 ........................ & 23 19 53.26 & $+$00 38 09 & 3.3 & 2.4 & 4.7 & $<$12 & ... \\

\hline
\end{tabular}
\begin{tablenotes}[normal]
$^{a}$Sources in italics are detected below 3.5$\sigma$ at 850\,\um\ and are therefore less reliable. In particular, we expect four of our sources to be spurious (see \S\ref{sec:false}).  \\
$^{b}$The measured 850\,\um\  flux, as described in \S\ref{sec:detection}.\\
$^{c}$The deboosted 850\,\um\  flux, which is a more accurate representation of the source (see \S\ref{sec:deboosting}).\\
$^{d}$Objects without a 450\,\um\ detection and those lying close to the edge of the usable 450\,\um\  field (where the noise is high) are given a $3\,\sigma$ upper limit on the 450\,\um\  flux.  The four objects that lie completely off the 450\,\um\  field of view are denoted by a solid dash (---).

\end{tablenotes}
\end{center}
\end{minipage}
\end{table*}

\subsubsection{450\,$\mu$m Detection}
\label{sec:450}
A unique advantage of SCUBA is its ability to simultaneously observe two submillimetre wavelengths using a dichroic beam splitter and two separate bolometer arrays, thus providing 450-\um\  data for free.  In principle, the smaller beam associated with the shorter wavelength should result in improved positional accuracy of sources; strong atmospheric emission at 450\,\um, however, causes deterioration of data quality.  Furthermore, telescope sensitivity to slight temperature changes cause high variability in the 450\,\um\  beam, thus complicating beam fitting.  We therefore resort to simple smoothing techniques, rather than convolving the raw data with a beam template, when performing source detection.

Despite this setback, our cluster sample contains fairly high quality data (i.e.\ less noisy than typical) at 450\,\um, allowing source extraction and counterpart identification to the 850\,\um\  emission.  We adopt a search radius of 12\,arcsec around an 850\,\um\  peak and a S/N cutoff of 2.5$\,\sigma$ at 450\,\um\  to locate any 450\,\um\  peak emission associated with the 850\,\um\  SMGs.  In the RCS 0224 field, \citet{Webb05} conclude that simulations of the 450\,\um\  positional uncertainties agree with those at 850\,\um\  (see \S\ref{sec:positional}), calculating a 13\,arcsec offset for the 95th percentile.  The large search area of 144 arcsec$^2$, therefore, accounts for errors on positions at both wavelengths.  The low detection limit (2.5$\,\sigma$) ensures that we will not miss any 450\,\um\  counterpart emission.

\begin{figure*}
   \centering
   \subfigure{\includegraphics[scale=0.35]{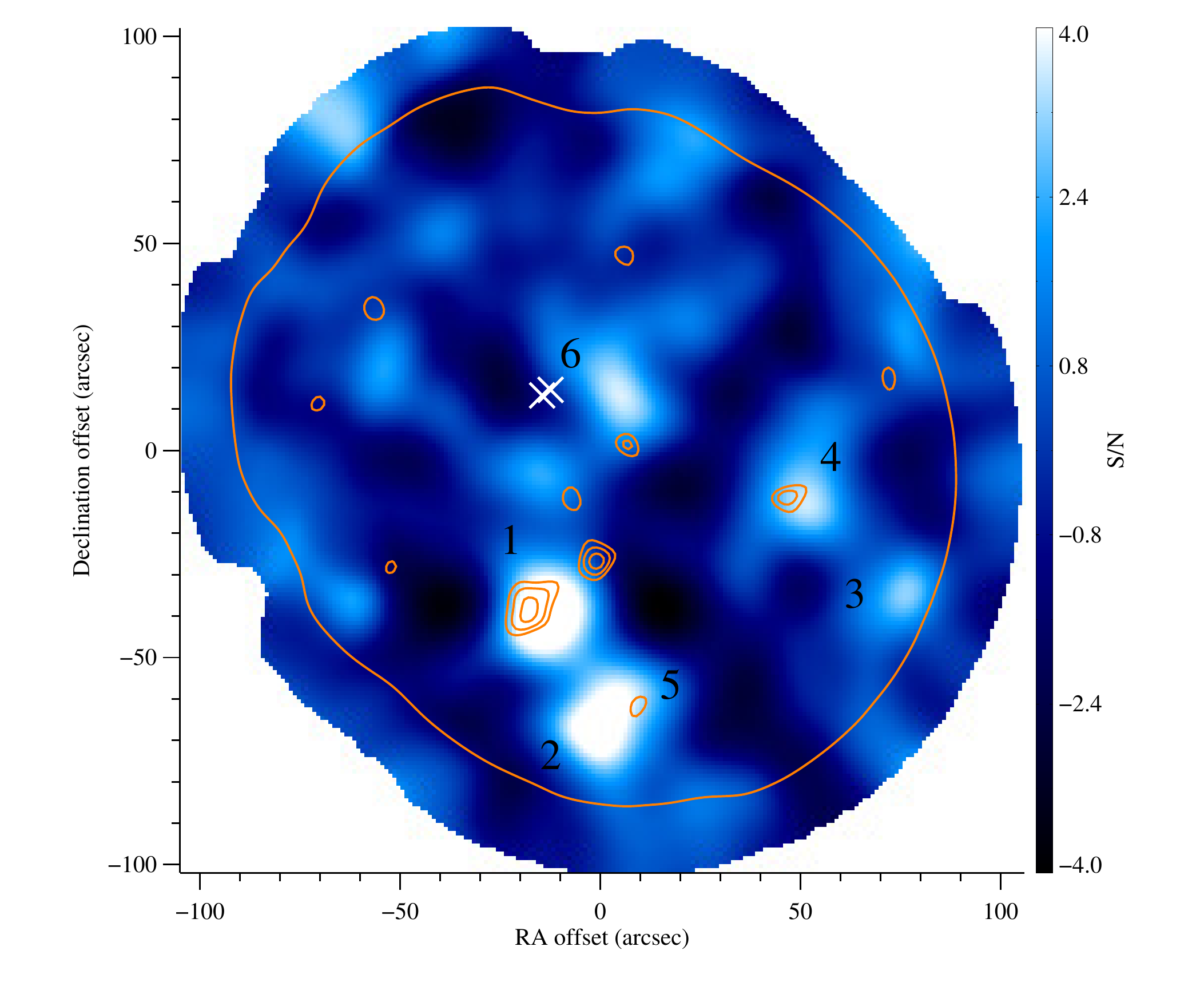}}
   \subfigure{\includegraphics[scale=0.35]{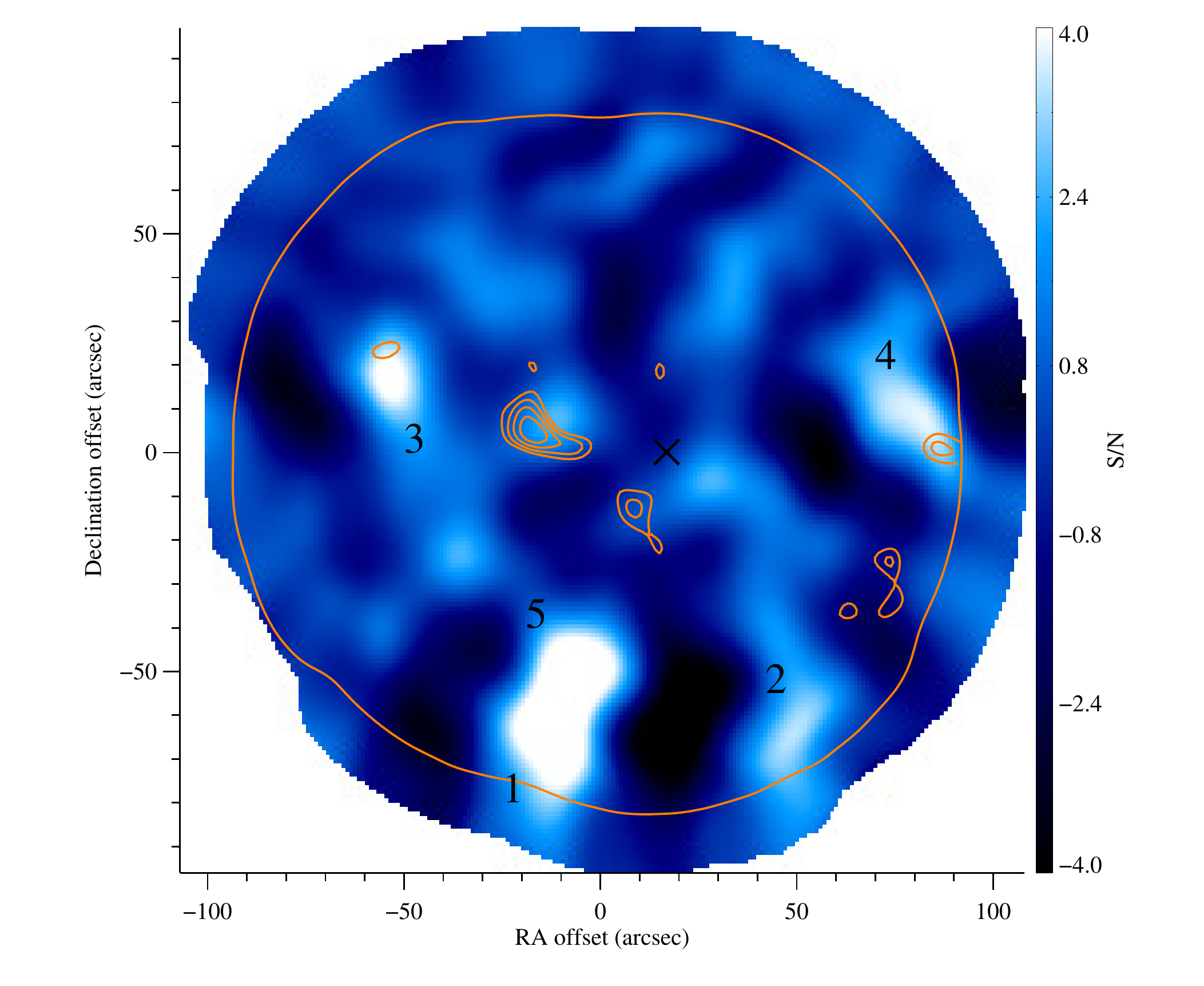}}
   \subfigure{\includegraphics[scale=0.35]{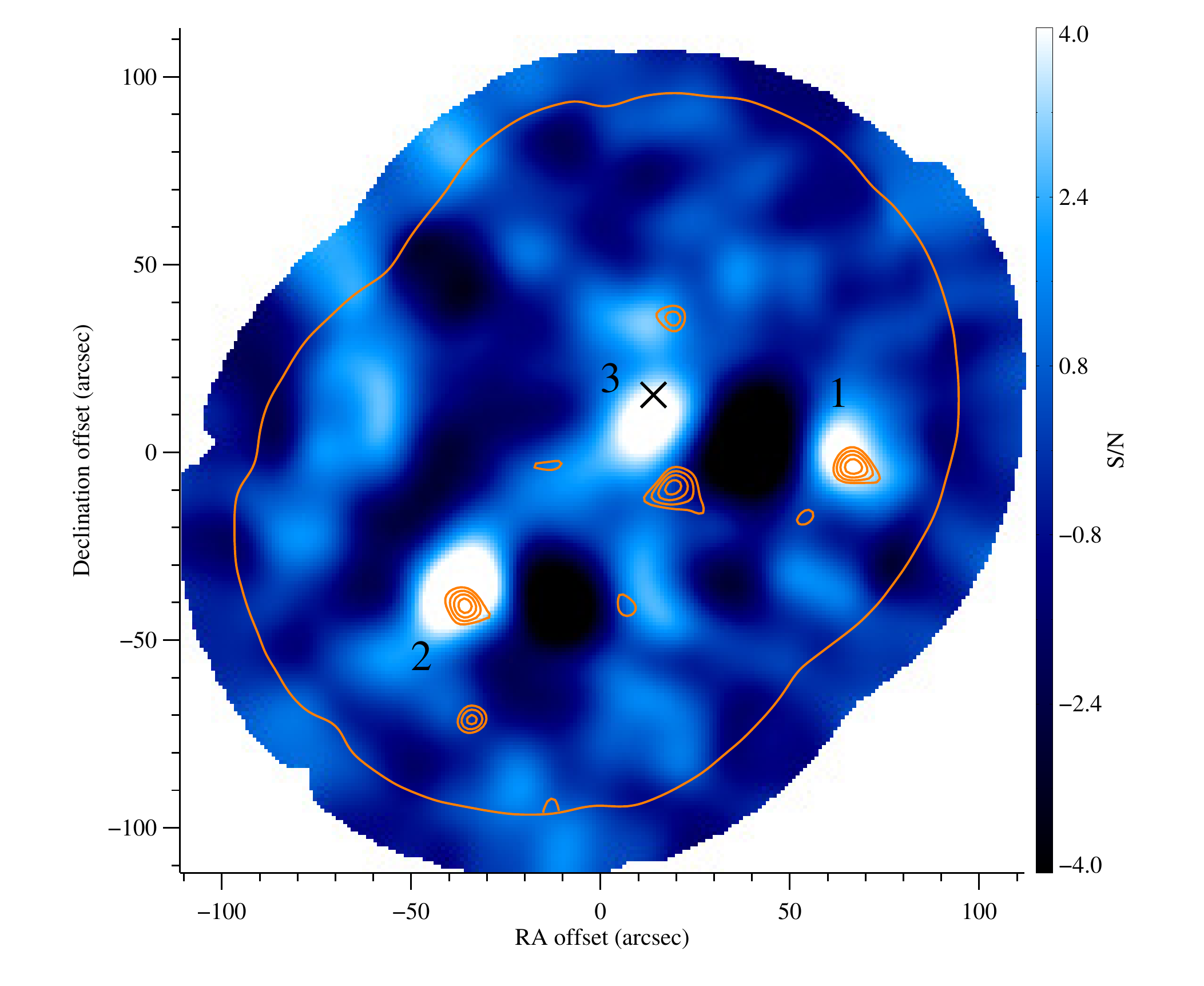}}
   \caption{850\,\um\  S/N maps (convolved with the beam template) for the three strong lensing clusters.  Overlaid are 450\,\um\  S/N contours (smoothed with an 8\,arcsec Gaussian) in steps of 0.5, beginning with 2.5$\,\sigma$ (the significance limit used for 450\,\um\  counterpart emission), and steps of 1.0 above 4.0$\,\sigma$.  Note the smaller field of view for 450\,\um, evident from the large orange borders.  Significant 850\,\um\  detections (within the usable map area---see \S\ref{sec:detection}) are evident as light blue or white areas (i.e.\ $\geq 3.0\,\sigma$) flanked by negative (black) lobes, and marked with numbers corresponding to their SMM IDs in Table \ref{tab:850cat}.  The location of the brightest cluster galaxy (BCG) is marked by an `x' in each field; note there are multiple BCGs in some clusters.  Clockwise from the upper left corner, the clusters are RCS 0224, 1419, and 2319.  In each map, north points up and east is towards the left; the map centres are given by the coordinates in the cluster names of Table \ref{tab:clusters}.}
   \label{fig:850mapslensing}
\end{figure*}

\begin{figure*}
   \centering
   \subfigure{\includegraphics[scale=0.35]{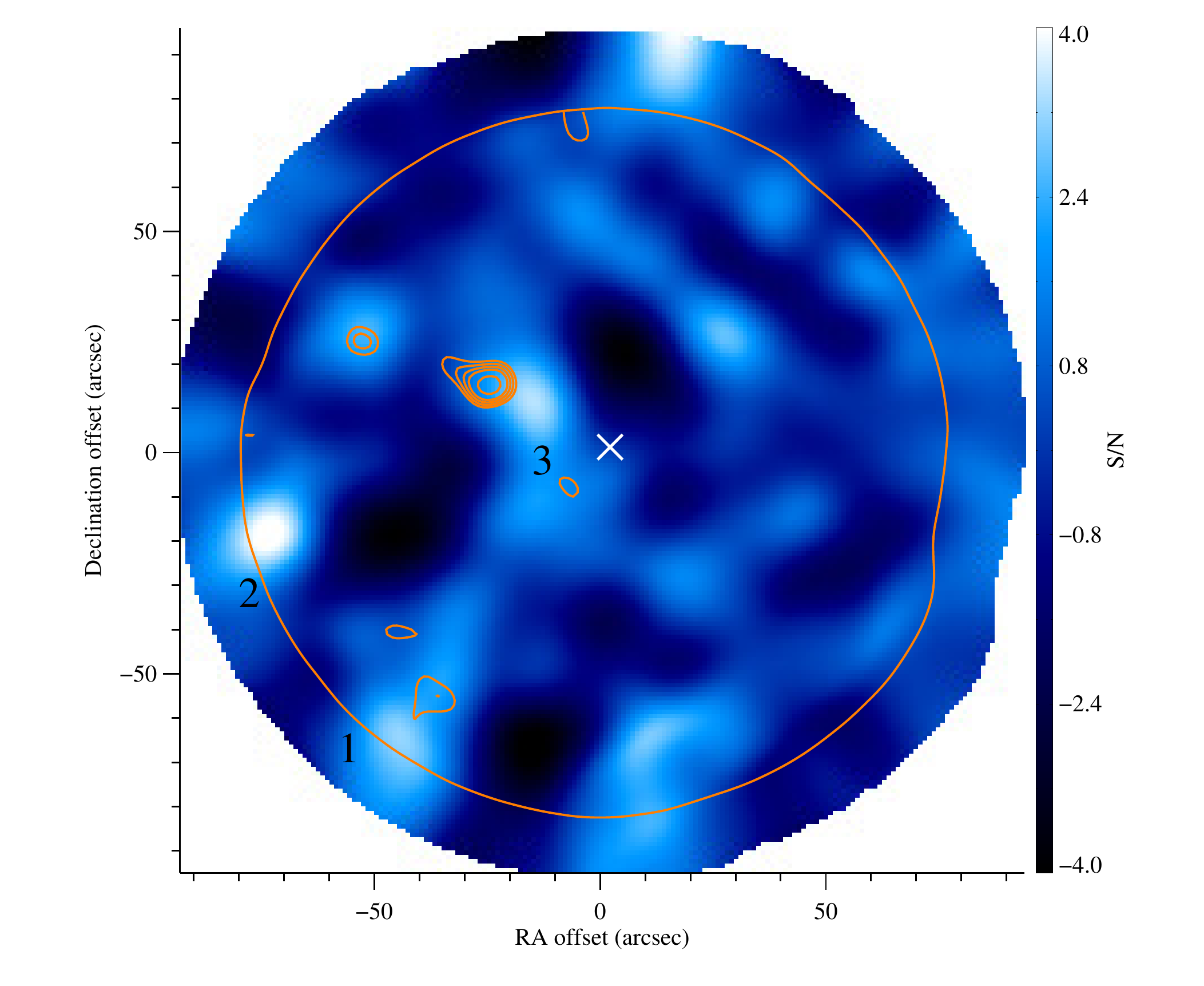}}
   \subfigure{\includegraphics[scale=0.35]{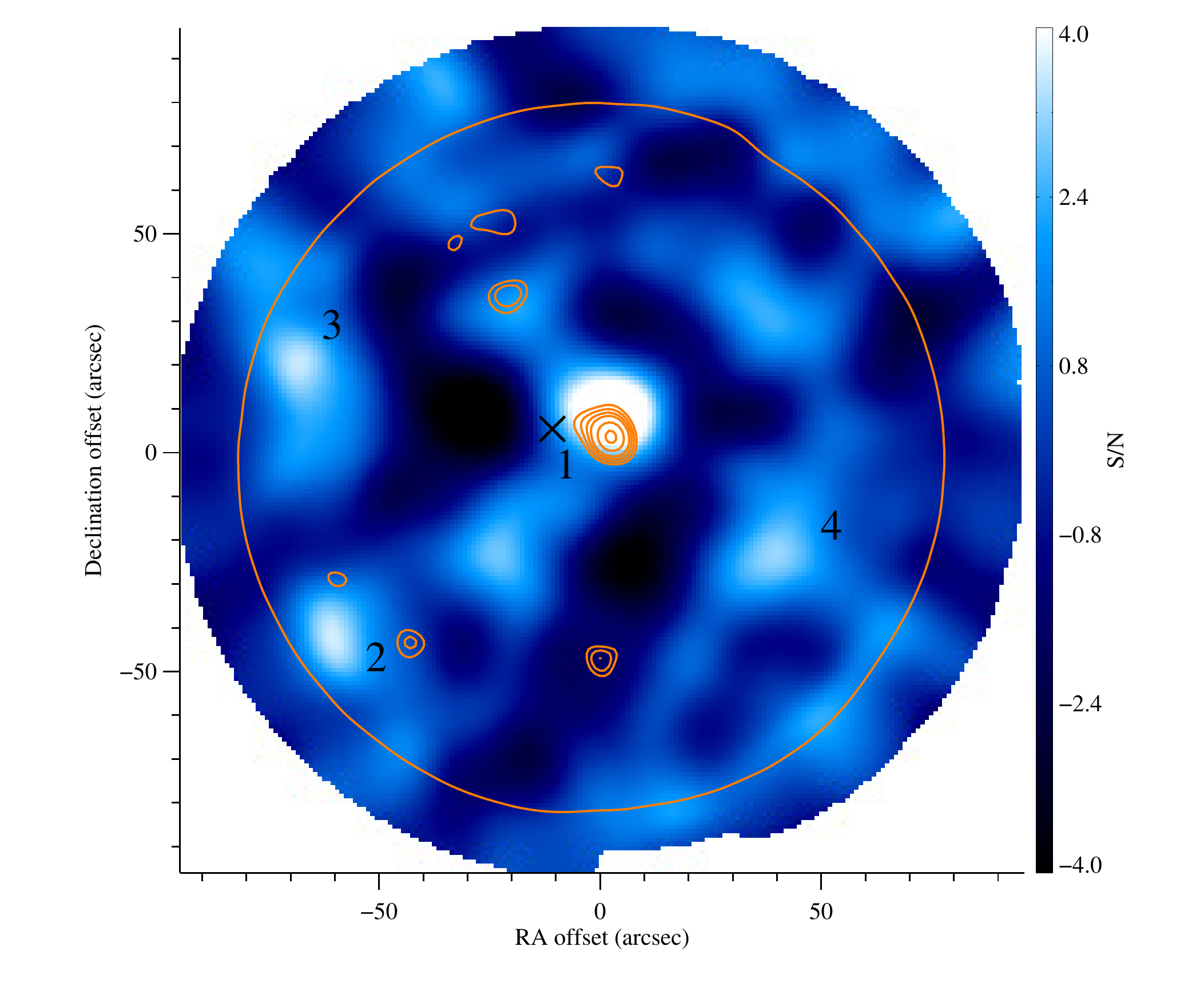}}
   \subfigure{\includegraphics[scale=0.35]{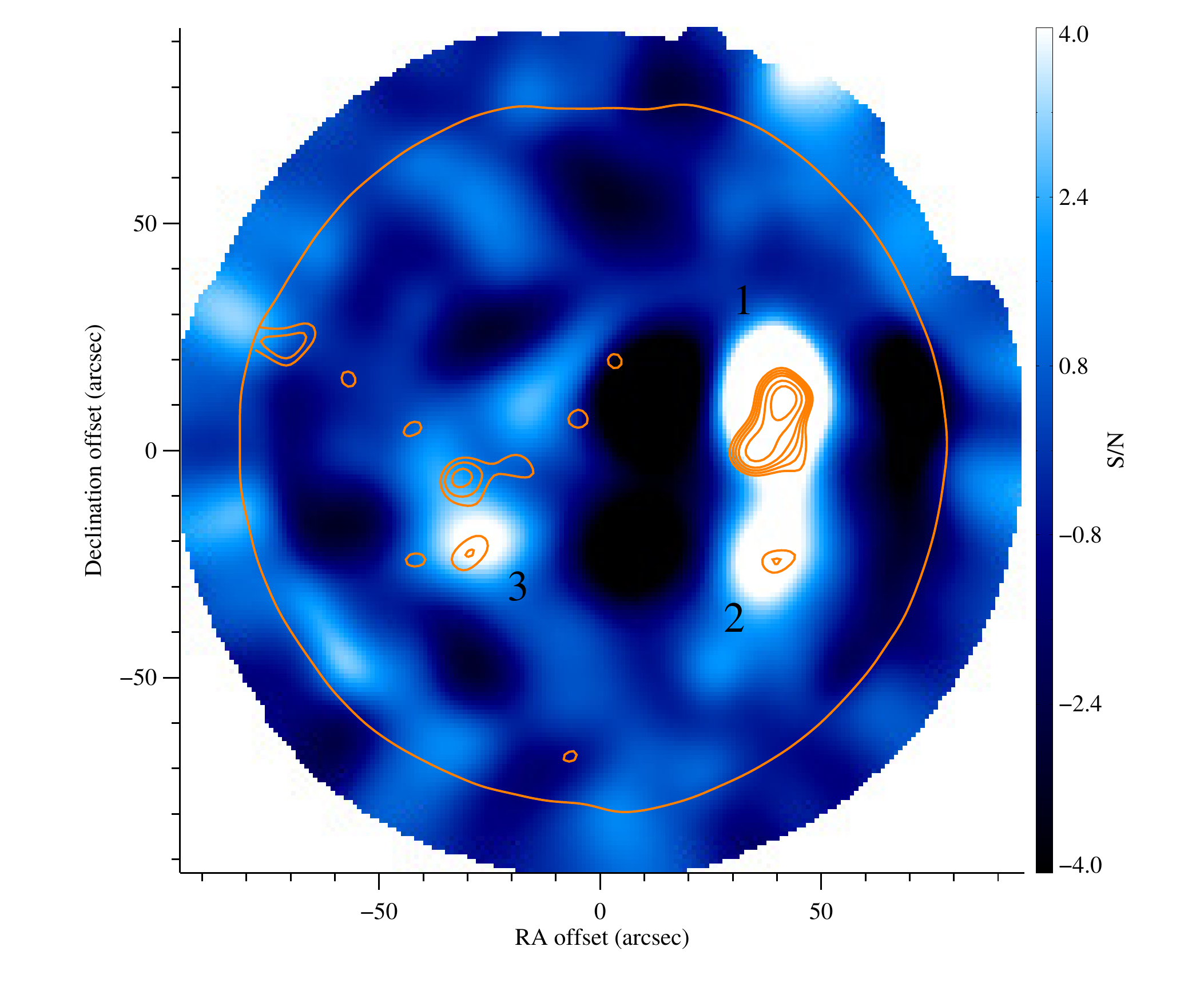}}
   \subfigure{\includegraphics[scale=0.35]{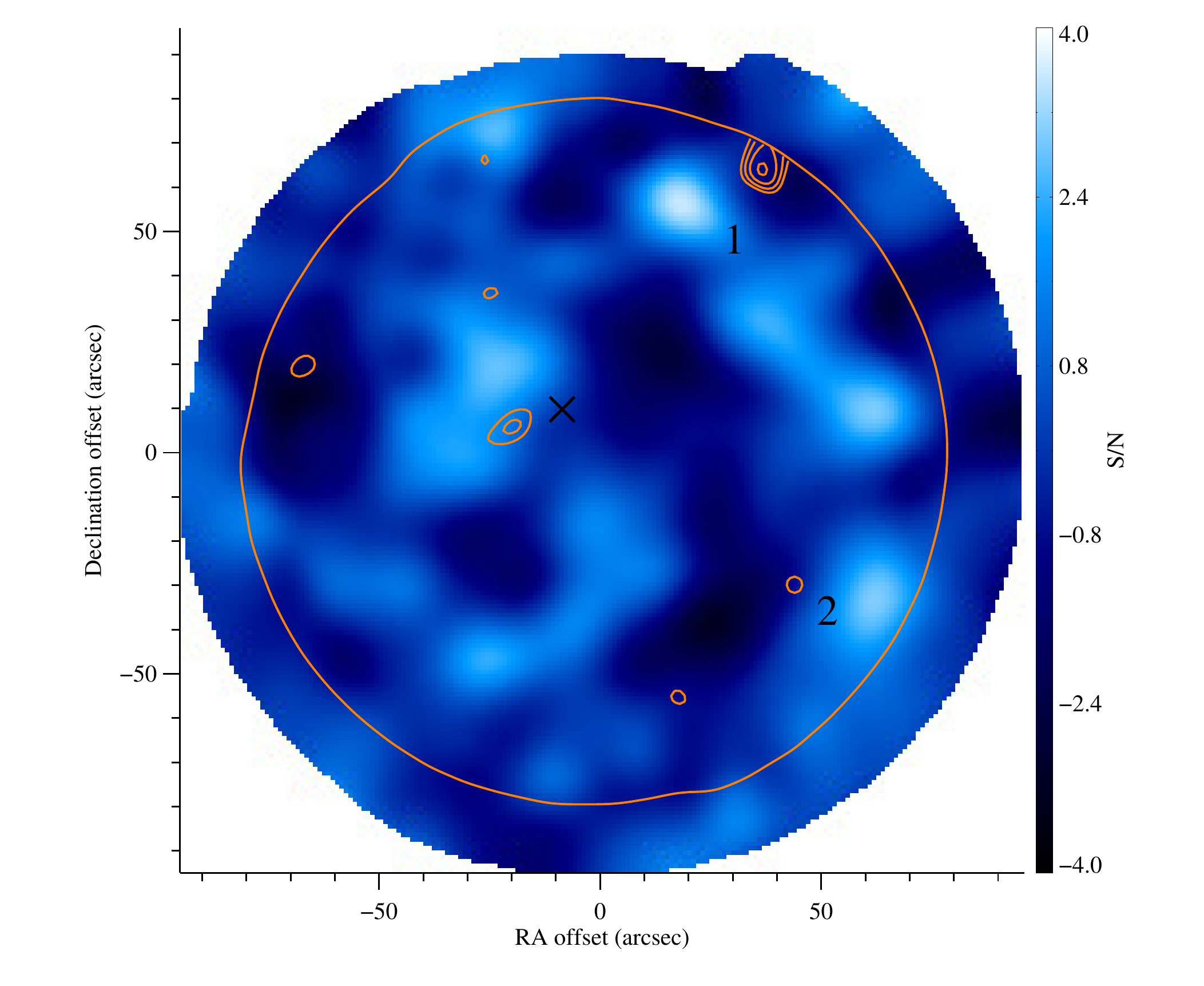}}
     \caption{850\,\um\  S/N maps (convolved with the beam template) for the four control clusters with the same explanation as Fig.~\ref{fig:850mapslensing}.  Clockwise from the upper left corner, the clusters are RCS 1122, 1326, 2152, and 2318.  Note that we have not been able to identify the BCG for RCS 2318. In each map, north points up and east is towards the left; the map centres are given by the coordinates in the cluster names of Table \ref{tab:clusters}.}
   \label{fig:850mapscontrol}
\end{figure*}

For the 26 SMGs, we find 450\,\um\  counterpart emission for 12 sources, 10 of which exceed a S/N of 3.0 at 450\,\um.   Due to the smaller 450\,\um\  field of view (FOV), four SMGs fall outside the shorter wavelength coverage, and seven lie on the noisy edge of the field (see Table \ref{tab:850cat}).  We plot 450\,\um\  S/N contours on the 850\,\um\  S/N map in Fig.~\ref{fig:850mapslensing} (lensing clusters) and Fig.~\ref{fig:850mapscontrol} (control clusters).

We compute the probability of detecting random 450\,\um\ emission within 12\,arcsec of the SMGs using the $P$-statistic from \cite{Downes86}, $P=1-$exp$(-\pi nr^2)$, where $n$ is the surface density of 450\,\um\ sources that have higher S/N than the candidate identifications.  In most of our cluster fields, we do not find any random detections with the required S/N.  In RCS 0224, 1419, and 2318, we calculate $P$-statistics of 0.05, 0.09, and 0.04, respectively.  This confirms that only 0--1 450\,\um\ candidate identifications per cluster are random associations. 

\subsection{Ancillary Data}
\label{sec:ancillary}
In addition to the SCUBA imaging, we have ancillary observations across the spectrum, including optical, mid-IR, radio, and X-ray.  Unfortunately, the coverage is not homogeneous over all the clusters.  A summary of the multi-wavelength data available for each cluster is provided in Table \ref{tab:clusters} and the observations are described below.

\subsubsection{Optical Data}
\label{sec:photoz}
Multi-band optical imaging for all clusters exists from the RCS survey with the CFH12K camera on the Canada-France-Hawaii Telescope (CFHT) and the Mosiac II camera on the Cerro Tololo Inter-American Observatory (CTIO) Blanco telescope in $z'$ and $R_c$ filters \citep{Gladders05}.  Follow-up observations in $V$ and $B$ using the CFH12K camera \citep{Hsieh05} have provided four-colour photometry for several optical sources within four cluster fields (RCS 0224, 1122, 2152, and 2319).  The data were used to calculate photometric redshifts.  For details on the observations and the photometric redshift fitting technique, see \cite{Hsieh05}.

\subsubsection{Spitzer MIPS Imaging}
\label{sec:mips}
The Multiband Imaging Photometer for \textit{Spitzer} (MIPS) at 24\,\um\  has optimal capabilities to detect any mid-IR emission associated with the 850\,\um\  SCUBA sources.  With a 5.4\,arcmin$^{2}$ field of view and point spread function of $\sim6$\,arcsec, MIPS provides improved source positions compared to the SCUBA maps with a 15\,arcsec beam.  These wavelengths typically trace the same populations since both probe dusty star formation. 

MIPS 24\,\um\  observations on the \textit{Spitzer Space Telescope} (proposal ID 30940) have yielded complementary data for six of the seven SCUBA clusters (RCS 2318 was inadvertently left out of the observing list), with integration times ranging from 2000--3300 seconds per pixel.  These data are discussed in more detail in Webb et al. (in preparation).  Briefly, the MIPS images were reduced using a combination of the Spitzer Science Center's MOPEX software and a suite of IDL routines we developed to further optimize background subtraction.  Source catalogues of the entire MIPS field contain flux densities estimated with DAOPHOT, the Dominion Astrophysical Observatory's stellar photometry package \citep{Stetson87}, and positions using a photometry pipeline \citep{Yee91}.  Monte Carlo simulations indicate 80 per cent completeness limits of $\sim$ 70\,\uJy, or L$_\mathrm{IR} \sim$ 10$^{11}$\,\Lsol, for all clusters.

\subsubsection{Spitzer IRAC Imaging}
\label{sec:irac}
The Infrared Array Camera (IRAC) on \textit{Spitzer} operates at 3.6, 4.5, 5.8, and 8.0\,\um.   We obtained IRAC data (program ID 20754) in all four channels for each SCUBA cluster, except RCS 2318, with 120\,second integrations for all clusters.  The images were processed using the Spitzer Science Center's {\it IRACproc} software suite \citep{Schuster06}.  The source finding was completed with a similar source detection and photometry pipeline that was used for the MIPS catalogues.  The images reach a 3$\,\sigma$ depth of $\sim$ 4.2, 6.0, 27, 26\,\uJy\ in channels 1--4, respectively.

\subsubsection{Radio Imaging}
\label{sec:radio}

Deep 1.4 GHz pseudo-continuum (L-band) imaging from the National Radio Astronomy Observatory's (NRAO)\footnote{NRAO is operated by the Associated Universities, Inc., under a cooperative agreement with the National Science Foundation.} Very Large Array (VLA) of the three lensing clusters exists from observations in 2003 through 2005.  Using the `A'  and `B' configurations we achieve a resolution of $\sim 4.5$\,arcsec at 1.4 GHz and reach down to a noise level of $\sim 15$\,\uJy\,beam$^{-1}$ after approximately 16 hours of integration time on each cluster.

Target data were recorded every 5\,seconds and periodically interrupted with observations of a nearby phase target in order to obtain calibration for phase, amplitude and bandpass.  Flux calibration was performed using standard unresolved sources with known flux densities; 3C48 was observed for RCS 0224 and 2319, while 3C286 was used for RCS 1419.  

Calibration and cleaning/imaging were completed using standard tasks in the Astronomical Imaging Processing System (AIPS) software.   The resulting clean images are $\sim$\,34 x 34\,arcmin$^2$ (for the central field), with RCS 1419 and RCS 2319 having slightly larger areas at each side due to mosiacing.  The RCS 2319 image also extends into the RCS 2318 field; the RCS 2319 radio catalogue therefore includes sources within the two SCUBA fields.

Source extraction and flux determination are achieved with the AIPS task `SAD' (Search And Destroy).  To reduce the detection of spurious sources, we first generate a S/N map from the clean image.  With the original clean image and noise map as inputs, the `SAD' task locates potential sources down to a specified level and uses a Gaussian fitting algorithm to determine the least square Gaussian fit to the source.  In the case of extended sources, up to four Gaussian components can be fit and integrated flux densities are summed.  We search down to a S/N of $3.5\,\sigma$ for each cluster.

\subsubsection{X-ray Data}
\label{sec:xray}
X-ray observations, described in \cite{Hicks08}, with \textit{Chandra's} Advanced CCD Imaging Spectrometer (ACIS) were taken between 2002 and 2005 of five (RCS 0224, 1326, 1419, 2318, and 2319) of the seven clusters with individual exposures ranging from $50 - 100$\,ks.  After initial cleaning, $0.3 - 0.7$\,keV flux images were created of the central 4 x 4\,arcmin$^2$ of each cluster.  The point source sensitivities range from $2.5 - 5.1\times10^{-15}$\,ergs\,s$^{-1}$\,cm$^{-2}$.

\section{Submillimetre Data Analysis and Results}
\label{sec:analysis}

\subsection{Monte Carlo Simulations}
\label{sec:monte}
Monte Carlo simulations allow for determination of the uncertainties on source properties, such as flux densities and positions (see \S \ref{sec:deboosting} and \S \ref{sec:positional}), and signal map characteristics, such as completeness, false-detection rates, and confusion (see \S \ref{sec:completeness}, \S \ref{sec:false}, and \S \ref{sec:detection}, respectively).

We populate the raw signal map for each cluster with simulated point sources of varying flux values at random positions and attempt to recover them with our standard source detection algorithm (\S \ref{sec:detection}).  The point sources are generated from the beam template and inserted one at a time to avoid significantly altering source density properties.  The total number of sources injected per signal map is dependent on the real sources in that cluster; the upper flux limit is roughly a factor of two greater than the brightest cluster source.  We inject 50 fake sources at each flux value, starting at 0.5\,mJy and stepping in 0.5\,mJy bins.  The positions are randomly chosen based on the noise mask criteria explained in \S \ref{sec:detection}.  After implementing our cleaning and detection routine, we measure the output flux and position of the point source.  If the source surpasses the 3$\,\sigma$ detection limit and is retrieved within 8\,arcsec (roughly half the width of the SCUBA beam) of the input position, it is considered recovered.

We also repeat the simulations on maps where all the real sources have already been removed (i.e.\ any real source signal has been cleaned out).  This is analogous to the criteria implemented by many groups (e.g.\ \citealp{Scott02, ScottK08}) in which any fake source injected or retrieved within a certain distance of a real source is excluded from further analysis.  The flux boosting estimate (see \S\ref{sec:deboosting}) utilizes these simulations.

\subsection{Completeness}
\label{sec:completeness}
\begin{figure}
   \centering
   \includegraphics[scale=0.45]{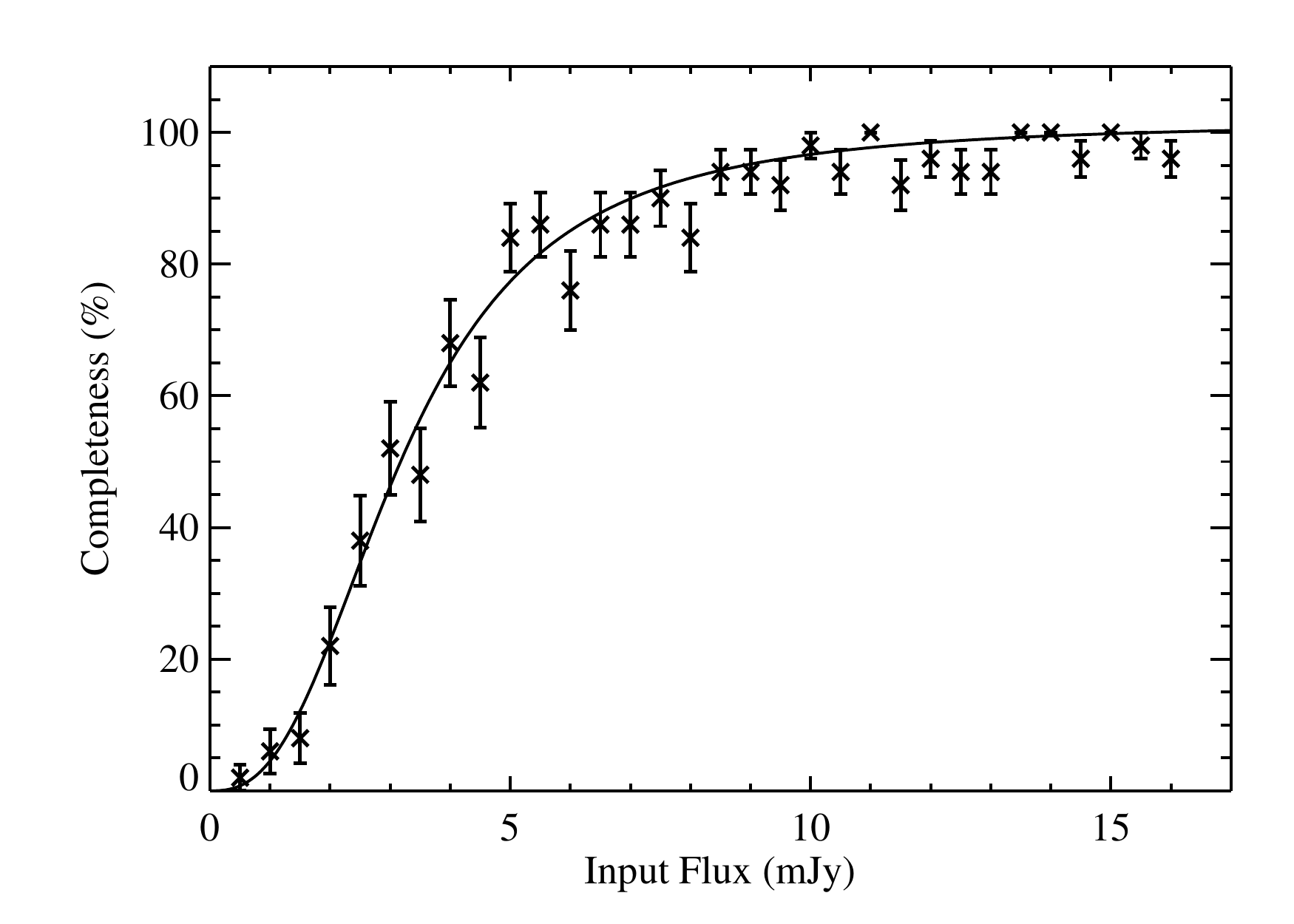} 
   \caption{Differential completeness versus intrinsic flux density for 3$\,\sigma$ detections in the real signal map of RCS 2319.  The error bars represent binomial uncertainties.  The points are fit to a curve of the form $(S^a)/(b+cS^a)$, where $S$ represents input flux.}
   \label{fig:completeness}
\end{figure}

The Monte Carlo simulations provide a means to estimate the completeness individually for each cluster map, using the aforementioned criteria for recovered sources.  In Fig.~\ref{fig:completeness}, we plot the differential completeness as a function of intrinsic source flux density for RCS 2319.  Following \citet{Coppin06}, we fit to a curve of the form $(S^a)/(b+cS^a)$, where $S$ represents input flux.  This empirically derived function provides a good fit to the calculated completeness.  For this specific cluster, we are $\sim$50 per cent complete at 3\,mJy and $\sim$95 per cent complete by 9\,mJy, which is consistent with the remaining six clusters.

\subsection{Flux Deboosting}
\label{sec:deboosting}
A well-established yet rather subtle bias, flux boosting describes the enhanced flux measurement for sources in low S/N ($\lesssim$10) maps that occurs when there is an increase in number counts of sources with decreasing flux density \citep{Hogg98}.  This produces an asymmetric distribution of sources available for scattering into another flux domain.  In other words, there is a greater probability that intrinsically faint sources are detected at a higher flux than bright sources are deboosted to a lower flux, which is known as the Eddington effect.  This measurement effect differs from a Malmquist bias, which arises in flux-limited samples due to greater detectibility of more luminous sources over a greater volume.  A second systematic flux measurement error for low S/N sources results from a tendency to be pulled to brighter fluxes due to random noise spikes in the map.  This is a consequence of the method of recovery of sources since peaks define the position and flux of the source.

To quantify this asymmetric flux error, many groups (e.g.\ \citealp{Coppin05,Coppin06,ScottK08}) have implemented a Bayesian approach by calculating a posterior flux distribution (PFD).  As an alternative to this analytical method, numerical simulations also provide a means to estimate the flux deboosting correction (e.g. \citealp{Scott02}).  A comparison of the output flux to the range of input fluxes that produced it should reveal a flux boosting factor that increases with decreasing measured flux density.

\begin{figure}
   \centering
   \includegraphics[scale=0.45]{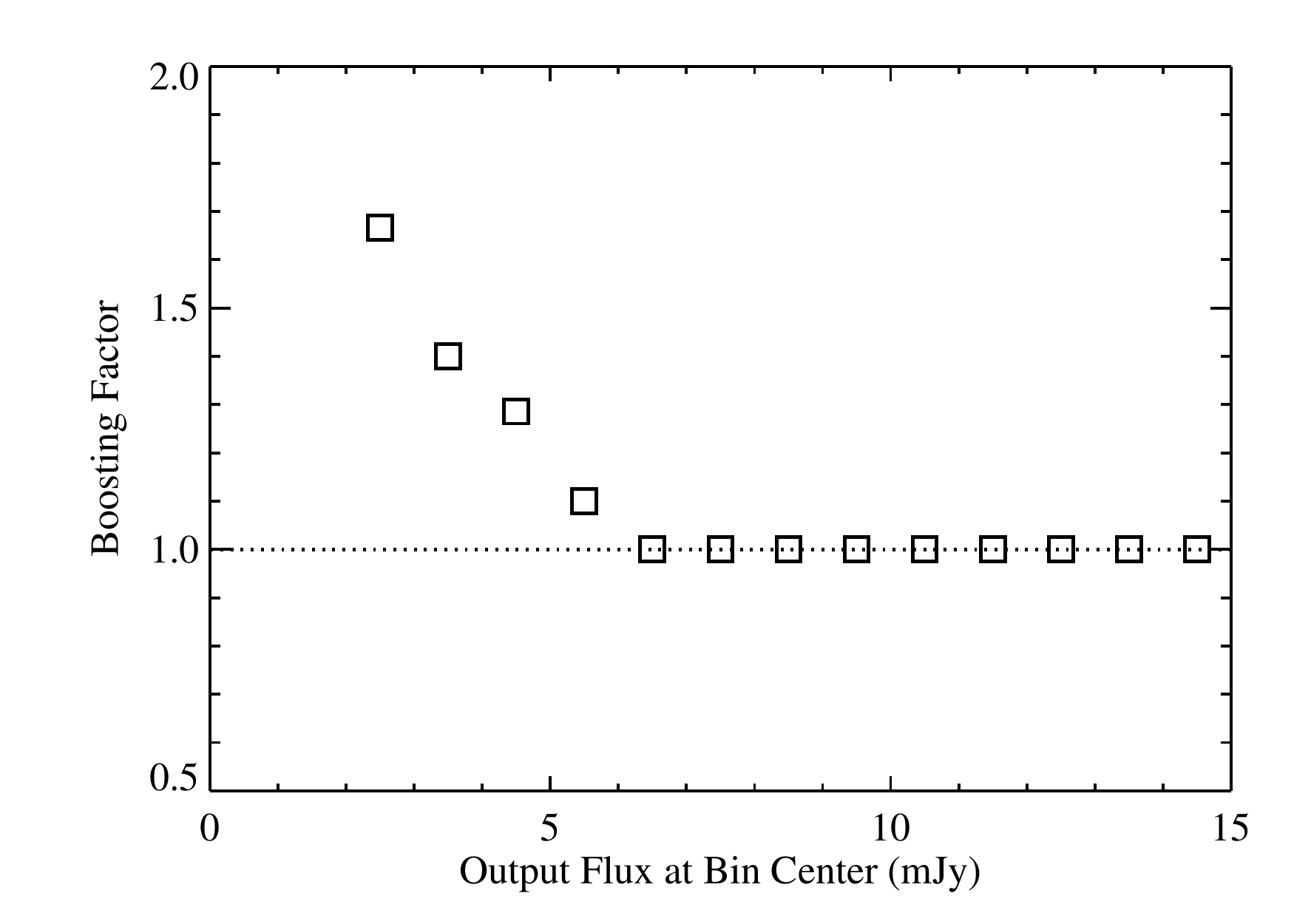} 
   \caption{The flux boosting factor as a function of output flux (i.e.\ measured flux) for all cluster fields combined.  All sources detected at a flux density where the boosting factor lies above the dotted line must be deboosted by that amount.}
   \label{fig:boost}
\end{figure}

Using the results from the injected point sources at each flux level within the uniform noise region, we bin all input fluxes (with 1\,mJy width) that yield a 1\,mJy range of output fluxes for each cluster field separately.  The centre of the output flux range is divided by the mode of input fluxes to calculate the flux boosting factor.  Each cluster produces approximately the same results within uncertainties, therefore the same procedure is repeated combining the simulations from all clusters to increase number statistics.  The final flux boosting factor as a function of output flux is shown in Fig.~\ref{fig:boost}.  This factor is then applied to the measured SMG fluxes within that output flux range, which are, consequently, deboosted.  The boosting factor ranges from 1.7 in the lowest flux bins to 1.0 at higher fluxes.  Sources detected above 6\,mJy are therefore left unaltered based on these simulations.  The final deboosted 850\,\um\  flux densities are recorded in Table \ref{tab:850cat}.

\subsection{Positional Uncertainty}
\label{sec:positional}
\begin{figure}
   \centering
   \includegraphics[scale=0.45]{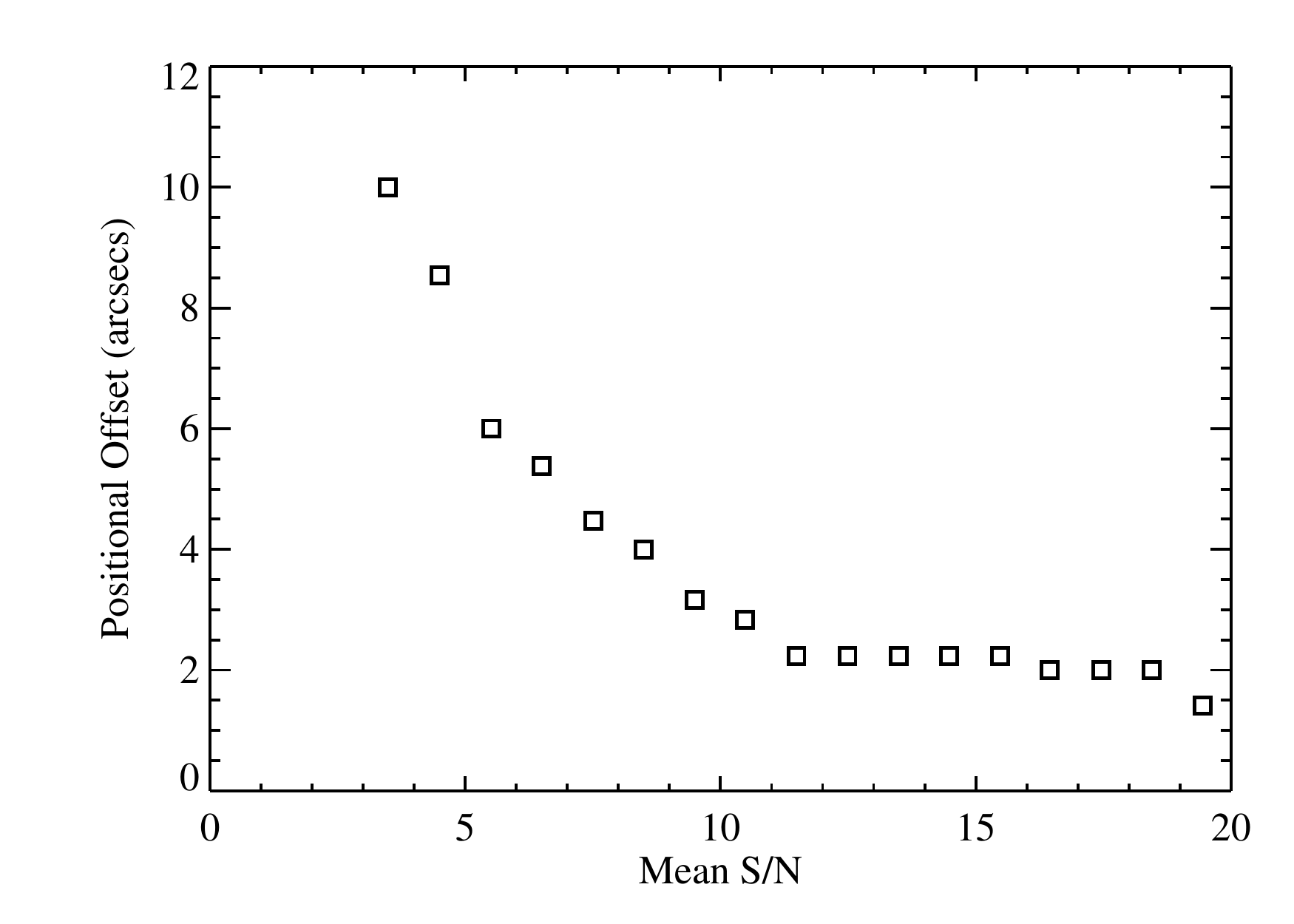} 
   \caption{The positional uncertainty of the SMG as a function of S/N based on our Monte Carlo simulations combined for all cluster fields.  The points represent the 95th percentile of all positional offsets at that S/N.}
   \label{fig:pos_offset}
\end{figure}

Lower S/N submillimetre sources are typically retrieved at larger positional offsets than more robust sources due to their susceptibility to being pulled off their intrinsic position by random noise spikes, or through confusion with brighter sources.  Based on our simulations, we calculate the 95th percentiles for the positional offsets after binning sources in S/N with 1$\,\sigma$ bin widths.  The resulting histograms of positional offsets for each cluster follow the same pattern, and we therefore combine all clusters to increase our number statistics.  We find maximum positional offsets ranging from 10 to 2\,arcsec for S/Ns between 3.5 and 7.5$\,\sigma$, respectively (see Fig.~\ref{fig:pos_offset}), which are consistent with other submillimetre cluster surveys (e.g.\ \citealp{Webb05,ScottK08}).  These offsets are subsequently used as search radii for counterpart identification (described in \S\ref{sec:counters}).

\subsection{False-Detection Rate}
\label{sec:false}
The inclusion of the negative beams from the chop in the convolved image renders the instrumental noise symmetric about zero.  However, the confusion noise could potentially produce an asymmetrical noise fluctuation distribution.  We quantify this by populating noise maps with sources distributed in flux according to the SHADES differential source counts \citep{Coppin06} fit to a Schechter function \citep{Schechter76}, up to three times the RMS noise in the map.  In other words, there should be no real, unblended sources with S/N $>3\sigma$.  We then count both the positive peaks ($>$3$\sigma$) and negative peaks ($<$-3$\sigma$), average over 100 realizations, and find a symmetric distribution.  For example, the RCS 2319 field yields $4.0\pm1.5$ positive peaks, and $3.5\pm1.6$ negative peaks. 

We therefore assume the noise is symmetric about zero, which allows for a direct method to quantify source reality.  After removing all the detected sources from the raw image, we invert the map.  We can then estimate the false-detection rate by counting the number of peaks $\ge 3\,\sigma$ within the usable map area on the inverted image.  To improve our statistics, we combine all the cluster fields in the calculation instead of considering them individually.  

\begin{figure}
   \centering
   \includegraphics[scale=0.45]{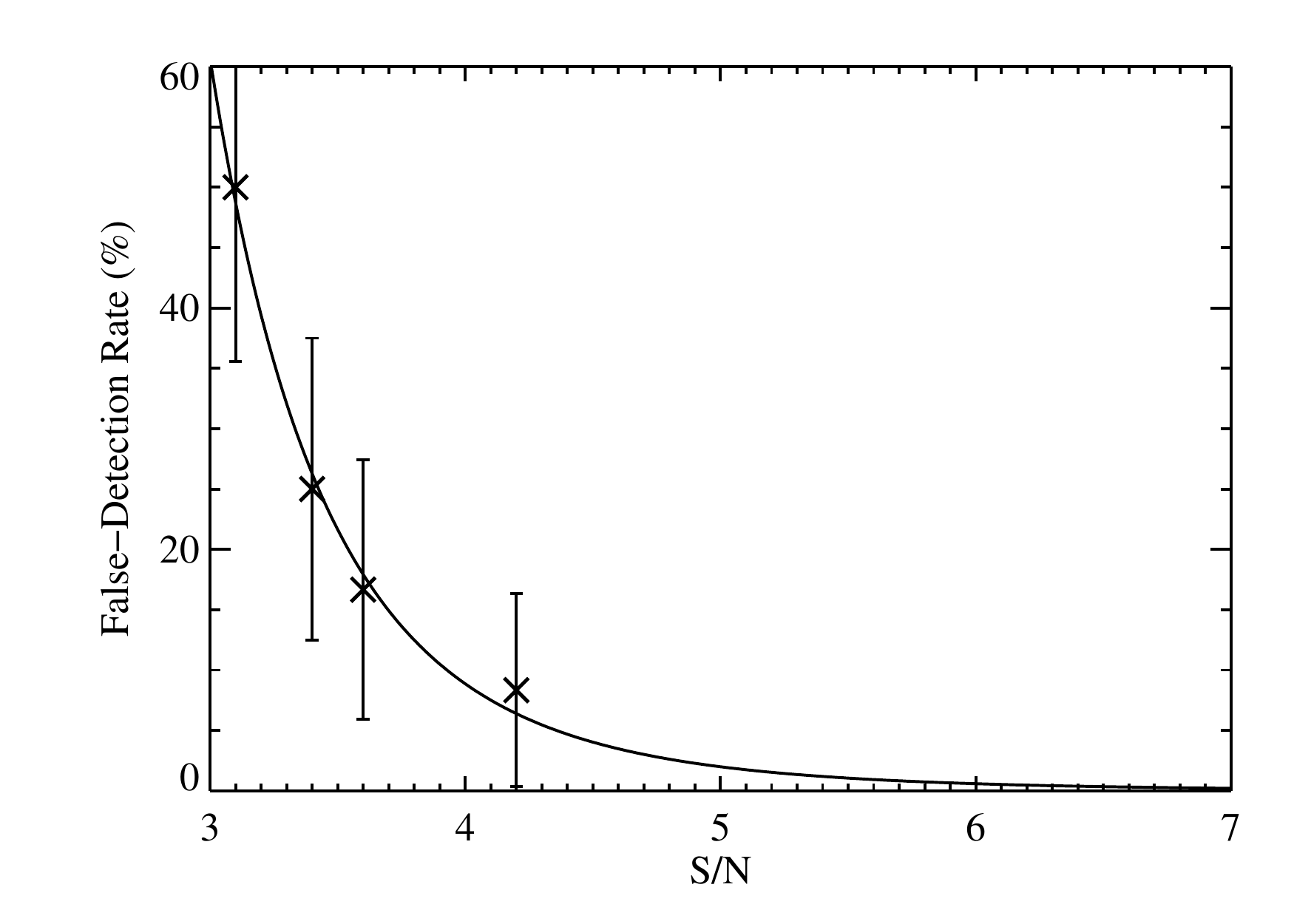} 
   \caption{The false-detection rate as a function of S/N in our combined cluster fields.  The points are empirically fit to a power-law and rates are determined from this curve for spurious source corrections (e.g.\ in source counts analyses---see \S\ref{sec:counts}).}
   \label{fig:spurious}
\end{figure}

Consistent with other groups (e.g.\ \citealp{Cowie02, Scott02}), we find a $3\,\sigma$ cut includes a high number of spurious sources, corresponding to a $\sim$46 per cent false-detection rate.  The situation remarkably improves for a threshold of 3.5$\,\sigma$, with only three (of 19) spurious sources detected, resulting in a 16 per cent contamination.  Fig.~\ref{fig:spurious} plots the false-detection rate at average S/N values including sources down to 3.0$\,\sigma$.  The points have binomial uncertainties and are empirically fit with a power-law of the form $y=ax^b$, where 941.5 and $-6.7$ are found for the amplitude and slope, respectively.  The false-detection rate for a given S/N is determined from this curve in subsequent analysis.

We can estimate our expected number of spurious sources by summing the false-detection rate equation over the total number of detected sources in our survey.  We calculate that four of the 26 sources in this survey should be false, given their range in S/N values (see Table \ref{tab:850cat}).

Despite the significant uncertainty in individual sources detected below 3.5$\,\sigma$, we decide to include them in the source counts analysis (as do other groups---e.g.\ \citealp{Cowie02,Coppin06}) as a means for comparison to previous surveys.  We correct the source counts for spurious sources (see \S\ref{sec:counts}), and  contend that the 3$\,\sigma$ threshold is reasonable for the counts analysis.  We do note, however, that counterpart identification at this level can be difficult to accurately interpret, as individual sources carry low significance.

\section{850 \lowercase{\um} Source Counts}
\label{sec:counts}
A crucial goal of most submillimetre surveys involves an understanding of the distribution of galaxies over a range of flux densities in order to constrain evolutionary models.  A plethora of studies have tackled this at 850\,\um\  across a high dynamic range of fluxes ($\sim 0.1$\,mJy $-$ 20\,mJy), utilizing massive cluster lenses to amplify the SMG population at the faint end (e.g.\ \citealp{Smail97, Cowie02, Smail02, Chapman02counts, Knudsen06}) and wide blank-field surveys to probe brighter SMGs (e.g.\ \citealp{Scott02, Borys03, Webb03VI, Coppin06}).

A comparison of the number of sources as a function of flux density (i.e.\ source counts) between cluster fields and unbiased regions of the sky (i.e.\ blank fields) can elucidate the influence of cluster environment on galaxy evolution through the verification of a statistical excess of SMGs in cluster fields.  Our survey offers a further analysis within those cluster fields by examining differences in number counts between clusters with and without an abundance of optical arcs from strong gravitational lensing and probes higher-redshift clusters, exposing excess counts as due to either lensing or contamination by cluster galaxies.

\subsection{Gravitational Lensing}
\label{sec:grav_lensing}
Strong gravitational lensing magnifies sources, and subsequently increases the observed
flux.  We therefore must assess the effect any flux magnification could have on the source counts.  The cluster mass inferred from the richness estimate (see Table \ref{tab:clusters}) can be used
to estimate the magnification caused by the smooth cluster mass
distribution. The amount of flux boosting depends on the source
redshift, and as these are unknown, we consider values between $z=1$ and
$z=4$.  To compute the magnification, we assume that the density profile
of the cluster is described by a Navarro, Frenk \& White profile \citep{Navarro97}, with a
mass-concentration relation given by \cite{Duffy08}.  Given that the control and lensing samples span the same range of cluster mass, we treat both cases with only a smooth cluster component, and do not attempt to reproduce the strong lensing features described in \S\ref{sec:strong_lensing}.

These calculations only consider the contribution of the smooth
density profile, but locally cluster members themselves can
contribute to the lensing signal as well. For SMGs in the outskirts of
clusters, the source would have to be located very close to the lens to
be magnified significantly.

We find that the magnification is modest for most sources, but in a
few cases the SMG is likely to be strongly lensed if it is indeed
at high redshift and the adopted mass is representative.  However,
it is difficult to quantify without being able to model
the cluster in more detail.  

Assuming $z=2.5$, a typical SMG redshift, we calculate an average magnification of $\sim1.5$, excluding the two SMGs that would be significantly lensed if confirmed to be at that redshift (SMMs 2319.3 and 1326.1).  In particular, the magnification values for SMM 2319.3 are unreliable as it is extremely close to the cluster centre (i.e.\ within a projected distance of 13\,arcsec).  This magnification value is fairly robust to $1\,\sigma$ increases/decreases in the cluster masses, yielding average values of $\sim1.8$ and 1.3, respectively, for a $z=2.5$ SMG population.  Even at the most extreme level, assuming a source plane redshift of $z=4$, we only find an average magnification of $\sim1.8$.  Due to the fairly low magnification values, we make no strong lensing correction in our source counts analysis (discussed further in \S\ref{sec:diff}) and note that none of these systems have actually been confirmed as lensed objects.

\subsection{Differential Counts}
\label{sec:diff}
An accurate estimate of 850\,\um\  source counts relies on properly correcting for several of the properties described above, such as incompleteness (\S \ref{sec:completeness}), flux-density boosting (\S\ref{sec:deboosting}), and false-detection contamination (\S\ref{sec:false}).  Following the procedure outlined in \cite{Scott02}, we exploit the results of our Monte Carlo simulations to account for these effects.

\begin{figure}
   \centering
   \includegraphics[scale=0.5]{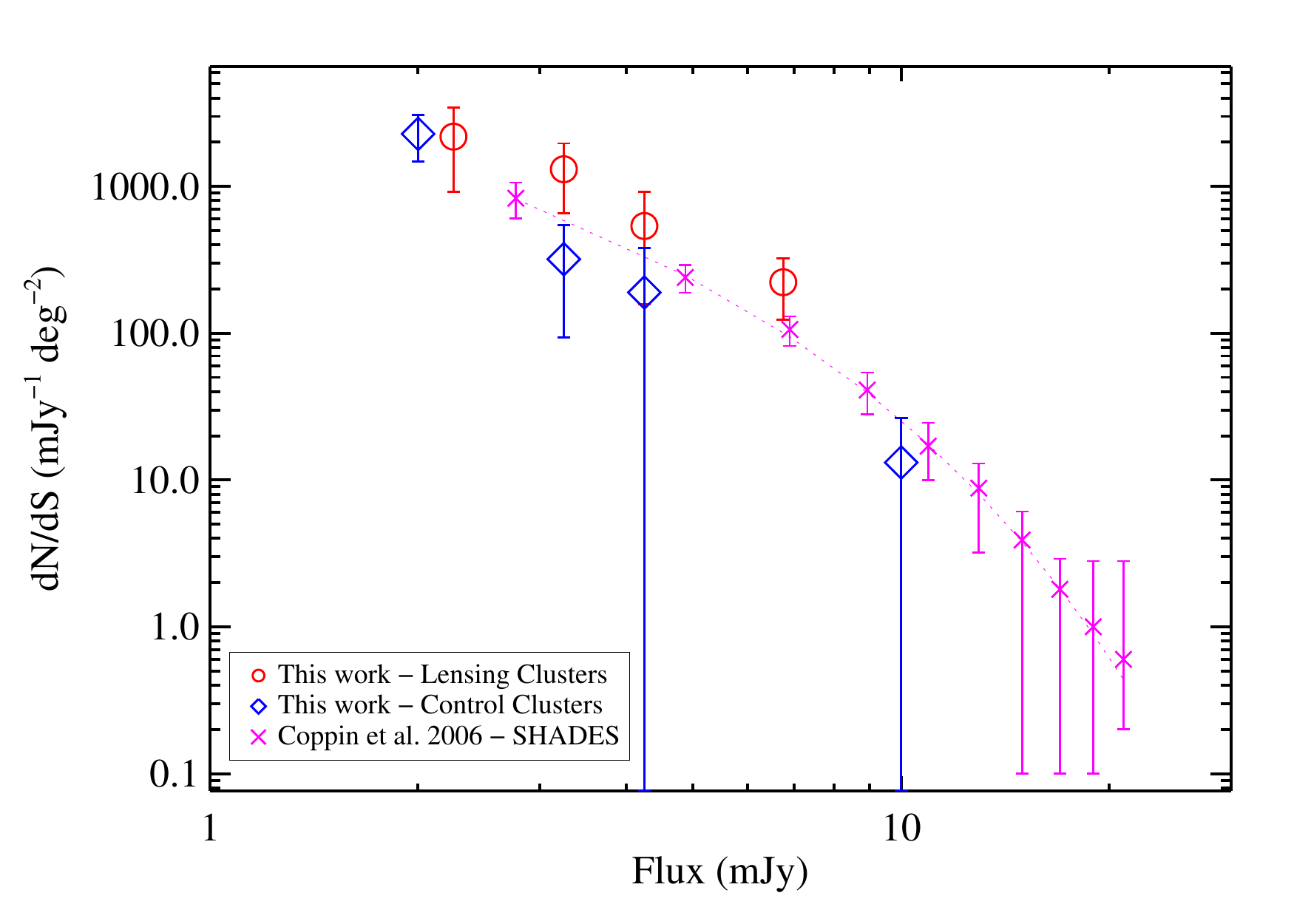} 
   \caption{Differential 850\,\um\  source counts.  Our results are denoted by the larger symbols, where the circles represent counts from the lensing clusters and the diamonds represent counts for the control group.  We also include differential counts from the largest blank-field survey---SHADES (crosses) \citep{Coppin06}---and a best-fitting Schechter function.  Our larger error bars are attributable to small number statistics.}
   \label{fig:diff_counts}
\end{figure}

To begin, we devise flux density bins based on the values used in the completeness calculation of \S\ref{sec:completeness} (i.e.\ 0.5\,mJy bins centred on integer and half-mJy fluxes).  We then calculate raw source counts for each bin, separately for each cluster, by adding the number of objects that fall within that bin using their statistically deboosted flux density.  To account for spuriously detected sources, each raw count is weighted by the probability that it is false, which is based on the S/N of the source.  The probability is estimated from the fit to the curve in Fig.~\ref{fig:spurious}.  This weighting scheme de-emphasizes counts from lower S/N sources which are more likely spurious.  Due to the observed incompleteness of each map, which increases with decreasing flux, we adjust the weighted counts by dividing them by the best-fitting completeness value at that flux determined from a curve analogous to that in Fig.~\ref{fig:completeness} (each cluster field has a slightly different completeness fit).  Bins with zero initial counts are left unaltered.  These corrected values are combined, preserving bins, for each subsample separately (i.e.\ lensing and control groups).  Based on the distribution of flux bins, we create larger bins to ensure all differential bins have a value greater than zero for display purposes.  This results in four new bins for each subsample, which are presented in Table \ref{tab:diff_counts}.  The highest flux bin in the control group is a result of one extremely bright source in RCS 2318, detected at $\sim$15\,mJy.  After adding the values from all the old bins that fit within the new bin boundaries, we divide each bin by its width for normalization purposes.  The final differential counts, shown in Fig.~\ref{fig:diff_counts}, are simply these values divided by the total area in square degrees covered by that subsample (see \S\ref{sec:detection} for a description of how usable map areas are determined and Table \ref{tab:clusters} for a list of areas).  The relative uncertainties are calculated from standard Poisson errors on the raw counts scaled by the bin size and map area.  The final counts and errors are tabulated in Table \ref{tab:diff_counts}.

In Fig.~\ref{fig:diff_counts}, we also plot the differential counts obtained from the SHADES blank-field survey, fit with the empirically motivated Schechter function \citep{Schechter76}
\begin{equation}
\label{eqn:schechter}
\frac{dN}{dS} = \frac{N^\prime}{S^\prime}S\left(\frac{S}{S^\prime}\right)^\alpha\exp(-S/S^\prime).
\end{equation}
The SHADES team has found the best-fitting form to the function using values of 3.3, 1599, and -2.0 for $S^\prime$, $N^\prime$, and $\alpha$, respectively.  The differential counts of our control sample agree quite well with the SHADES results; the lensing sample tends toward a slight excess in source counts compared with both the control sample and SHADES survey counts, hinting at an inherent difference between the cluster environments of the two subsamples in our survey.  We note that demagnifying the fluxes of the lensing cluster sample by a minimum of $\sim1.5$ does bring the differential counts in line with SHADES.  However, if we assume the field is lensed, we also need to decrease the total survey area, since massive foreground lenses cause the background area to look expanded \citep{Broadhurst95}.  This brings the counts back up, and thus we emphasize that this is only a lower limit on the required magnification.  Since the control sample covers the same range of magnifications, the same adjustments would be made for both samples, and the counts discrepancy between the two would remain.  As none of these galaxies have been confirmed as lensed systems, it is difficult to fully assess the magnification correction.

\begin{table}
\begin{center}
\caption{Differential 850\,\um\  source counts for the control and lensing samples.}
\label{tab:diff_counts}
\begin{tabular}{cccc}
\hline 
\multicolumn{1}{c}{Cluster} &
\multicolumn{1}{c}{Flux Density$^{a}$} &
\multicolumn{1}{c}{Bin Width} &
\multicolumn{1}{c}{$dN/dS$} \\
\multicolumn{1}{c}{Sample} &
\multicolumn{1}{c}{(mJy)} &
\multicolumn{1}{c}{(mJy)} &
\multicolumn{1}{c}{(mJy$^{-1}$ deg$^{-2}$)} \\
\hline
Lensing  & 2.25 & 1.0 & 2183$\pm$1260	\\
& 3.25 & 1.0 & 1309$\pm$654	\\
& 4.25 & 1.0 & 537$\pm$380 	 \\
& 6.75 & 4.0 & 223$\pm$100 	\\
\hline
Control  & 2.0 & 1.5 & 2277$\pm$805\\
& 3.25 & 1.0 & 319$\pm$226\\
& 4.25 & 1.0 & 190$\pm$190\\
& 10.0 & 10.5 & 13$\pm$13\\
\hline
\end{tabular}
\begin{tablenotes}[normal]
$^{a}$The flux density given is for the bin centre.
\end{tablenotes}
\end{center}
\end{table}

\subsection{Cumulative Counts}
\label{sec:cumul}
\begin{figure*}  
 \centering
   \includegraphics[scale=0.94]{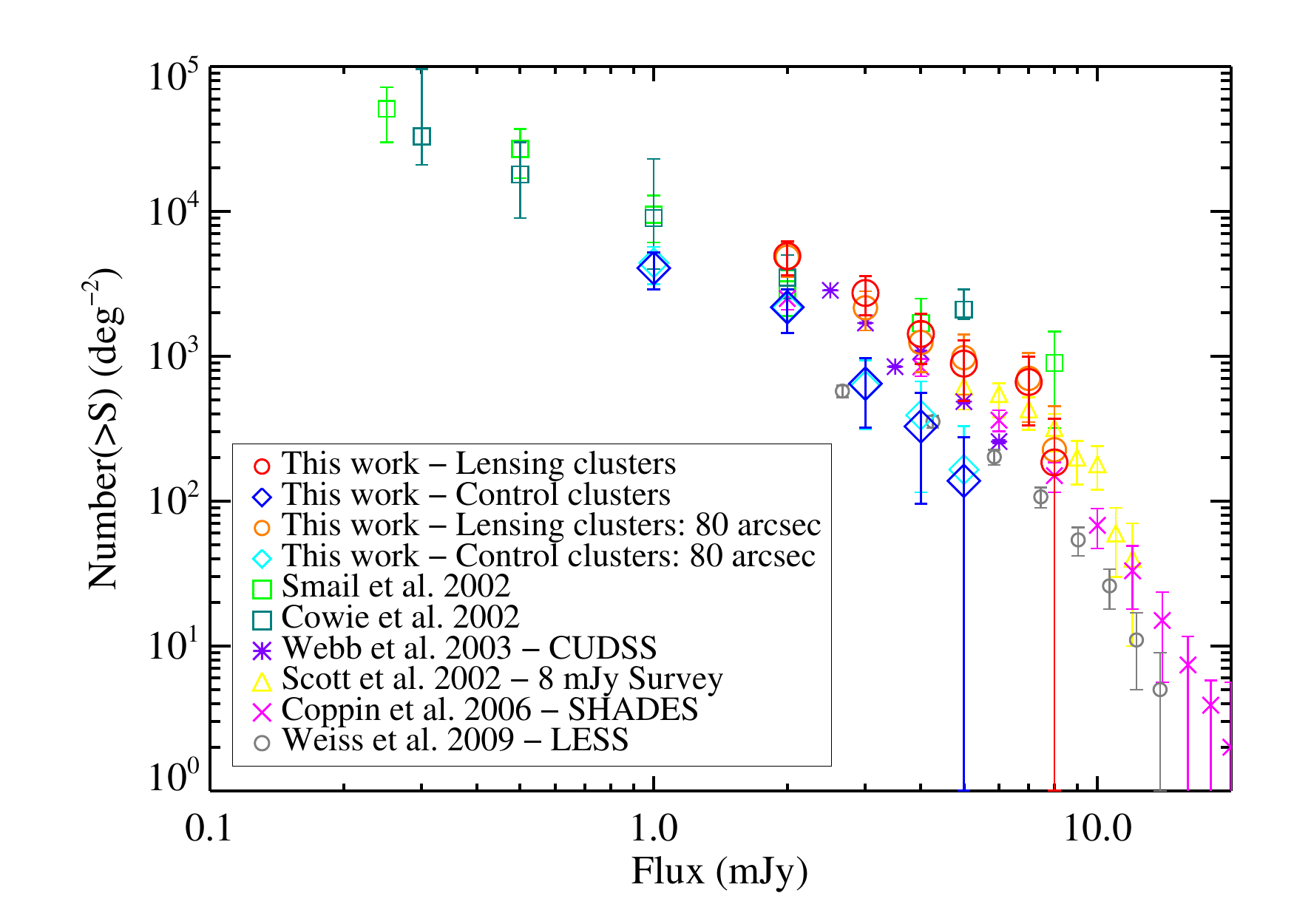} 
   \caption{Cumulative 850\,\um\  source counts for the lensing clusters (large red circles) and the control clusters (large blue diamonds).  Also shown are counts determined from other cluster surveys \citep{Smail02, Cowie02} and blank-field surveys \citep{Scott02, Webb03VI, Coppin06, Weib09}.  We have scaled the 870\,\um\  LESS counts to 850\,\um\  using a submillimetre spectral index of 2.7.  Note the tentative excess of source counts for the two cluster surveys mentioned, along with the counts derived from the strong lensing sample.}
   \label{fig:counts}
\end{figure*}

The same corrections (i.e.\ flux deboosting, false-detection rates, and completeness adjustments) are applied to properly compute source counts at each 0.5\,mJy flux density bin.  The cumulative counts are recorded at each discrete flux density as the total number of sources per square degree at that flux and higher, meaning each bin is correlated with all higher flux density bins.  The control sample ranges from 1.0 to 5.0\,mJy, while the lensing group encompasses fluxes between 2.0 and 8.0\,mJy.  The resulting counts are given in Table \ref{tab:counts} and shown in Fig.~\ref{fig:counts} along with various blank-field and cluster surveys.  

We test the validity of our source count results by randomly eliminating one cluster from each sample and re-calculating the cumulative counts.  Separately, we also transfer one control cluster to the lensing sample to ensure the discrepancy in source counts between the two samples does not flip.  The cumulative counts prove to be robust to all test cases as the lensing sample counts remain above the control sample.  We also attempt to account for any human biases (mentioned in \S\ref{sec:IRintro}) that may enhance the source counts in the lensing sample.  We note that the sample was selected in an unbiased way, and no clusters have been left out of the analysis.  As evident in Figures \ref{fig:850mapslensing} and \ref{fig:850mapscontrol}, the lensing clusters have larger map areas as the result of multiple pointings.  These three clusters were all mapped in classical observing time, and therefore observations were designed in real-time to maximize the S/N of detected sources; the control clusters, on the other hand, were all observed later in service time.  To determine the potential effect this bias might have on our results, we compute the cumulative source counts within the same projected radius from the original pointing for each cluster.  We adopt a radius of 80\,arcsec to ensure the search area is within the uniform noise for each field.  The resulting counts are plotted alongside the original counts in Fig.~\ref{fig:counts}.  We find our results are robust to any potential bias resulting from multiple pointings, and we therefore continue our analysis with the original source counts from the larger map areas.  

Although just below the SHADES results, the control sample is consistent with the counts from the Large Apex Bolometer Camera (LABOCA) Extended Chandra Deep Field South (ECDFS) Submillimetre Survey (LESS); the ECDFS has a deficit of bright SMGs and is known to be an underabundant ULIRG field \citep{Weib09}.  We note that we have plotted the LESS counts estimated directly from the catalogue, rather than those estimated from the $P(D)$ analysis, although both methods show a deficit of counts compared to other deep fields.  The lower counts for both LESS (with the direct source count method) and our control group could be the result of faint multiple sources in the beam identified as a single brighter source, which would lead to an underestimate of fainter sources \citep{Weib09}. 

With the exception of the highest flux bin, the lensing sample shows an excess of counts compared to blank-field surveys, consistent with other lensing cluster surveys (e.g.\ \citealp{Cowie02, Smail02}).  This is a $\sim2.7\,\sigma$ overabundance relative to the control sample (at the lowest flux point, which accounts for all sources with $S\ge2$\,mJy), suggesting that enhanced source counts are unique to strong lensing cluster fields.  We limit our quantitative comparison to the two samples presented in this work, as we can be assured that the counts were calculated identically, and do not suffer from any systematic differences.

\begin{table}
\begin{center}
\caption{Integral 850\,\um\  sources counts for the control and lensing samples.}
\label{tab:counts}
\begin{tabular}{ccc}
\hline
\multicolumn{1}{c}{} &
\multicolumn{2}{c}{Integral Counts} \\
\multicolumn{1}{c}{} &
\multicolumn{2}{c}{$N(>S)$ (deg$^{-2}$)} \\
\multicolumn{1}{c}{Flux Density (mJy)} &
\multicolumn{1}{c}{Lensing Sample} &
\multicolumn{1}{c}{Control Sample} \\
\hline
1.0 & --- & 4063$\pm$1173\\
2.0 & 4919$\pm$1315 & 2181$\pm727$\\
3.0 & 2736$\pm$825 & 647$\pm324$\\
4.0 & 1427$\pm$540 & 328$\pm232$\\
5.0 & 890$\pm$398 & 138$\pm138$\\
6.0 & 890$\pm$398 & ---\\
7.0 & 664$\pm$332 & ---\\
8.0 &185$\pm$185 & ---\\
\hline
\end{tabular}
\end{center}
\end{table}

\section{Counterpart Identification}
\label{sec:counters}
\begin{figure}
   \centering
   \includegraphics[scale=0.45]{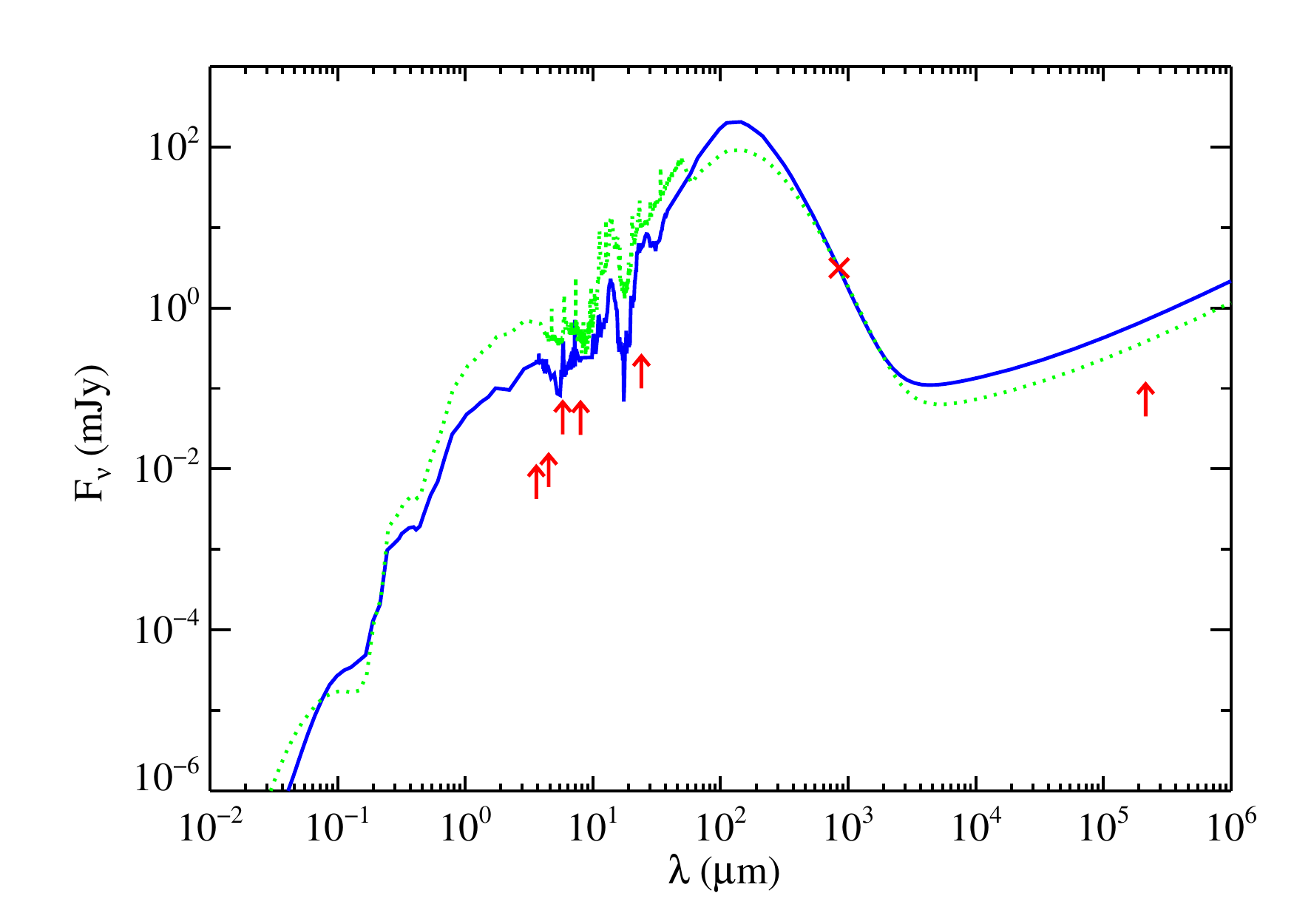} 
   \caption{SED templates for Arp 220 (solid line) and M82 (dotted line) redshifted to $z=1.0$, the highest redshift of our submillimetre cluster sample.  The templates have been normalized in flux density to 3.1\,mJy (roughly the 3$\,\sigma$ detection limit in our SCUBA maps) at 850\,\um.  The 3$\,\sigma$ detection limits for the MIPS, IRAC, and radio data are also shown.  Note that these local galaxy SED templates are not representative for high-redshift SMGs at rest-frame optical and UV wavelengths (i.e.\ below $\sim$1\,\um). }
   \label{fig:sed_detect}
\end{figure}

A wealth of ancillary data to the submillimetre observations exists at mid-infrared (mid-IR), radio, optical, and x-ray wavelengths (see \S\ref{sec:ancillary}), allowing for the identification of counterparts to the SMGs which improves positional accuracy and helps constrain redshifts.  A possible complication with regards to identifying counterpart emission with ancillary observations involves the detection limit at various wavelengths.  To ensure we are sensitive to detecting counterparts, we plot the $3\,\sigma$ depths for the MIPS, IRAC, and radio fields in Fig.~\ref{fig:sed_detect} overlaid with two spectral energy distribution templates for Arp 220, a typical star-forming ULIRG at low redshift, and for M82, a star-forming galaxy with irregular morphology.  Each SED has been converted to the highest redshift for the entire sample, $z\sim1$, and normalized to the SCUBA detection limit at 850\,\um\  in our cluster fields, $3\,\sigma\simeq 3$\,mJy.  The detection limits for 3.6, 4.5, 5.8, 8.0, and 24\,\um\  along with 1.4\,GHz are below the SEDs, thereby allowing sensitive counterpart detection for cluster members.  A non-detection may by indicative of either a high-redshift background source or a spurious SMG, particularly for low S/N cases.  We reiterate that although we expect four spurious sources given our false-detection rate equation (see \S\ref{sec:false}), we search for counterparts in the entire source catalogue since we do not know which of these are in fact spurious. However, we limit our potential cluster members (see \S\ref{sec:tent_mems}) to sources with S/N$>$3.5.

Given the fairly homogeneous data coverage with MIPS (imaging over six of the seven cluster fields), its higher positional accuracy (5.7\,arcsec point spread function) compared to SCUBA, and the low density of mid-IR sources compared to optical, we take the MIPS imaging to be our primary dataset for counterpart identification.  The radio data, which covers four cluster fields, is used as a secondary indicator.   We classify our identifications as secure (robust 24\,\um\  detection with supporting radio and/or IRAC emission), ambiguous (the 24\,\um\  emission present suffers from high confusion or there are two equally robust identifications), or tentative (there are a few possible candidates, but none are overwhelmingly convincing).  The optical and X-ray data are only used to help estimate redshifts and confirm the astrometry; they do not contribute to the strength of the classification. 

\subsection{MIPS Counterparts}
\label{sec:mipscounters}
The utility of mid-IR follow-up observations in the identification of SMG counterparts has been exploited by many groups (e.g.\ \citealp{Ivison04, Pope06, Biggs10}).  Like the submillimetre regime, mid-IR imaging probes dusty star-formation and AGN activity by sampling thermally emitting dust grains.  Observations of high-redshift clusters at 24\,\um\  also provide a rough estimate of the star formation rate through an extension of an empirical correlation between it and the infrared luminosity of a starburst galaxy \citep{Kennicutt98, Bell03, Calzetti07, Kennicutt09}.  

To locate counterpart 24\,\um\  emission, we employ a signal-to-noise dependent search radius using our positional uncertainty simulations described in \S\ref{sec:positional}, which represent the 95th percentile of SMG offsets.  We limit the maximum radius to 8.0\,arcsec (since this defines the source recovery cut-off in our simulations from \S\ref{sec:monte}) and the minimum radius to 2\,arcsec, where the positional offset tapers off in Fig.~\ref{fig:pos_offset}.  We also introduce an additional 2.0\,arcsec to account for uncertainties in the MIPS source position, resulting in a final search range up to 4.0\,arcsec for the highest S/N SMGs and 10.0\,arcsec around the lowest S/N SMGs.  

We find single MIPS counterparts for 13 ($\sim57$ per cent) of the 23 SMGs with 24\,\um\  coverage.  Three SMGs lack any 24\,\um\  emission, and seven have multiple (two or three) MIPS identifications within the search radius (see \S\ref{sec:multiple}).  The complete counterpart identification results are shown in Table \ref{tab:counters}. 

\subsubsection{Multiple MIPS Counterparts}
\label{sec:multiple}
In the case where multiple  24\,\um\  sources are found within the search radius (35 per cent of the 20 sources with mid-IR emission), we investigate whether one source in particular dominates the infrared flux, and if so, take this to be the more likely counterpart.  \cite{Pope06} conclude that typically only one source significantly contributes to the submillimetre flux, while the secondary sources within the search area are incapable of producing the flux due to faintness.  

For the purposes of this work, we take the brightest MIPS source within the SCUBA beam to be the dominant emitter, but note that this could potentially bias us towards foreground galaxies.  In all cases, the brightest MIPS source is coincident with emission at another wavelength if there exists any within the SCUBA beam, and therefore provides evidence in favour of our counterpart assumption.  There is one source in RCS 1419 that has two MIPS sources of roughly equal fluxes (and no emission at any other wavelength) for which we take the closer of the two (for both 850 and 450\,\um) as the correct identification.

\subsubsection{Assigning MIPS Likelihoods}
\label{sec:likeli}
\begin{figure}
   \centering
      \includegraphics[scale=0.45]{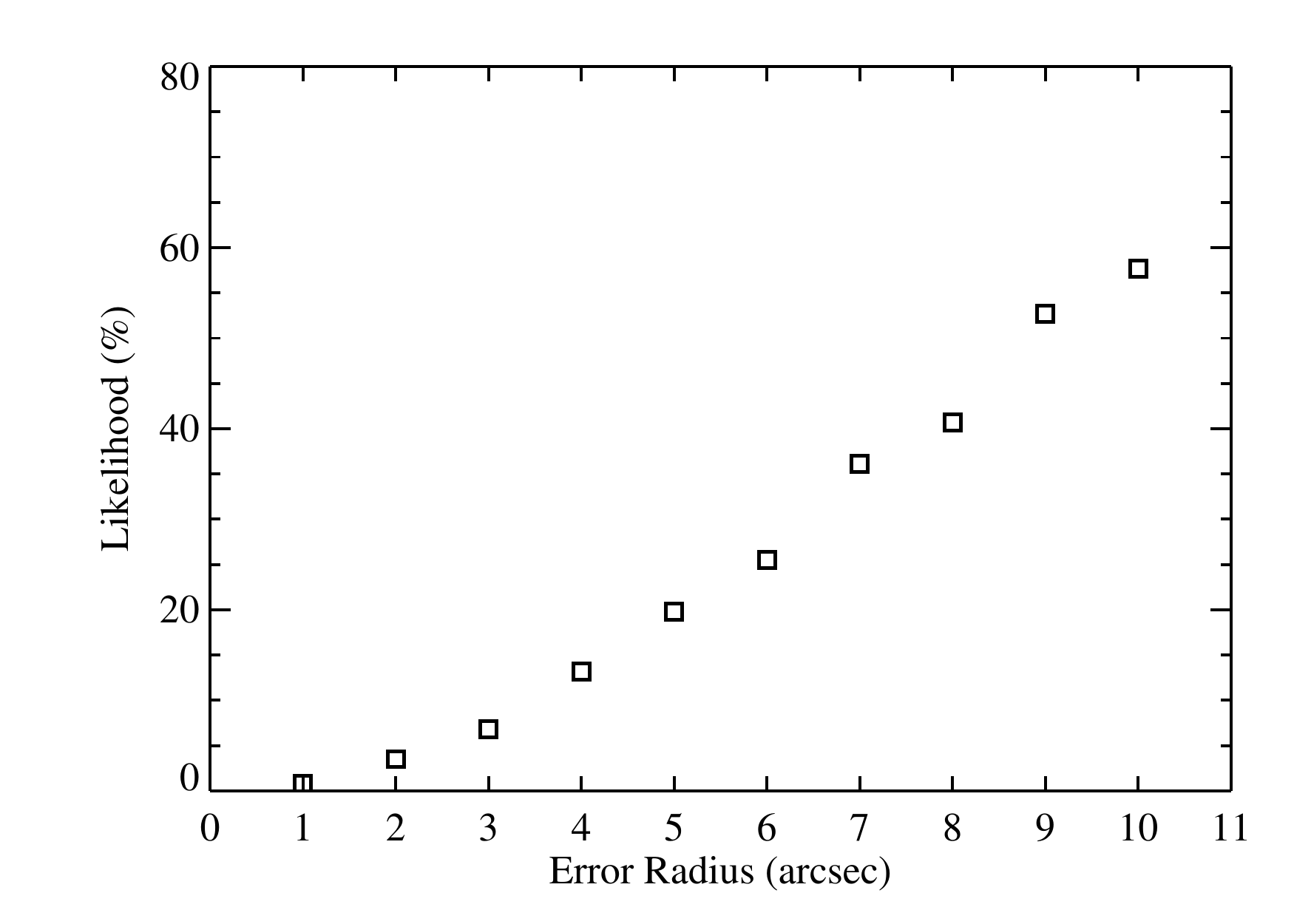}
   \caption{A plot of the likelihood of finding one or more source within an aperture of the given radius.  We linearly interpolate between these values to assign likelihoods to MIPS counterparts based on its offset from the SMG.}
   \label{fig:aperture}
\end{figure}

To quantify the probability that the MIPS detection is associated with the 850\,\um\  emission, we perform Monte Carlo simulations, placing apertures of various radii on each 24\,\um\  cluster field and counting the number of sources detected within that area.  The importance of using our own data for the simulations, rather than a pure statistical method such as the $P$-statistic (e.g.\ \citealp{Downes86}) which requires an estimate of the source density, derives from the prior expectation of an overdensity in our cluster fields.  We drop 1000 apertures at each radius from 1.0 to 10.0\,arcsec in steps of 1.0\,arcsec on the MIPS images (see Fig.~\ref{fig:aperture}).  A likelihood for each radius is assigned based on the number of times we detect one or more MIPS sources within that aperture.  Using the positional offset between the SCUBA source and the most robust MIPS counterpart (in the case of multiple 24\,\um\  sources within the search radius), we interpolate between the adjacent aperture radii (see Fig.~\ref{fig:aperture}) to determine an individual likelihood that each MIPS counterpart is not just a chance alignment. 

\begin{figure}
   \centering
   \includegraphics[scale=0.45]{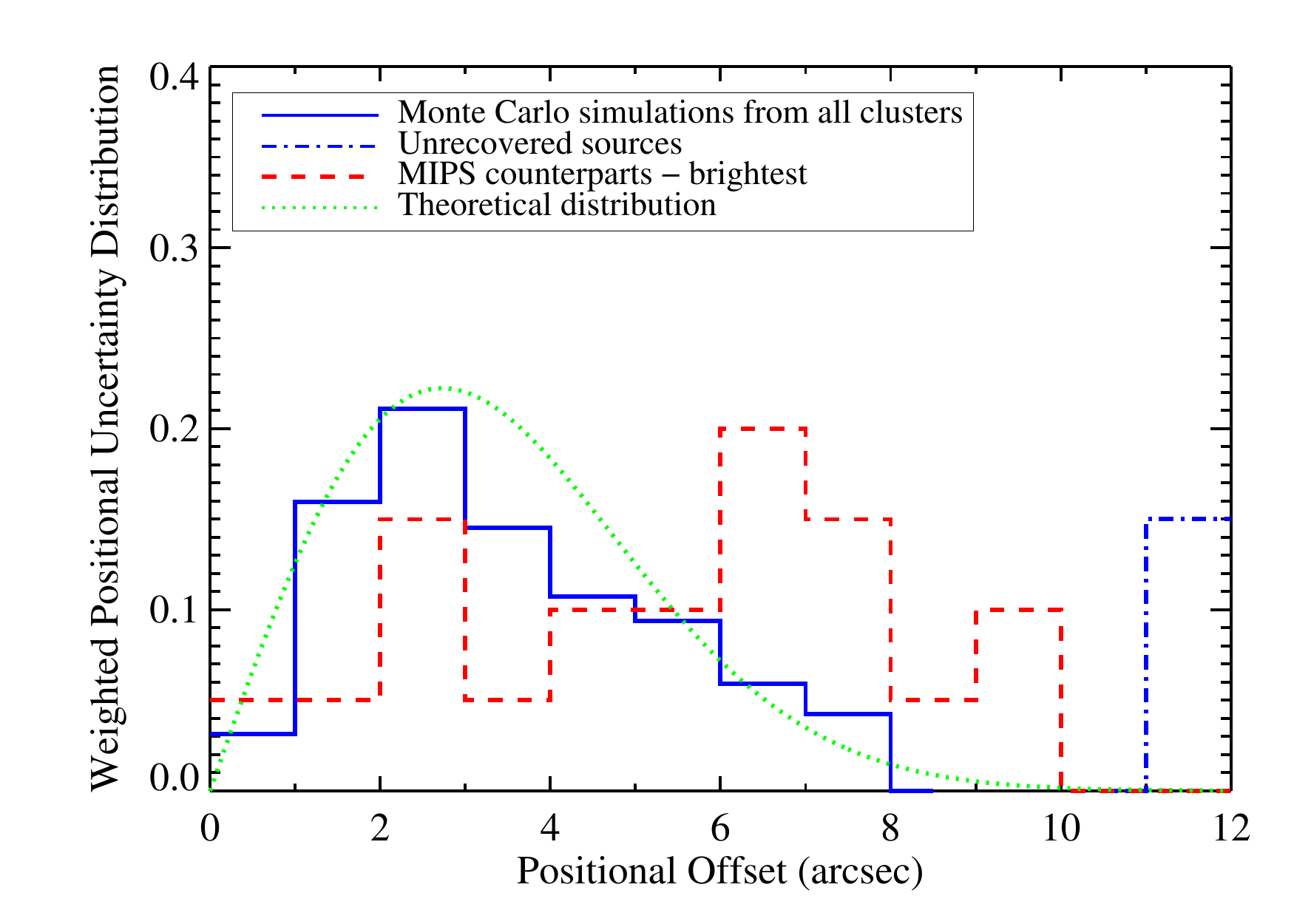} 
 \caption{The distribution of positional offsets for the brightest MIPS counterparts in our sample (dashed histogram) compared to Monte Carlo simulations of the SMG positional uncertainties (solid histogram).  We also overplot the theoretical distribution of radial offsets for flux-boosted SMGs (dotted curve), as defined in \citealp{Ivison07}, using the mean S/N of the SMG sample.  The dash-dotted histogram at 11--12\,arcsec denotes the fraction of unrecovered submillimetre sources (recovered at offsets $>8$ arcsec) in our simulations, meaning 15 per cent of our sources should not have counterpart identifications.  Note that this distribution is for all 850\,\um\  sources with S/N $\geq$ 3.0.}
   \label{fig:mips_offsets}
\end{figure}

To investigate the importance of random associations that contaminate our identifications, we compare the positional offsets between secure counterparts and their associated SMG to our Monte Carlo simulations of submillimetre positional uncertainties (\S\ref{sec:positional}) combined for all clusters (see Fig.~\ref{fig:mips_offsets}).  The simulations have been weighted proportionally by S/N based on the percentage of SMGs with MIPS counterparts in each S/N bin (with 1$\,\sigma$ width).  This places more emphasis on lower S/N (i.e.\ 3--4$\,\sigma$) positional uncertainty simulations as 60 per cent of our submillimetre detections fall within this range.  We also include the theoretical distribution of radial offsets for flux-boosted SMGs from \cite{Ivison07}.  Assuming a Gaussian distribution of errors in right ascension (RA) and declination, this is given by $r{\rm e}^{-r^2/2\sigma^2}$, where $\sigma=0.6 \theta[$S/N$^2_{\rm{app}}-(2\beta+4)]^{-1/2}$ and the S/N$_{\rm{app}}$ has not been corrected for flux boosting.

The simulated and theoretical distributions agree quite well and both peak around 2.5\,arcsec, while our broader distribution of MIPS counterpart offsets peaks further out at $\sim6$\,arcsec.  This suggests that some of our counterparts are falsely identified, primarily those with offsets beyond 6\,arcsec, where 50 per cent of the histogram of MIPS counterparts lies above the simulated histogram (including unrecovered sources).  Of the ten MIPS sources in this high offset region, five are securely identified (see \S\ref{sec:counters} for a definition) as the correct SMG counterpart, while the remaining 50 per cent are only tentative or ambiguous counterparts (see Appendix A for details on classifications of counterparts).  Our classifications are therefore consistent with the distributions at large positional offsets, with tentative counterparts more likely being wrongly identified.  Alternatively, these sources with large MIPS offsets could also represent blends of multiple submillimetre sources, or spurious sources (see \S\ref{sec:false}).  In fact, four of the sources with MIPS offsets above 6\,arcsec have S/N$<$3.5, increasing the likelihood that they are either spurious or blended sources.  Moreover, the fraction of unrecovered submillimetre sources in our simulations (dot-dashed histogram) provides a handle on how many real sources should remain unidentified, roughly 15 per cent, which is congruous with our findings of three SMGs that lack MIPS counterpart emission within the search radius.

\subsection{Radio Counterparts}
\label{sec:radiocounters}
The powerful benefits associated with complementary radio data at 1.4\,GHz (21\,cm) with the VLA are primarily two-fold.  First, the sub-arcsecond spatial resolution attainable with radio interferometry provides improved astrometric precision compared to SCUBA and MIPS positions.  Second, radio sources are fairly sparse in the sky \citep{Richards00}, making coincident detections of 850\,\um\  and 1.4 GHz emission highly unlikely.  Furthermore, radio emission is sensitive to synchrotron radiation from relativistic electrons associated with supernovae, and thus traces recent star formation, resulting in a correlation between radio and submillimetre emission.  Any radio emission detected within the 850\,\um\  search area, therefore, offers likely counterpart identifications to the SCUBA sources \citep{Ivison00}.  In principle, the simulations for the MIPS likelihoods (\S\ref{sec:likeli}) could be applied to the radio maps as well, but because the number of radio sources is even sparser, we automatically assign the 1.4 GHz emission as the correct identification.  Note that one exception to this occurs in source SMM 2319.3, in which two radio galaxies are found; further discussion is provided in \S\ref{sec:gglensing}.    

Radio counterparts are searched for within a maximum of 8\,arcsec radius around the SMG, depending on the S/N of the source.  No additional margin is given due to the high astrometric precision of the radio data.  For the four clusters (17 SMGs in total) with complementary 1.4\,GHz imaging, we detect six radio sources within the search radii of five SMGs, although one source is suspected to be unassociated with the SMG (discussed in detail in \S\ref{sec:members})  In all radio-detected cases, a secure 24\,\um\  counterpart has already been identified, and the MIPS and radio positions are within 3\,arcsec of each other.

\subsection{IRAC Counterparts}
\label{sec:iraccounters}
Many groups (e.g.\ \citealp{Smail99, Ivison02, Webb03VII, Ashby06, Pope06, Yun08, Hainline09, Wardlow10}) have investigated the efficacy of infrared colours of SMGs as a rough proxy for redshift as well as a counterpart selection technique.  As these colour criteria can be biased towards higher redshift SMGs and in some cases, are perhaps too general (as found by \citealp{Hainline09}), we employ them merely as a means to secure one identification over another in the case of multiple MIPS detections within the search radius.  

\begin{figure*}
\centering
\includegraphics[scale=0.49]{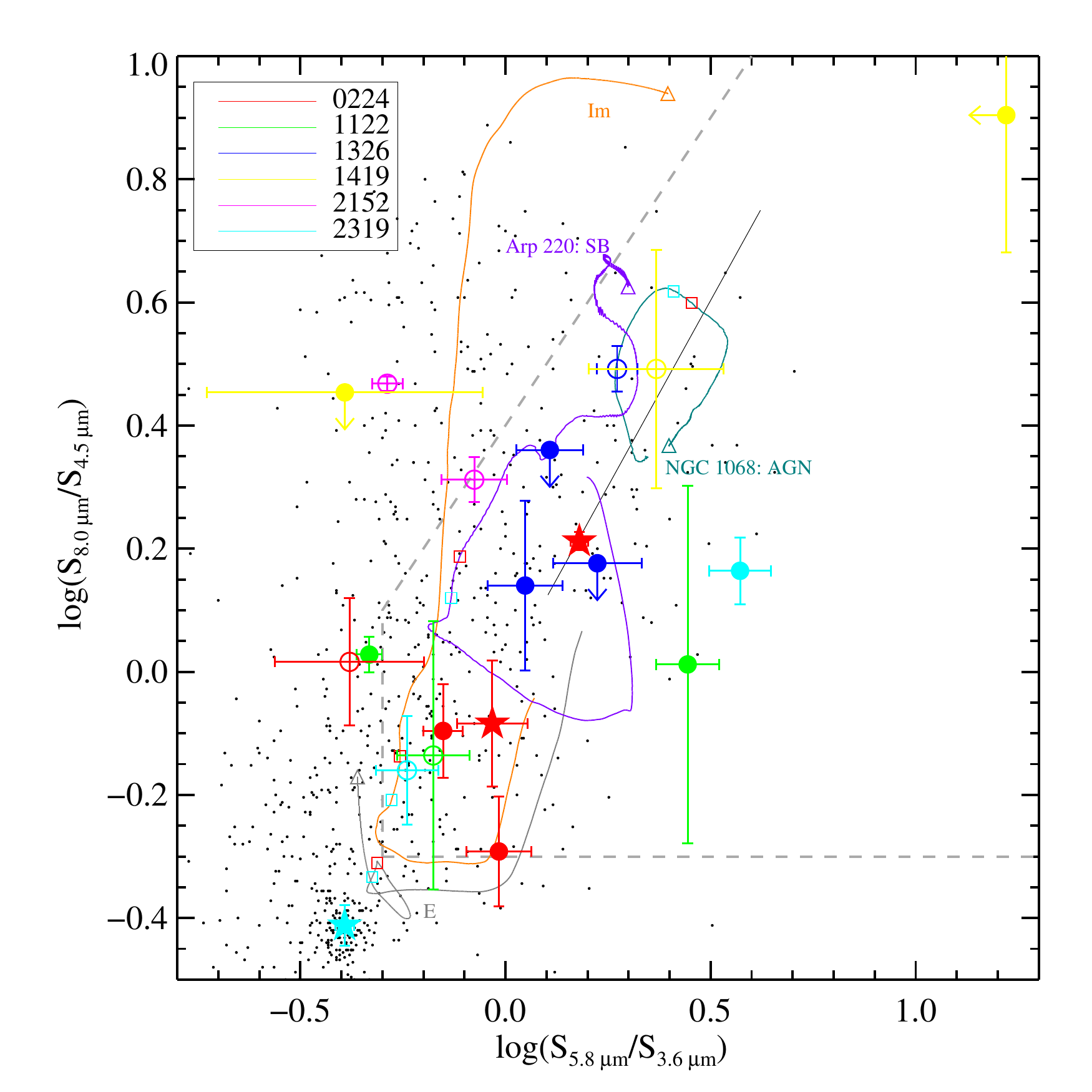}
\includegraphics[scale=0.49]{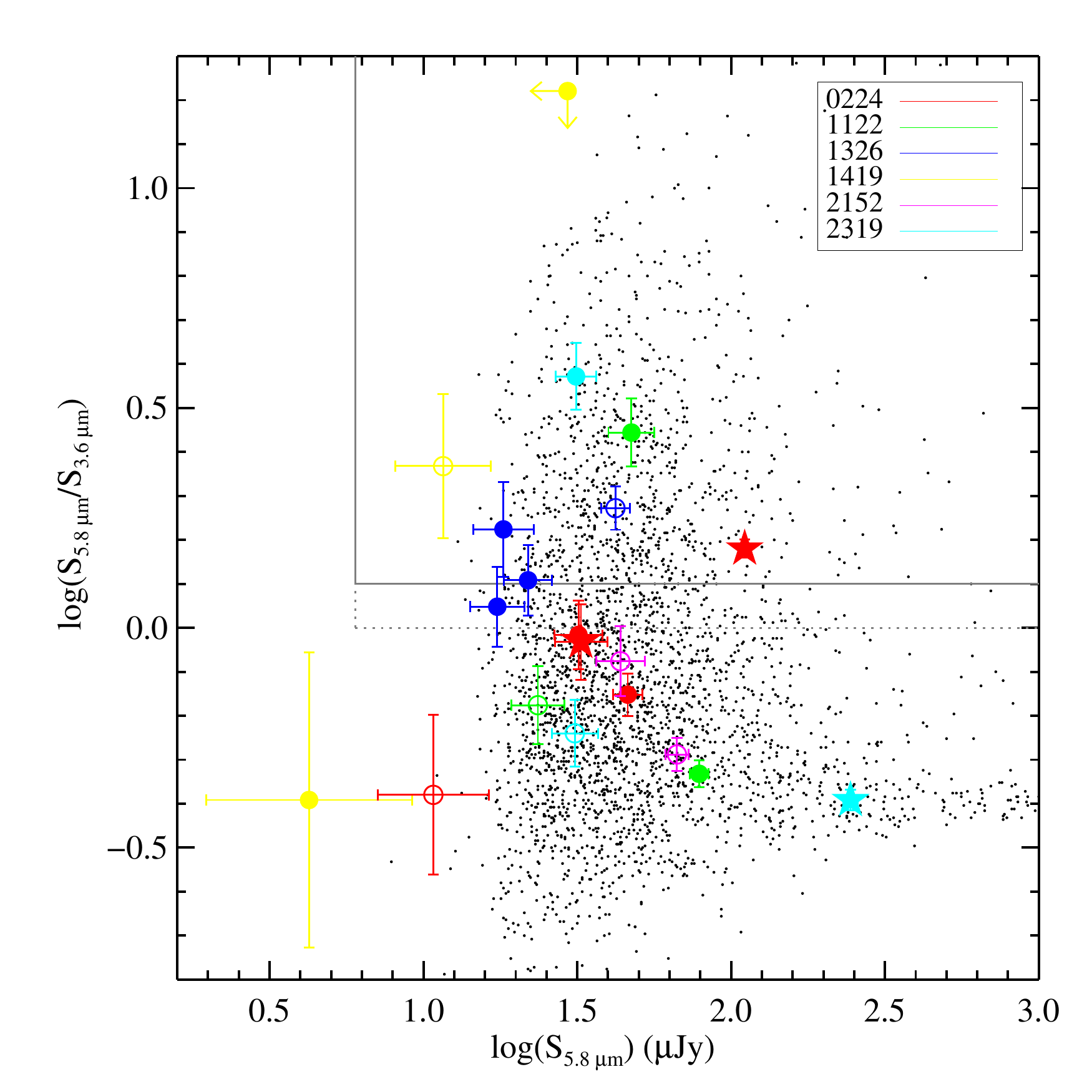}
\caption{[Left]$S_{8.0\,\mu m}/S_{4.5\,\mu m}$ versus $S_{5.8\,\mu m}/S_{3.6\,\mu m}$ colour-colour diagram for both secure (large filled circles) and tentative/ambiguous (large empty circles) IRAC counterparts to the SMGs.  The filled stars denote likely cluster members (see \S\ref{sec:members}).  Small black dots correspond to field galaxies in the six clusters detected above 5$\,\sigma$ in all four IRAC channels.  SED tracks for representative galaxies are overlaid from $z=0$ to $z=3.1$, with the triangle indicating low redshift.  Also shown is a pure power-law galaxy, $S\propto\nu^{\alpha}$, with a spectral slope ranging from $-0.5$ to $-3.0$.  The cluster redshifts of RCS 0224 (0.773) and 2319 (0.9024) are displayed along the tracks as red and cyan squares, respectively. The dotted gray line represents the boundary for separation of $z>1$ SMG counterparts from field galaxies as suggested by \citealp{Yun08}. [Right] A colour-magnitude diagram for the secure and tentative IRAC counterparts.  The symbols are equivalent to those in the colour-colour diagram to the left.  The solid black line denotes the region proposed by \citealp{Pope06} for SMG counterpart selection, while the dotted gray line traces the more relaxed criteria given by \citealp{Hainline09}. } 
\label{fig:colmag}
\end{figure*}

Previous studies (e.g.\ \citealp{Yun08}) have used a $S_{8.0\,\mu m}/S_{4.5\,\mu m}$ versus $S_{5.8\,\mu m}/S_{3.6\,\mu m}$ colour-colour diagram (see Fig.~\ref{fig:colmag}) to identify mid-IR counterparts to SMGs based on the claim that their colours occupy a different region of the plot than those from stellar photospheric emission (the cloud in the blue region) and of low-$z$ late-type galaxies (the vertical branch emerging from the cloud).  These colour criteria derive from the region populated by AGN identified by \cite{Lacy04}.  In Fig.~\ref{fig:colmag} we show the infrared colours for both secure and tentative IRAC counterparts to the SMGs against a backdrop of galaxies in the six cluster fields.  Overlaid are SED tracks for representative galaxies, including the ULIRG Arp 220, an irregular starburst galaxy, an elliptical galaxy, NGC 1068 (an AGN), and a pure power-law galaxy.  Of our secure counterparts, 11 of 12 ($92$ per cent) fall, within the errors, inside the marked SMG domain.  The remaining secure counterpart falls along the low-$z$ late-type sequence and has only a single MIPS detection in the entire search radius; it is therefore still identified as a secure detection.  Thus, the IRAC observations support all of our previous conclusions based solely on the MIPS and radio data.

We also plot a colour-magnitude diagram (Fig.~\ref{fig:colmag}) for the same counterparts described above using the criteria suggested by \cite{Pope06} and \cite{Hainline09}.  This domain is used to select $z>1.5$ SMGs with a $90$ per cent success rate and $6$ per cent contamination rate by $z<1.5$ galaxies (based on the more relaxed \citealp{Hainline09} criteria).  This diagram also serves as a selection cut for SMG counterparts as there is a low probability of randomly displaying these colours.  Regarding our secure counterparts, $75$ per cent are within error ($1\,\sigma$ error bars on infrared colours) of the $z>1.5$ region, suggesting the majority of the SMGs in the sample lie at higher redshifts and are not cluster members.

\subsection{X-ray Counterparts}
\label{sec:xraycounters}
The X-ray regime provides a window into obscured AGN and star formation; many groups (e.g.\ \citealp{Almaini99, Fabian00}) have exploited hard (i.e.\ 2--8\,keV) X-ray observations to locate any AGN associated with submillimetre emission which might contribute to the extreme luminosities produced by these sources.

X-ray counterparts must adhere to the same distance criteria as radio counterparts---within a maximum of 8\,arcsec of the SMG (depending on the S/N) and 3\,arcsec of a secure MIPS counterpart.  Only five X-ray sources are found within the search radii, leaving 16 SMGs without any X-ray association.

\begin{landscape}
\begin{table}
\begin{center}
\begin{scriptsize}
\caption{A summary of the multi-wavelength counterpart emission for each SMG.}
\label{tab:counters}
\begin{tabular}{cllccccccccccccccccc}
\hline 
\multicolumn{1}{c}{} &
\multicolumn{1}{c}{} &
\multicolumn{1}{c}{} &
\multicolumn{3}{c}{Submillimetre} &
\multicolumn{3}{c}{MIPS} &
\multicolumn{4}{c}{IRAC} &
\multicolumn{2}{c}{Radio} &
\multicolumn{1}{c}{X-ray} &
\multicolumn{1}{c}{Optical} \\
\multicolumn{1}{c}{SMM} &
\multicolumn{1}{c}{RA} &
\multicolumn{1}{c}{Dec} &
\multicolumn{1}{c}{850\,\um} &
\multicolumn{2}{c}{450\,\um} &
\multicolumn{3}{c}{24\,\um} &
\multicolumn{4}{c}{3.6, 4.5, 5.8, 8.0\,\um\ } &
\multicolumn{2}{c}{1.4 GHz} &
\multicolumn{1}{c}{0.3--0.7\,keV} &
\multicolumn{1}{c}{$z$, $R_c$, $V$, $B$} \\
\multicolumn{1}{c}{ID} &
\multicolumn{1}{c}{(J2000)} &
\multicolumn{1}{c}{(J2000)} &
\multicolumn{1}{c}{S/N} &
\multicolumn{1}{c}{S/N} &
\multicolumn{1}{c}{Offset$^{a}$} &
\multicolumn{1}{c}{S$_{24\, \mu\text{m}}$$^{b}$} &
\multicolumn{1}{c}{Offset$^{a}$} &
\multicolumn{1}{c}{$\%$$^{c}$} &
\multicolumn{1}{c}{S$_{3.6\, \mu\text{m}}$$^{b}$} &
\multicolumn{1}{c}{S$_{4.5\, \mu\text{m}}$$^{b}$} &
\multicolumn{1}{c}{S$_{5.8\, \mu\text{m}}$$^{b}$} &
\multicolumn{1}{c}{S$_{8.0\, \mu\text{m}}$$^{b}$} &
\multicolumn{1}{c}{S$_{1.4\,\text{GHz}}$$^{b}$} &
\multicolumn{1}{c}{Offset$^{d}$} &
\multicolumn{1}{c}{S/N} &
\multicolumn{1}{c}{Redshift$^{e}$} \\
\hline
\textcolor[gray]{0.5}{0224.1}	&	02 24 34.19			&	$-$00 03 25	&	6.1	&	3.7	&	4.1	&	x			&	x	&	x	&	x	&	x	&	x	&	x	&	x	&	x	&	x	&	x	\\
\bf{*0224.2*} 	&	02 24 33.30$^{\ast}$ 	&	$-$00 04 06	&	4.3	&	x	&	x	&	617$\pm$24	&	9.3	&	46	&	73.1$\pm$0.9	&	115.9$\pm$1.2	&	110.7$\pm$5.2	&	188.8$\pm$6.1	&	68$\pm$29	&	0.5	&	5.5	&	0.677$\pm$0.074	\\
\bf {*0224.3*} 	&	02 24 28.67$^{\dagger}$	&	$-$00 03 18	&	3.5	&	x	&	x	&	404$\pm$41	&	9.1	&	47	&	35.0$\pm$0.8	&	30.2$\pm$1.0	&	32.5$\pm$6.4	&	24.9$\pm$5.8	&	x	&	x	&	5.6	&	0.688$\pm$0.071	\\
\bf{0224.4}&	02 24 29.97$^{\ast}$ 	&	$-$00 03 04	&	3.9	&	3.3	&	5.0	&	412$\pm$25	&	5.6	&	77	&	33.1$\pm$0.9	&	39.1$\pm$1.2	&	31.9$\pm$5.7	&	20.0$\pm$4.1	&	66$\pm$21	&	0.4	&	x	&	x	\\
0224.5	&	02 24 33.18$^{\dagger}$	&	$-$00 03 42	&	3.4	&	2.8	&	5.1	&	225$\pm$23	&	7.7	&	61	&	25.8$\pm$1.0	&	20.7$\pm$1.1	&	10.8$\pm$4.5	&	21.5$\pm$5.0	&	x	&	x	&	x	&	0.943$\pm$0.074	\\
\bf{0224.6}&	02 24 32.87$^{\ast}$ 	&	$-$00 02 31	&	3.7	&	3.1	&	11.4	&	313$\pm$24	&	1.6	&	98	&	65.5$\pm$1.2	&	47.4$\pm$1.3	&	46.1$\pm$5.0	&	38.0$\pm$6.6	&	90$\pm$28	&	0.8	&	x	&	\emph{1.0995	}	\\
\hline
1122.1	&	11 22 29.57$^{\dagger\dagger}$	&	$+$24 21 20	&	3.2	&	x	&	x	&	\emph{326$\pm$24}	&	3.5	&	92	&	35.3$\pm$1.4	&	33.4$\pm$1.6	&	23.6$\pm$4.7	&	24.4$\pm$12.2	&	---	&	---	&	---	&	x	\\
\bf{1122.2}&	11 22 31.74$^{\dagger}$	&	$+$24 22 09	&	4.3	&	x	&	x	&	272$\pm$24	&	0.7	&	98	&	17.1$\pm$0.8	&	21.9$\pm$1.2	&	47.4$\pm$8.2	&	22.5$\pm$15.0	&	---	&	---	&	---	&	1.073$\pm$0.803	\\
\bf{1122.3}&	11 22 27.47$^{\dagger}$	&	$+$24 22 46	&	3.4	&	5.6	&	3.9	&	1096$\pm$23	&	7.5	&	69	&	169.1$\pm$1.0	&	129.4$\pm$1.3	&	78.7$\pm$5.5	&	138.1$\pm$9.2	&	---	&	---	&	---	&	99$\pm$99	\\
\hline
\bf{1326.1}&	13 26 28.08$^{\dagger}$	&	$+$29 03 09	&	6.5	&	6.2	&	4.1	&	343$\pm$21	&	4.8	&	83	&	15.6$\pm$0.7	&	19.4$\pm$1.0	&	17.4$\pm$3.5	&	26.8$\pm$8.4	&	---	&	---	&	x	&	---	\\
\bf{1326.2}&	13 26 33.43$^{\dagger}$	&	$+$29 02 15	&	3.6	&	x	&	x	&	249$\pm$22	&	6.9	&	67	&	10.8$\pm$1.1	&	14.6$\pm$1.5	&	18.2$\pm$4.1	&	$<21.9$		&	---	&	---	&	5.6	&	---	\\
\bf{1326.3}&	13 26 33.60$^{\dagger}$	&	$+$29 03 25	&	3.4	&	x	&	x	&	110$\pm$20	&	2.6	&	95	&	17.1$\pm$0.9	&	9.6$\pm$1.3	&	21.9$\pm$3.9	&	$<21.9$		&	---	&	---	&	x	&	---	\\
1326.4 &	13 26 25.80$^{\dagger\dagger}$	&	$+$29 02 43	&	3.2	&	x	&	x	&	\emph{174$\pm$21}	&	7.3	&	64	&	22.5$\pm$0.9	&	33.1$\pm$1.2	&	42.1$\pm$4.5	&	102.8$\pm$7.9	&	---	&	---	&	5.7	&	---	\\
\hline
\bf{1419.1}&	14 19 15.33$^{\dagger}$	&	$+$53 24 51	&	5.2	&	x	&	x	&	63$\pm$20	&	4.8	&	83	&	1.8$\pm$0.6	&	1.9$\pm$0.7	&	$<29.4$		&	15.0$\pm$5.6	&	x	&	x	&	x	&	---	\\
\textcolor[gray]{0.5}{1419.2}	&	14 19 08.76			&	$+$53 24 56	&	3.0	&	x	&	x	&	x			&	x	&	x	&	x	&	x	&	x	&	x	&	x	&	x	&	x	&	---	\\
1419.3	&	14 19 19.48$^{\dagger}$	&	$+$53 26 31	&	4.5	&	2.7	&	7.1	&	54$\pm$19	&	6.6	&	69	&	5.0$\pm$0.6	&	4.7$\pm$0.9	&	11.6$\pm$4.1	&	14.6$\pm$5.8	&	x	&	x	&	x	&	---	\\
\bf{1419.4}&	14 19 04.99$^{\dagger}$	&	$+$53 26 25	&	3.8	&	x	&	x	&	111$\pm$22	&	6.4	&	71	&	10.5$\pm$0.9	&	7.5$\pm$1.1	&	4.3$\pm$3.3	&	$<21.4$		&	x	&	x	&	x	&	---	\\
\textcolor[gray]{0.5}{1419.5}	&	14 19 14.69			&	$+$53 25 21	&	5.9	&	x	&	x	&	x			&	x	&	x	&	x	&	x	&	x	&	x	&	x	&	x	&	x	&	---	\\
\hline
2152.1	&	21 52 46.35$^{\dagger}$	&	$-$06 08 55	&	3.5	&	x	&	x	&	\emph{477$\pm$25}	&	6.9	&	70	&	\emph{129.4$\pm$1.4}	&	\emph{100.9$\pm$1.6}	&	\emph{66.7$\pm$5.7}	&	\emph{296.5$\pm$6.2}	&	---	&	---	&	---	&	\emph{ND	}\\
2152.2	&	21 52 44.19$^{\ddagger}$	&	$-$06 10 15	&	3.0	&	x	&	x	&	\emph{382$\pm$23}	&	8.5	&	57	&	\emph{52.0$\pm$1.1}	&	\emph{57.0$\pm$1.9}	&	\emph{43.7$\pm$7.9}	&	\emph{117.0$\pm$9.0}	&	---	&	---	&	---	&	0.827$\pm$0.468	\\
\hline
2318.1	&	23 18 27.89	&	$+$00 34 55	&	17.0	&	7.1	&	2.2	&	---	&	---	&	---	&	---	&	---	&	---	&	---	&	x	&	x	&	x	&	---	\\
2318.2	&	23 18 28.03	&	$+$00 34 18	&	5.6	&	3.0	&	2.2	&	---	&	---	&	---	&	---	&	---	&	---	&	---	&	x	&	x	&	x	&	---	\\
2318.3	&	23 18 32.63	&	$+$00 34 22	&	4.0	&	3.3	&	12.0	&	---	&	---	&	---	&	---	&	---	&	---	&	---	&	x	&	x	&	x	&	---	\\
\hline
2319.1 &	23 19 49.70$^{\ast}$ 	&	$+$00 37 56	&	5.5	&	4.4	&	2.8	&	628$\pm$23	&	2.1	&	97	&	54.0$\pm$1.4	&	59.2$\pm$2.1	&	31.1$\pm$5.4	&	40.9$\pm$8.2	&	80$\pm$27	&	1.5	&	x	&	\emph{ND}\\
\bf{2319.2}&	23 19 56.73$^{\dagger}$	&	$+$00 37 19	&	7.5	&	4.4	&	3.6	&	372$\pm$21	&	2.5	&	95	&	8.4$\pm$0.7	&	31.3$\pm$1.2	&	31.3$\pm$4.7	&	45.7$\pm$5.4	&	x	&	x	&	x	&	\emph{ND}	\\
*2319.3*	&	23 19 53.42$^{\ast}$ 	&	$+$00 38 13	&	4.7	&	x	&	x	&	71$\pm$22	&	5.5	&	76	&	602.7$\pm$3.0	&	280.6$\pm$2.4	&	244.4$\pm$13.0	&	108.7$\pm$8.2	&	1159$\pm$32	&	1.9	&	3.9	&	\emph{0.90126}	\\
\hline
\end{tabular}
\begin{tablenotes}[normal]
\textit{Notes.} Radio and MIPS positions, when available, are primarily used for the SMG astrometry (instead of the 850\,\um\  position) and denoted by $\ast$ and $\dagger$, respectively, in the RA column.   In a few cases, the MIPS source is highly blended and either the IRAC ($\dagger\dagger$) or optical ($\ddagger$) position is given instead.  SMGs lacking coverage at certain wavelengths is marked by ---, while x signifies that no emission exists within the search radius despite coverage in that band.  SMM IDs shown in bold, normal, or gray text signify secure, tentative/ambigiguous, or non-identifications, respectively.  The three likely cluster members are marked by asterisks on either side of their SMM ID.    \\
$^{a}$The positional offset in arcsec with respect to the 850\,\um\  position.\\
$^{b}$The flux densities are given in \uJy\,beam$^{-1}$ with $1\,\sigma$ errors.  Flux limits on the IRAC detections are determined from the $3\,\sigma$ depths for each channel within that cluster field.  MIPS and IRAC fluxes written in italics are blended.  \\
$^{c}$The likelihood that the MIPS source is the correct counterpart identification, based solely on the offset between the MIPS and SCUBA positions.\\
$^{d}$The positional offset in arcseconds between the MIPS and radio positions.\\
$^{e}$Italicized redshifts are spectroscopic, and all others are photometric based on the multi-band optical imaging.  A redshift of $99\pm99$ indicates the redshift was not constrained with the four-band photometry, while \emph{ND} indicates that the source was not detected in the optical filters, but faintly revealed in \textit{HST} imaging. 
\end{tablenotes}
\end{scriptsize}
\end{center}
\end{table}
\end{landscape}

\section{Discussion}
\label{sec:disc}

\subsection{Tentative Cluster Members}
\label{sec:tent_mems}
After careful inspection of all counterpart identifications, we infer that four (SMMs 0224.2, 0224.3, 2152.2, and 2319.3) of the 26 SMGs are potential cluster members based on their photometric or spectroscopic redshift.  We note that this conclusion is biased towards cluster fields with radio and/or multi-band optical data, as these wavelength regimes provide positional accuracy and redshift estimates.  With more consistent wavelength coverage for all clusters, the fraction of possible members could be higher.  

For the remainder of this paper, however, we choose to limit our discussion to the three most likely cluster members and eliminate one candidate (SMM 2152.2) based on its low ($<3.5$) S/N at 850\,\um.  As determined in \S\ref{sec:false}, a detection threshold of 3.0\,$\sigma$ carries a high false-detection rate and therefore these sources are not reliable as individual objects. Three remaining candidates (SMMs 0224.2, 0224.3, and 2319.3) form the basis of our cluster member sample and are discussed in detail here.  Multi-wavelength postage stamps of the counterpart emission for these three SMGs are shown in Fig.~\ref{fig:stamps}.  A discussion of counterparts for the remaining SMGs is deferred to the appendices.

\subsubsection{Notes on Individual Cluster Members}
\label{sec:members}
\begin{figure*}
   \centering
      \subfigure{\includegraphics[scale=0.42]{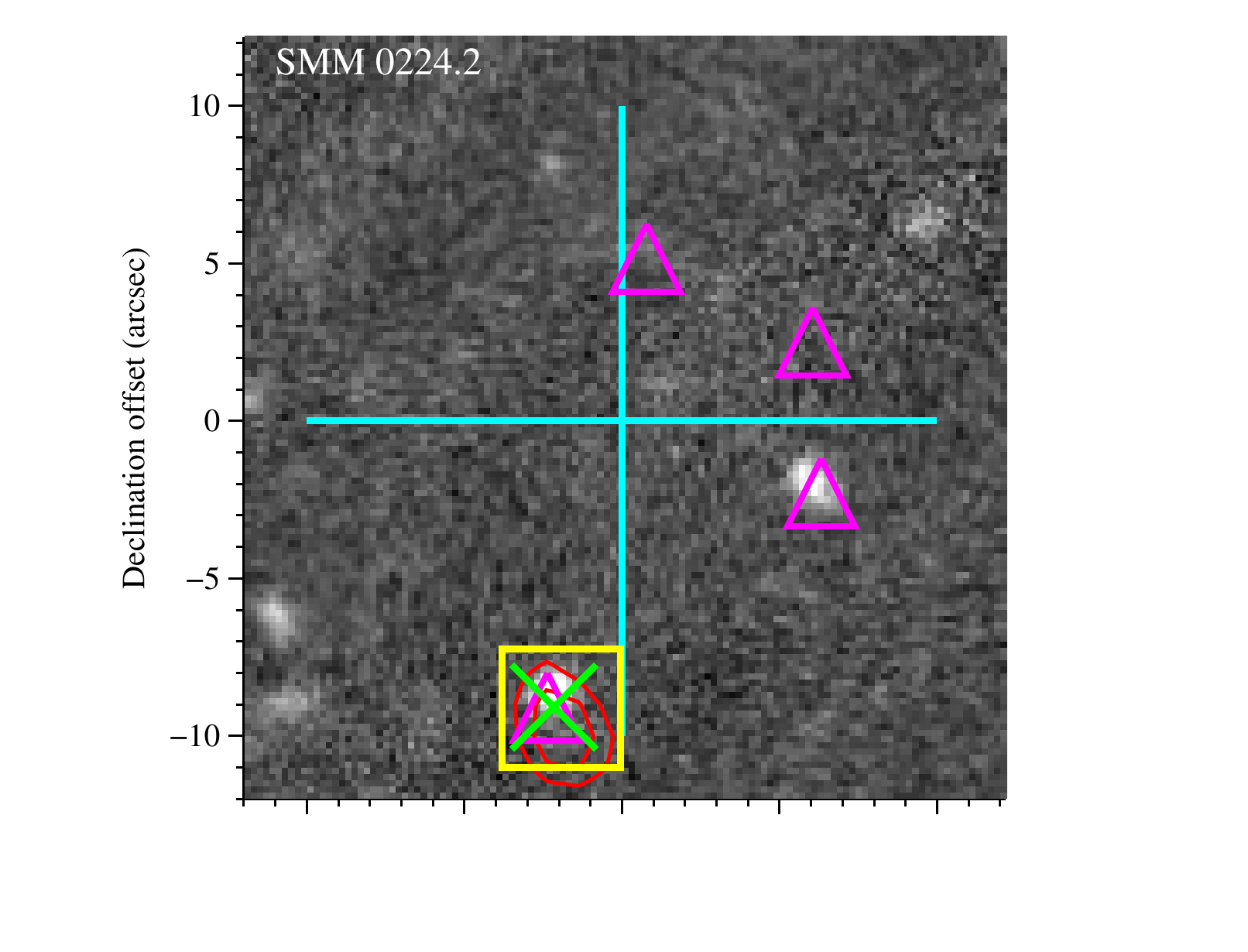}}
   		\hspace{-27mm}
   \subfigure{\includegraphics[scale=0.42]{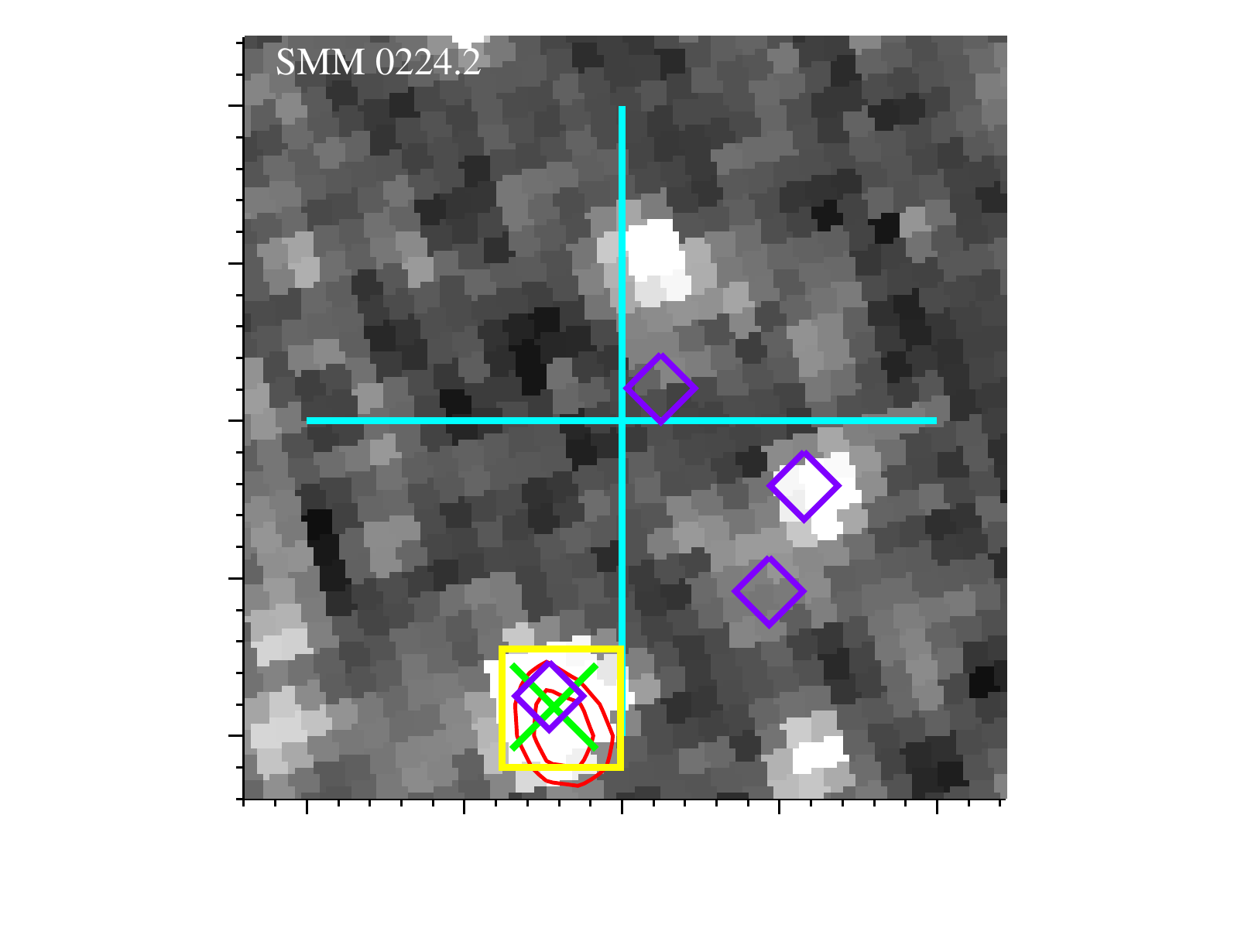}}
   		\hspace{-27mm}
  	   	 \vspace{-12mm}
   \subfigure{\includegraphics[scale=0.42]{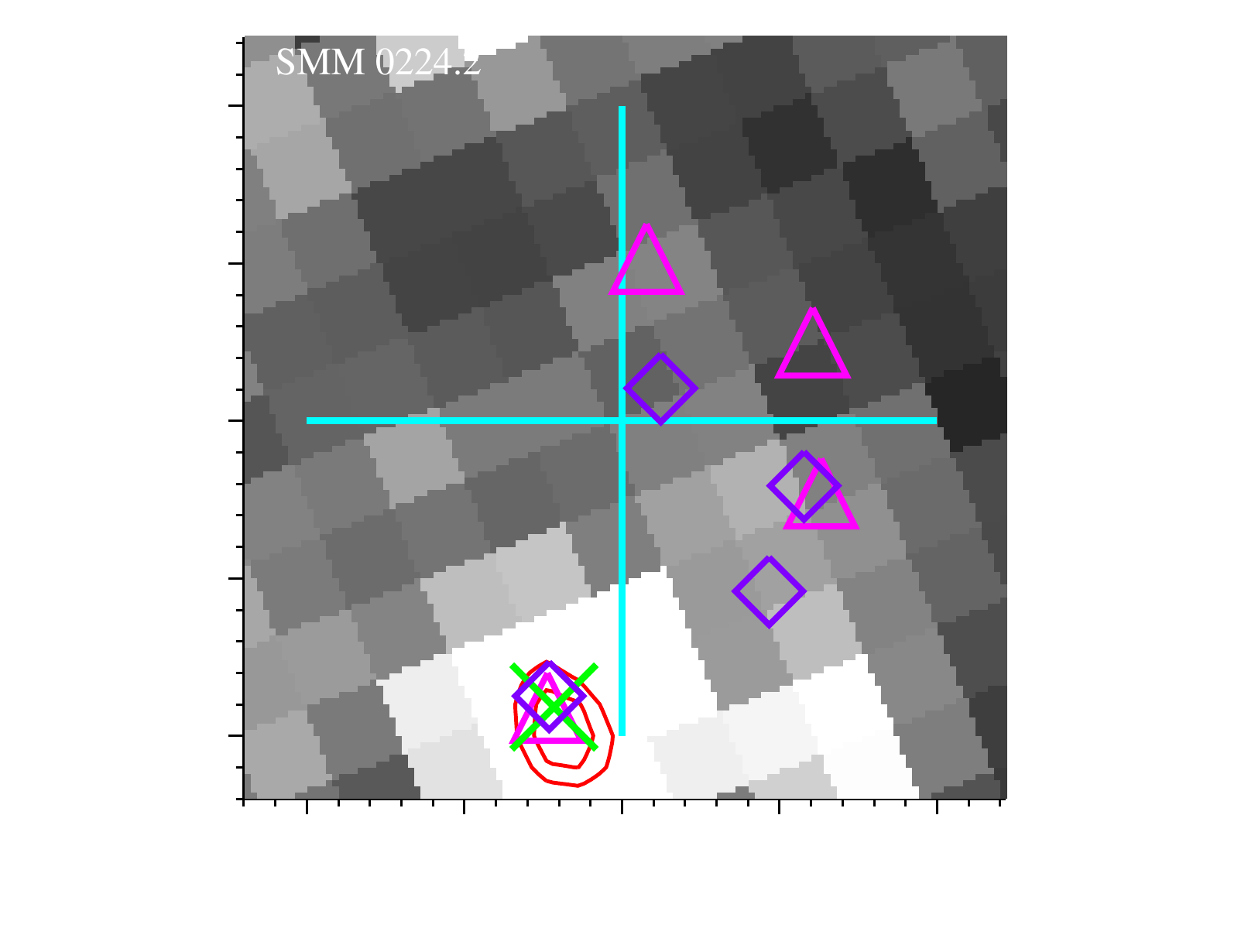}}
     \subfigure{\includegraphics[scale=0.42]{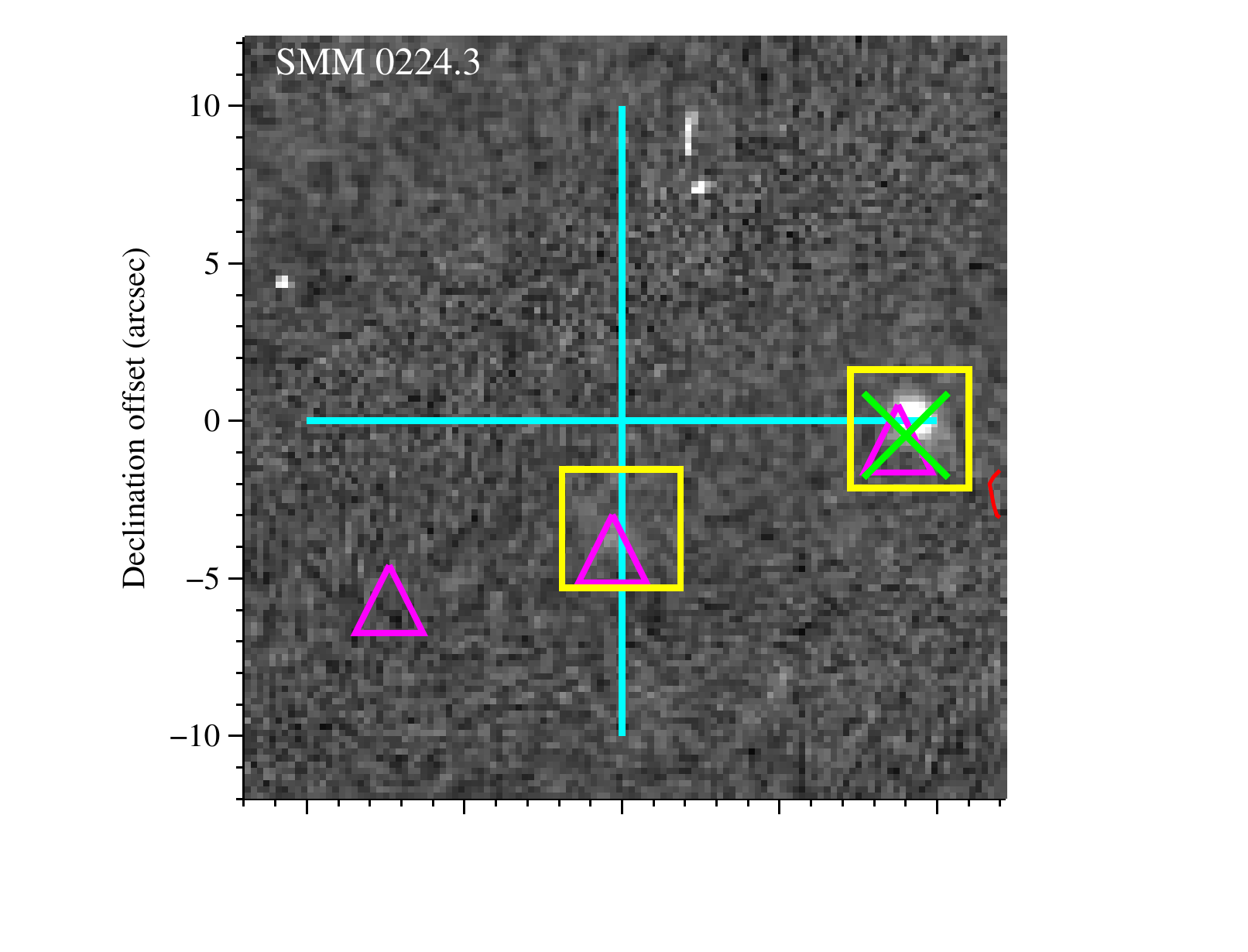}}
        		\hspace{-27mm}
   \subfigure{\includegraphics[scale=0.42]{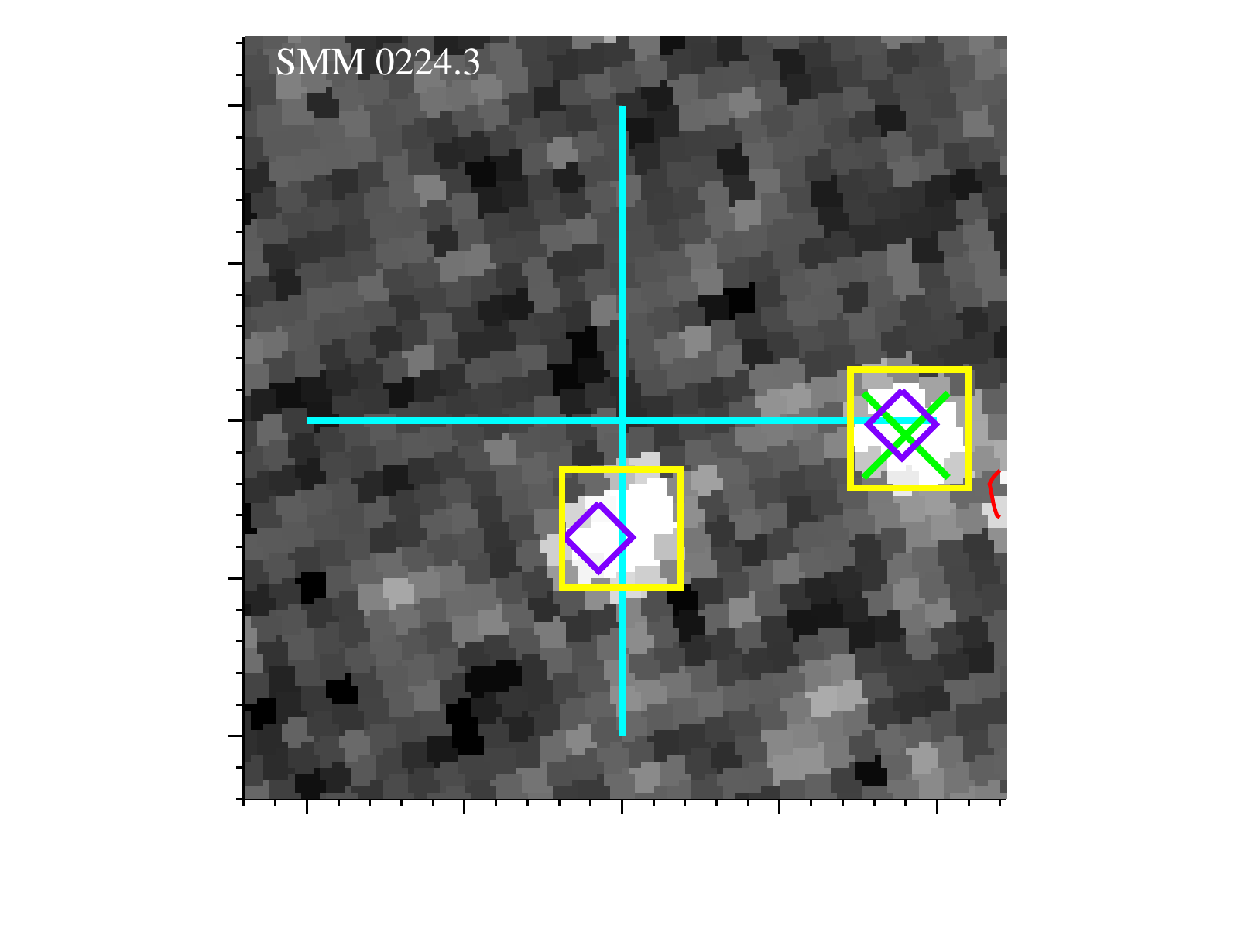}}
      		\hspace{-27mm}
  	   	 \vspace{-12mm}
   \subfigure{\includegraphics[scale=0.42]{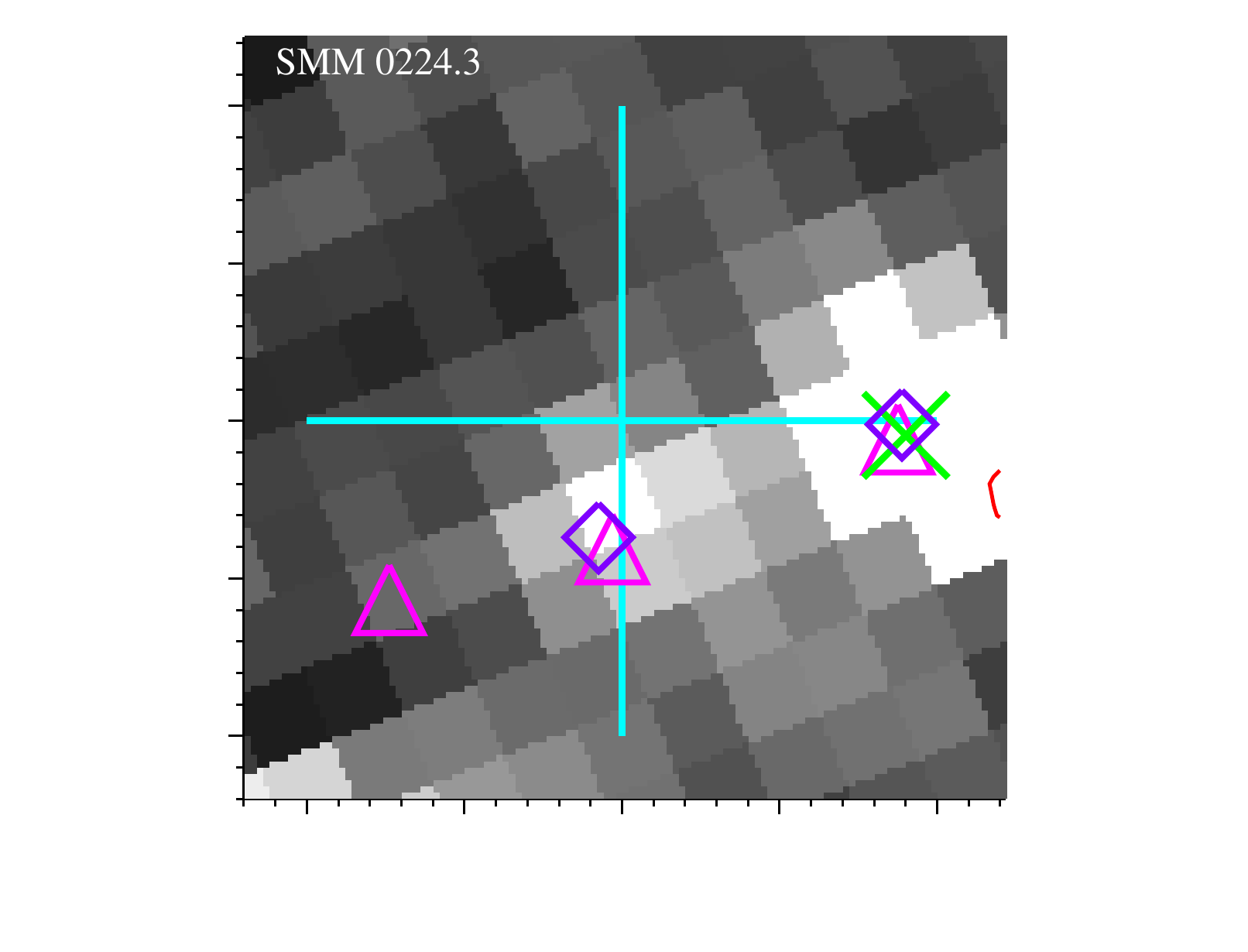}}
      \subfigure{\includegraphics[scale=0.42]{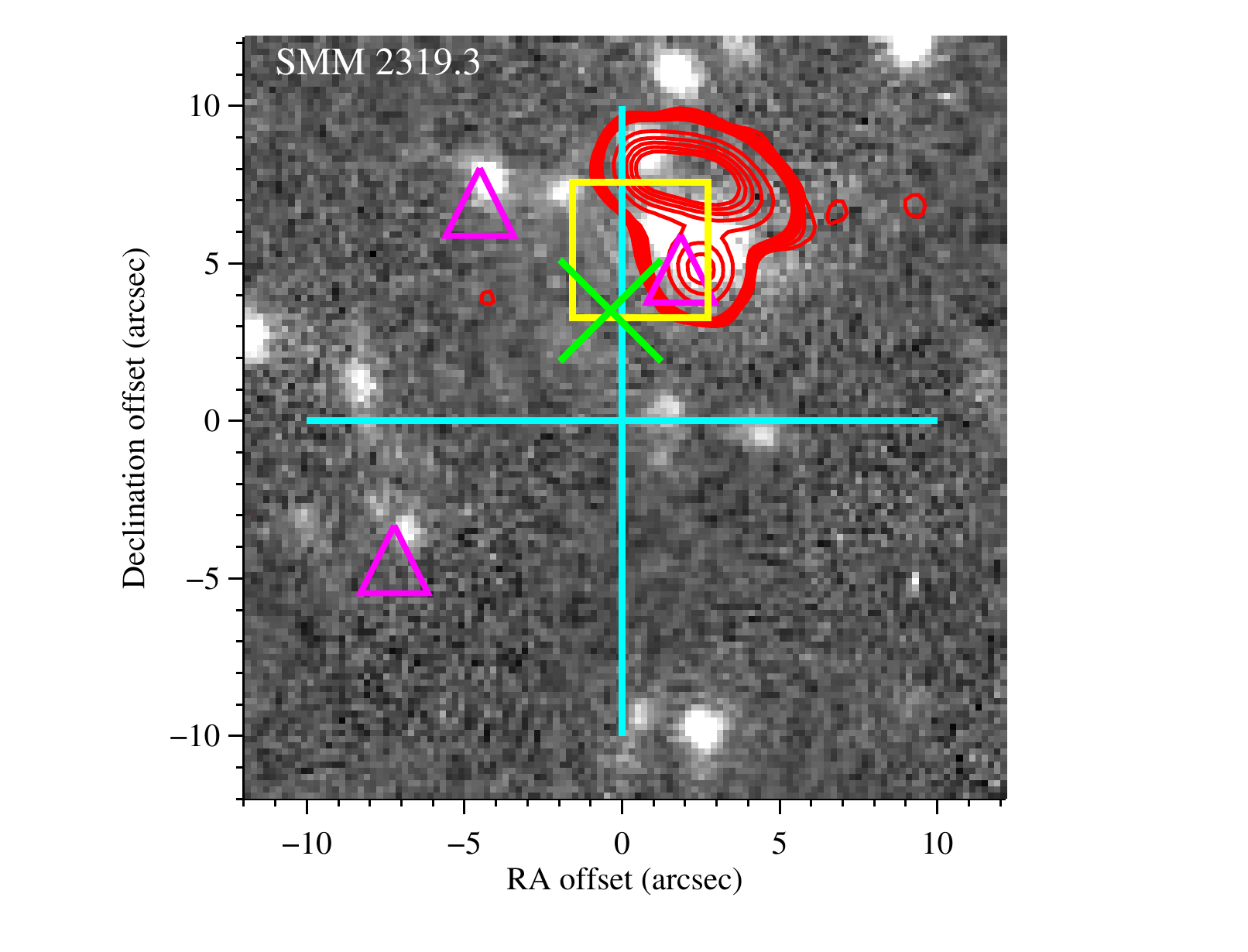}}
         		\hspace{-27mm}
     \subfigure{\includegraphics[scale=0.42]{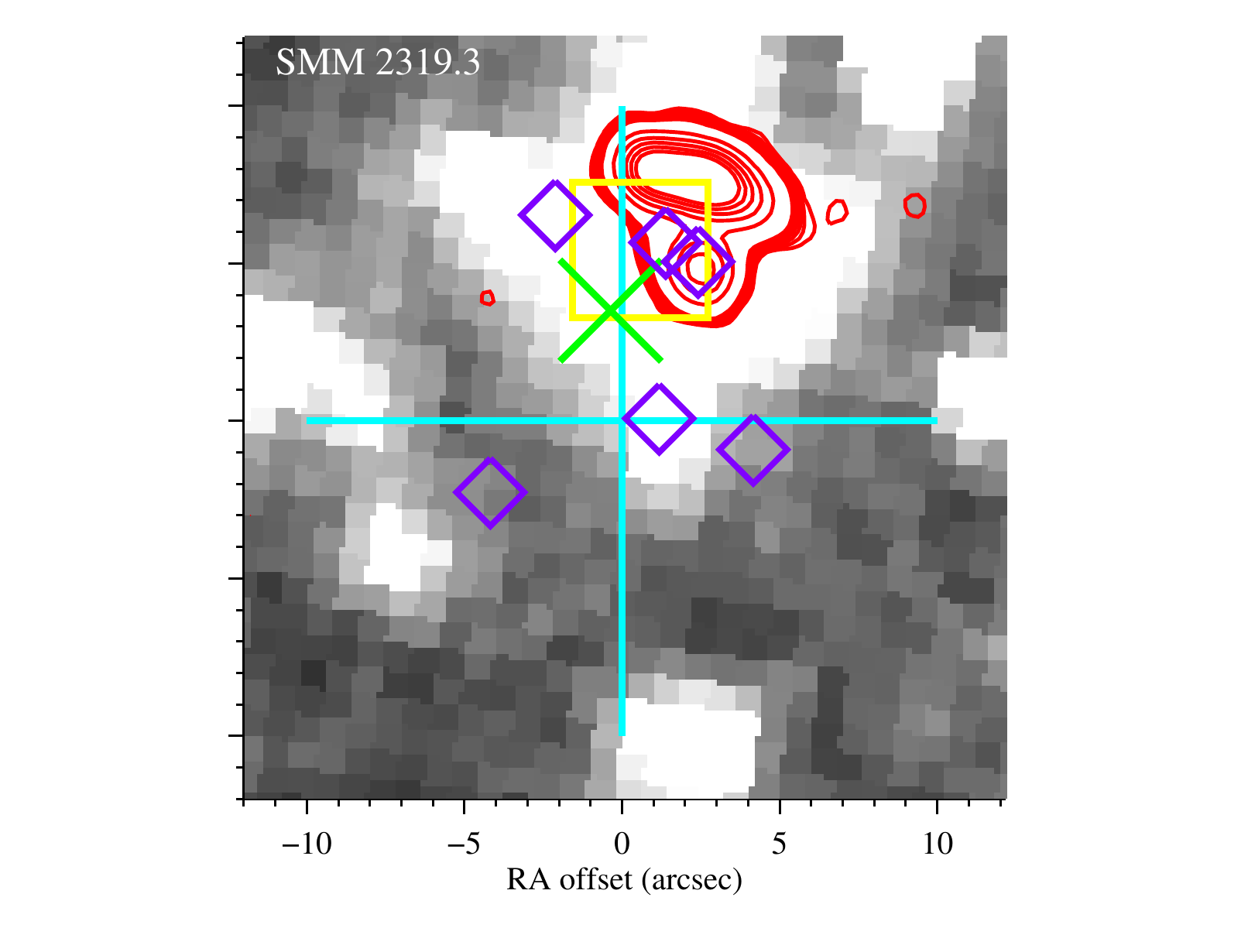}}
        		\hspace{-27mm}
  	   	 \vspace{0mm}
      \subfigure{\includegraphics[scale=0.42]{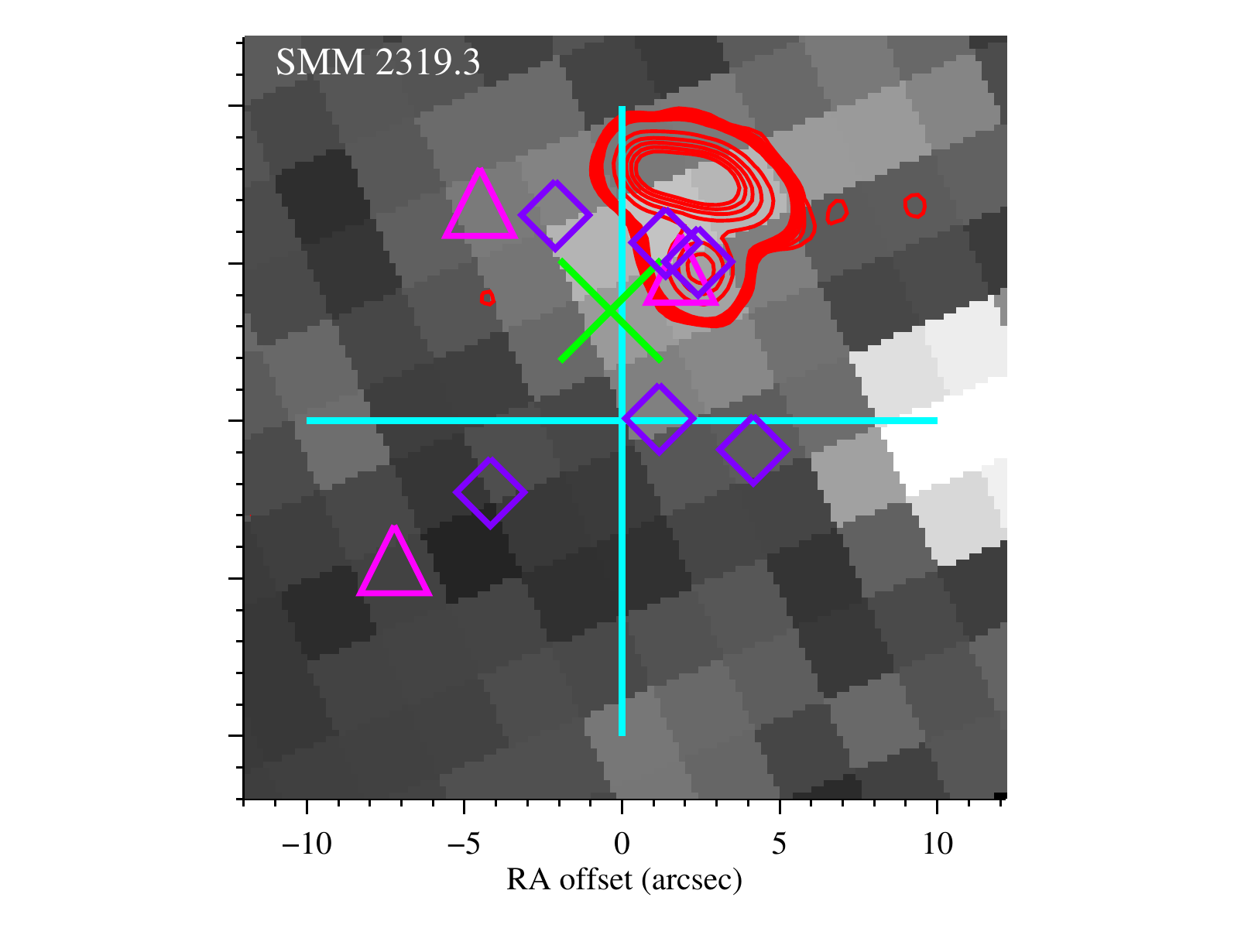}}
   \caption{24 $\times$ 24 arcsec$^2$ postage stamps for the three cluster members:  SMGs 0224.2 (top), 0224.3 (middle), and 2319.3 (bottom).  The grey-scale images are optical $z^{\prime}$ band, IRAC 3.6\,\um, and MIPS 24\,\um, from left to right, where north points up and east is to the right.  Each image is centred on the SMG position, denoted with a large blue cross.  The size of the cross represents the search radius for MIPS counterparts.  The rest of the symbols are as follows: a yellow square is a MIPS source, magenta triangles signify IRAC detections, the red contours show radio emission, an orange cross is 450\,\um\ emission (although none is present in the candidate members), purple diamonds show optical sources with photometric redshifts, and a green `x' denotes an X-ray detection.  Note that the radio contours cover the entire field, but the symbols are only shown for sources within the search radius.  The radio contours start at 45\,\uJy\ and increase in 10\,\uJy\ steps up to 105\,\uJy, after which they increase by 200\,\uJy.}
   \label{fig:stamps}
\end{figure*}

\textbf{SMM 0224.2}.  This SMG has a secure identification based on a robust 24\,\um\  detection.  Coincident radio, X-ray, IRAC, and optical emission all fall within 0.5\,arcsec of the MIPS position and greatly strengthen the security of this identification (see Fig.~\ref{fig:stamps}).  The absence of 450\,\um\  emission is likely due to the high noise at the edges of the 450\,\um\  FOV where this SMG is located.  The slight offset between the optical and radio centroids is probably due to the low S/N (3.8\,$\sigma$) of the radio detection.  The infrared emission is significantly detected in all four IRAC channels and falls within the SMG counterpart region of the $S_{5.8\,\mu m}/S_{3.6\,\mu m}$ versus $S_{5.8\,\mu m}$ plot, making it a likely SMG counterpart (see \citealp{Pope06,Hainline09}).  Although this region of the colour-magnitude plot typically accounts for $z > 1.5$ SMGs, the $S_{5.8\,\mu m}/S_{3.6\,\mu m}$ colour could be reddened due to AGN contamination.  Indeed, the source is along the power-law AGN region in $S_{8.0\,\mu m}/S_{4.5\,\mu m}$ versus $S_{5.8\,\mu m}/S_{3.6\,\mu m}$ colour-colour plot, and has MIPS-IRAC colours indicative of a starburst (SB) $ + $ AGN component (see \S\ref{sec:agn}).  The photometric redshift places the object at $z=0.677\pm0.074$, consistent with the cluster redshift at the 1.3$\,\sigma$ level (see also \citealp{Webb05}).  
\newline
\\
\textbf{SMM 0224.3}.  Although two sources of 24\,\um\  emission are found within the 10\,arcsec search radius, their flux densities differ by a factor of 2.5.  At 404\,\uJy, the brighter (but further) source coincides with X-ray emission and an IRAC source robustly detected in all four channels, further securing the correct identification.  The infrared colours of $S_{8.0\,\mu m}/S_{4.5\,\mu m}$ and $S_{5.8\,\mu m}/S_{3.6\,\mu m}$ fall within the likely SMG counterpart region of the colour-colour plot as denoted by \cite{Yun08} and are marginally consistent with the cluster redshift on the irregular starbursting galaxy track.  It also falls within the appropriate SMG domain proposed by \cite{Hainline09} on the colour-magnitude diagram. We note there is no 1.4\,GHz detection for this SMG; weak radio emission is found to the southeast, but with too large an offset from the MIPS position to be associated.  The lack of 450\,\um\  emission is a result of the noisy edges on the map.  The photometric redshift for a coincident optical source places this galaxy at the redshift of the cluster (at the 1.2$\,\sigma$ level), with $z=0.688\pm0.071$.  
\newline
\\
\textbf{SMM 2319.3}.  SMM 2319.3, the most intriguing SMG of the sample, represents the only field where two radio sources have been detected.  This is not necessarily indicative of false radio detections as 10 per cent of SMGs are expected to contain two radio sources within the search radius, both of which are related to the 850\,\um\  emission \citep{Ivison02, Chapman05, Pope06, Ivison07}.  Furthermore, this SMG is positioned very close to the cluster centre ($\sim10$\,arcsec) where the density of galaxies is high.  

Both radio sources are extremely bright at 1159 and 5656\,\uJy.  Closer investigation reveals a possible head-tail morphology for the brighter object; it is an extended radio source with a single bright optical source at one end.  It is most likely the result of radio jets bending due to the bulk flows in the intracluster medium, and possibly suggestive of merger-induced activity.  It has a spectroscopic redshift consistent with the cluster redshift ($z=0.90490$).  However, the spectrum shape is indicative of an old elliptical galaxy, which is not typical for SMGs.  The fainter radio source is also a spectroscopically confirmed member ($z=0.90126$) and most likely associated with the brightest cluster galaxy (BCG).  Spectroscopic observations were obtained with the Visible Multi-Object Sectrograph (VIMOS; \citealp{LeFevre03}) on the 8.2-m VLT UT3 telescope, and the Inamori Magellan Areal Camera and Spectrograph (IMACS; \citealp{Dressler06}) on the 6.5-m Baade telescope, for the head-tail source and BCG, respectively.

The radio sources lie on opposite sides of a 24\,\um\  source, within 1.9 and 2.7\,\,arcsec, for the fainter and brighter source respectively.  Given the above description on the spectrum shapes and positional offsets, the MIPS emission is likely associated with the fainter radio source.  The IRAC emission is highly confused, complicating the identification process.  X-ray emission is detected as well, but at a distance of 3.0\,arcsec from the fainter radio source; it is more closely aligned with the MIPS emission.  The slightly lower S/N of the X-ray source (3.9$\,\sigma$) might explain the larger offset between it and the more accurate radio position.  Another possible interpretation could involve an ongoing merger event between the MIPS/X-ray and radio sources, as there are only $\sim$15\,kpc between the MIPS and radio centroids.  Based on the proximity to the cluster centre, the submillimetre emission could also be a background lensed source; however we have no evidence to substantiate this possibility.  The SMG search radius also has three optical sources with photometric redshifts consistent with the cluster.  Although the crowded field complicates the process of identifying coincident emission, it seems clear that this is a submillimetre detection of the cluster centre, and therefore a likely member.
\newline
 
Thus far, only a handful of SMG studies have detected submillimetre cluster galaxies, some of which where discovered serendipitously as a result of utilizing massive clusters as gravitational telescopes to probe the background population (e.g.\ \citealp{Smail02,Cowie02}).  Other studies have indeed focused on the cluster galaxies, but either rely on unconfirmed members (e.g.\ \citealp{Best02}) or study lower redshift clusters than presented here, for example two clusters at $z=0.23$ and $z=0.25$ that each contain a central galaxy with submillimetre emission  \citep{Edge99}, and a rich cluster at $z=0.54$ with two likely 1.1\,mm detected cluster members \citep{Wardlow10}.  Our detection of three likely submillimetre cluster members at  the higher redshifts of $z=0.773$ and $z=0.9024$ not only has implications for galaxy cluster evolution, but also serves as an important probe of some of the most intense starbursts in clusters.  

\subsubsection{AGN Contamination}
\label{sec:agn}
Infrared emission is a well-established tracer of star formation activity, in particular for very luminous and dusty galaxies.  An exception to this involves the contamination to the infrared luminosity due to AGN activity.  

In order to isolate sources contaminated by reprocessed AGN emission, we employ various infrared colour selection techniques (e.g.\ \citealp{Ivison04, Lacy04, Ashby06, Yun08, Hainline09}).  For instance, the left panel in Fig.~\ref{fig:colmag} displays the redshift track of an AGN in colour-colour space along with that of a pure power-law galaxy, $S\propto \nu^{\alpha}$.  Of the three cluster members, SMM 0224.2 is the most likely AGN candidate, inhabiting the power-law AGN region, with log$(S_{8.0\,\mu m}/S_{4.5\,\mu m}) = 0.21$, which is consistent with AGN colours and redder than the majority of SMGs which typically have a ratio $<0.2$ (see \citealp{Yun08}).  Furthermore, according to \cite{Ivison04} and \cite{Hainline09}, the colour ratios of $S_{24\,\mu m}/S_{8\,\mu m}$ (3.3) versus $S_{8.0\,\mu m}/S_{4.5\,\mu m}$ (1.6) for RCS 0224.2 are indicative of a starburst (SB) $ + $ AGN component.

In contrast, SMM 0224.3 shows no evidence for AGN contamination.  A comparison of the MIPS and IRAC fluxes yields a starbursting system with no AGN component (i.e.\ $S_{24\,\mu m}/S_{8\,\mu m}=16.2$ versus $S_{8.0\,\mu m}/S_{4.5\,\mu m} = 0.8$; see \citealp{Ivison04}).  In the IRAC colour-colour plot, this SMG lies directly between the two starbursting tracks.  SMM 2319.3 is too confused in IRAC to accurately interpret its infrared colours.

Our AGN analysis is further supported by empirically fitting the IRAC SEDs of the two cluster members with good infrared detections to a power law (e.g.\ \citealp{Ashby06, Barmby06}).  The sole AGN candidate, SMM 0224.2, can be fit to a red power law with $\alpha \approx -1.2$, and monotonically increases in flux with wavelength through the IRAC bands; these features are both indicative of AGN-dominated emission.  SMM 0224.3 has a blue power law ($\alpha \approx +0.4$), suggesting stellar-dominated emission, and SMM 2319.3 is too blended in the IRAC channels to analyse.

\subsubsection{Cluster Member Properties}
We estimate the infrared luminosities (over 8--1000\,\um) of our cluster members following the prescription outlined in \cite{Elbaz02} and \cite{Sanders96}.  Using a galaxy SED template for Arp 220, we shift the distribution to the respective redshifts of the host clusters, normalize to the 850\,\um\  flux of each SMG, and convert to a total infrared luminosity.  This yields $L_{\rm{IR}} = 6.3\times 10^{12}$\,\Lsol,  $5.6\times 10^{12}$\,\Lsol, and $3.9\times 10^{12}$\,\Lsol\ for SMMs 0224.2, 0224.3, and 2319.3, respectively, which are consistent with ULIRG level luminosities, albeit on the higher end of the distribution.

Luminous galaxies (such as SMGs) powered by dense starbursts output the bulk of their luminosity in the infrared as young stars heat the dust, which thermally re-emits the radiation.  The infrared luminosity, therefore, traces young stellar populations and provides an estimate of the star formation rate (SFR).  Using the relation derived in \cite{Kennicutt98a} and assuming pure starbursting systems, we calculate star formation rates corresponding to 1082, 968, and 665\,\myr\ for SMM 0224.2, 0224.3, and 2319.3, respectively.  This is in good agreement with rates derived in larger surveys with spectroscopically (or photometrically) identified SMGs (e.g.\ \citealp{Chapman05, Pope06}), where they calculate a median SFR of 1100\,\myr.  We note that our estimates are highly sensitive to the dust temperature \citep{Webb05}, and could also suffer from AGN contamination, in particular SMM 0224.2, which would bring down our estimates.

The intense star formation in these likely cluster members prompts the broader question: what is the trigger for the creation of these luminous systems?  We can attempt to draw clues from their radial distribution.  For instance, the potential member within RCS 2319 is located in the cluster core; the MIPS emission is $\sim8$\,kpc from the brightest cluster galaxy (BCG; see Fig.~\ref{fig:850mapslensing}).  Such high activity in the proximity of the core is indicative of a major merger-induced episode of star formation as a bound pair falls through the cluster centre and is tidally torqued by the potential (e.g.\ \citealp{Oemler09}).  Moreover, recent evidence suggests that the build-up of BCG mass occurs via major mergers \citep{Tran08}.  Either scenario could explain the extreme star formation in the core, though the high-density of sources complicates identification of counterpart emission.

In contrast, the two members of RCS 0224 are both further out at $\sim95$\,arcsec from the BCG, corresponding to $\sim700$\,kpc at $z=0.773$, and therefore could belong to the infalling population.  Indeed, galaxy infall events have been observed to increase at higher redshifts (e.g.\ \citealp{Ellingson01,Loh08}).  In this scenario, the increased star formation of these accreted systems might merely reflect the increased activity in the field, as suggested by numerous studies (e.g.\ \citealp{Tran05, Saintonge08}).

\subsection{Redshift Estimates}
\label{sec:redshift}
For the remaining 23 SMGs in the sample, the counterpart identification process revealed only two secure identifications with either an optical spectroscopic (SMM 0224.6; see \S\ref{sec:gglensing}) or photometric (SMM 1122.2) redshift, however the uncertainties are fairly large on the latter source.  Two additional ambiguous/tentative sources (SMMs 0224.5 and 2152.2) have photometric redshifts for their most probable identification (i.e.\ the brightest MIPS emitter) in the SMG search radius.  Therefore, for the remaining 16 SMGs (not including potential cluster members or those without any counterpart identifications), we must rely on more crude redshift estimates based on flux ratio trends.

In principle, we can calculate a redshift for each radio identification based on the strong evolution of the spectral index between 850\,\um\  and 1.4\,GHz emission in AGN and star-forming galaxies.  However, there is significant uncertainty in this technique (see fig. 4 in \citealp{Webb05}), as it is highly sensitive to the dust temperature and to AGN contamination of the radio emission \citep{Carilli99}.  We therefore do not use this redshift estimate for individual sources, but instead calculate a median redshift for the radio-identified sample (5 SMGs) of  $z\gtrsim 2$ (assuming a dust temperature of $\sim 40$\,K), which is consistent with other submillimetre surveys with deep radio imaging (e.g.\ \citealp{Smail00,Ivison02,Chapman03,Chapman05}).

\begin{figure}
\centering
\includegraphics[scale=0.45]{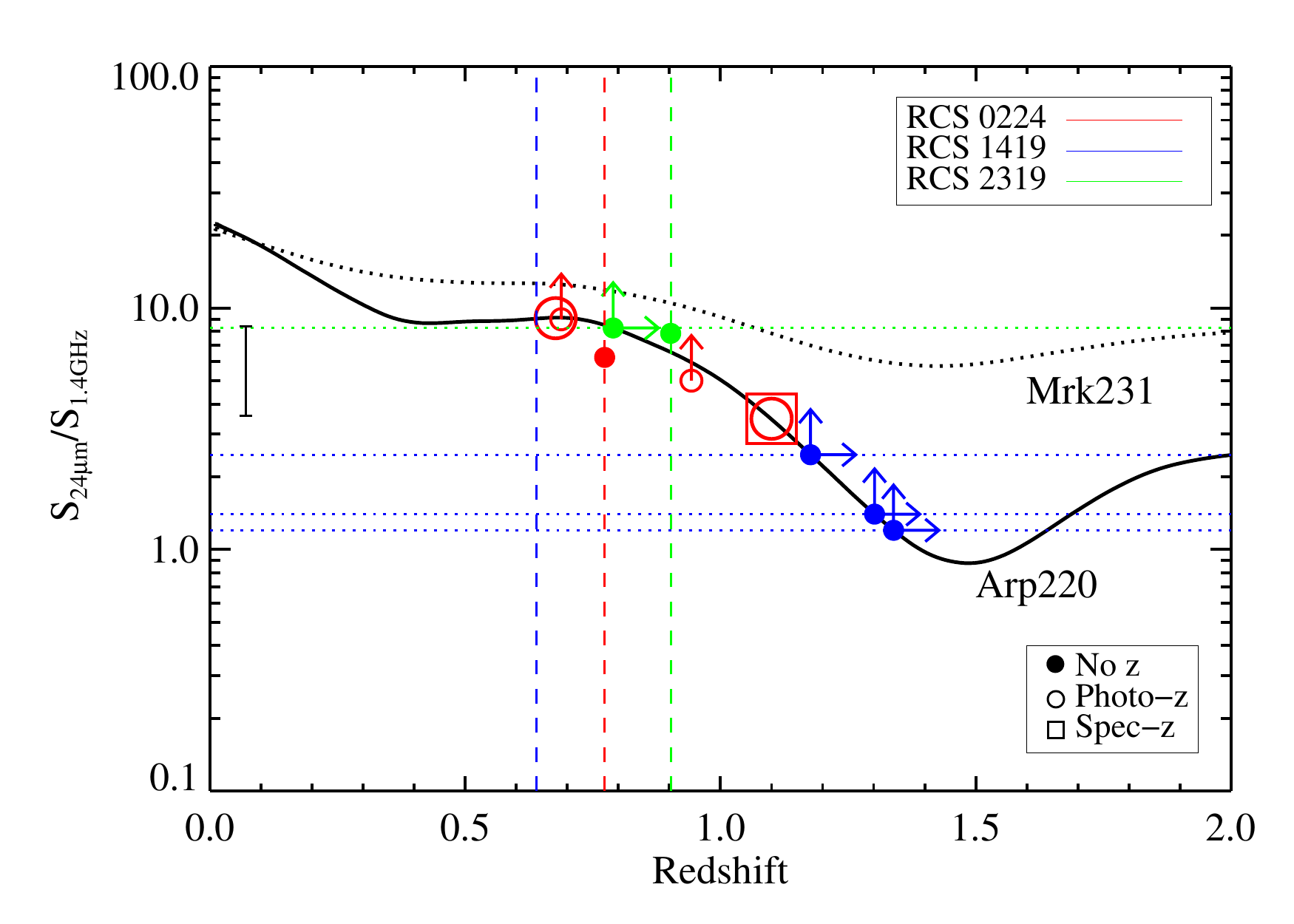}
\caption{The ratio of $S_{24\mu m}/S_{1.4GHz}$ versus redshift for Arp 220 (solid line), Mrk231 (dotted line; an AGN-dominated ULIRG), and for the ten SMGs with counterpart identifications in the three fields with radio and MIPS imaging.  The four radio-identified SMGs (SMMs 0224.2, 0224.4, 0224.6, and 2319.1) are plotted as open and filled circles without arrows; SMM 2319.3 is omitted since the 24\,\um\ flux is severely blended. The remaining six SMGs with MIPS emission but no radio detection are given a 3$\,\sigma$ upper flux limit at 1.4\,GHz and shown as circles with arrows.  SMGs with (without) optical redshifts are plotted as open (filled) symbols; the spectroscopic redshift for SMM 0224.6 is highlighted by a box.   The two larger open circles denote the only two SMGs with detections at both 24\,\um\  and 1.4\,GHz and redshift estimates (SMMs 0224.2 and 0224.6).  The four SMGs that are only detected at 24\,\um\  and lack redshift estimates (SMMs 1419.1, 1419.3, 1419.4, and 2319.2) are plotted as horizontal dotted lines and at redshifts interpolated onto Arp 220 given their flux ratio limit; they are given both lower limits on their flux ratio and redshift (filled circles with double arrows).  The dashed vertical lines represent the three cluster redshifts. Error bars on each point have been omitted to avoid crowding the plot; instead we show a representative error bar on the left side of the plot.}
\label{fig:mip_radio_z}
\end{figure}

We also investigate the redshift distribution of the SMG sample through a comparison of 24\,\um\  and 1.4\,GHz fluxes \citep{Ivison07} in the three cluster fields with radio and MIPS imaging (RCS 0224, 1419, and 2319), as shown in Fig.~\ref{fig:mip_radio_z}.  The SMGs with photometric (SMMs 0224.2, 0224.3, 0224.5; open circles) or spectroscopic redshifts (SMM 0224.6; open circle with box) all are consistent with the track for Arp 220, supporting their identification and redshift.  The two SMGs with robust mid-IR and radio detections but no redshift estimate (SMMs 0224.4 and 2319.1; filled circles) are also consistent with the Arp 220 track at their respective cluster redshift, and therefore may also be cluster members (but see \S\ref{sec:gglensing} for an alternative explanation regarding SMM 2319.1).  The three SMGs of RCS 1419 with counterpart identifications but no radio detections (SMMs 1419.1, 1419.3, and 1419.4; dotted blue lines) are given a 3$\,\sigma$ upper flux limit at 1.4 GHz and have $S_{24\mu m}/S_{1.4GHz}$ ratios consistent with higher redshifts ($z>1$) assuming an Arp 220-like SED.  Based on the lower limit of their flux ratio, we also plot these sources (filled blue circles with double arrows) interpolated onto the Arp 220 track.  Although the remaining SMG (SMM 2319.2; dotted green line) lacks a radio detection and redshift estimate, it is not inconsistent with a lower redshift source based on the interpolated value (filled green circle with double arrows) to Arp 220 and could in fact be a cluster member.

\begin{figure}
\centering
\includegraphics[scale=0.45]{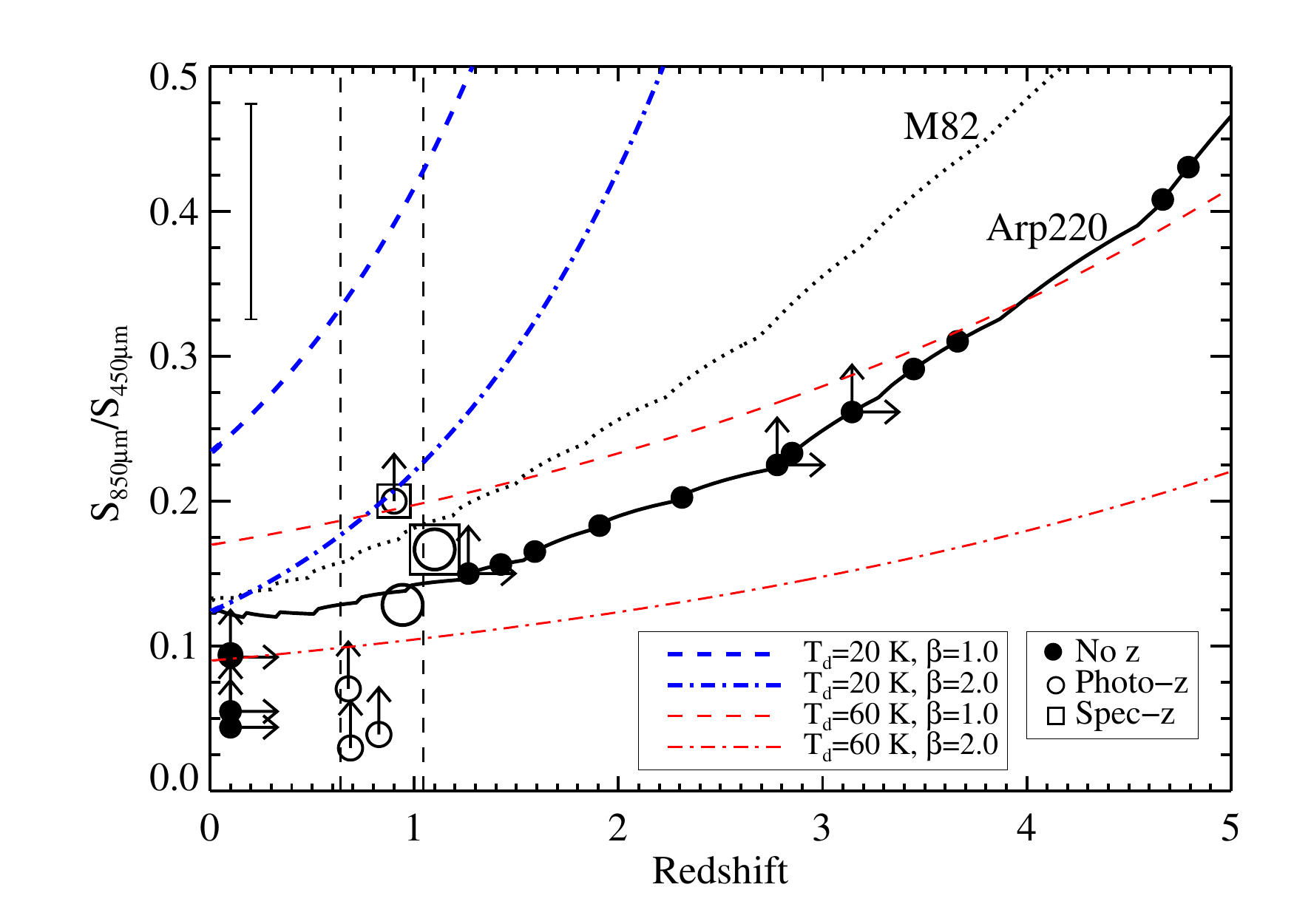}
\caption{The ratio of $S_{850\mu m}/S_{450\mu m}$ as a function for redshift for all SMGs within the 450\,\um\  FOV.  The solid (dotted) line is the redshift track for Arp 220 (M82) and the dashed lines represent the range of redshifts for all seven clusters.  We also plot isothermal grey bodies over a range of plausible dust temperatures and dust emissivities.  The symbols are the same as in Fig.~\ref{fig:mip_radio_z}.  Non-detections at 450\,\um\  are given 3$\,\sigma$ upper flux limits (open and filled circles with arrows) and interpolated onto the Arp 220 track when possible.  The two larger open circles denote the only two SMGs with detections at 450\,\um\  and redshift estimates (SMMs 0224.5 and 0224.6).  Error bars on each point have been omitted to avoid crowding the plot; instead we show a representative error bar in the upper left corner.}
\label{fig:submm_z}
\end{figure}

The submillimetre flux-density ratio provides another means of constraining redshifts (e.g.\ \citealp{Hughes98, Smail99, Fox02, Coppin08}).  In Fig.~\ref{fig:submm_z} we plot $S_{850\,\mu m}/S_{450\,\mu m}$ as a function of redshift for Arp 220, M82, our entire SMG sample within the 450\,\um\  field of view (22 SMGs total), and four isothermal grey bodies with a range of dust temperatures ($T_{\rm{d}}$) and dust emissivities ($\epsilon\propto\nu^{\beta}$).   Given the monotonic rise in the Arp 220 flux ratio with redshift, we plot the $3\,\sigma$ upper flux limit for 450\,\um\  non-detections as a lower limit in redshift space (the ten cases of circles with arrows).  SMGs with optical redshifts are shown as open circles (boxes represent spectroscopic redshifts), while those lacking redshift information (filled circles) are given interpolated values to the Arp 220 curve based on either flux ratios or limits.  Ratios below the lowest value from Arp 220 are plotted arbitrarily as lower limits at $z=0.1$.  The plot clearly illustrates the sensitivity of redshift with the chosen SED for an individual SMG, and therefore, the colour ratio should only be used to estimate the redshift distribution of the entire sample (see also \citealp{Amblard10}).  Assuming an Arp 220-like trend, our submillimetre flux ratio predicts that roughly over half our sample lies at $z\gtrsim1$ and the rest within $z\lesssim1$.

We note that for individual galaxies there is a fairly large scatter in the redshifts estimated from the techniques discussed above.  Since the MIPS flux could suffer from confusion and the submillimetre colour ratio is highly dependent on dust temperatures, the uncertainty on an individual redshift is typically $\sim0.5$.  However, we attempt to extract broad redshift trends for the sample; overall, we find that the combination of flux ratios between radio, MIPS, and 450\,\um\  indicates that the majority of the our SMGs lie at $z>1$.  Even after accounting for individual redshift uncertainties, this implies that the bulk of our sources are in the background of the clusters.  

\subsection{Galaxy-Galaxy Lensing}
\label{sec:gglensing}
Given the steep submillimetre source counts at the bright tail, recent studies have investigated the role of gravitational lensing in the SMG population (e.g.\ \citealp{Perrotta03, Almaini03, Almaini05, Austermann09, Negrello10, Vieira10, Aretxaga11}).  Theoretical models predict that $\sim30$ per cent of SMGs could be gravitationally lensed at fluxes above 10\,mJy at 850\,\um\  \citep{Perrotta03}, with the fraction increasing at even brighter fluxes \citep{Negrello07}.  

Current observational studies support this claim.  For instance, \cite{Almaini05} found a positive angular correlation between SCUBA sources and low-redshift optical galaxies, with $20-25$ per cent of submillimetre sources associated with foreground overdensities within a 30\,arcsec aperture. They argue that a possible cause of this correlation could be gravitational lensing.  More recently, \cite{Vieira10} discovered 20 dusty sources above $10$\,mJy at 1.4\,mm within 87 deg$^{2}$ with the South Pole Telescope (SPT), and postulated that they represent a population of gravitationally lensed, dust-dominated sources. Indeed, \cite{Negrello10} claim that bright 500\,\um-selected sources are almost all lensed.  Moreover, \cite{Austermann09} report a $\sim3\,\sigma$ excess of SMGs compared to the blank-field population with the AzTEC camera in the Cosmological Evolution Survey (COSMOS), a 0.15 deg$^2$ field that includes a massive $z=0.73$ cluster and an overdensity of optical-IR galaxies at $z\lesssim1.1$.  They similarly calculate a spatial correlation between bright SMGs and low-redshift mass structure, attributable to lensing of the background submillimetre population by individual galaxies.  \cite{Aretxaga11} corroborate this claim with a larger AzTEC survey of the COSMOS field (0.72 deg$^2$) and find an excess of number counts compared to those of SHADES at 1.1\,mm \citep{Austermann10} due to moderate magnification levels from galaxy-galaxy and galaxy-group lensing.

In light of these recent observations, we explore the possibility of galaxy-galaxy gravitational lensing within our sample between an optical cluster member and background SMG.  Although lensing events are expected to be fairly common with the submillimetre population due to its high median redshift ($z\gtrsim1$) \citep{Blain96}, large-area surveys are essential to obtain a statistically significant sample of lensed SMGs since they have a low surface density at higher fluxes where they are expected to dominate over un-lensed SMGs \citep{Negrello10}.  Given the increased density of optical sources in cluster fields, there could be a larger correlation in our sample.  However, even a handful of gravitationally lensed SMGs in our relatively small sample, although difficult to statistically confirm, is noteworthy, as it is a possible complication for future observations of cluster fields.  

\begin{figure*}
   \centering
      \subfigure{\includegraphics[scale=0.42]{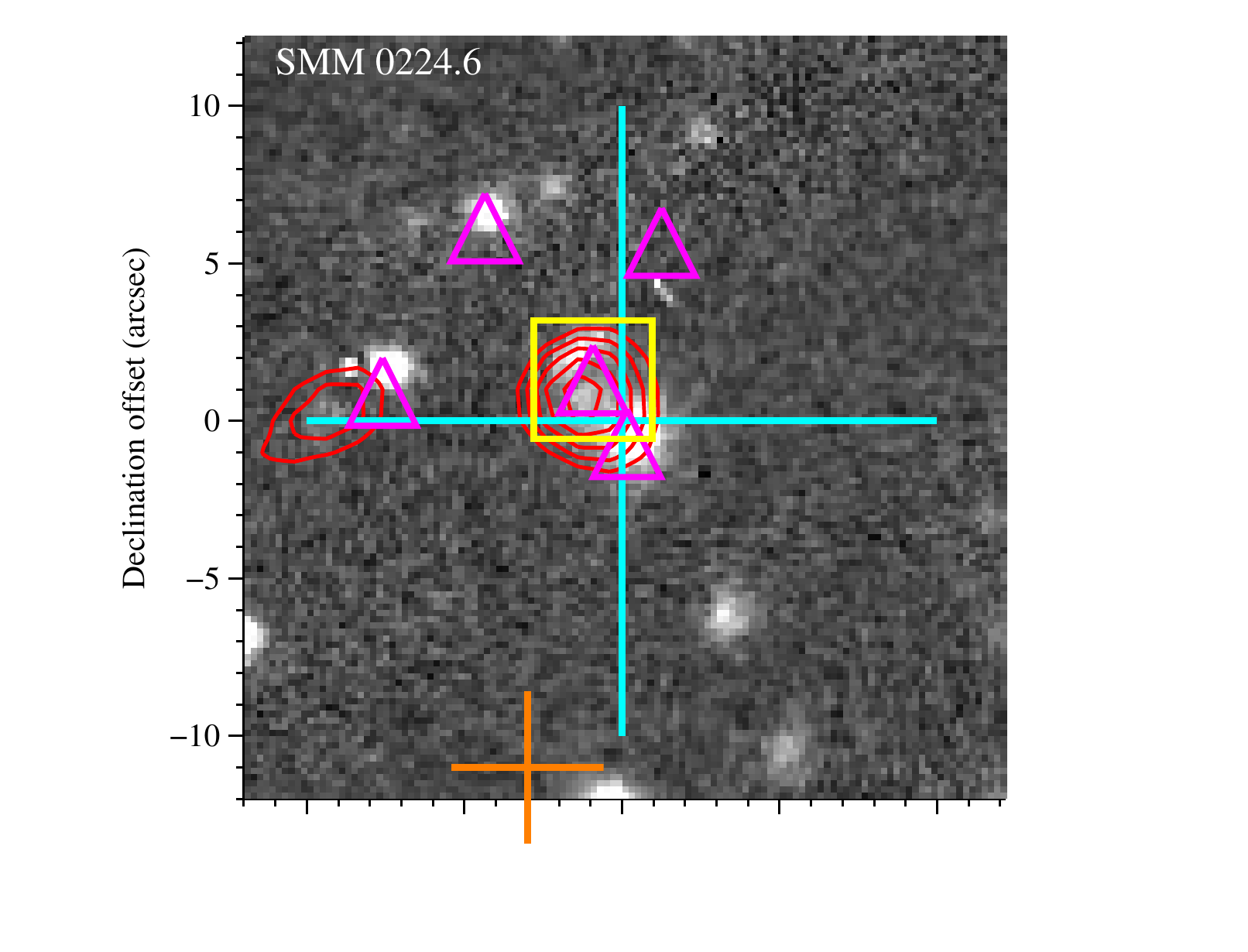}}
   		\hspace{-27mm}
   \subfigure{\includegraphics[scale=0.42]{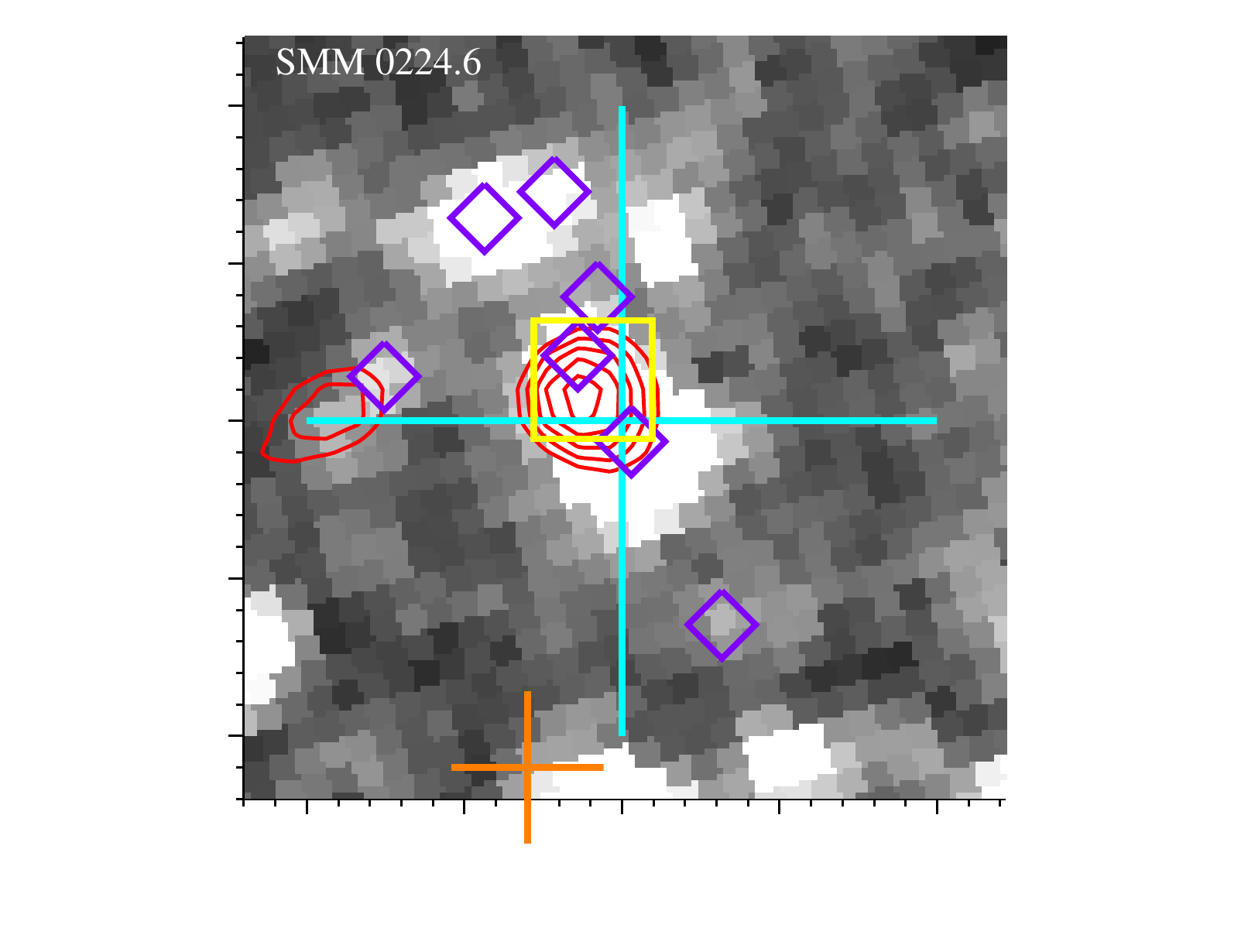}}
   		\hspace{-27mm}
  	   	 \vspace{-12mm}
   \subfigure{\includegraphics[scale=0.42]{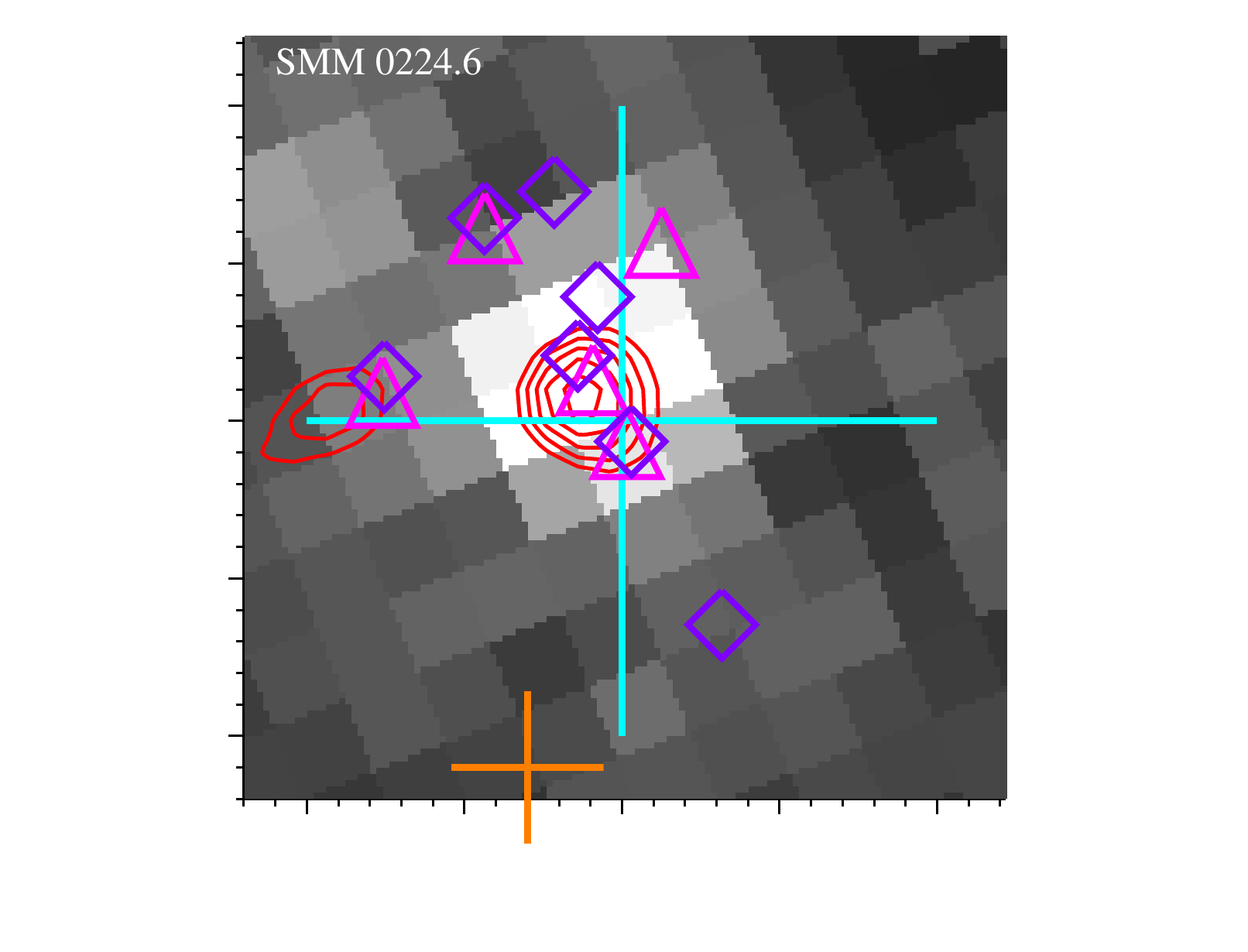}}
      \subfigure{\includegraphics[scale=0.42]{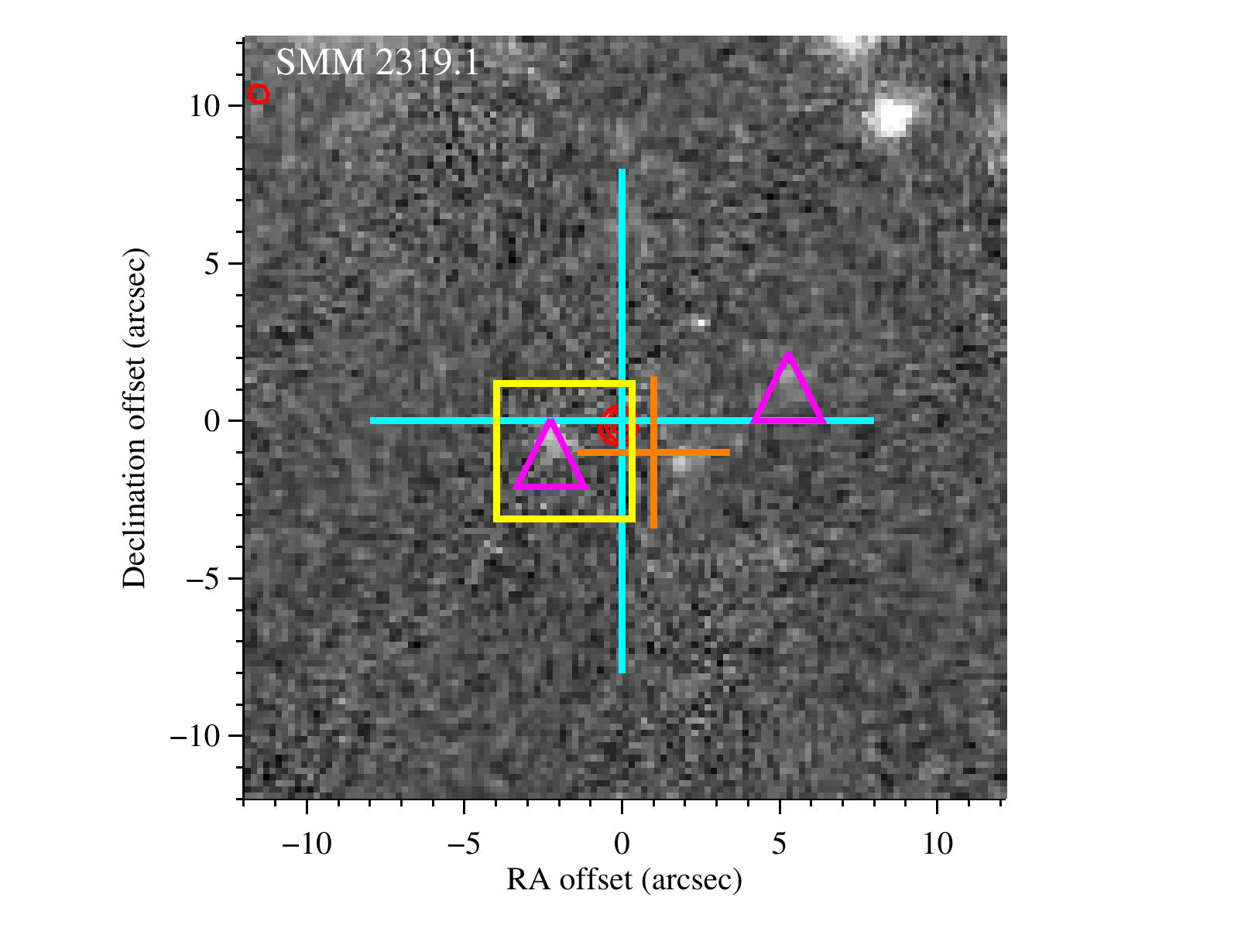}}
         		\hspace{-27mm}
     \subfigure{\includegraphics[scale=0.42]{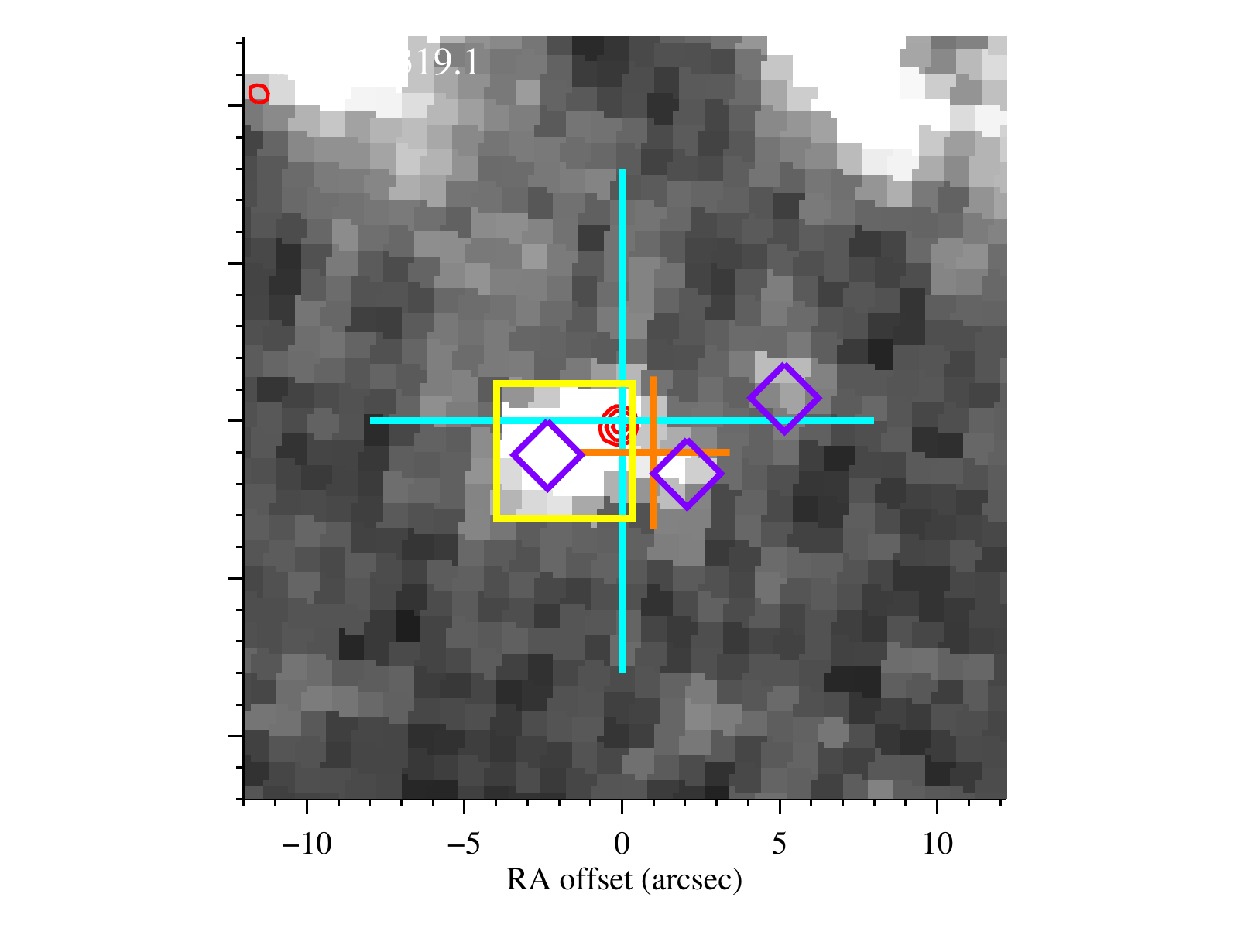}}
        		\hspace{-27mm}
  	   	 \vspace{0mm}
      \subfigure{\includegraphics[scale=0.42]{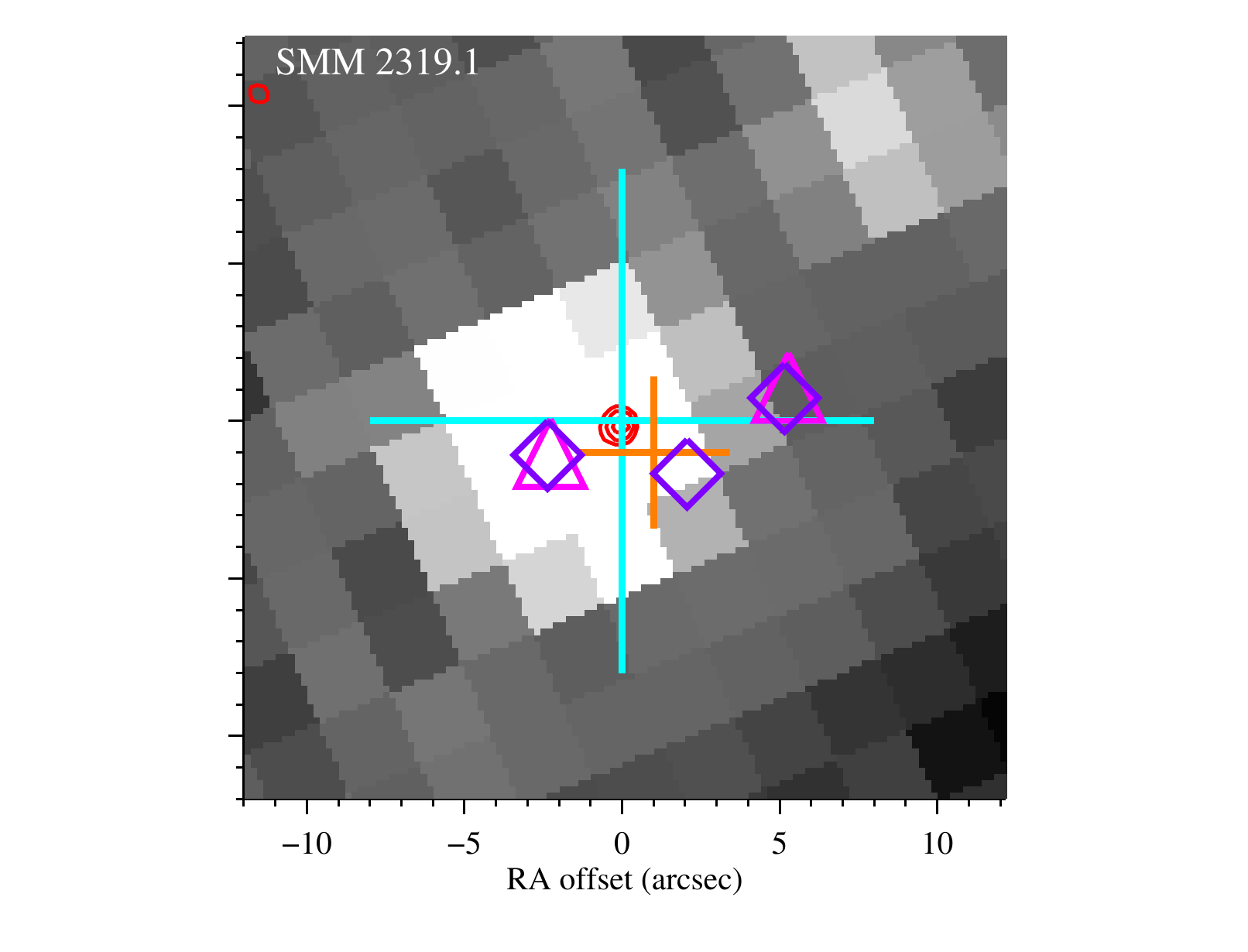}}
   \caption{24 $\times$ 24 arcsec$^2$ postage stamps for the two cases of possible galaxy-galaxy lensing:  SMGs 0224.6 (top) and 2319.1 (bottom).  The grey-scale images and symbols are the same as those in Fig.~\ref{fig:stamps}.}
   \label{fig:gglensing}
\end{figure*}

\begin{figure*}
   \centering
      \subfigure{\includegraphics[scale=0.42]{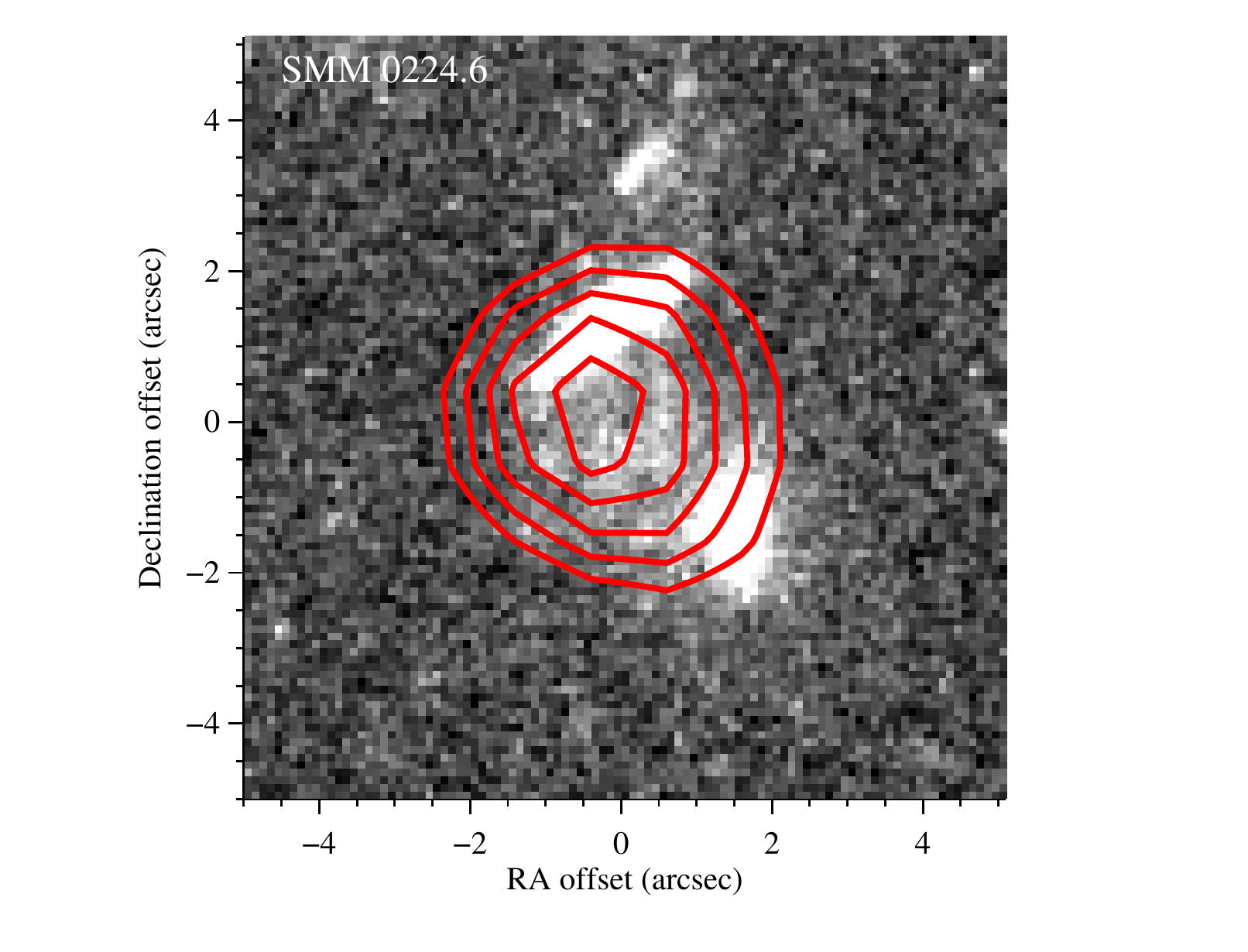}}
      \hspace{-27mm}
       \subfigure{\includegraphics[scale=0.42]{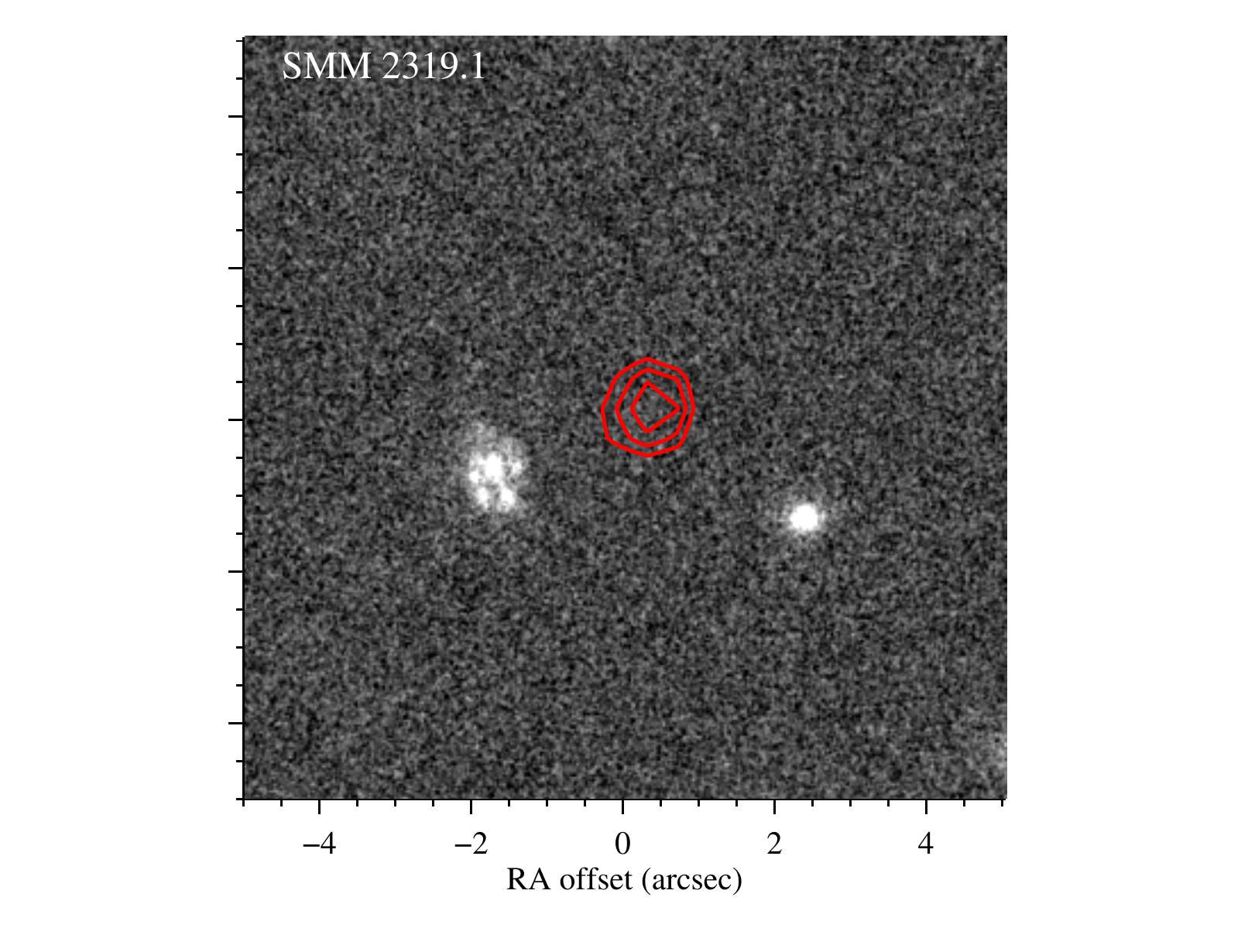}}   
       \hspace{-27mm}
      \subfigure{\includegraphics[scale=0.42]{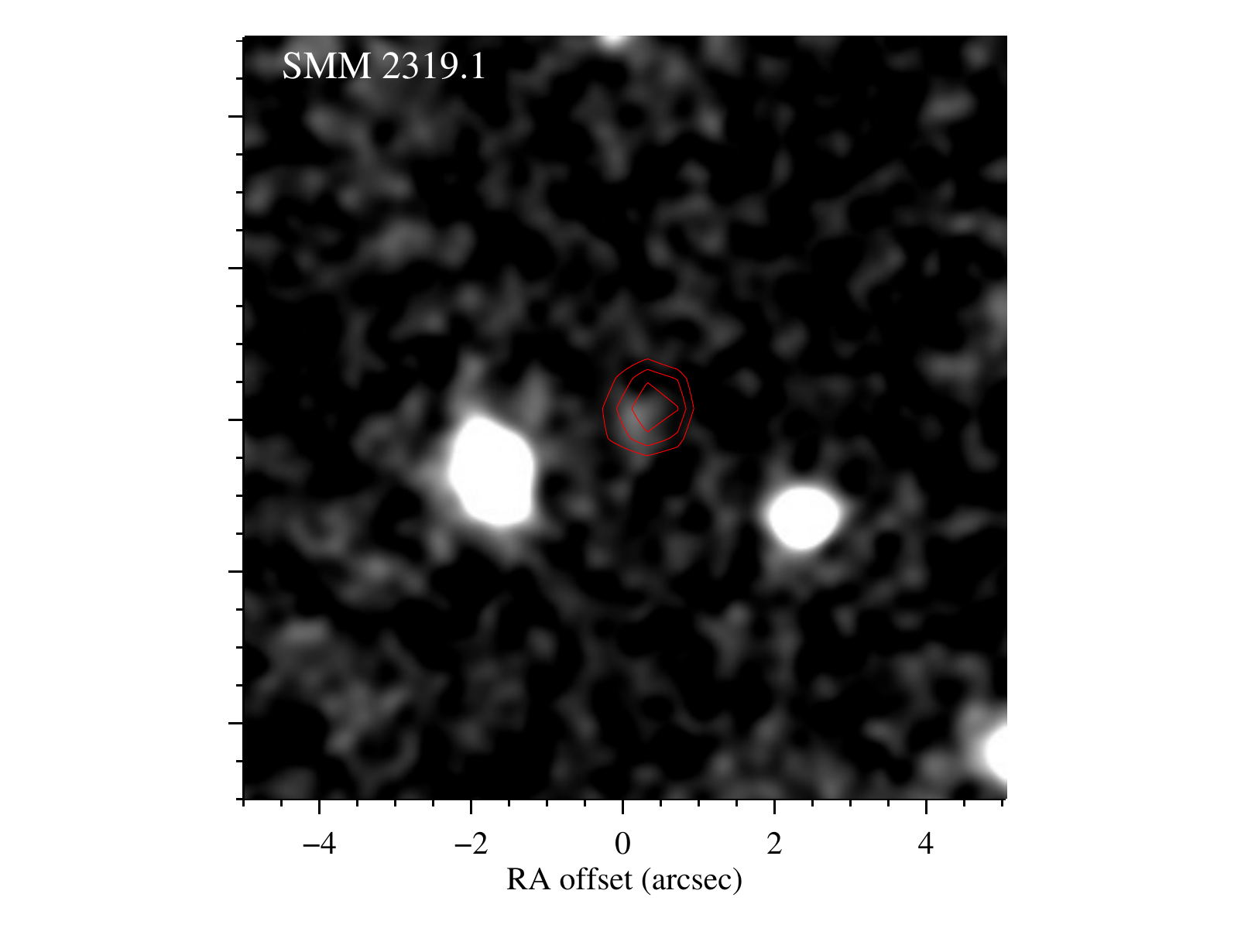}}      
     \caption{10 $\times$ 10 arcsec$^2$ \textit{HST} postage stamps for the two cases of possible galaxy-galaxy lensing.  The rightmost stamp is the  \textit{HST} stamp of SMM 2319.1 smoothed with a 0.3\,arcsec Gaussian to highlight seemingly faint optical emission at the radio centroid.  The red contours show 1.4\,GHz emission starting at 45\,\uJy\ and increasing in 10\,\uJy\ steps.}
   \label{fig:hstlensing}
\end{figure*}

Again, we are biased towards our fields with radio and optical coverage, as accurate positions are necessary to differentiate the foreground optical lensing galaxy from the background SMG with radio emission.  We find two possible cases of galaxy-galaxy lensing out of our five radio-identified SMGs.  We note there are other cases of blended IRAC sources elongated in the direction of a possible background source (SMM 2152.1 and 2152.2), but they lack follow-up radio observations to verify the submillimetre position.  Here we describe the counterpart identifications for the two possible anomalous lensed systems and include their postage stamps for visual reference (Fig.~\ref{fig:gglensing}) along with archival \textit{HST} imaging\footnote{Based on observations made with the NASA/ESA \textit{Hubble Space Telescope}, and obtained from the Hubble Legacy Archive, which is a collaboration between the Space Telescope Science Institute (STScI/NASA), the Space Telescope European Coordinating Facility (ST-ECF/ESA) and the Canadian Astronomy Data Centre (CADC/NRC/CSA).} (Fig.~\ref{fig:hstlensing}) to support this idea.
\newline
\\
\textbf{SMM 0224.6}.  Counterpart emission is detected at mid-IR, radio, and optical wavelengths, all coinciding within $\sim2$\,arcsec of the 850\,\um\  emission.  Distant 450\,\um\  emission at $3.1\,\sigma$ is detected 11.4\,arcsec away from the 850\,\um\  source.  The radio centroid lies between two bright optical sources and directly on top of diffuse  \textit{HST} emission (see Fig.~\ref{fig:hstlensing}) and a faint $K$-band detection (see \citealp{Webb05}).  The northern optical source has a foreground photometric redshift estimate of $z=0.351\pm0.116$, while the southern source has a spectroscopic redshift of $z=0.7832$, consistent with the redshift of the cluster. 

Spectroscopic observations were obtained with the DEep Imaging Multi-Object Spectrograph (DEIMOS; \citealp{Faber03}) on the Keck II 10-m telescope as part of the Observations of Redshift Evolution of Large Scale Environment survey (ORELSE; \citealp{Lubin09}). The diffuse \textit{HST} emission, which is likely the submillimetre counterpart based on the radio position, was serendipitously observed with DEIMOS and has a spectroscopic redshift of $z=1.0995$. This system, therefore, represents a possible case of a background galaxy that may be lensed by either a foreground galaxy or cluster member.  Assuming a singular isothermal sphere (SIS) profile and either a massive lens for the cluster member ($z=0.7832$) or a lower mass ($\sim60$ per cent of the cluster member mass) for the $z=0.351$ source results in magnifications of $\sim1.3$ and $\sim1.7$, respectively.  This range of values, depending on the lens, is comparable to the possible magnification due to the cluster potential.  As this is only one of two SMGs with a spectroscopic redshift in our sample, we briefly outline its properties.  

The spectrum of the submillimetre counterpart contains a faint continuum and a single emission line detection of strong [OII] $\lambda3727\textup{\AA}$ emission.  The line flux of the [OII] emission was measured using bandpass measurement techniques performed on the rest-frame spectrum, adopting the [OII] bandpass of \cite{Fisher98} and a constant slit throughput of $\omega_{slit}=0.37$ \citep{Lemaux10}. This measurement resulted in a line luminosity of $L(\rm{[OII]})=2.75\times10^{41}$\,ergs s$^{-1}$, which translates to an [OII]-derived SFR of 3.85\,\myr\ using the relationship of \cite{Kennicutt98a}. Combining this measurement with the 24\,\um-derived SFR of 123\,\myr\ from the \cite{Chary01} SED templates results in an $E(B-V)_{s}=1.13$. This value is consistent with other measurements of IR-luminous star-forming galaxies at similar redshifts (e.g.\ \citealp{Kawara10}), thus increasing our confidence that this galaxy is correctly matched to the submillimetre emission.  As noted before in the literature (e.g.\ \citealp{Geach09}), the optical star formation estimate is drastically lower than that from the mid-IR.
\newline
\\
\textbf{SMM 2319.1}.  Radio, mid-IR, 450\,\um, and optical emission are all within 2.5\,arcsec of the 850\,\um\  (5.5$\,\sigma$) source (see Fig.~\ref{fig:gglensing}), though the correct ID is ambiguous.  The radio centroid lies slightly offset from the optical galaxy, as seen in the  \textit{HST} stamp (Fig.~\ref{fig:hstlensing}), and therefore might be a distinct source. Indeed, the IRAC emission looks like two blended sources: one coincident with the optical source, and the other associated with the radio source.  The bright optical galaxy has a photometric redshift consistent with the cluster at $z=0.868\pm0.139$ .  The IRAC colours further support membership as the source is consistent with the cluster redshift along the irregular starbursting galaxy track on the colour-colour diagram.  The MIPS emission is closer to the optical source than radio centroid, but could be from both given the large PSF.  A smoothed  \textit{HST} image (Fig.~\ref{fig:hstlensing}) reveals tentative evidence for faint optical emission at the radio position, which could indicate a background galaxy.  This therefore might represent another case of galaxy-galaxy lensing between a cluster member in the outskirts and a background radio source.  Conservatively assuming a redshift of $z=4$, a fairly massive lens with a velocity dispersion of $\sigma \sim 200$ km\,s$^{-1}$, and an SIS profile, we estimate a magnification of $\sim1.3$.  This modest magnification would be low compared to that from the cluster potential ($\sim2.3$) assuming the same source plane redshift of $z=4$.  We note, however, that the combination of the cluster mass distribution and galaxy-galaxy lensing can enhance the magnification further, as they do not add linearly.
\\

\subsection{The Possible Origin of Excess Counts in Cluster Fields}
Contamination by both submillimetre cluster members and strong lensing by individual massive galaxies (either within the cluster or along the line-of-sight) might drive excess source counts in cluster fields, given the observed evolution of cluster galaxies and the recent detection of lensed SMGs, respectively.  We describe each of these below.

The growth and evolution of galaxy clusters, through the accretion of galaxies, galaxy group infall, and cluster-cluster mergers, have drastic influences on populations within clusters and may temporarily spark extreme starbursts and fuel AGN activity \citep{BlainJames99, Owen99, Miller03a}.  Moreover, since clusters accrete populations from the surrounding field, any evolution in cluster galaxies may trace trends in the field (e.g.\ \citealp{Tran05}).

The first evidence for a rapid evolution of cluster galaxies emerged at optical wavelengths when \cite{Butcher78,Butcher84} observed a larger fraction of blue star-forming galaxies in moderate redshift galaxy clusters ($z\sim0.5$) compared to their local counterparts; this discovery has since been confirmed through many complementary techniques (e.g.\ \citealp{Dressler82, Couch87, Ellingson01}).  Local clusters, however, are dominated by passively evolving elliptical galaxies, suggesting that star-forming galaxies are transforming into the early-type population prevalent in clusters at $z\sim0$ \citep{Poggianti99}. Moreover, the field population, which feeds the infall, parallels this trend with an increasing star formation density with redshift (e.g.\ \citealp{Lilly96, Ellis96}).  

Since the most extreme activity in field and cluster galaxies may be enshrouded by dust (e.g.\ \citealp{Duc02}), recent studies have transitioned to longer wavelengths in the infrared and submillimetre to penetrate the heavy obscuration, and to more accurately measure the total star formation.  Thus, the now well-established increase in the number of dust-obscured submillimetre luminous field galaxies with redshift \citep{Smail97} may similarly be mirrored in cluster environments. Indeed, evidence for enhanced star formation activity in moderate and higher redshift clusters has recently been reported at mid-IR wavelengths \citep{Fadda00, Geach06, Bai07, Saintonge08, Haines09}. If some of these systems reach ULIRG-level luminosities they will be detectable with SCUBA at 850\,\um, and the trend should be even more apparent at the higher redshifts in our sample.  Therefore, any submillimetre cluster member should be removed from the source counts for direct comparison to the blank-field.

Given the recent detection of a population of gravitationally lensed dusty objects \citep{Negrello10, Vieira10} and the correlation between observed SMGs and low-redshift galaxies \citep{Almaini05, Austermann09}, the excess of counts at higher fluxes could also be due to contamination from lensing effects.  Although our sample only contains a few potential cases, they could considerably affect the shape due to the steepness of the counts curve \citep{Negrello10}.  

\begin{figure*}
   \centering
   \includegraphics[scale=0.5]{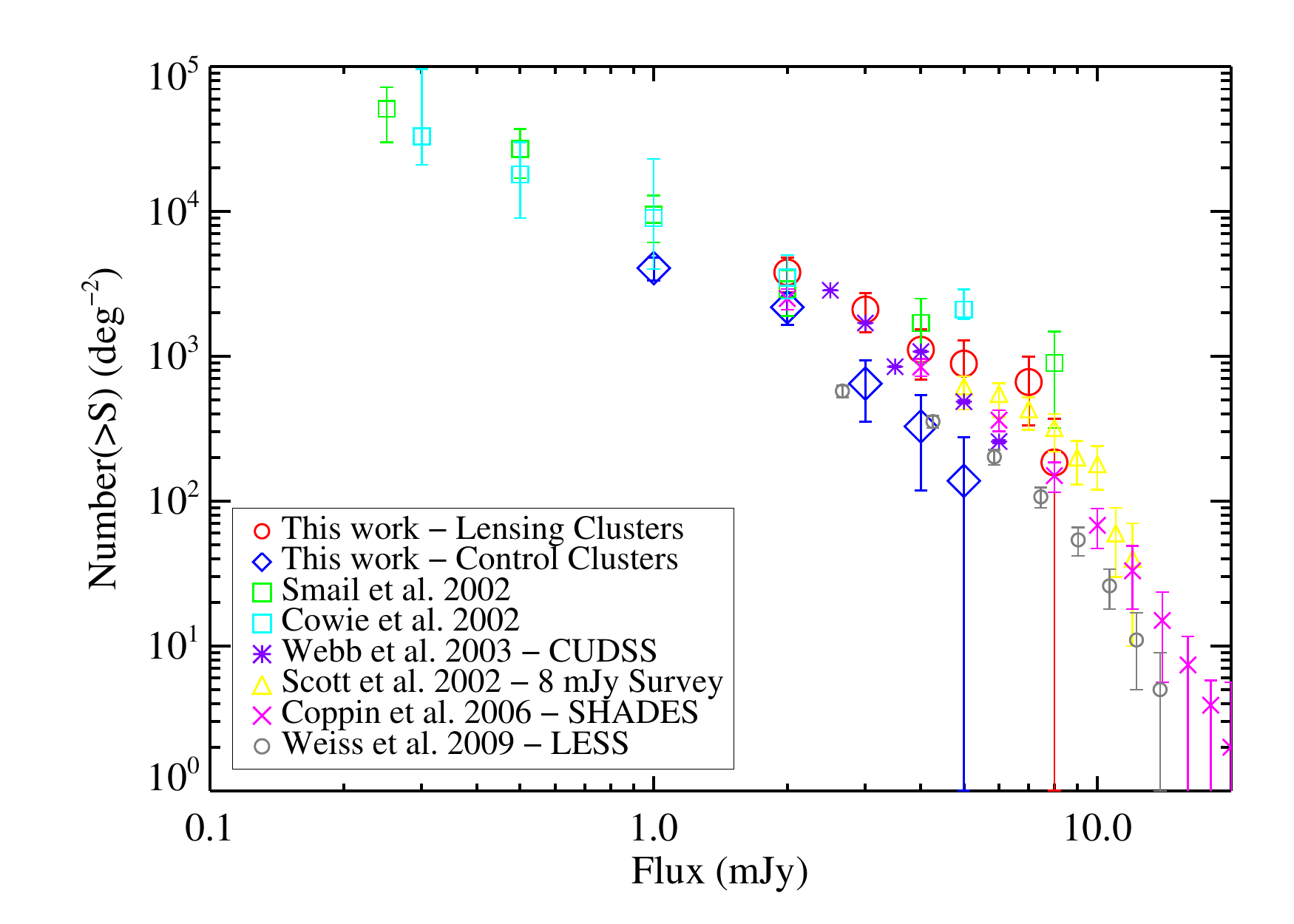} 
   \hspace{-5mm}
     \includegraphics[scale=0.5]{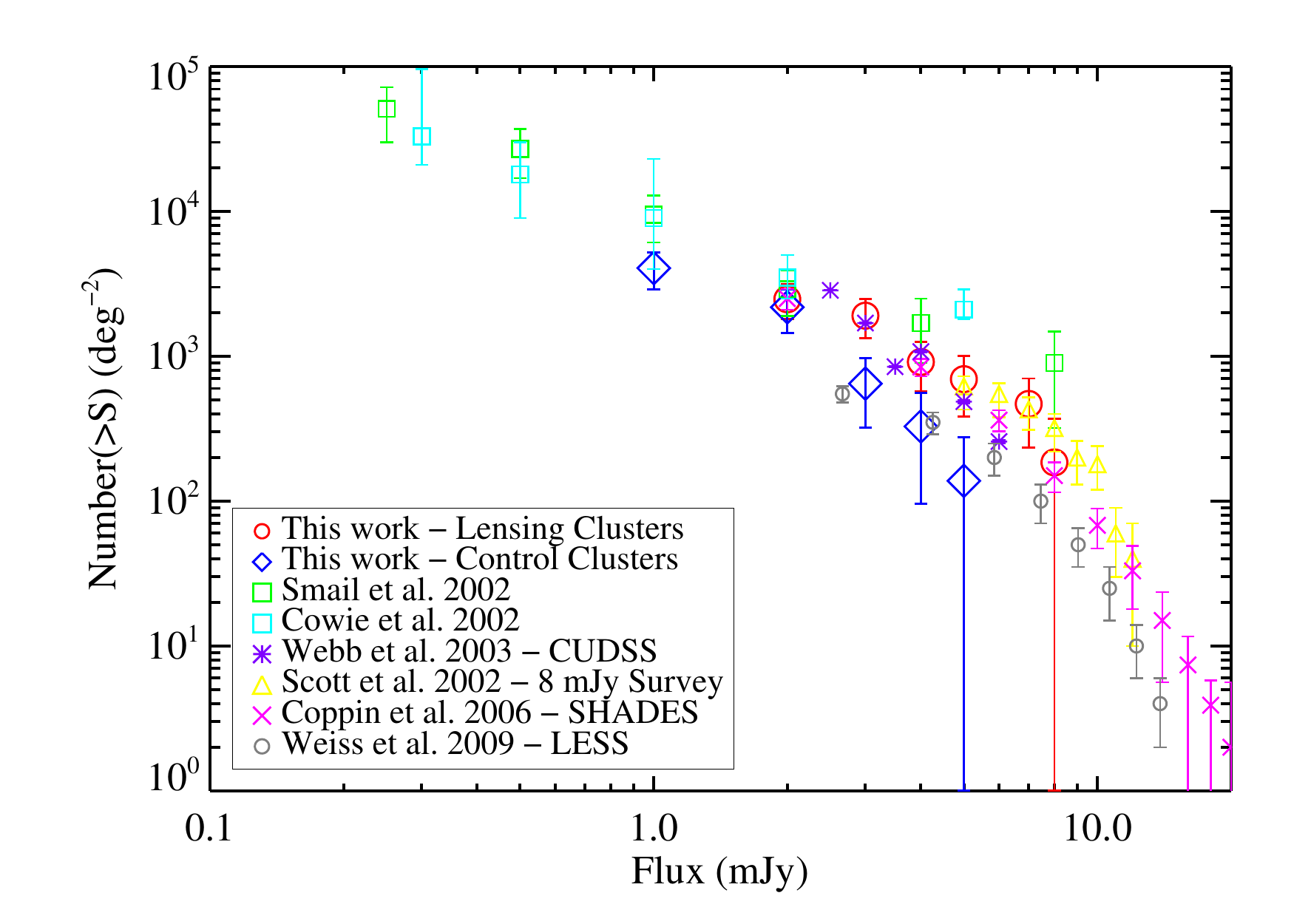} 
   \caption{[Left] Cumulative 850\,\um\  source counts, excluding the three likely cluster members.  Our results are denoted by the larger symbols, where the red circles represent counts from the strong lensing clusters and the blue diamonds represent counts for the control group. [Right] Cumulative 850\,\um\  source counts, excluding the three cluster members and two possible cases of galaxy-galaxy lensing.}
   \label{fig:counts_nonmemb}
\end{figure*}

To determine the potential origin of excess source counts in cluster fields, we have attempted to isolate these two possible drivers.  First, using the strongest member candidates (SMMs 0224.2, 0224.3, and 2319.3), we investigate the effects of cluster member contamination in our source counts by removing these three galaxies from the calculation.  The resulting cumulative counts (Fig.~\ref{fig:counts_nonmemb}) display minimal differences from the original counts.  As members are only found within the lensing sample (but, again, this may be biased by the more extensive follow-up data), the source counts for the control cluster remain unaffected.

We next exclude all possible cases of contamination from the source counts, including cluster members (SMMs 0224.2, 0224.3, and 2319.3) and possible examples of lensed SMGs (SMMs 0224.6 and 2319.1).  While this does bring down the cumulative counts of the lensing sample to blank-field levels, it is not sufficient to entirely remove the discrepancy between the two cluster samples.  Further high-resolution observations could isolate lensed from un-lensed SMGs by providing clear positional information and possibly verify the source of discrepancy between the counts in clusters and blank fields, and moreover between the two cluster samples.

We note that the other cluster surveys plotted \citep{Cowie02, Smail02} have removed central cluster galaxies from their source counts analysis and corrected for strong lensing due to the cluster potential, yet still show an excess compared to the blank field.  We do not attempt a lensing correction from the cluster potential, as the average lensing magnification is expected to be $\lesssim2$ beyond the central $\sim30$\,arcsec, even at the most extreme source plane redshifts.  In our lensing sample, the two SMGs most centrally located in the clusters are SMMs 0224.6 and 2319.3, both of which have already been excluded in the right panel of Fig.~\ref{fig:counts_nonmemb}.  Moreover, in the control group, only SMMs 1122.3 and 1326.1 (both within $\sim20$\,arcsec, projected) are likely to be significantly lensed (assuming source plane redshifts $z\gtrsim2.5$); removing these from the counts analysis would only amplify the discrepancy between the two samples.  Future cluster surveys should ensure all possible members are accounted for and possible lensing effects due to individual galaxies and the cluster potential are noted.  

The fact that all of our likely cluster members originate from lensing clusters might be indicative of dynamically unrelaxed states or physical mechanisms (e.g.\ mergers of cluster galaxies) that enhance activity and induce star formation, while temporarily boosting the lensing cross section.  Indeed, RCS 0224 represents one of highest known redshift systems with such bright and abundant lensed sources \citep{Gladders02,Gladders03}, and contains at least three optical arcs, the largest of which originates from a background source at $z=4.88$.  Moreover, preliminary measurements of the dynamical state from 117 member galaxies observed in the ORELSE survey show evidence for an unrelaxed cluster with significant substructure.  RCS 2319, a rare and distant supercluster, comprises three spectroscopically confirmed bound clusters, separated by a projected distance of $<3$\,Mpc, which are expected to merge by $z\sim0.5$ to form a massive structure with mass $\ge10^{15}$\,\Msol\ \citep{Gilbank08}.  The cluster with SCUBA coverage has three optical arcs \citep{Gladders03}, one of which belongs to a high-redshift Lyman break galaxy at $z=3.86$.  

However, the lensing clusters seem to be fairly relaxed systems in terms of their X-ray properties; in particular, RCS 1419 is compact with very regular X-ray morphology.  On the other hand, RCS 2318 of the control sample shows significant substructure at X-ray wavelengths \citep{Hicks08}.  Perhaps then, the dynamical distinction between the samples is that of a relaxed cluster with a sub-structured collapsed core (i.e.\ the super-lenses) versus a wholly unrelaxed cluster.  The former may therefore be a signpost for both enhanced lensing due to a boosted central lensing cross-section and increased star formation in the form of excess SMGs, for example.

\section{Conclusions}
\label{sec:conclusions}
We have presented a study of seven galaxy clusters between $0.640<z<1.045$, three of which display strong gravitational arcs.  Within these fields, which cover a total of 47 arcmin$^2$, we have detected 26 SMGs and SMG candidates at 850\,\um\  with S/N $\ge 3.0$.  We summarize our results as follows:
\begin{enumerate}

\item{As with other cluster surveys, we see an excess of SMG source counts compared to the blank field; however, this trend is limited to our strong lensing sample.  The control sample actually shows a slight deficit of counts, though consistent with the results from the LABOCA ECDFS Submillimetre Survey (LESS); the ECDFS has been shown to be an underabundant field at many wavelengths (\citealp{Weib09} and references therein).  There is a $2.7\,\sigma$ excess of 850\,\um\  source counts in the lensing sample compared to the control clusters.}

\item{We utilize mid-IR, radio, optical, and X-ray catalogues to determine counterpart emission to the SMGs, thereby improving the SMG astrometry and, in some cases, providing redshift estimates.  We classify our identifications as secure, ambiguous, or tentative, and identify 12, 6, and 5 SMGs for each classification, respectively.  Three SMGs lack any plausible counterpart emission and are therefore indicative of either spurious sources or high-redshift galaxies.}

\item{Employing infrared colour criteria and locating optical counterparts with photometric redshifts allows us to identify three likely cluster members: SMMs 0224.2, 0224.3, and 2319.3.  We calculate infrared luminosities of $\sim 10^{12}$\,\Lsol, consistent with ULIRG levels.  Assuming no AGN contamination, all three members have extreme star formation rates of $\sim 700$--$1000$\,\myr, but we argue that the most luminous of the three probably contains an AGN.  The extreme luminosities and star formation rates of these ULIRG-level sources are not typical in cluster environments, which are usually populated with gas-poor spirals or old elliptical galaxies \citep{Sanders96} at low redshift.  Therefore, these cluster members may reflect the increased infall and activity that is seen in clusters at higher redshift with the Butcher-Oemler effect and similar mid-IR studies. All three likely cluster members derive from strong lensing clusters that display evidence of additional cluster-wide activity: the RCS 2319 supercluster comprises three merging systems that give rise to disturbed galaxy morphologies;  it is conceivable that RCS 0224 might be undergoing similar activity, although thus far, current measurements in the ORELSE survey only hint at enhanced substructure.}

\item{We observe two possible cases (SMMs 0224.6 and 2319.1) of galaxy-galaxy lensing between an optical foreground galaxy (or cluster member) and a background SMG.  Follow-up high-resolution observations or the detection of CO emission lines are essential to verify this claim, but beyond the scope of our current work.}

\item{While there is some evidence for cluster member contamination in the source counts, it is not sufficient to remove the excess from the lensing fields.  We note, however, that the effect of cluster contamination is only a lower limit as we are biased by the quantity of the multi-wavelength follow-up.  Galaxy-galaxy lensing can also contribute to source count contamination at higher fluxes, but the effect on our counts must be verified through follow-up observations on the possible lensed systems.  We suggest that the dynamical state of the cluster might account for any additional source count discrepancies between the two cluster samples, as a sub-structured core, of an otherwise relaxed cluster, could boost central lensing efficiency while triggering extreme starbursts and driving the formation of ULIRGs within cluster fields.}

\item{While we have not unambiguously accounted for the excess source counts between the two cluster samples, we have highlighted possible problems for cluster surveys, including increased cluster member contamination at higher redshifts and potentially lensed SMGs.  Further work is required to fully isolate the driver of contamination, such as high-resolution Submillimetre Array imaging, CO emission line observations, and homogeneous radio coverage.  Although this study suffers from both small number statistics and difficulties associated with the SMG counterpart identification process, it still has important ramifications for intrinsic submillimetre source counts and SZ cosmological studies that are concerned with point source contamination.  We therefore urge caution for future large-scale cluster surveys, for instance with the Herschel telescope and the forthcoming SCUBA-2 instrument, and stress the importance of accurate lens models and proper subtraction of cluster members for counts analyses.}
\end{enumerate}

\section*{Acknowledgments}
We thank the referee, Stephen Serjeant, for thorough comments and suggestions which improved the manuscript.  We also thank Kristen Coppin for many useful discussions, particularly on SMG counterpart identification.  Tracy Webb acknowledges the support of the Natural Sciences and Engineering Research Council of Canada's Discovery Grant program, and \'Etablissement de Nouveaux Chercheurs program of Le Fondes Qu\'ebecois de la Recherche sur la Nature et les echnologies.  This work is based (in part) on observations made with the \textit{Spitzer Space Telescope}, which is operated by the Jet Propulsion Laboratory, California Institute of Technology under a contract with NASA.  The JCMT is operated by The Joint Astronomy Centre on behalf of the Science and Technology Facilities Council of the United Kingdom, the Netherlands Organisation for Scientific Research, and the National Research Council of Canada.

\bibliography{mnemonic,references}
\bibliographystyle{apj}

\appendix
\section{A discussion of individual sources}
\label{app:cases}
Here we provide a discussion of each submillimetre source detected in the survey.  We reiterate that individual sources detected below 3.5$\,\sigma$ have a high probability of being spurious, making counterpart identification rather uncertain.  Nevertheless, these sources are included in the discussion for completeness.  Table \ref{tab:counters} summarizes our multi-wavelength counterpart analysis and Figures \ref{fig:secure}-\ref{fig:hstall} provide visual support for each description.  

\subsection{Secure counterparts}
\noindent\textit{\textbf{SMM 0224.2}}.\  See \S\ref{sec:members}.

\noindent\textit{\textbf{SMM 0224.3}}.\  See \S\ref{sec:members}.

\noindent\textit{\textbf{SMM 0224.4}}.\  We find secure MIPS, radio, IRAC and 450\,\um\  counterparts within the 10\,arcsec search radius.  The 450\,\um\ emission (3.3$\,\sigma$) is isolated, offset in declination from the other three sources.  The MIPS source is 5.6\,arcsec away from the 850\,\um\  emission but within 0.5\,arcsec of the radio and IRAC positions.  Based on positional offsets to the MIPS emission and infrared colours, the more northern of the two blended IRAC sources (shown in the 3.6\,\um\  grayscale image) is the more likely SMG counterpart.  There is also a faint $K$-band detection at the radio centroid with extremely red colours (see \cite{Webb05}), consistent with a high-redshift galaxy.  It is too faint to for sufficient detection in optical bands, and we must therefore rely on the redshift deduced from the radio flux, $z=2.56$, making the SMG consistent with a background galaxy.   \textit{HST} imaging (see Fig.~\ref{fig:hstall}) reveals two optical sources a few arcseconds south of the radio centroid (also seen in the $z$' image).  The western source is likely a star, while the eastern detection could be a foreground galaxy acting as a lens to the background radio source, but further observations, beyond the scope of this paper, are required.

\noindent\textit{\textbf{SMM 0224.6}}.\ See \S\ref{sec:gglensing}.

\noindent\textit{\textbf{SMM 1122.2}}.\  Within 10 arcsec, only one 24\,\um\  source and one optical source are detected.  The MIPS source is offset by 0.7\,arcsec and therefore provides a secure identification.  The optical source lacks a reliable photometric redshift ($z=1.073\pm0.803$), so we cannot determine cluster membership for this SMG, but  \textit{HST} imaging (see Fig.~\ref{fig:hstall}) reveals a faint compact source.  Of the four IRAC detections, only one is significant at 3.6\,\um\  and coincides with the MIPS and optical emission.  This source lies within the likely SMG counterpart region of the colour-magnitude diagram ($z>1.5$) and within the colour-colour boundary for SMGs.

\noindent\textit{\textbf{SMM 1122.3}}.\  An extremely bright MIPS source (1096.4\,\uJy) lies 7.5\,arcsec away with coincident optical and IRAC sources.  Slightly further out (9.9\,arcsec from the SMG) is 450\,\um\  emission detected at a S/N of 5.63.  The optical counterpart lacks a photometric estimate, but the IRAC colours suggest this SMGs lies at $z<1.5$.  The  \textit{HST} image (Fig.~\ref{fig:hstall}) shows a likely foreground galaxy, possibly a face-on spiral with a central nucleus.  We therefore conclude this is a secure ID, and most likely at lower redshift.

\noindent\textit{\textbf{SMM 1326.1}}.\  MIPS, IRAC, and 450\,\um\  (6.2$\,\sigma$) emission all exist within 5\,arcsec of the submillimetre source, offering a robust counterpart identification, but there is no visible optical companion.  The IRAC source falls within the colour-colour boundary from \cite{Yun08} and within the colour-magnitude selection criteria for SMG counterparts proposed by \cite{Hainline09}, and therefore likely at $z>1.5$, which would also explain the lack of an optical detection.

\noindent\textit{\textbf{SMM 1326.2}}.\ The total 24\,\um\  flux density within this search radius is significantly dominated by one MIPS source (249\,\uJy).  We also detect X-ray emission at 5.6$\,\sigma$, coincident (offset by 1.1 arcsec) with the brighter MIPS source.  As we do not have any photometric redshifts for this entire cluster field, we cannot determine cluster membership, but there is a faint optical counterpart coincident with the secure ID.  The two IRAC sources associated with each MIPS object are strongly detected in every channel except at 8.0\,\um, so we can only place an upper limit on the $S_{8.0\,\mu m}/S_{4.5\,\mu m}$ colour.  However, the redder $S_{5.8\,\mu m}/S_{3.6\,\mu m}$ colour of the IRAC source associated with the X-ray and brighter MIPS emission places the object in the likely SMG counterpart region, making this a secure identification, likely at high redshift.

\noindent\textit{\textbf{SMM 1326.3}}.\  We find a single MIPS source (coincident with IRAC and optical emission) with a 2.6\,arcsec offset from the SMG, providing improved positional accuracy for the submillimetre emission.  The associated IRAC source secures the identification as it falls within the colour-magnitude selection criteria given by \citealp{Hainline09}.

\noindent\textit{\textbf{SMM 1419.1}}.  A single faint MIPS source (63\,\uJy) is located 4.8\,arcsec away and offers better astrometry for the submillimetre source.  The IRAC source associated with the MIPS emission is weakly detected, so we can only estimate a limit for its infrared colours.  While the other IRAC source in the field is brighter in all channels (except at 8.0\,\um\ ), its infrared colours are inconsistent with an SMG, so we classify the MIPS/IRAC detection as the more likely counterpart.  A smoothed  \textit{HST} image (see Fig.~\ref{fig:hstall}) reveals a two possible optical counterparts: one just southwest of the MIPS centroid, and another fainter source aligned with the IRAC position.  We note that there are no photometric redshifts available for this entire cluster.  

\noindent\textit{\textbf{SMM 1419.4}}.\  We detect a 111\,\uJy\ MIPS source within the search area, altering the submillimetre astrometry by 6.4 arcsec.  There is also a coincident IRAC source, but weakly detected at 5.8\,\um\  and without a detection at 8\,\um, preventing a reliable colour analysis.  The  \textit{HST} image (Fig.~\ref{fig:hstall}) shows a slight extension in the optical emission.  As there is only one MIPS source in the entire search radius, we classify this as a secure identification.

\noindent\textit{\textbf{SMM 2319.2}}.\ We locate 24\,\um\  emission at 2.5\,arcsec from the SMG.  Corresponding 450\,\um\  emission (4.4$\,\sigma$) lies 1.1\,arcsec beyond the MIPS source, yielding a secure detection for this 7.5$\,\sigma$ SMG.  The associated IRAC source to the MIPS emission falls within the SMG counterpart boundaries on the colour-magnitude diagram (which typically indicates $z>1.5$) and colour-colour diagram. Besides the star to the northwest, there is only a faint optical source (see Fig.~\ref{fig:hstall}) slightly offset from the IRAC position, possibly indicative of a high-redshift source.

\subsection{Ambiguous counterparts}

\noindent\textit{\textbf{SMM 1122.1}}.\  We discover two MIPS sources, each with a coincident optical and IRAC source, and both within 5\,arcsec of the 850\,\um\  emission, but this identification is complicated by a third MIPS/IRAC detection, blended with the northern MIPS source.  Based on IRAC colours (although somewhat unreliable due to the confusion), the central IRAC source is the most likely counterpart.  Deep  \textit{HST} imaging reveals faint optical emission between the two bright optical sources; this source becomes even more obvious after smoothing the image with a 0.3\,arcsec Gaussian (see Fig.~\ref{fig:hstall}), and is probably a high-redshift source.  Photometric redshifts for the two bright optical sources place them in the foreground at $z=0.282\pm0.023$ (northern) and $z=0.30\pm0.032$ (southern).  We therefore classify this ID as ambiguous, but a possible example of galaxy-galaxy lensing between the northern foreground optical source and the central IRAC/MIPS/optical emission (our most likely SMG ID).

\noindent\textit{\textbf{SMM 1326.4}}.\  Confused MIPS sources populate the area surrounding this SMG, complicating the ID.  However, there is only one distinct IRAC source with coincident X-ray emission as well.  The red $S_{5.8\,\mu m}/S_{3.6\,\mu m}$ colour places this source in a SMG region of the colour-magnitude plot ($z>1.5$), supporting the MIPS and X-ray emission as the most likely identification.  The infrared colours are within the colour-colour boundary, specifically along the power-law AGN region.  Furthermore, the MIPS-to-IRAC colours are indicative of an AGN component with $S_{24\,\mu m}/S_{8\,\mu m}=1.7$ and $S_{8.0\,\mu m}/S_{4.5\,\mu m} =3.1$ \citep{Ivison04}.  Due to the blended MIPS emission, we classify this ID as ambiguous, but most likely a background source as it lacks optical emission.

\noindent\textit{\textbf{SMM 2152.1}}.\  Although we find two MIPS sources within the 10\,arcsec search radius, one is clearly a few sources blended together.  While slightly less confused, the IRAC emission can also not be separated into distinct sources for reliable colour estimates.  The sole IRAC source associated with the fainter MIPS emission in the northeast has colours consistent with an elliptical galaxy, making it an unlikely SMG candidate. \textit{HST} imaging (see Fig.~\ref{fig:hstall}) reveals faint emission to the southwest of the dominating foreground object.  This therefore might be another case of galaxy-galaxy lensing, making the most likely submillimetre ID the background lensed galaxy, but still rather ambiguous.

\noindent\textit{\textbf{SMM 2152.2}}.\  Severely blended MIPS emission is coincident with optical and blended IRAC sources.  The bright optical source is a foreground galaxy ($z=0.301\pm0.030$), but a fainter source evident in the \textit{HST} image (Fig.~\ref{fig:hstall}) has a more uncertain redshift of  $0.827\pm0.468$, which tentatively agrees with the cluster redshift.  The IRAC and MIPS emission are too confused for reliable photometry.  We classify this as a possible galaxy-galaxy lensing pair, with the background source as the more likely SMG counterpart, but the ID remains ambiguous.

\noindent\textit{\textbf{SMM 2319.1}}.\ See \S\ref{sec:gglensing}.

\noindent\textit{\textbf{SMM 2319.3}}.\ See \S\ref{sec:members}.

\subsection{Tentative counterparts}
\noindent\textit{\textbf{SMM 0224.5}}.\  Three MIPS sources are detected within the 10\,arcsec search radius of this SMG.  The brightest source (225\,\uJy) dominates the other two (both are $<90\,\mu$Jy) by a factor $\sim$2.5 and we therefore adopt this source as the primary contributer to the 850\,\um\  emission. It is offset in right ascension from the SMG position by 7.7\,arcsec and in proximity of two optical sources.  The closest (0.8\,arcsec away) has a photometric redshift estimate of $z=0.943\pm0.074$, likely indicating a background galaxy, but consistent with the cluster redshift at just over the 2$\,\sigma$ level.  The corresponding IRAC source lies within error of SMG counterpart region on the colour-colour plot.  Emission at 450\,\um\  also exists, closest to a weaker 24\,\um\  source, but we emphasize the high positional uncertainty of such a low 450\,\um\  detection (2.8$\,\sigma$) and assume it too is associated with the brightest MIPS source.  Due to the multiple MIPS detections, we label this identification as tentative, but place a higher counterpart likelihood on the brightest 24\,\um\  source (most eastern).  Given its low S/N at 850\,\um\ (3.4$\,\sigma$), it could also be a blend of multiple submillimetre sources.

\noindent\textit{\textbf{SMM 1419.3}}.\  We find two MIPS sources, each with an IRAC detection, and 450\,\um\  emission (at 2.7$\,\sigma$) neighbouringing the SMG.  The MIPS sources are of roughly equal flux densities ($\sim55\,\mu$Jy), but one source is closer to both the 450 and 850\,\um\  emission.  The IRAC source associated with the closer MIPS position also has infrared colours more appropriate for an SMG.  The source is firmly within the colour-magnitude boundary and situated along the power-law AGN track in the colour-colour diagram.  The \textit{HST} image (Fig.~\ref{fig:hstall}) reveals a compact optical object associated with the MIPS/IRAC emission, making this a likely, but tentative counterpart.

\noindent\textit{\textbf{SMM 2318.1}}.\  We detect strong 450\,\um\  emission (7.1$\,\sigma$) coincident (2.2\,arcsec offset) with this extremely bright 850\,\um\  SMG (14.9\,mJy with a S/N of 17.0).  There is also a tentative optical counterpart close to the 450\,\um\  emission in the \textit{HST} image with a disturbed morphology (see Fig.~\ref{fig:hstall}).  Given the steep drop in SMG source counts at flux levels greater than $\sim10$\,mJy, such a bright source is fairly unique and could represent a strongly gravitational lensed SMG.  Alternatively, it might be multiple detections of distinct objects; indeed, the 450\,\um\  emission is fairly elongated (see Fig.~\ref{fig:850mapscontrol}) and possibly double-peaked.  Further high-resolution observations are needed to verify either possibility.  

\noindent\textit{\textbf{SMM 2318.2}}.\  Emission at 450\,\um\  (3.0$\,\sigma$) lies within 2.2\,arcsec of the submillimetre 850\,\um\  source (5.6$\,\sigma$).  The \textit{HST} image (Fig.~\ref{fig:hstall}) displays a few possible counterparts.

\noindent\textit{\textbf{SMM 2318.3}}.\  This SMG has 450\,\um\  emission, but is detected just within the search radius for 450\,\um\  sources at 12.0 arcsec.  \textit{HST} imaging (see Fig.~\ref{fig:hstall}) reveals several tentative optical counterparts, but this SMG lacks any secure identification.

\subsection{Insecure counterparts or possible spurious sources}

\noindent\textit{\textbf{SMM 0224.1}}.\  The robust 850\,\um\  detection of this source (6.1$\,\sigma$) means the assigned search radius is fairly small (7.4 arcsec).  This source only has candidate counterpart emission at 450\,\um\  (3.7$\,\sigma$ at 4.1\,arcsec away) and two possible optical counterparts.  Due to the high probability of detecting optical sources within this search area, we are unable to assign it a unique identification without a supporting radio or MIPS detection.  We note, however, that there are radio and mid-IR sources (each with coincident IRAC and optical emission) just beyond the search radius with offsets of 9.9 and 8.0 arcsec, respectively, but on opposite sides of the SMG.  Given that the search radius is a 95 per cent confidence interval, it is possible, though unlikely, that one of these objects is the counterpart.   For now, however, we conclude that this source has no secure counterpart, perhaps due to a high redshift.

\noindent\textit{\textbf{SMM 1419.2}}.\  Weakly detected at 850\,\um\  (3.0$\,\sigma$), this source only has counterpart emission at IRAC and optical wavelengths (but no photometric redshifts available).  The brightest IRAC source at 3.6\,\um\  is an unlikely SMG candidate based on its infrared colours, and most likely a star based on the \textit{HST} image (Fig.~\ref{fig:hstall}).  The best suited IRAC source is the most southern detection and is located within both appropriate SMG boundaries on the colour-colour and colour-magnitude diagrams.  There is an \textit{HST} source nearby that might be associated with the faint IRAC detection, however, without an associated MIPS source, we classify this identification as insecure.

\noindent\textit{\textbf{SMM 1419.5}}.\  This source evades counterpart emission at all other wavelengths, making it a possible candidate for one of the spurious sources.   We note there is MIPS, IRAC and faint radio emission in the southwest corner, but beyond the assigned search radius of 8.0\,arcsec.  Given the high detection at 850\,\um\ ($6\,\sigma$), this source could instead be a high redshift SMG, and therefore not sensitive to counterpart emission.

\begin{figure*}
   \centering
   \vspace{-4mm}
      \subfigure{\includegraphics[scale=0.42]{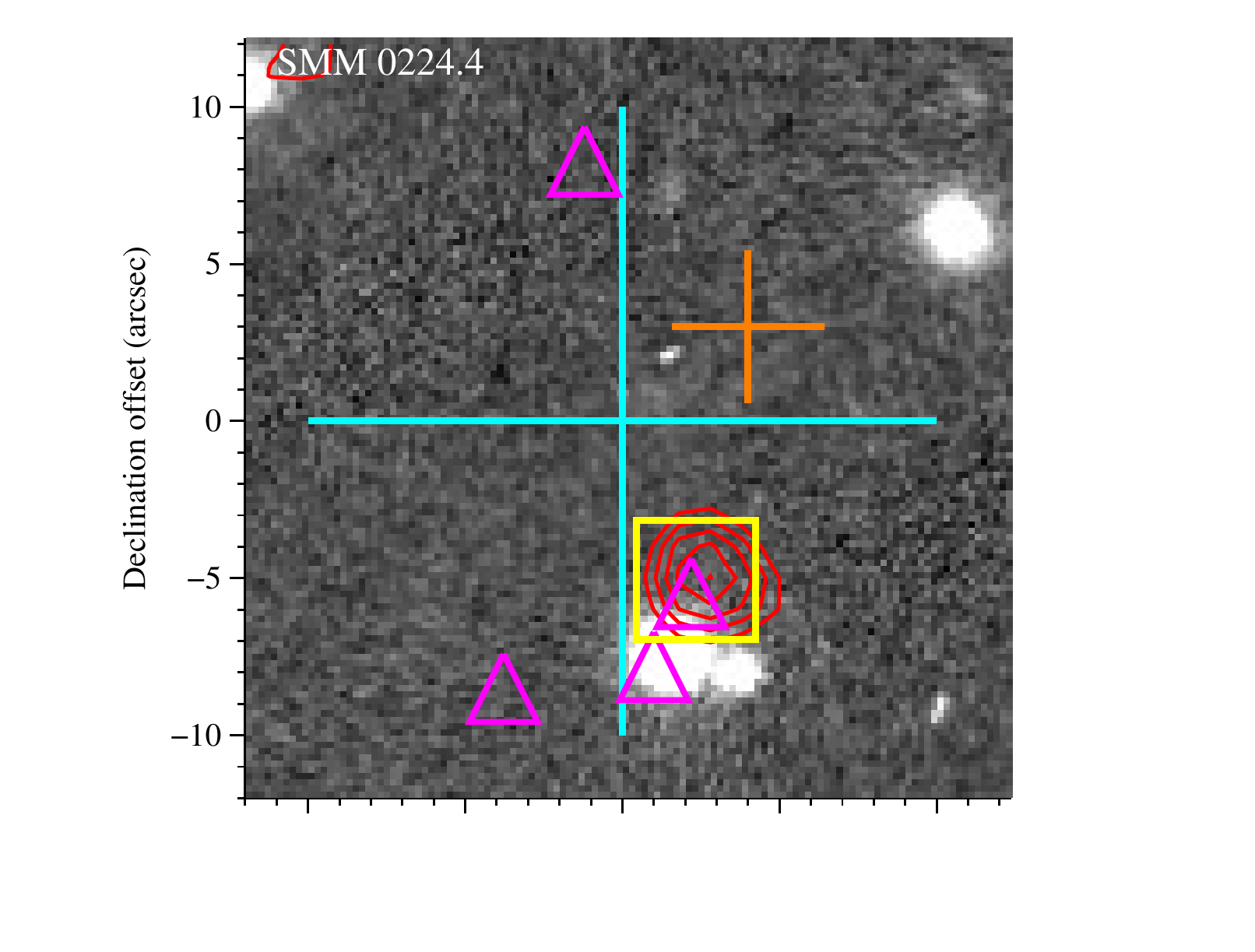}}
   		\hspace{-27mm}
   \subfigure{\includegraphics[scale=0.42]{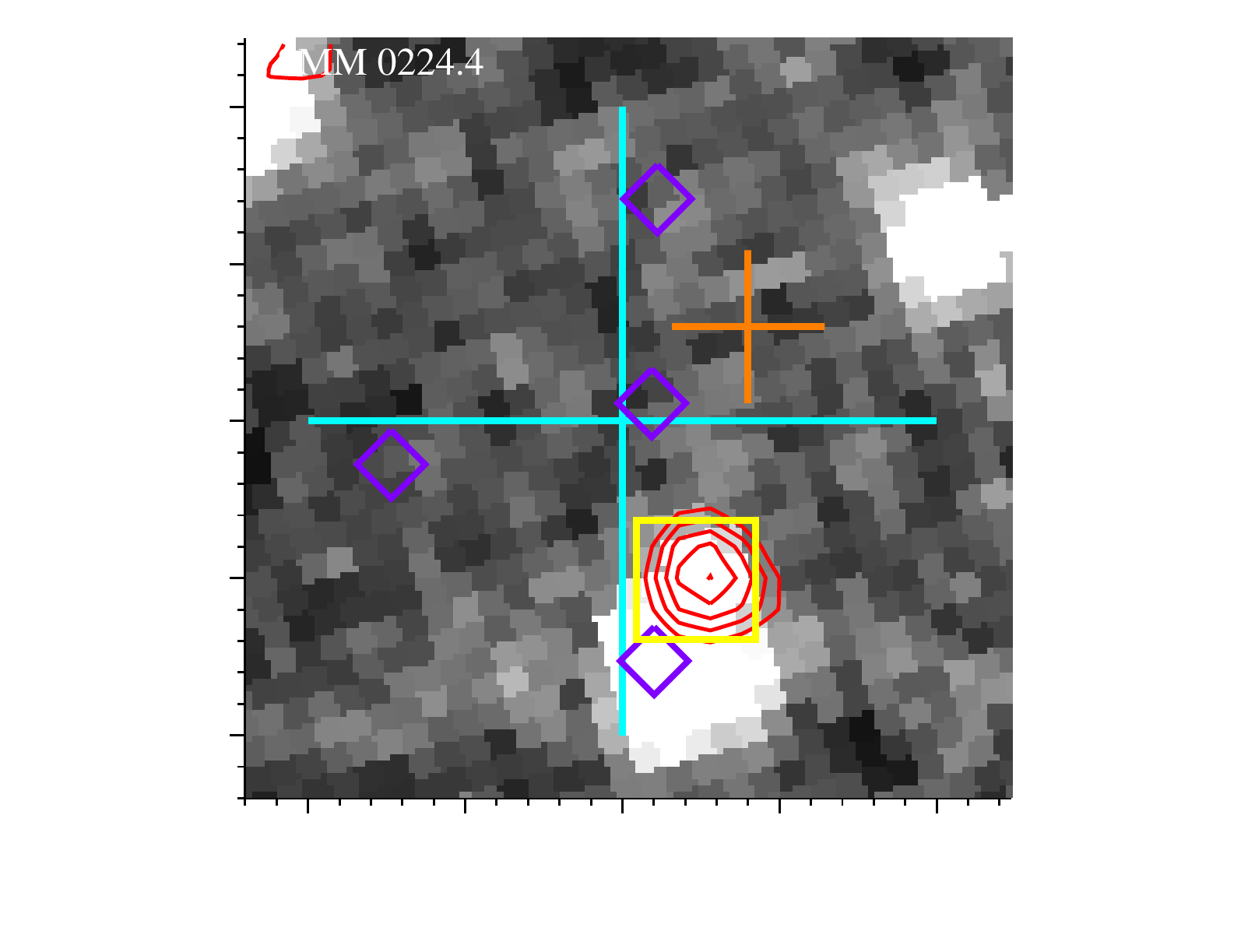}}
		 \hspace{-27mm}
  	   	 \vspace{-12mm}
     \subfigure{\includegraphics[scale=0.42]{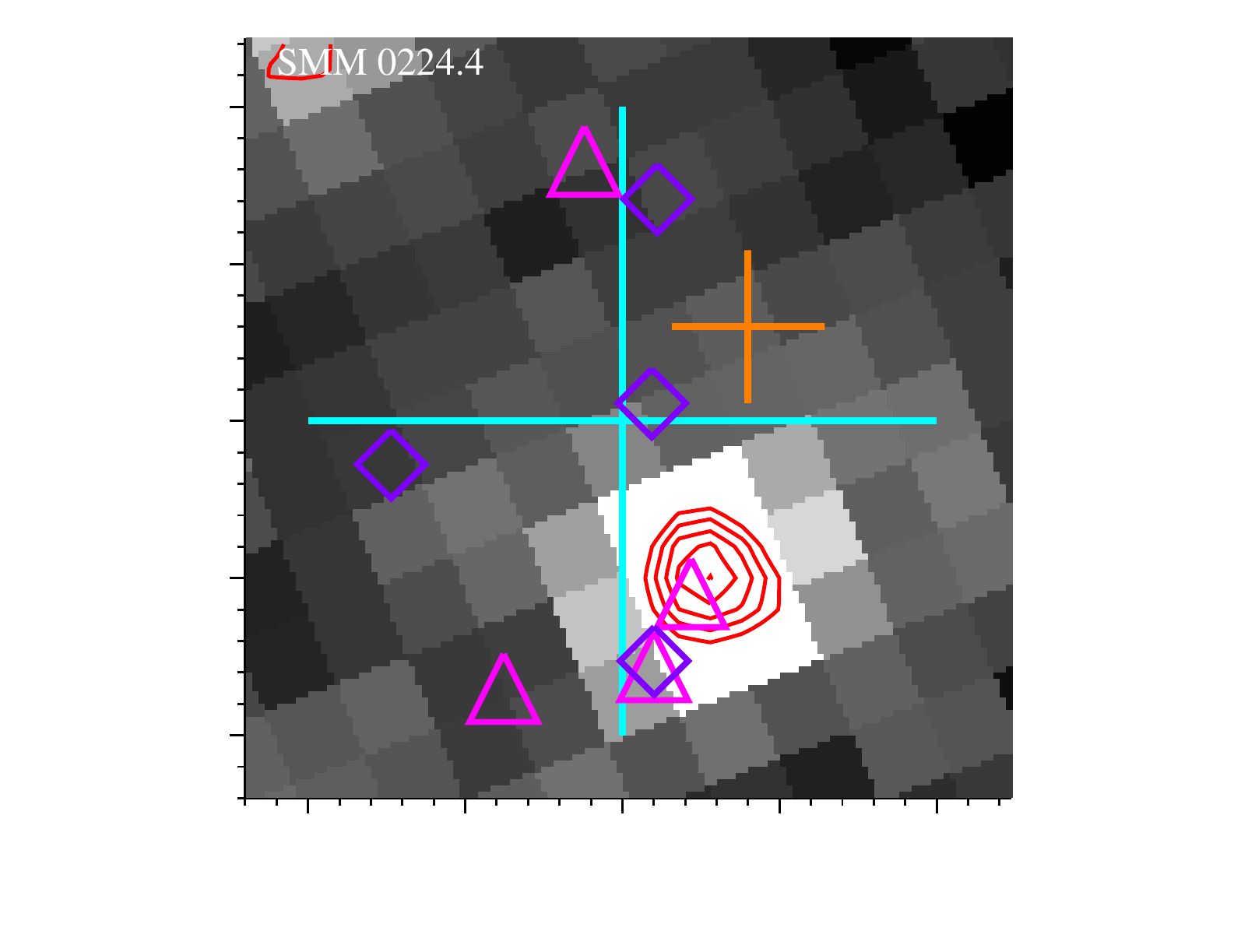}}
       \subfigure{\includegraphics[scale=0.42]{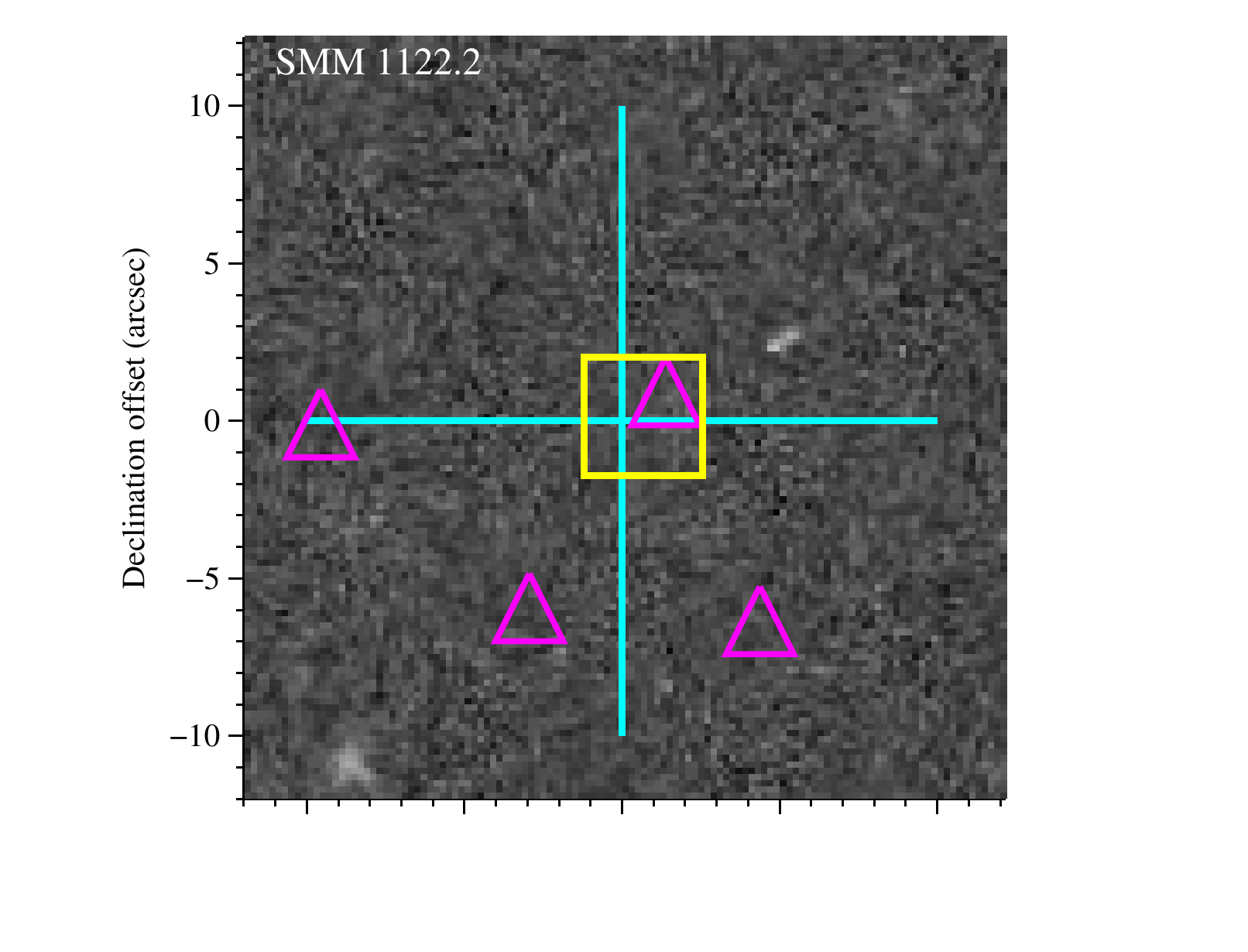}}
   		\hspace{-27mm}
     \subfigure{\includegraphics[scale=0.42]{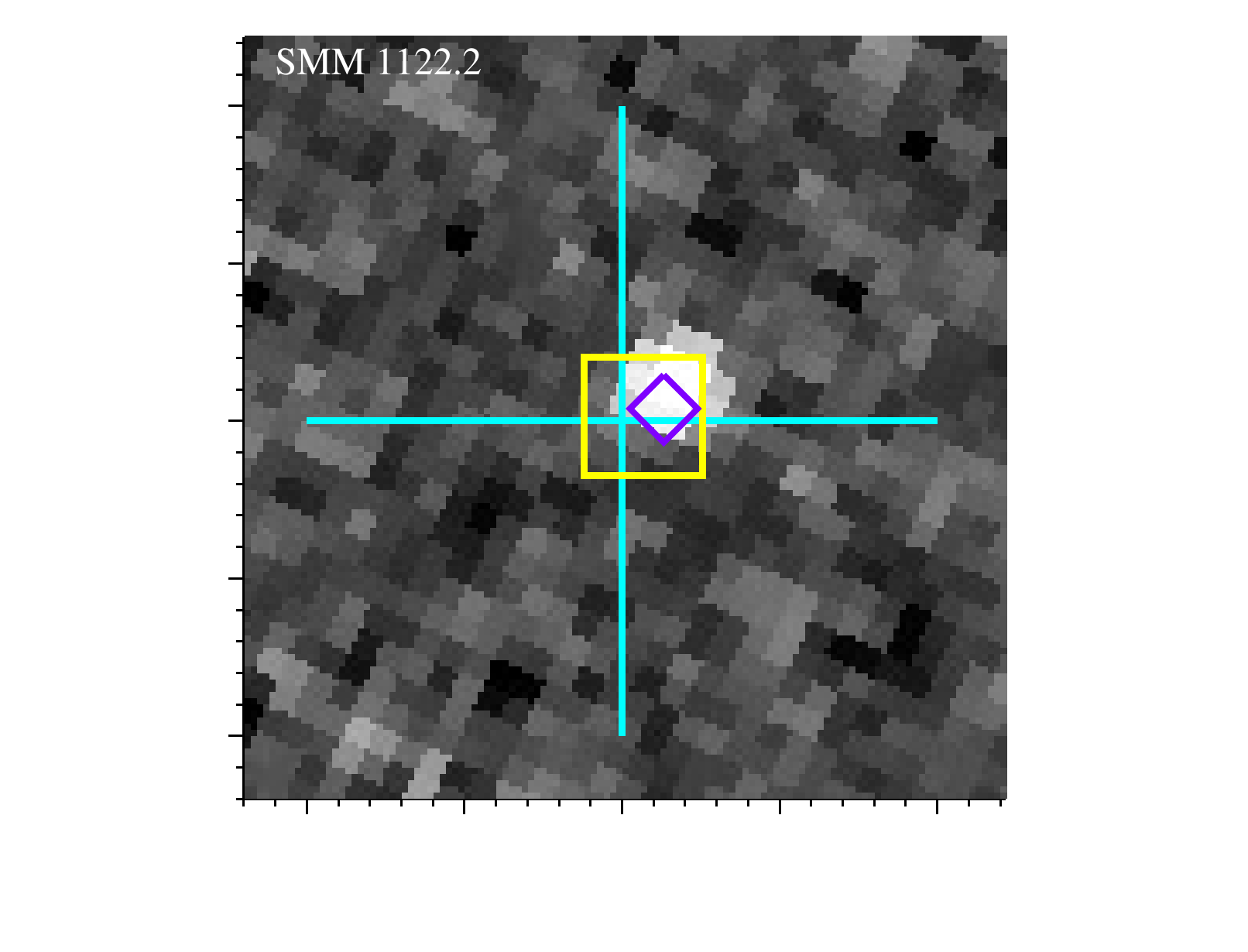}}
    		\hspace{-27mm}
  	   	 \vspace{-12mm}
      \subfigure{\includegraphics[scale=0.42]{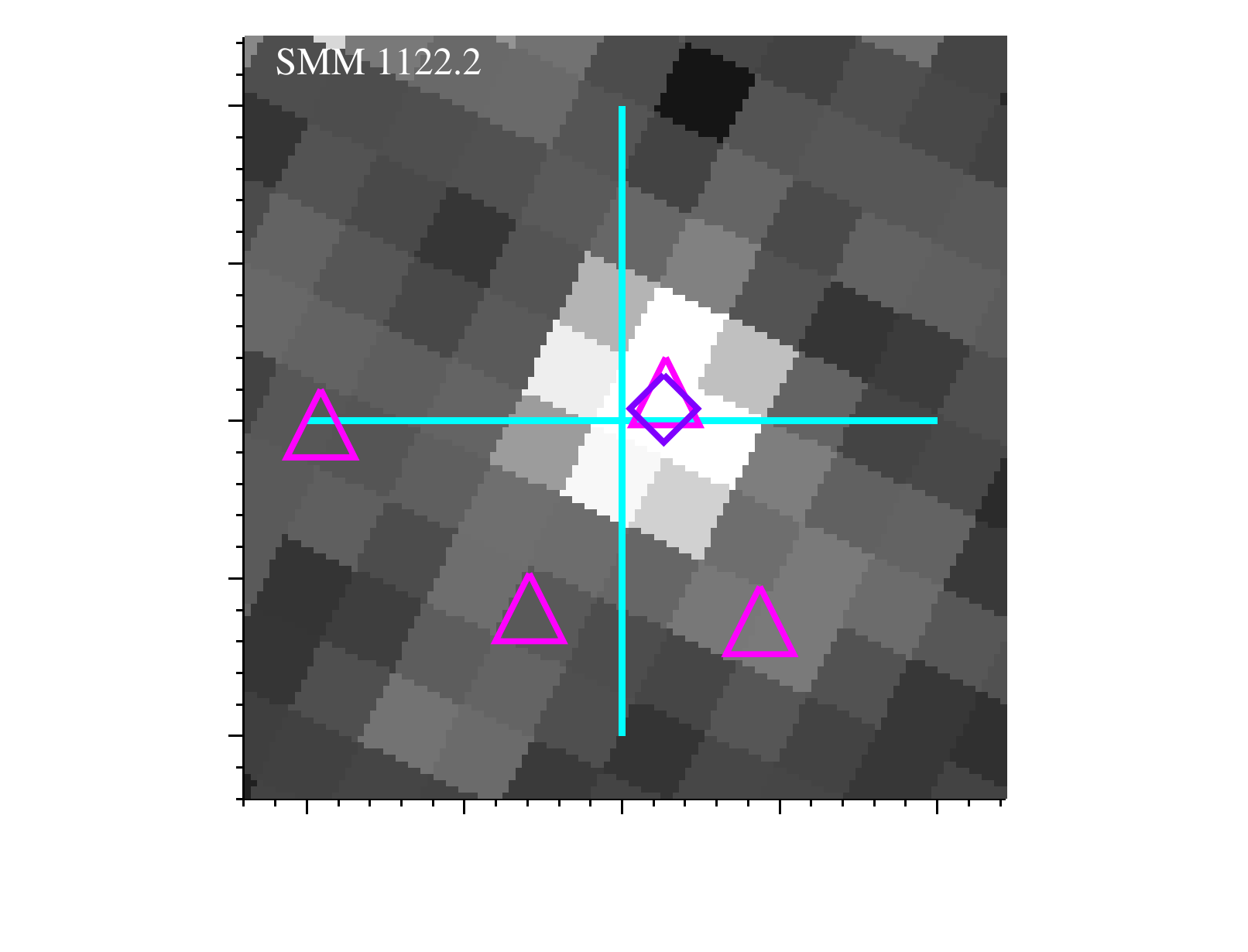}}
        \subfigure{\includegraphics[scale=0.42]{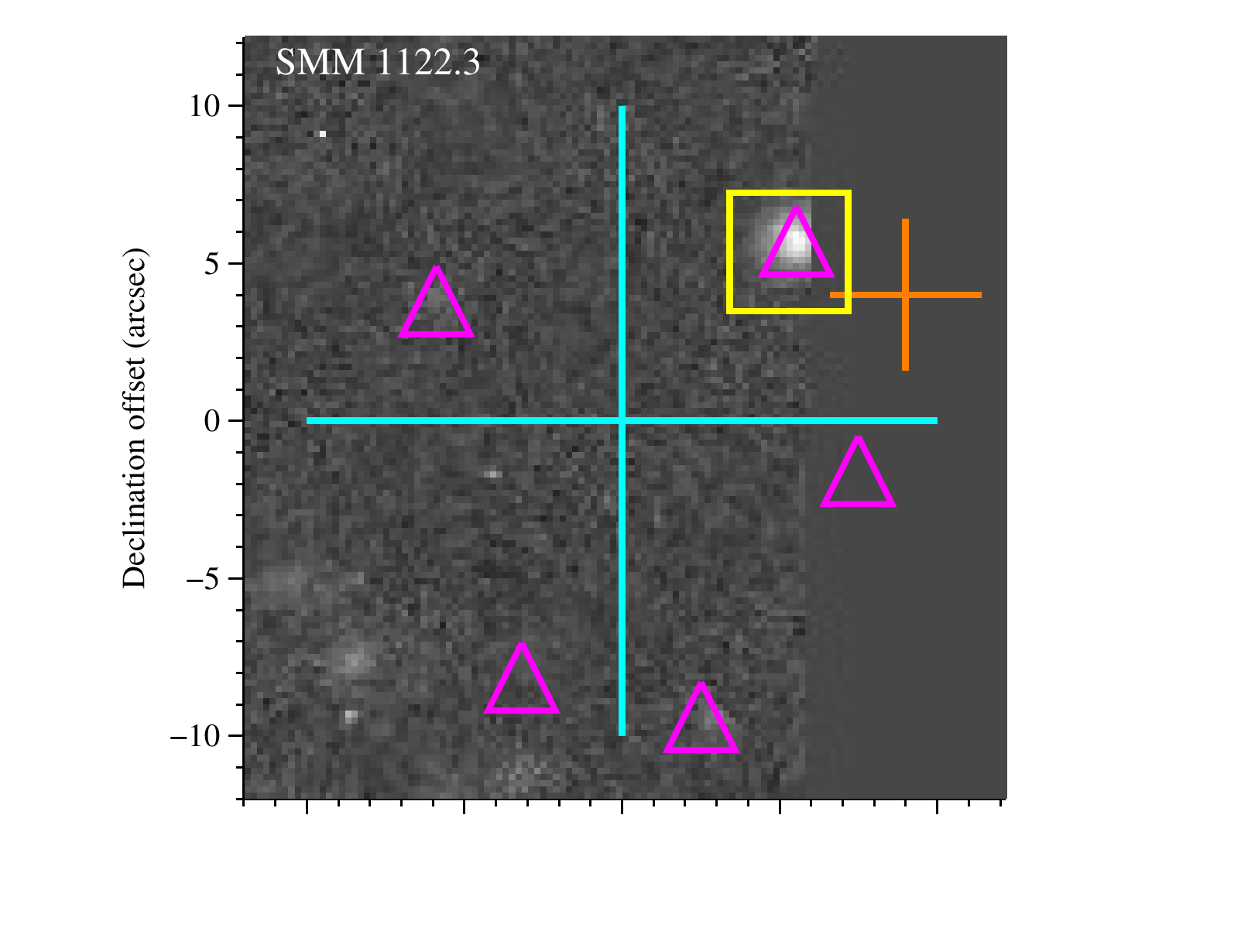}}
   		\hspace{-27mm}
     \subfigure{\includegraphics[scale=0.42]{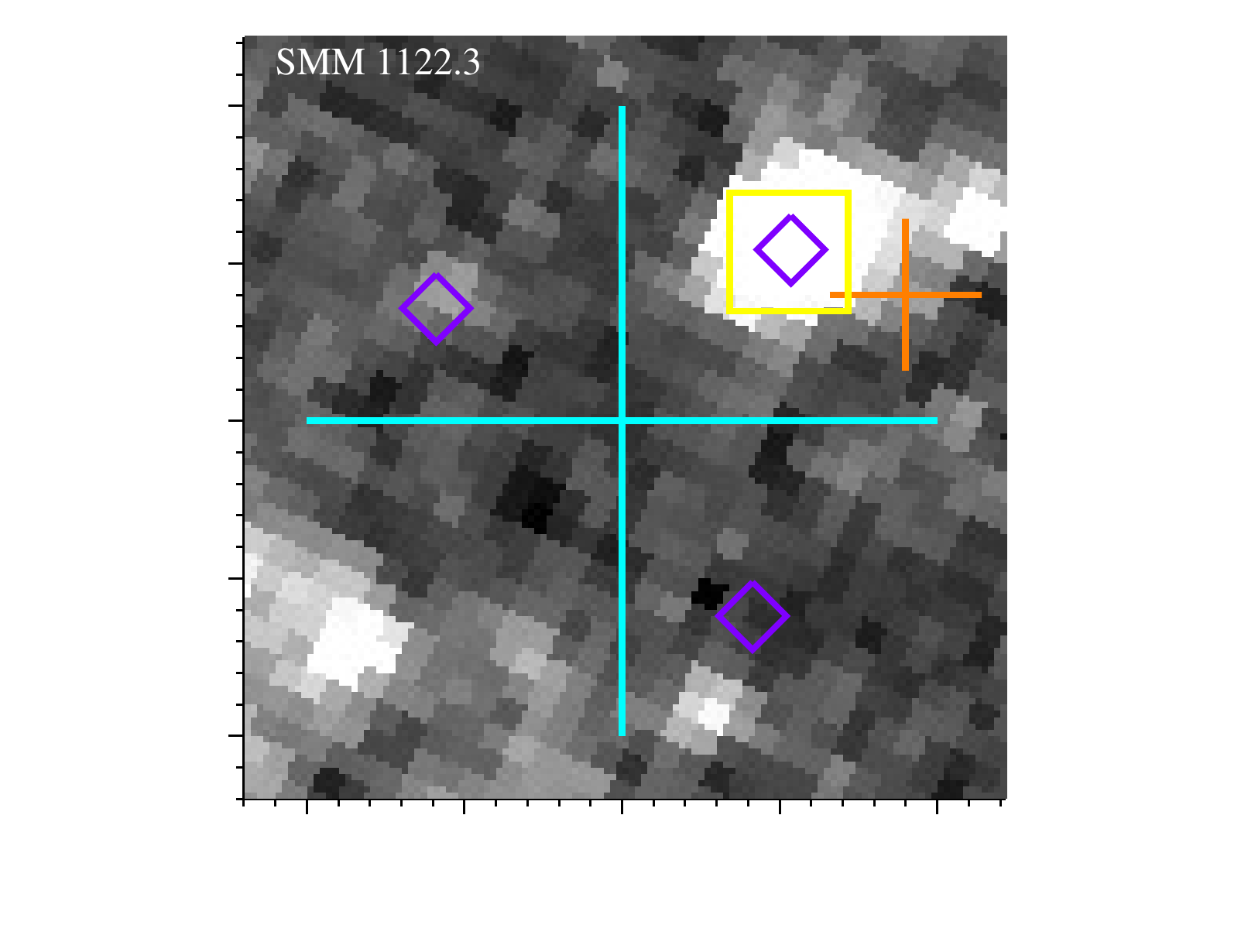}}
   		\hspace{-27mm}
  	   	 \vspace{-12mm}
      \subfigure{\includegraphics[scale=0.42]{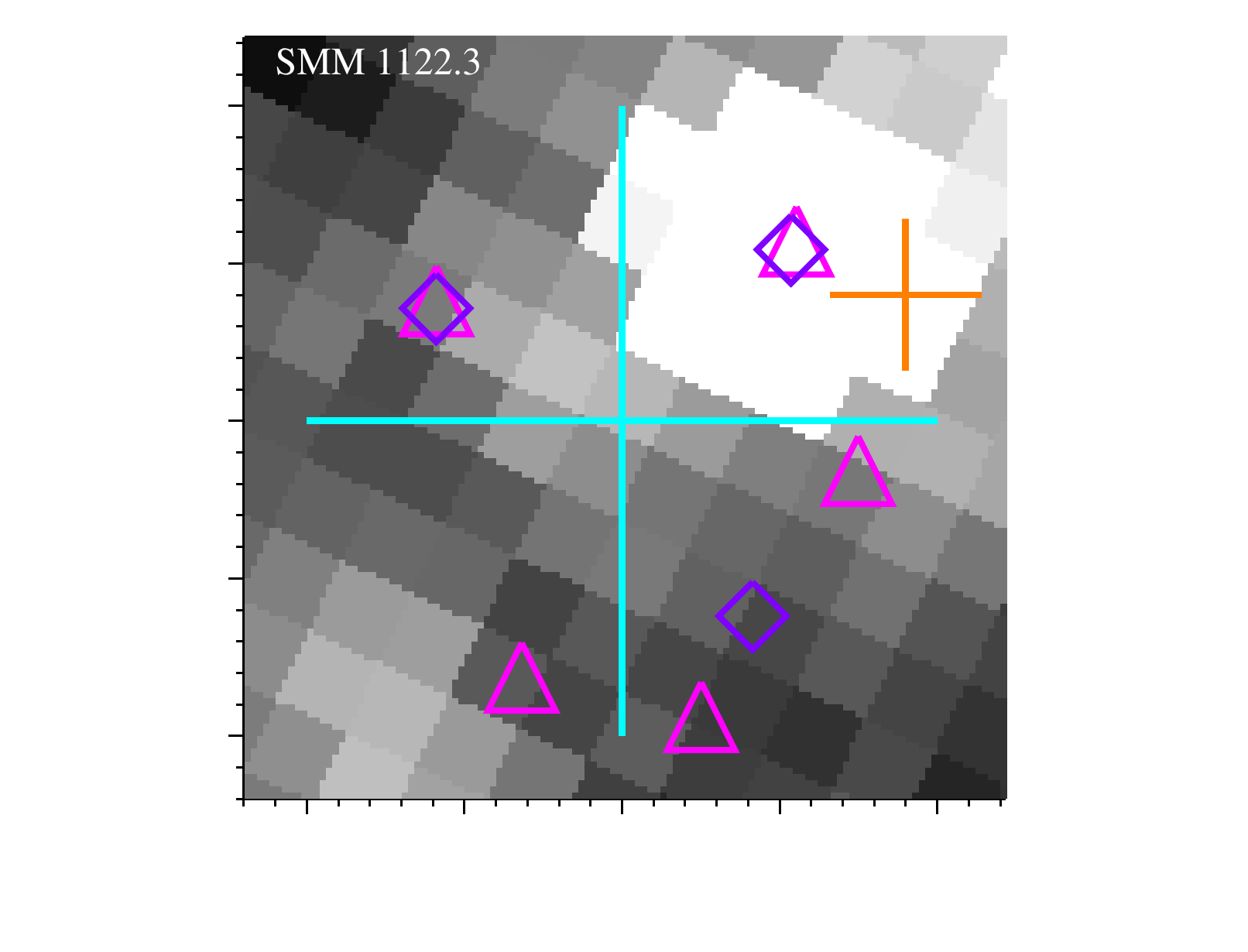}}
       \subfigure{\includegraphics[scale=0.42]{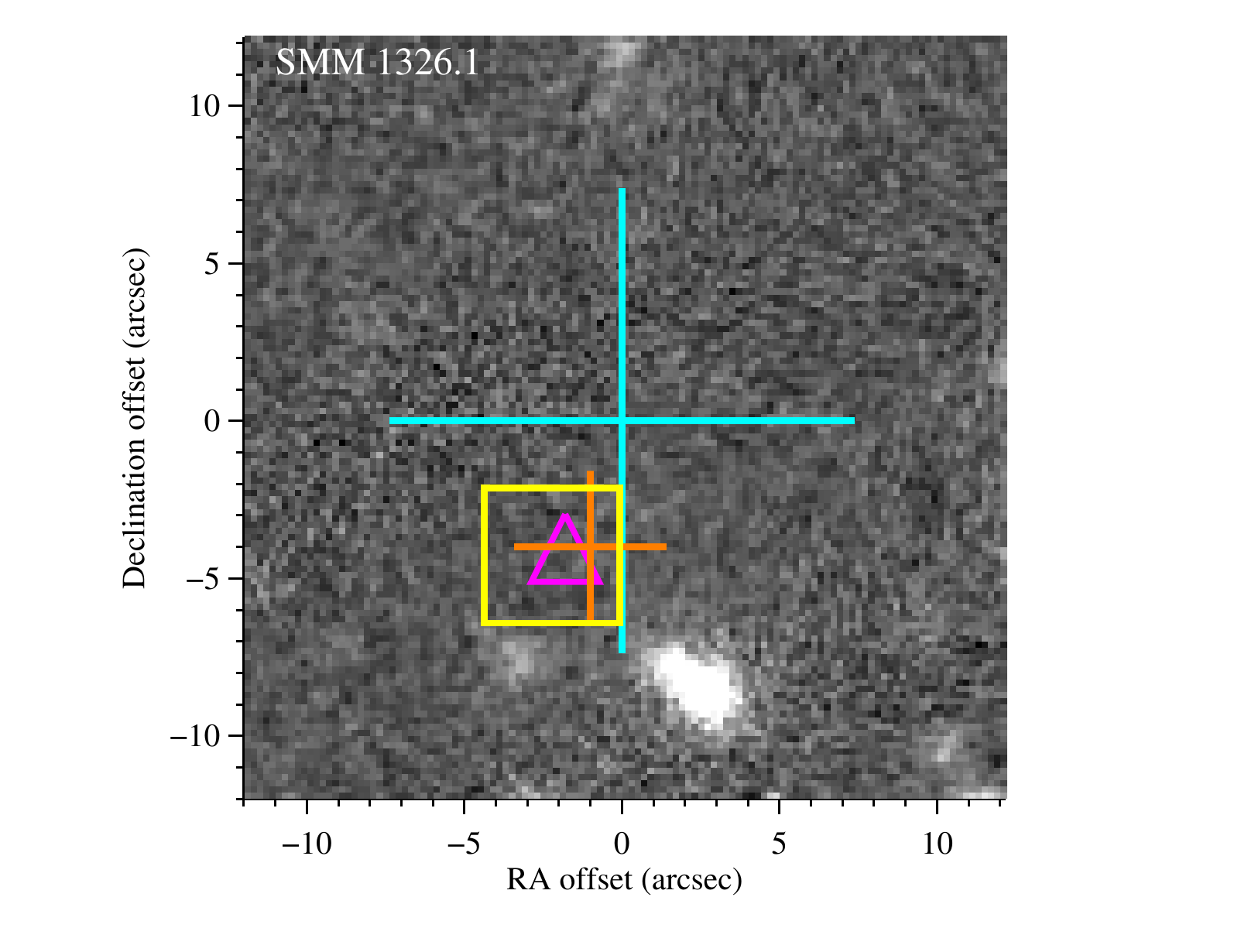}}
   		\hspace{-27mm}
     \subfigure{\includegraphics[scale=0.42]{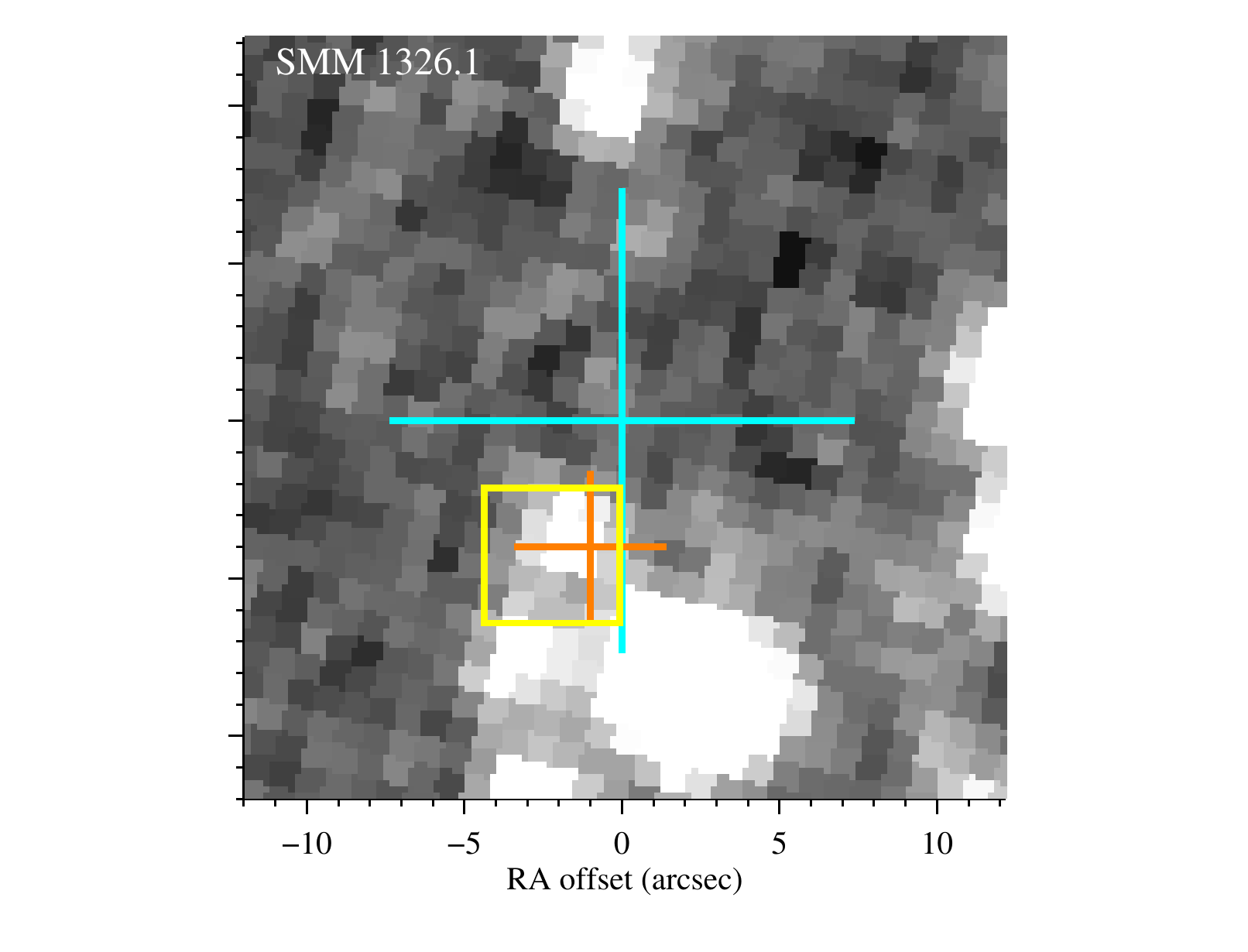}}
   		\hspace{-27mm}
  	   	 \vspace{-4mm}
      \subfigure{\includegraphics[scale=0.42]{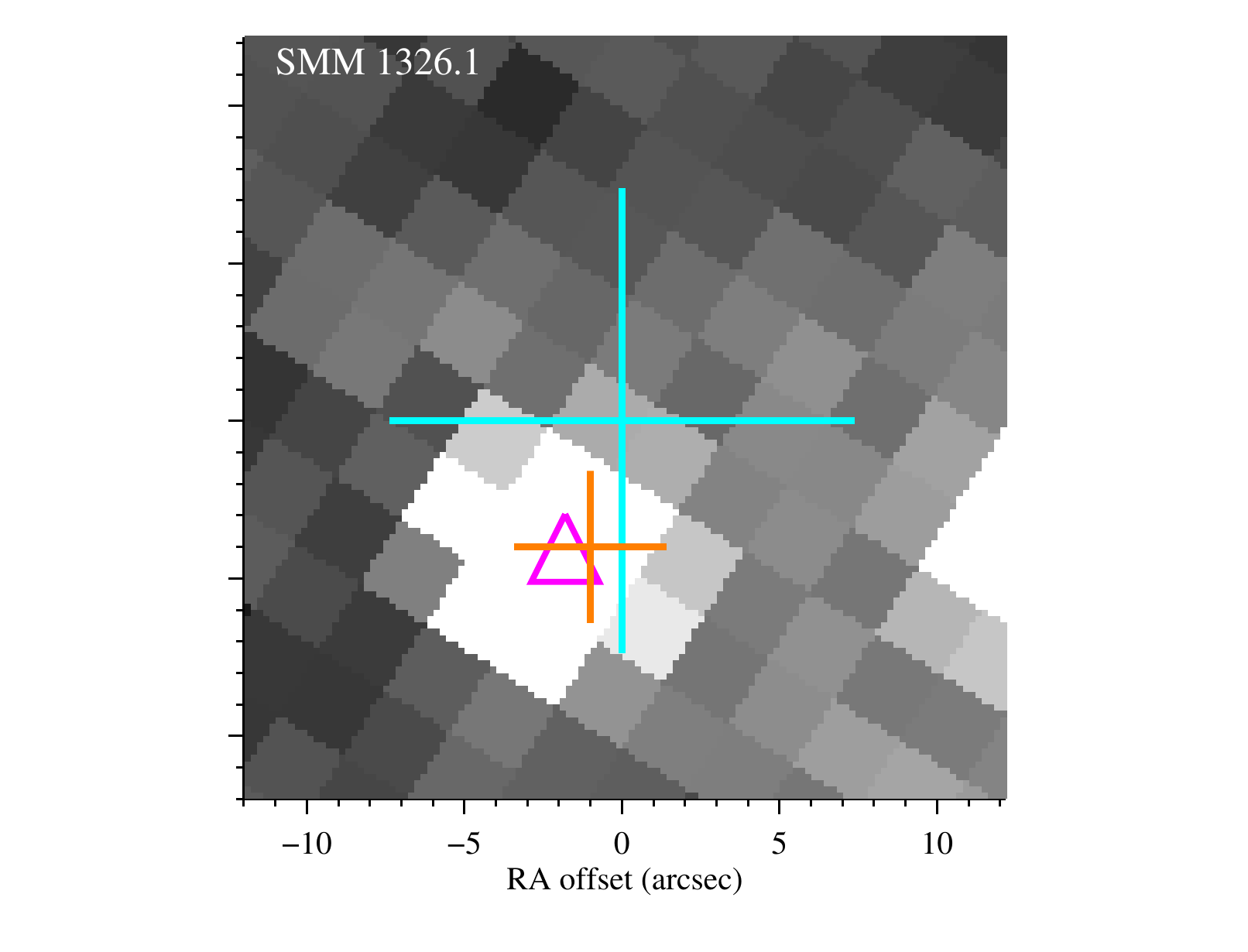}}
\caption{24 $\times$ 24 arcsec$^2$ postage stamps for SMGs with secure counterparts.  The grey-scale images are optical $z^{\prime}$ band, IRAC 3.6\,\um, and MIPS 24\,\um, from left to right, where north points up and east is to the right.  Each image is centred on the SMG position, denoted with a large blue cross.  The size of the cross represents the search radius for MIPS counterparts.  The rest of the symbols are as follows: an orange cross is 450\,\um\  emission, a yellow square is a MIPS source, magenta triangles signify IRAC detections, the red contours show radio emission, purple diamonds show optical sources with photometric redshifts, and a green `x' denotes an X-ray detection.  Note that the radio contours cover the entire field, but the symbols are only shown for sources within the search radius.  The radio contours start at 45\,\uJy\ and increase in 10\,\uJy\ steps up to 105\,\uJy, after which they increase by 200\,\uJy.}
   \label{fig:secure}
\end{figure*}

\addtocounter{figure}{-1}
\begin{figure*}
   \centering
      \subfigure{\includegraphics[scale=0.42]{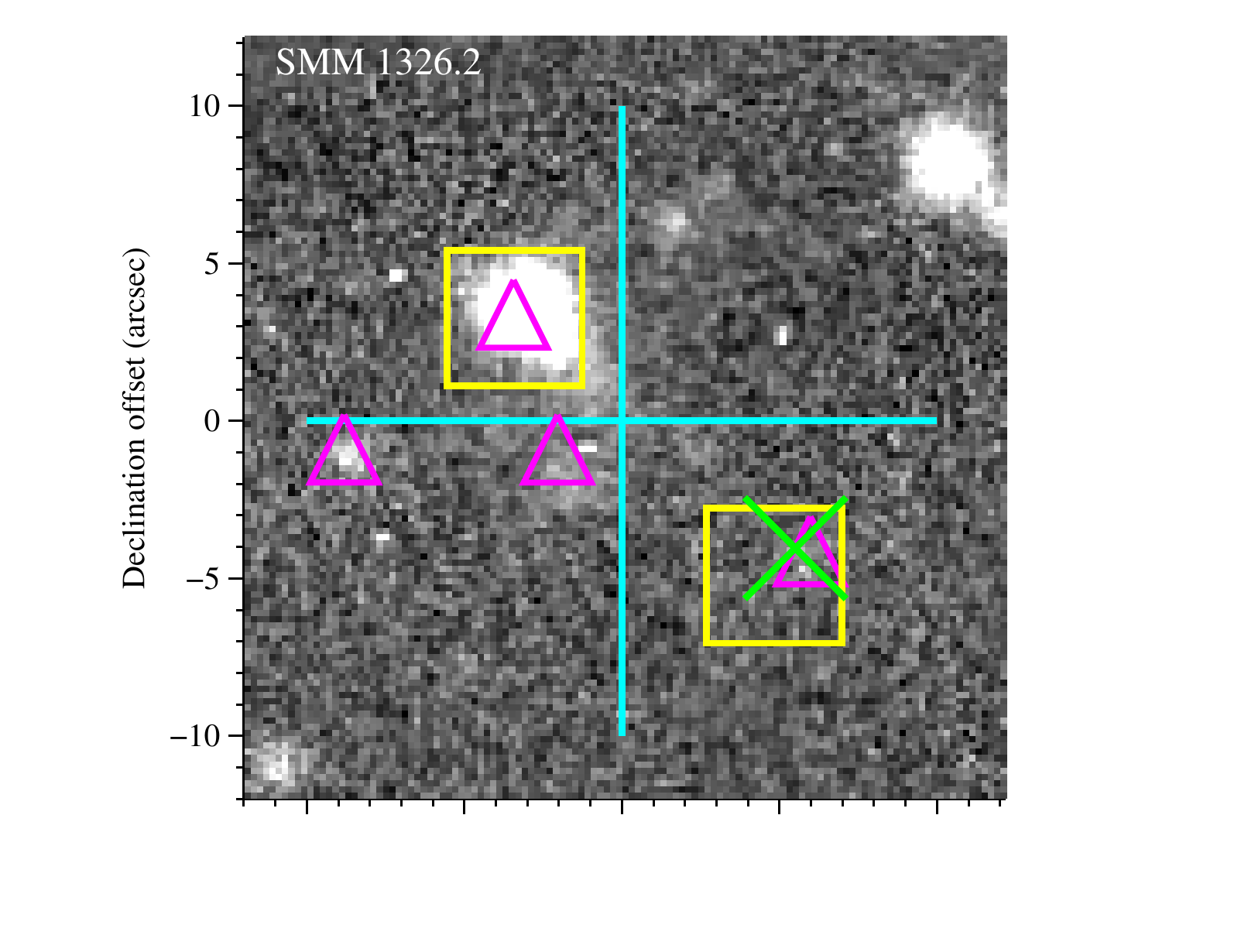}}
   		\hspace{-27mm}
   \subfigure{\includegraphics[scale=0.42]{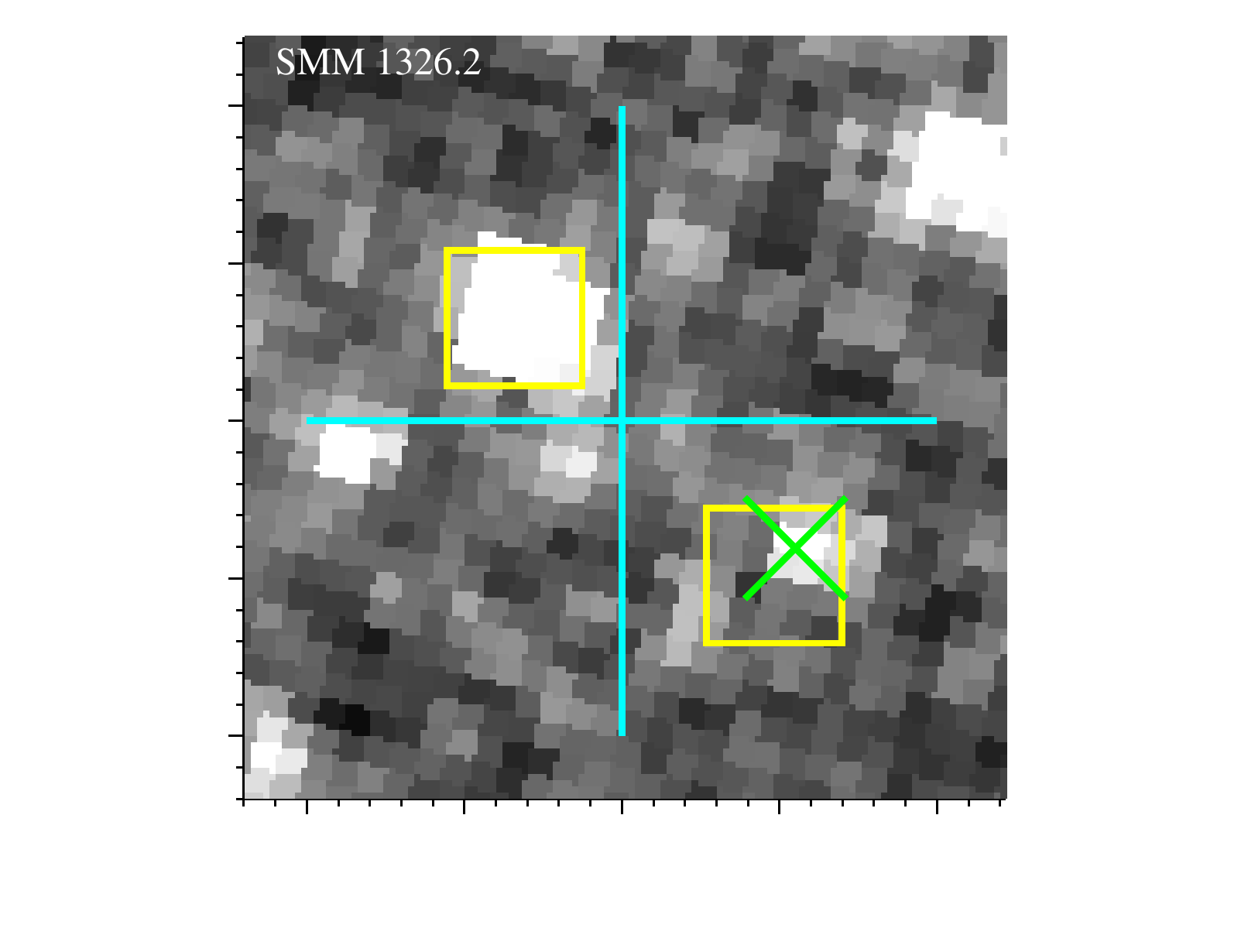}}
		 \hspace{-27mm}
  	   	 \vspace{-12mm}
     \subfigure{\includegraphics[scale=0.42]{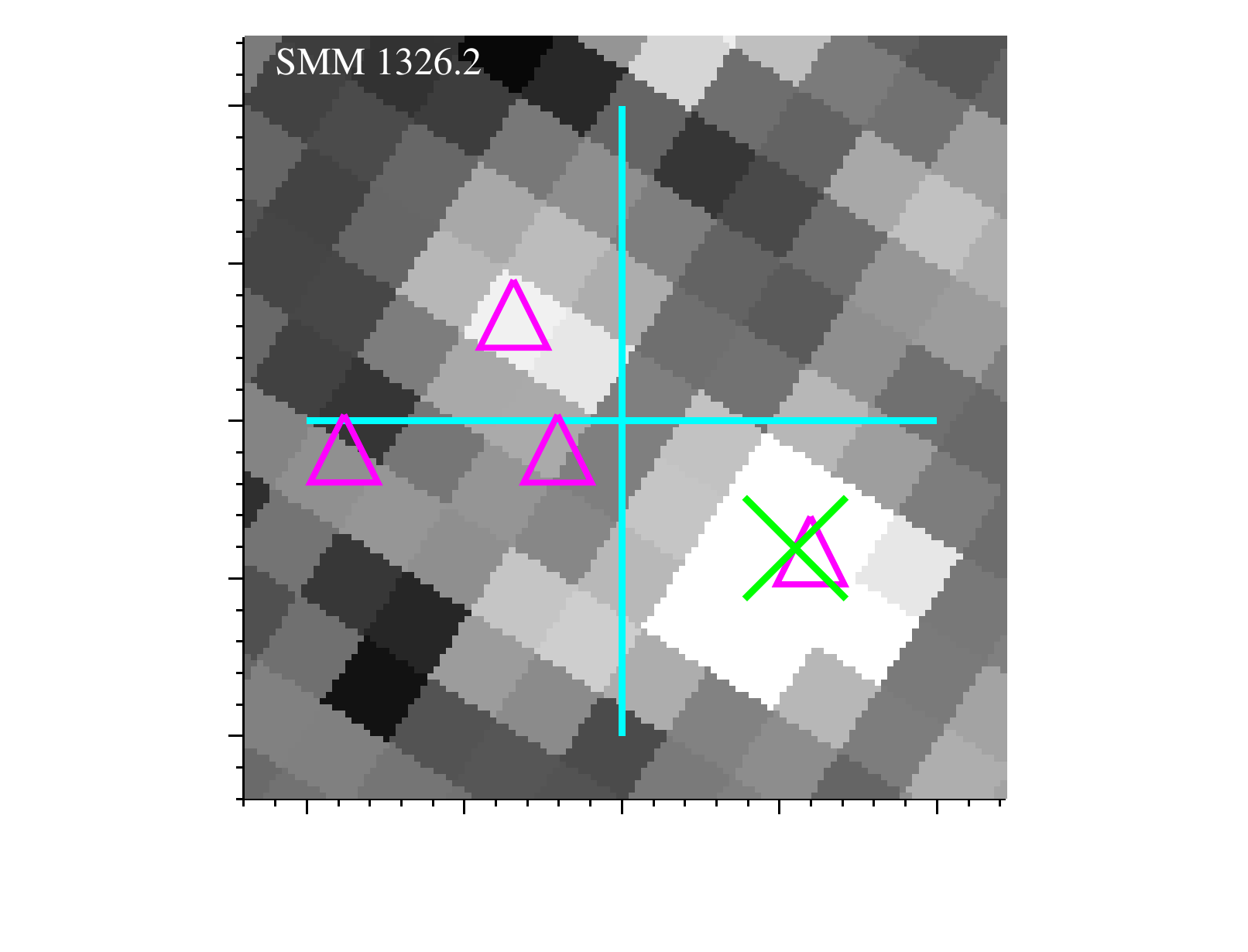}}
       \subfigure{\includegraphics[scale=0.42]{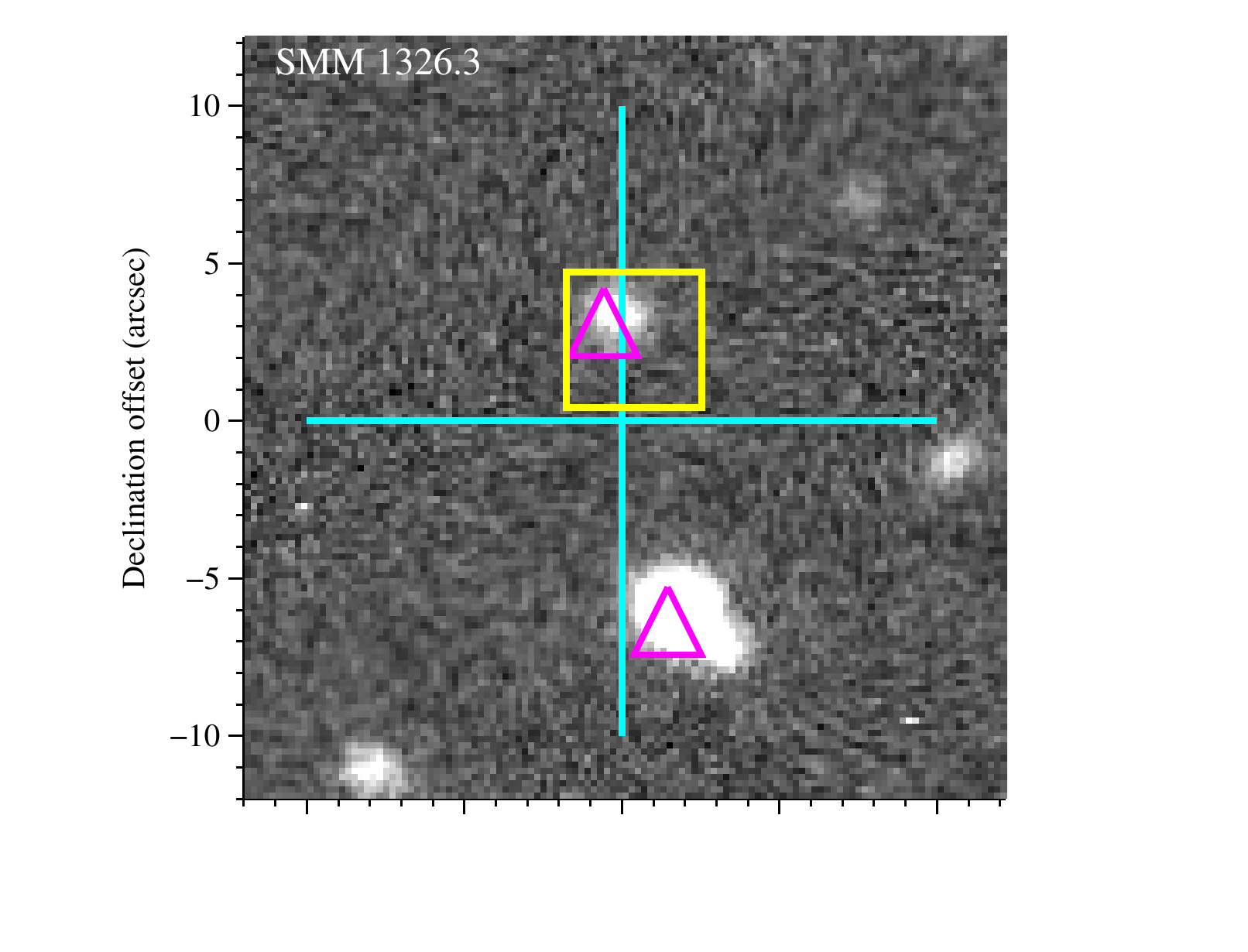}}
   		\hspace{-27mm}
     \subfigure{\includegraphics[scale=0.42]{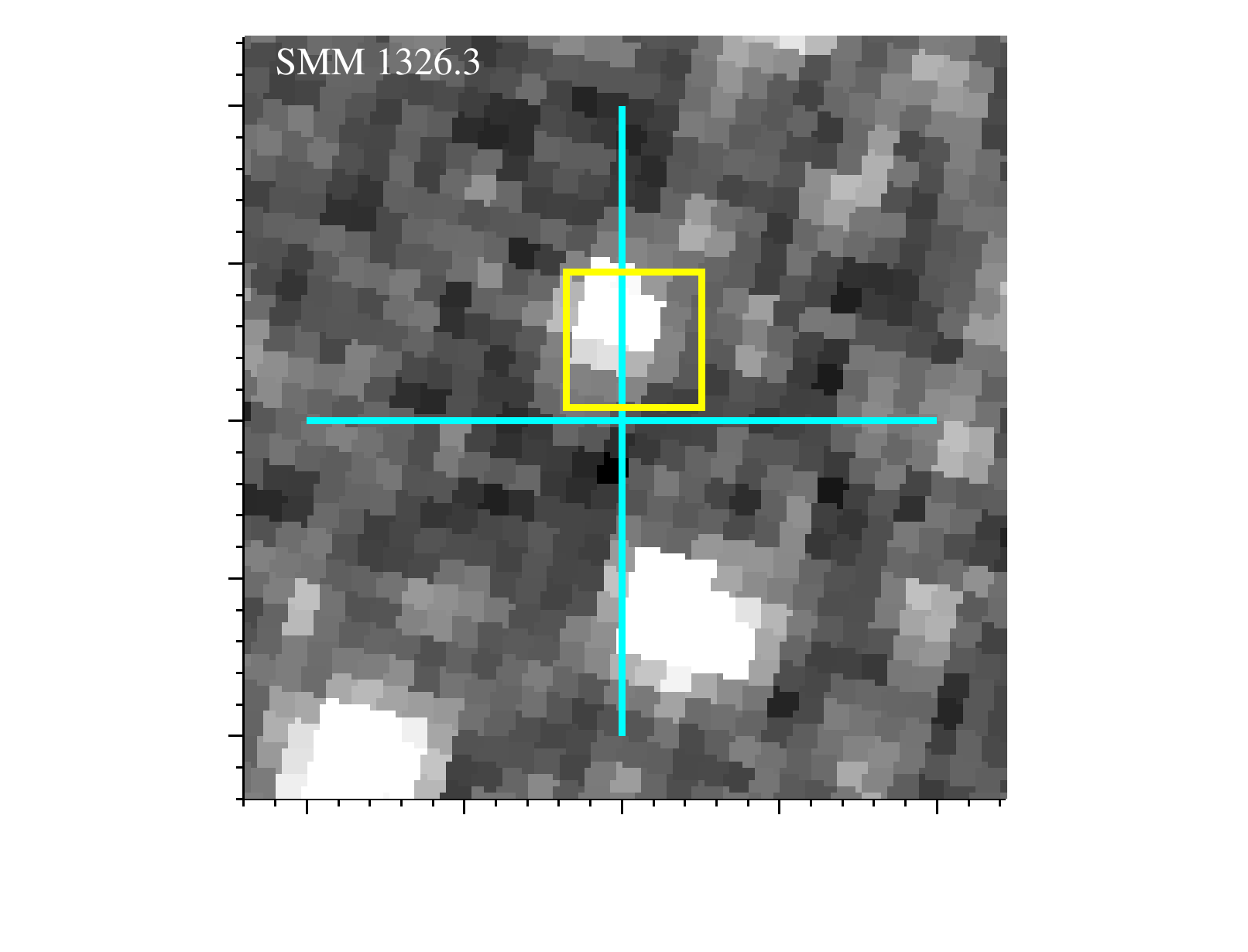}}
    		\hspace{-27mm}
  	   	 \vspace{-12mm}
      \subfigure{\includegraphics[scale=0.42]{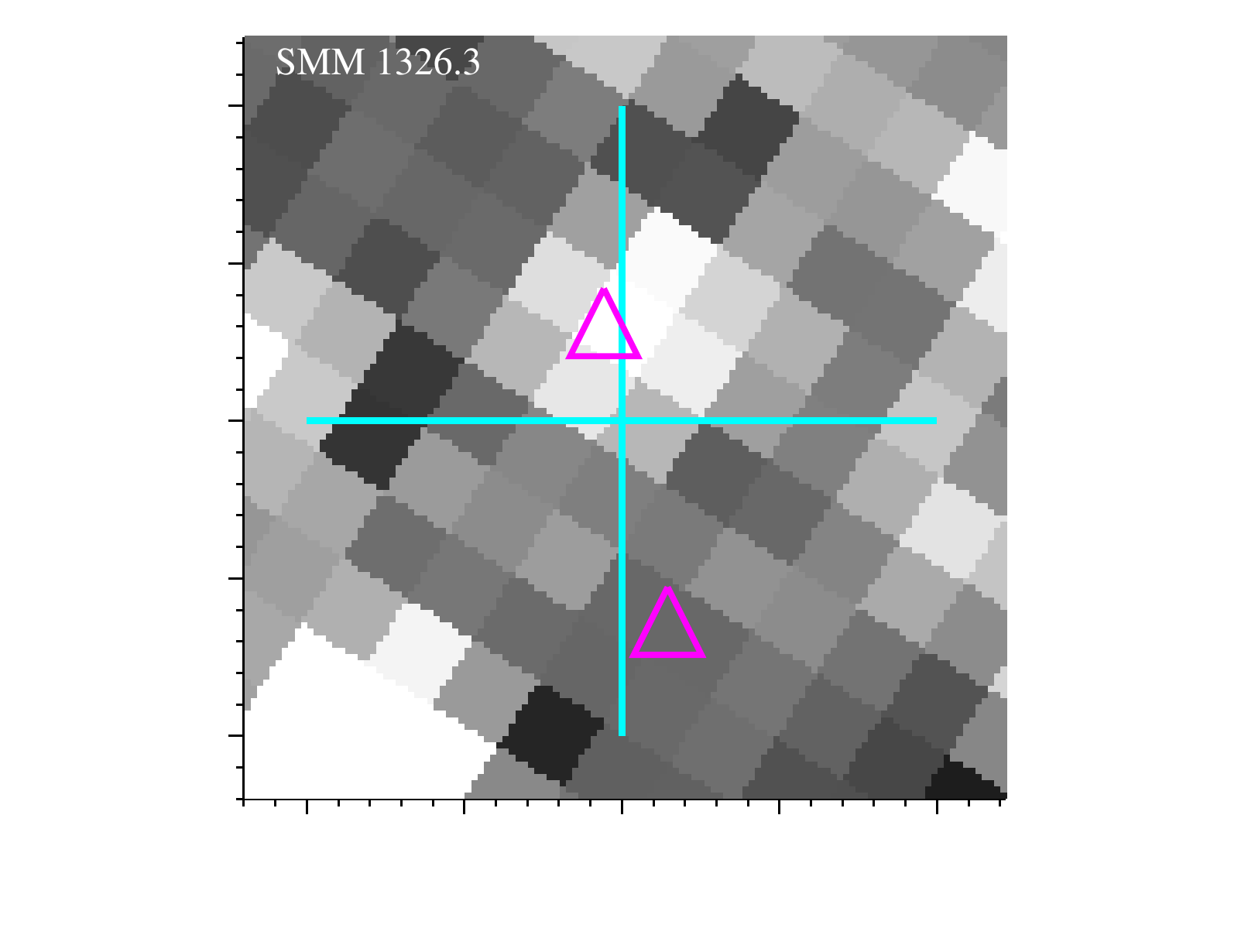}}
        \subfigure{\includegraphics[scale=0.42]{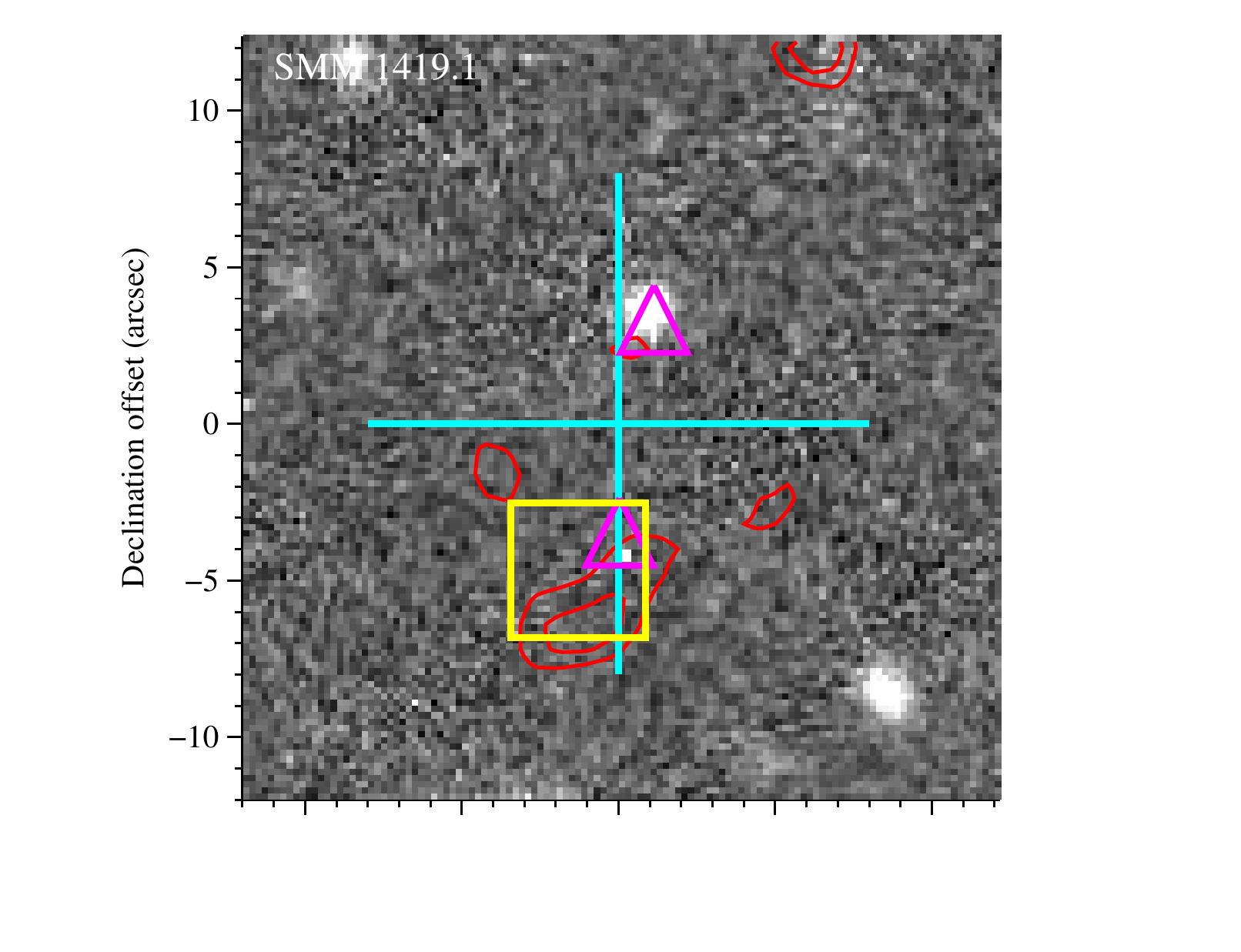}}
   		\hspace{-27mm}
     \subfigure{\includegraphics[scale=0.42]{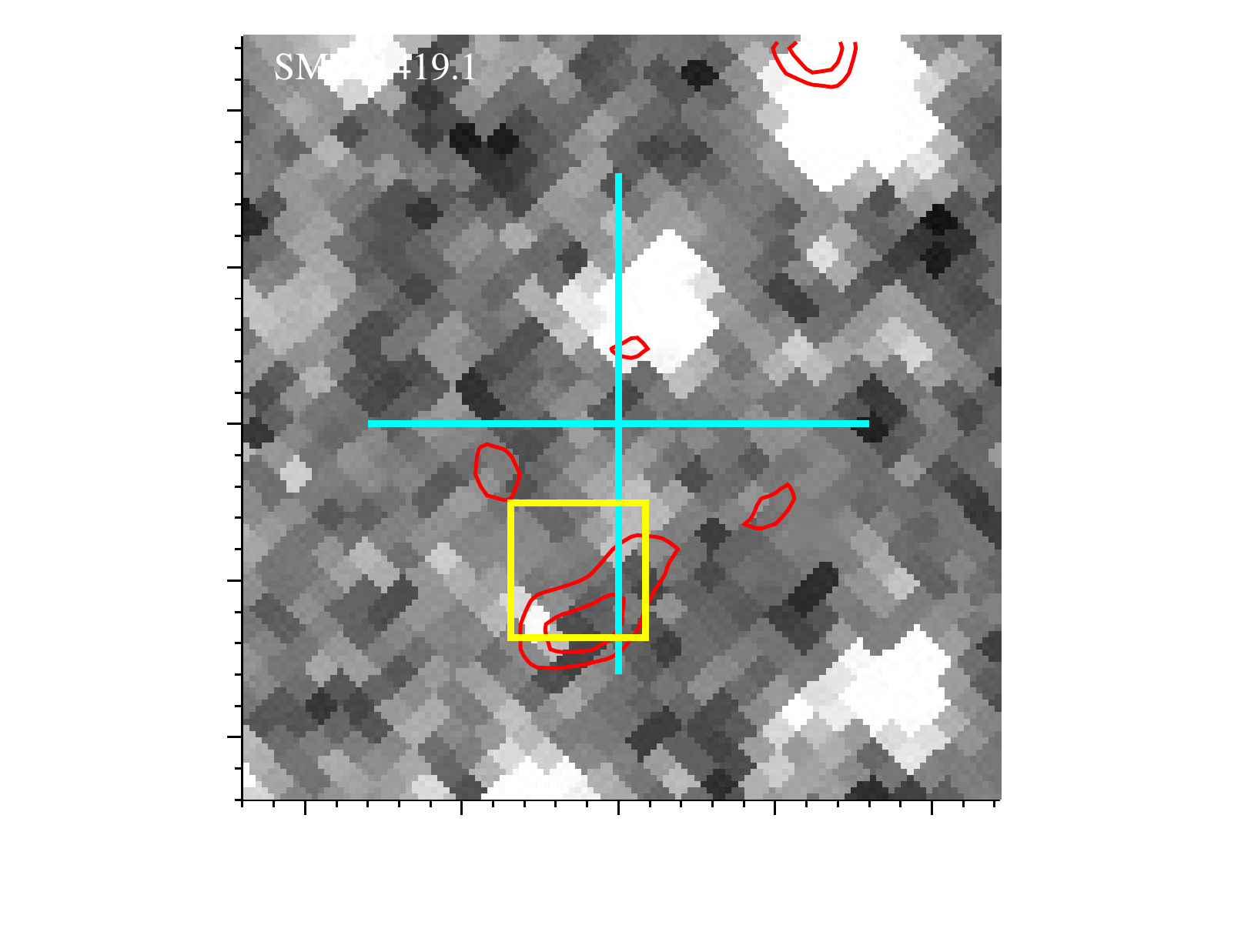}}
   		\hspace{-27mm}
  	   	 \vspace{-12mm}
      \subfigure{\includegraphics[scale=0.42]{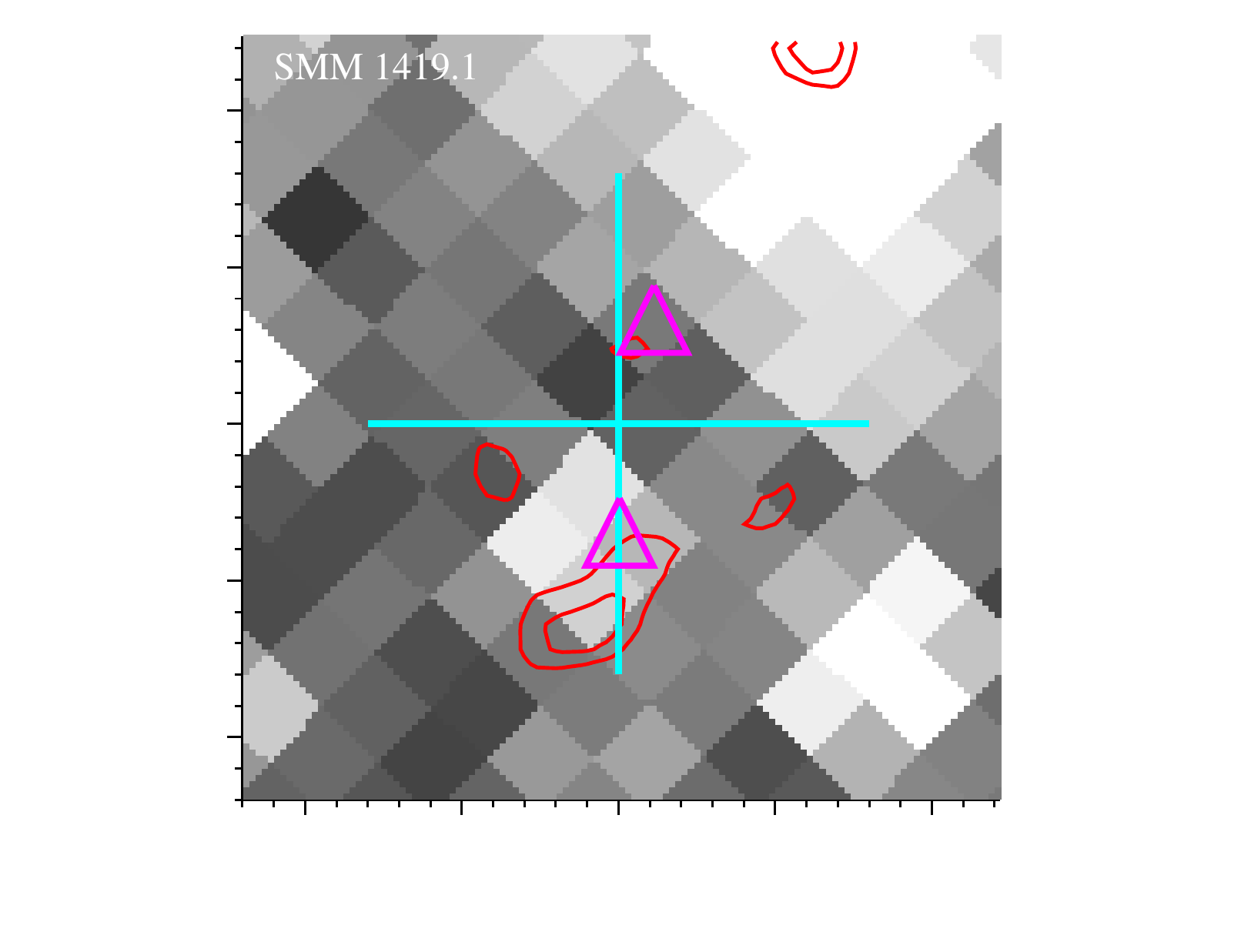}}
       \subfigure{\includegraphics[scale=0.42]{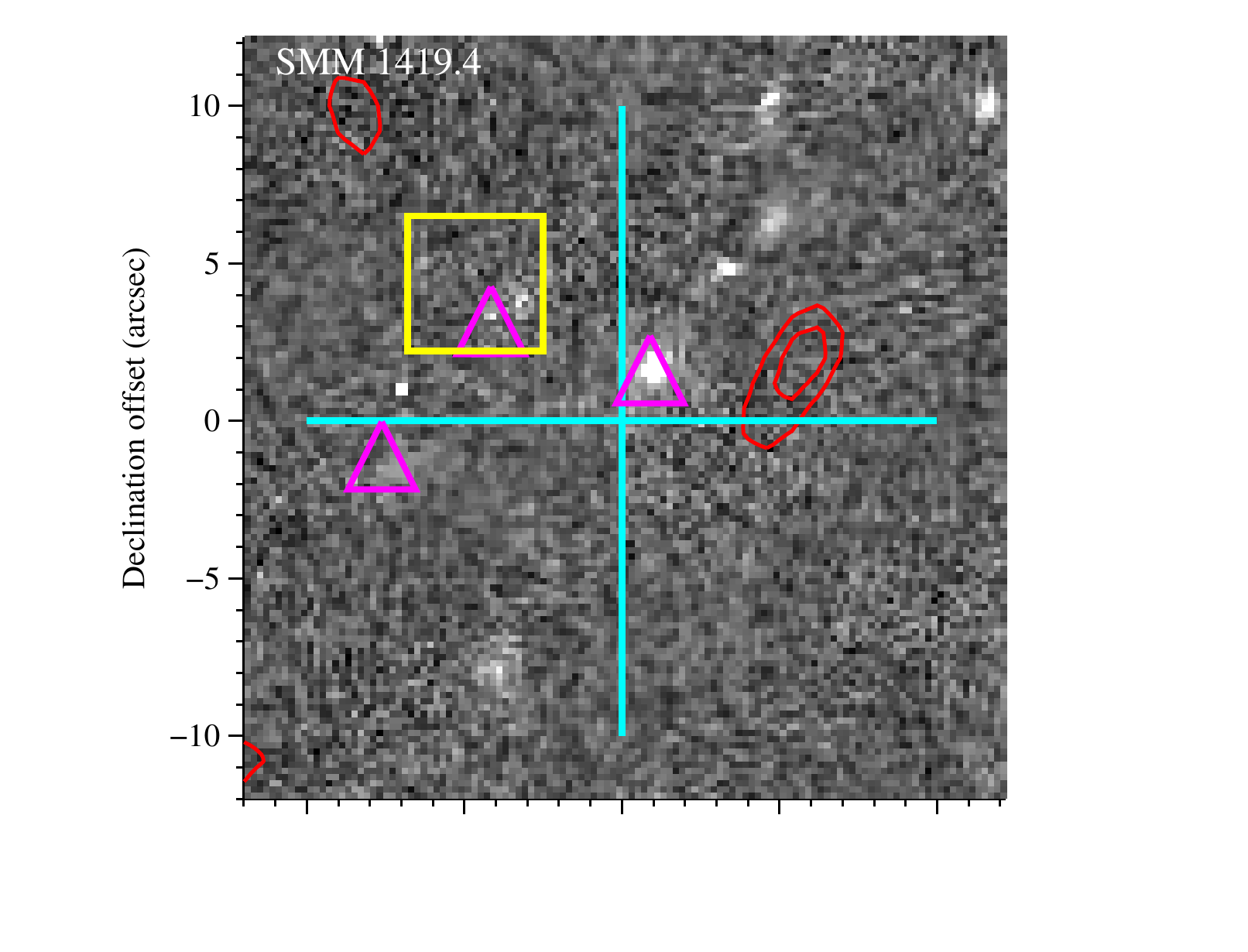}}
   		\hspace{-27mm}
     \subfigure{\includegraphics[scale=0.42]{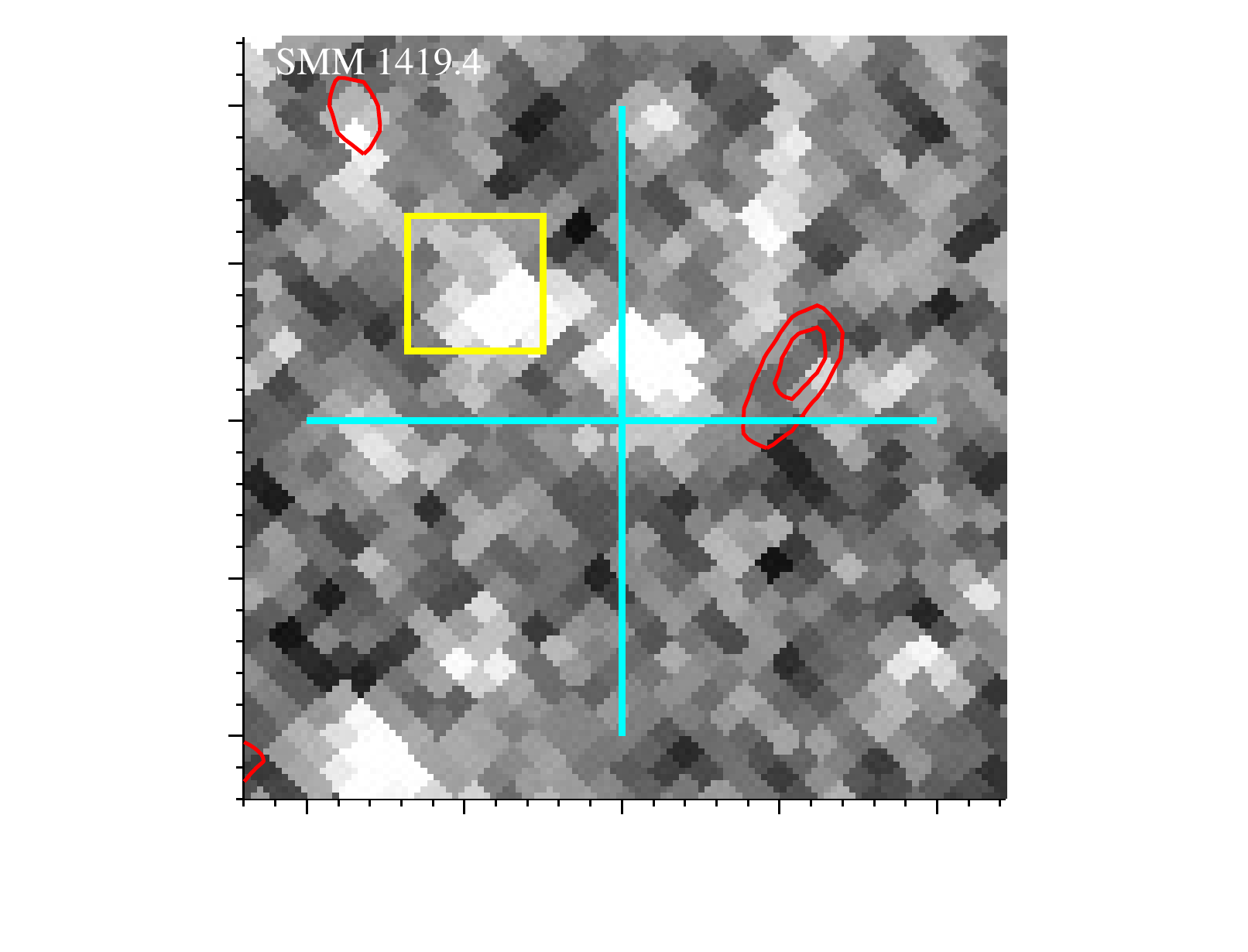}}
   		\hspace{-27mm}
  	   	 \vspace{-12mm}
      \subfigure{\includegraphics[scale=0.42]{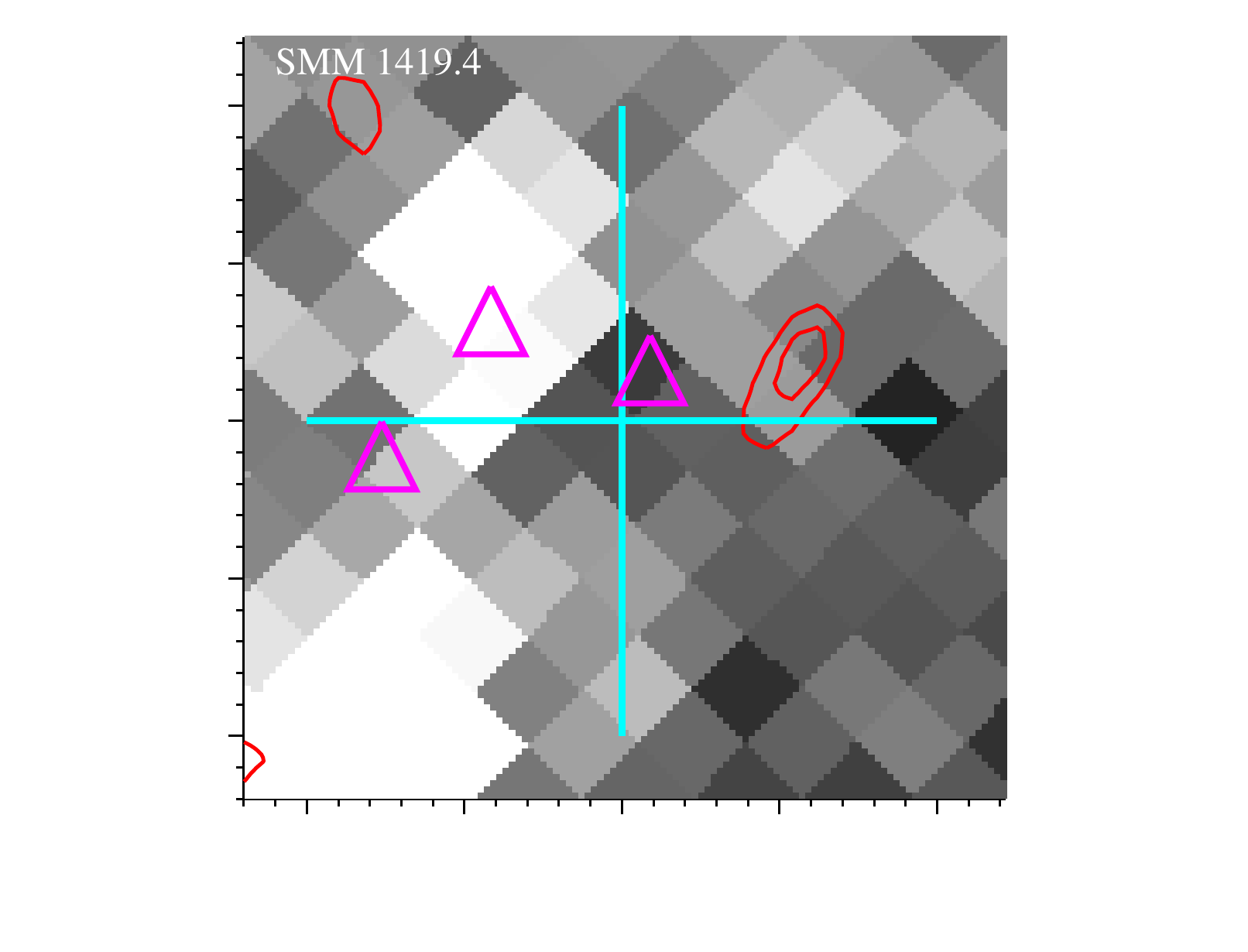}}
         \subfigure{\includegraphics[scale=0.42]{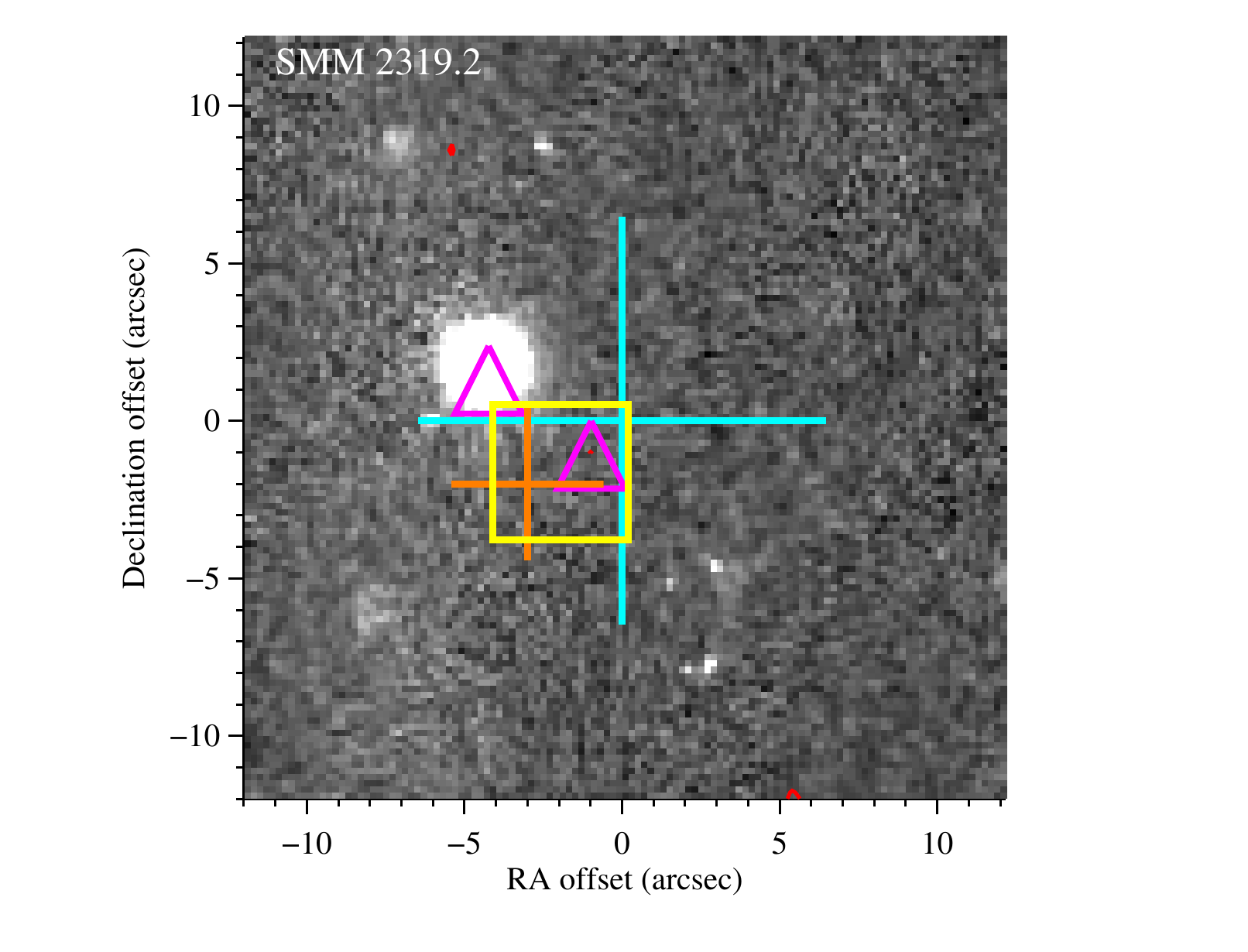}}
   		\hspace{-27mm}
     \subfigure{\includegraphics[scale=0.42]{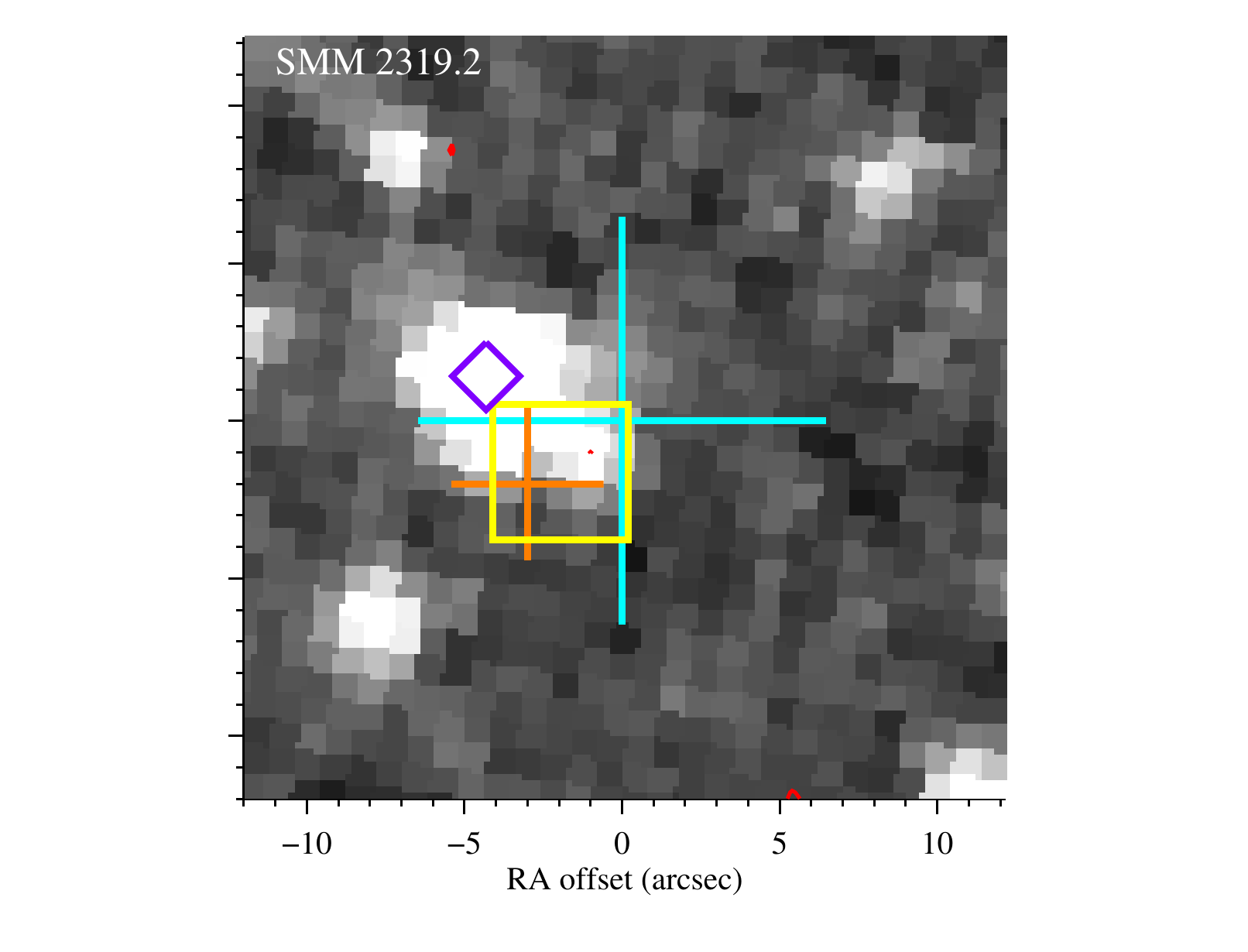}}
   		\hspace{-27mm}
  	   	 \vspace{-4mm}
      \subfigure{\includegraphics[scale=0.42]{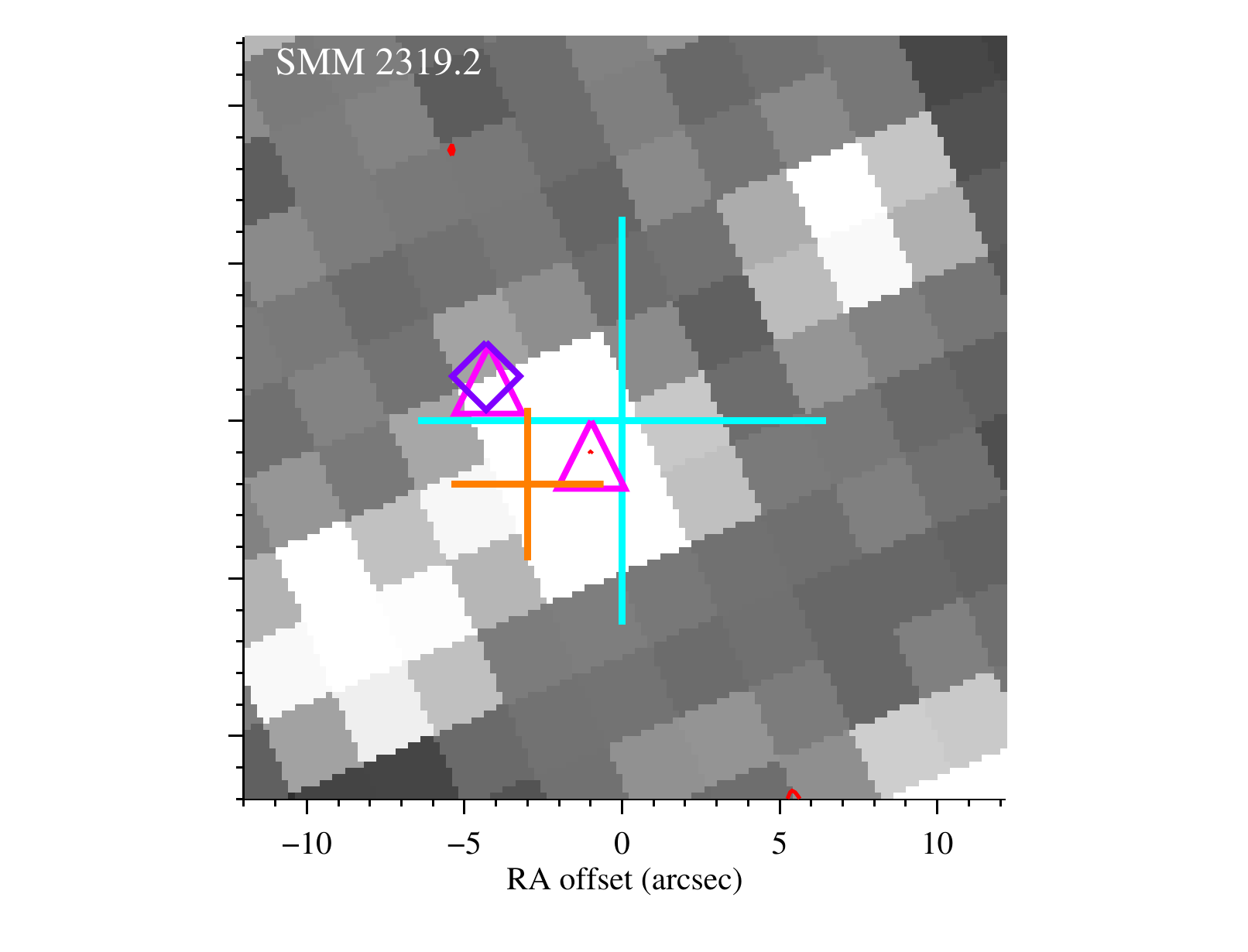}}
        \caption{\it{continued}}
\end{figure*}

\begin{figure*}
   \centering
           \subfigure{\includegraphics[scale=0.42]{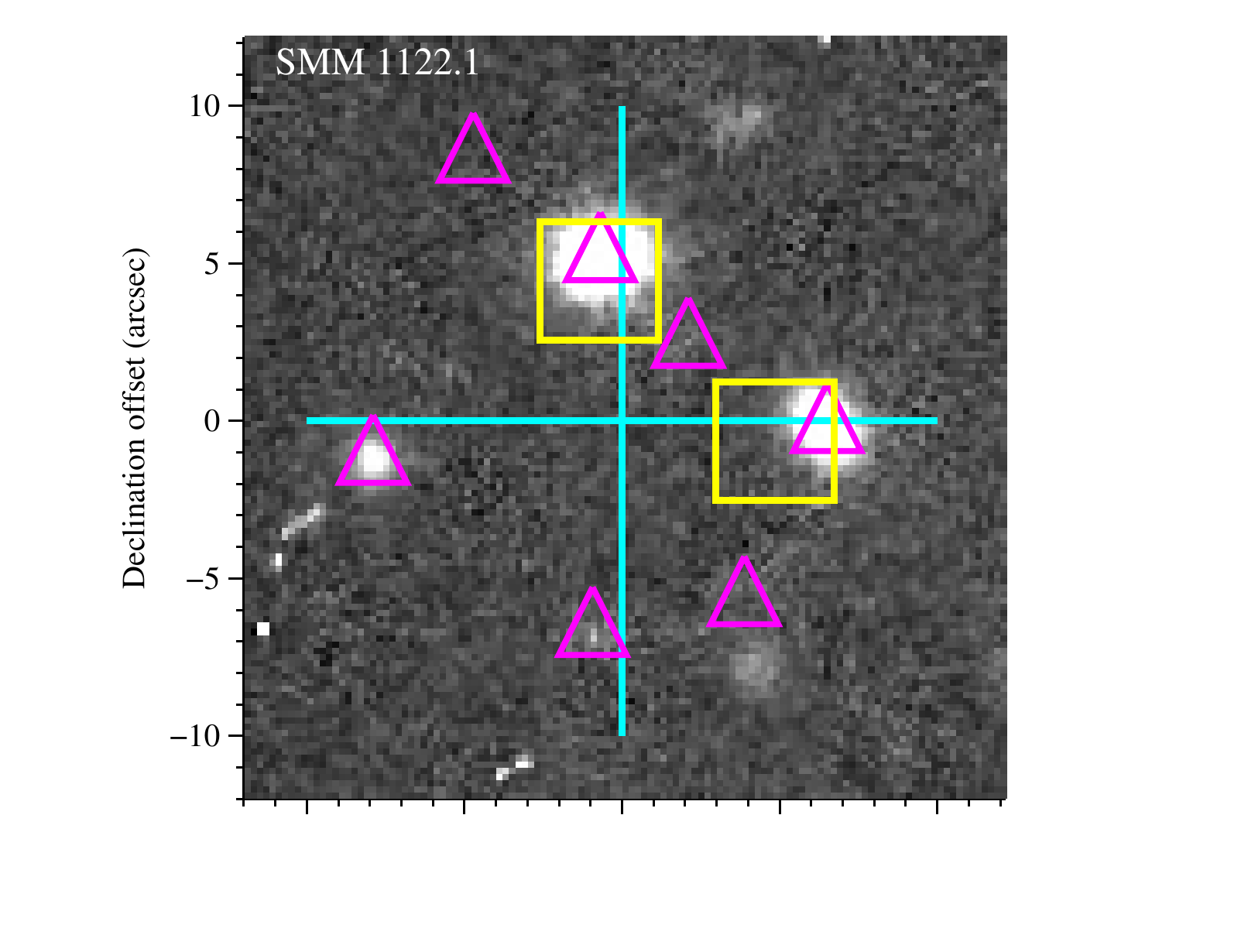}}
              	\hspace{-27mm}
     \subfigure{\includegraphics[scale=0.42]{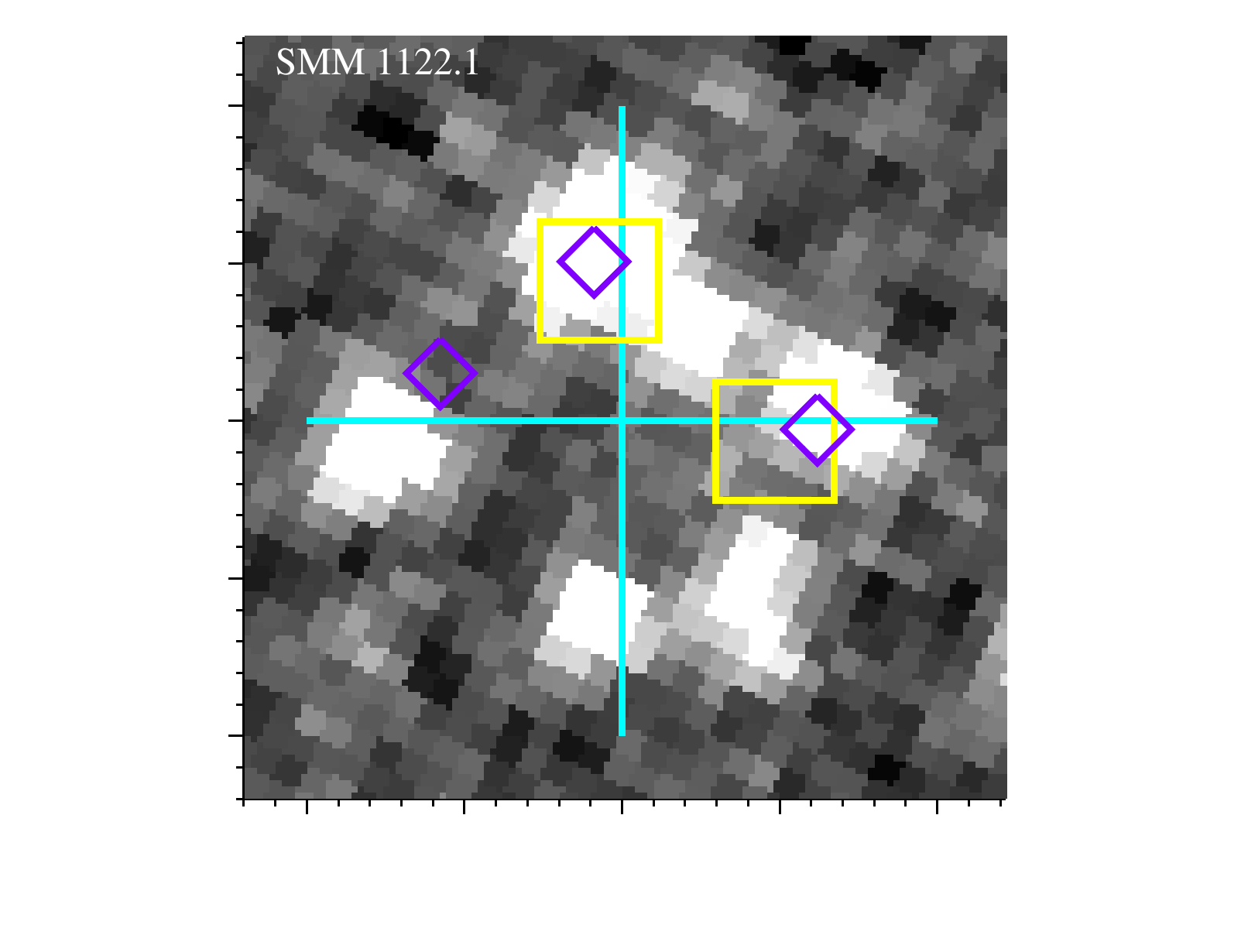}}
     		 \hspace{-27mm}
  	   	 \vspace{-12mm}
      \subfigure{\includegraphics[scale=0.42]{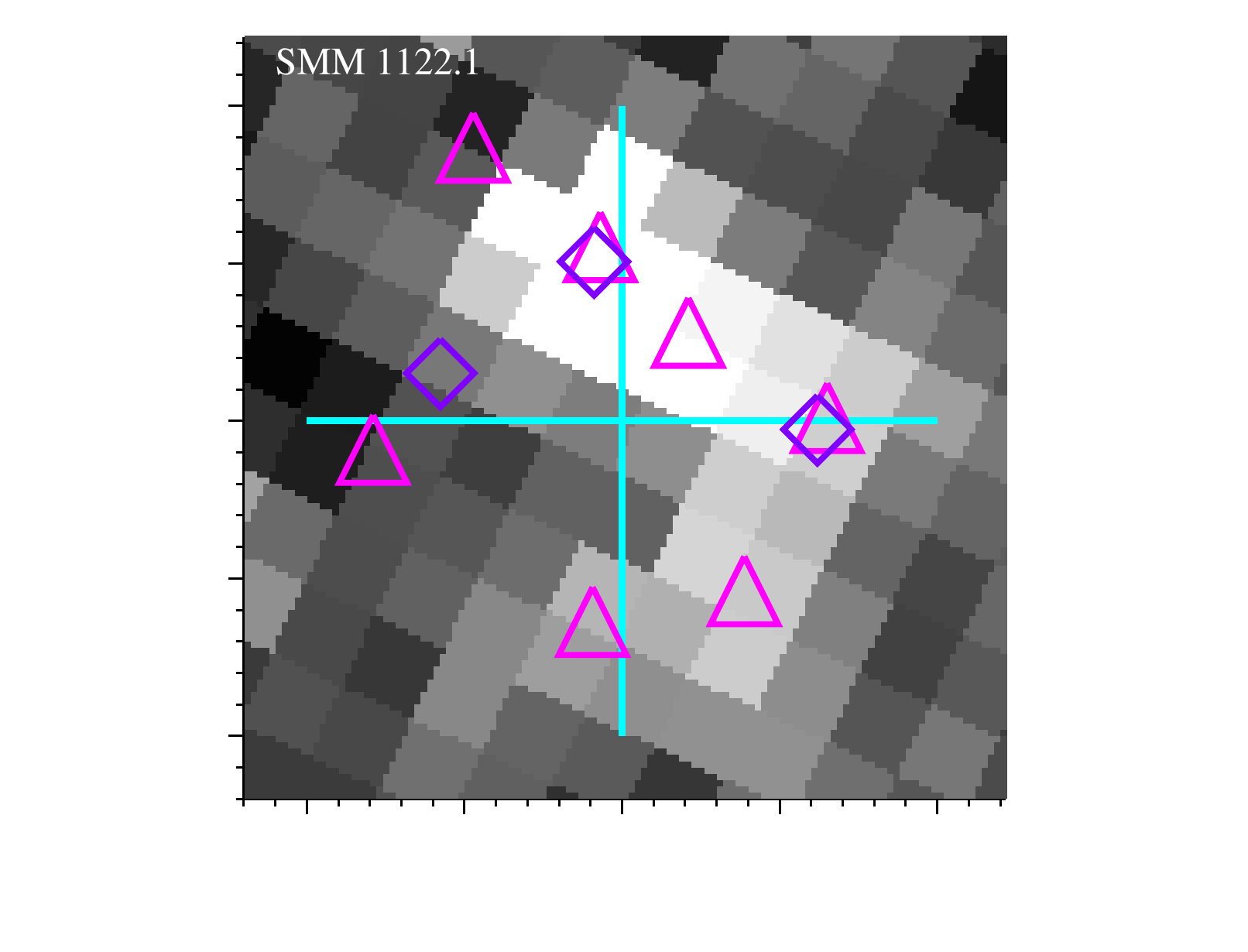}}
         \subfigure{\includegraphics[scale=0.42]{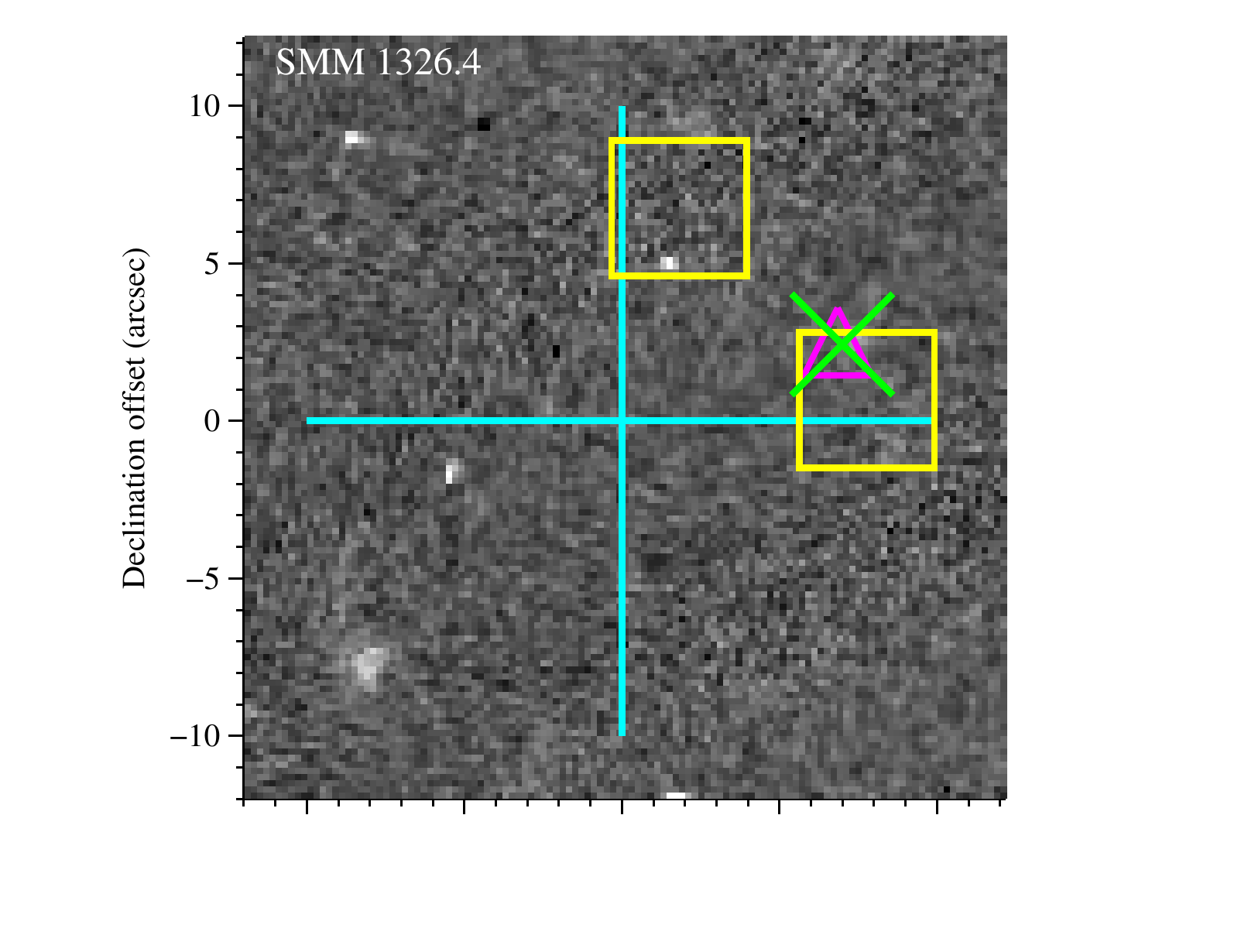}}
                       	\hspace{-27mm}
     \subfigure{\includegraphics[scale=0.42]{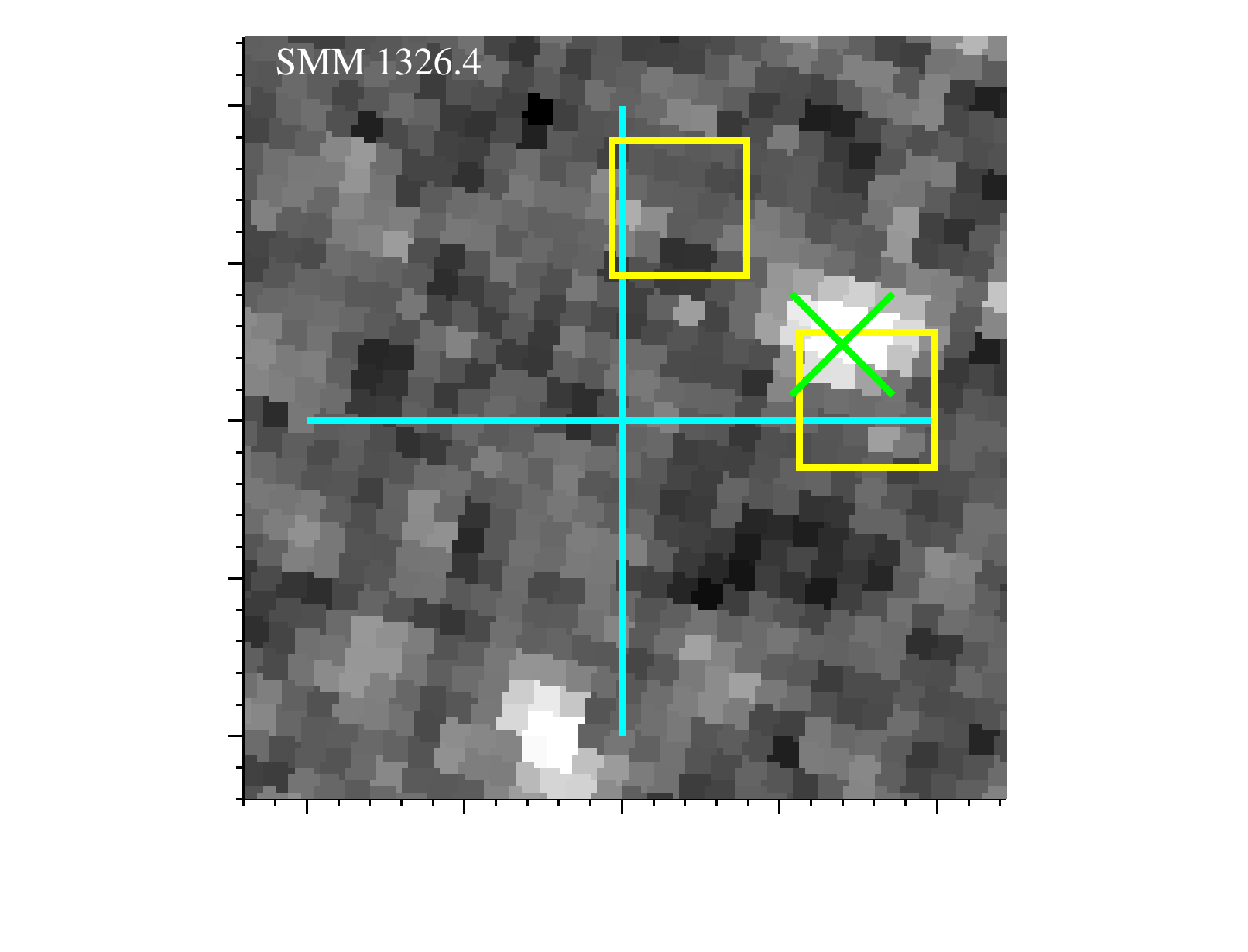}}
     		 \hspace{-27mm}
  	   	 \vspace{-12mm}
      \subfigure{\includegraphics[scale=0.42]{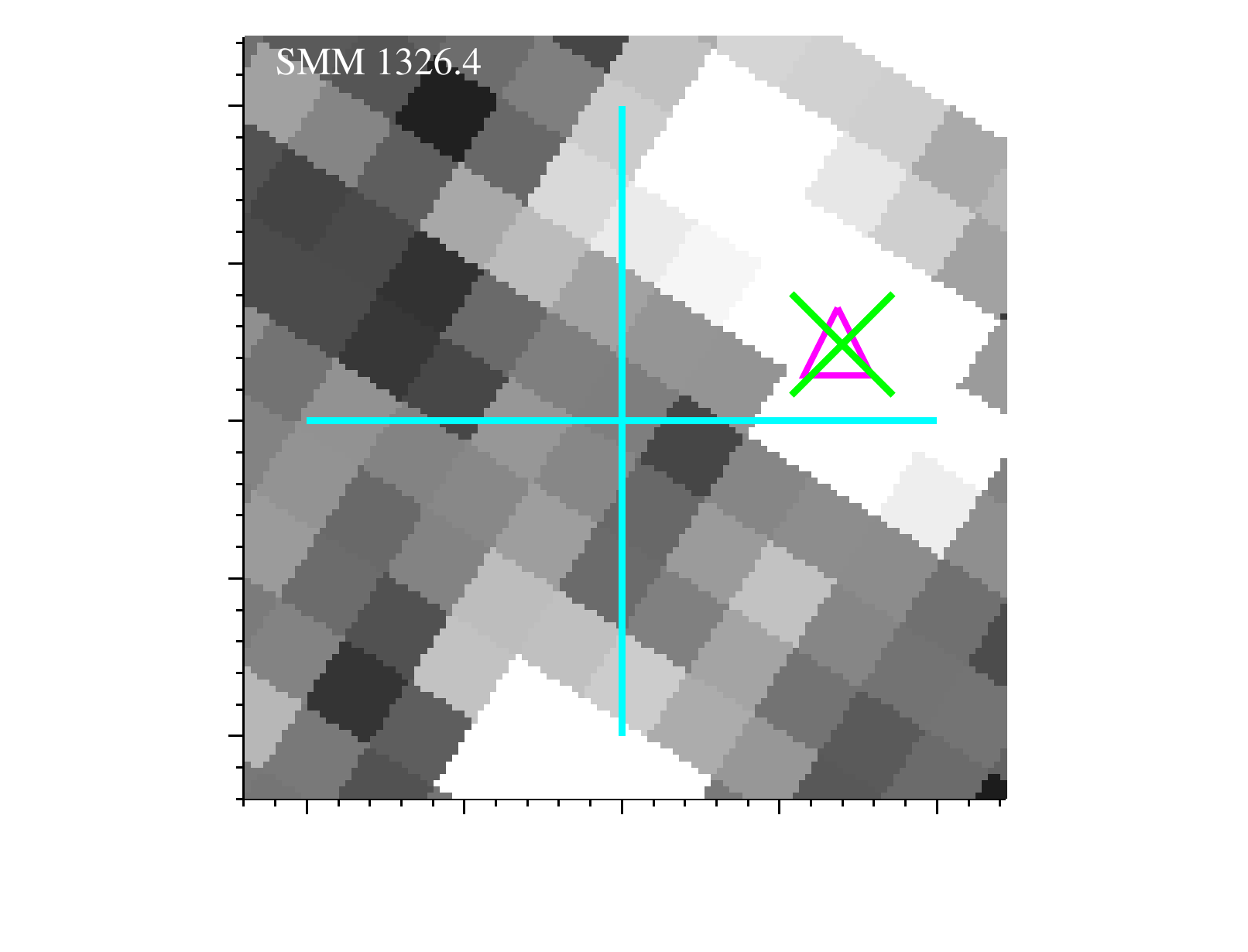}}
       \subfigure{\includegraphics[scale=0.42]{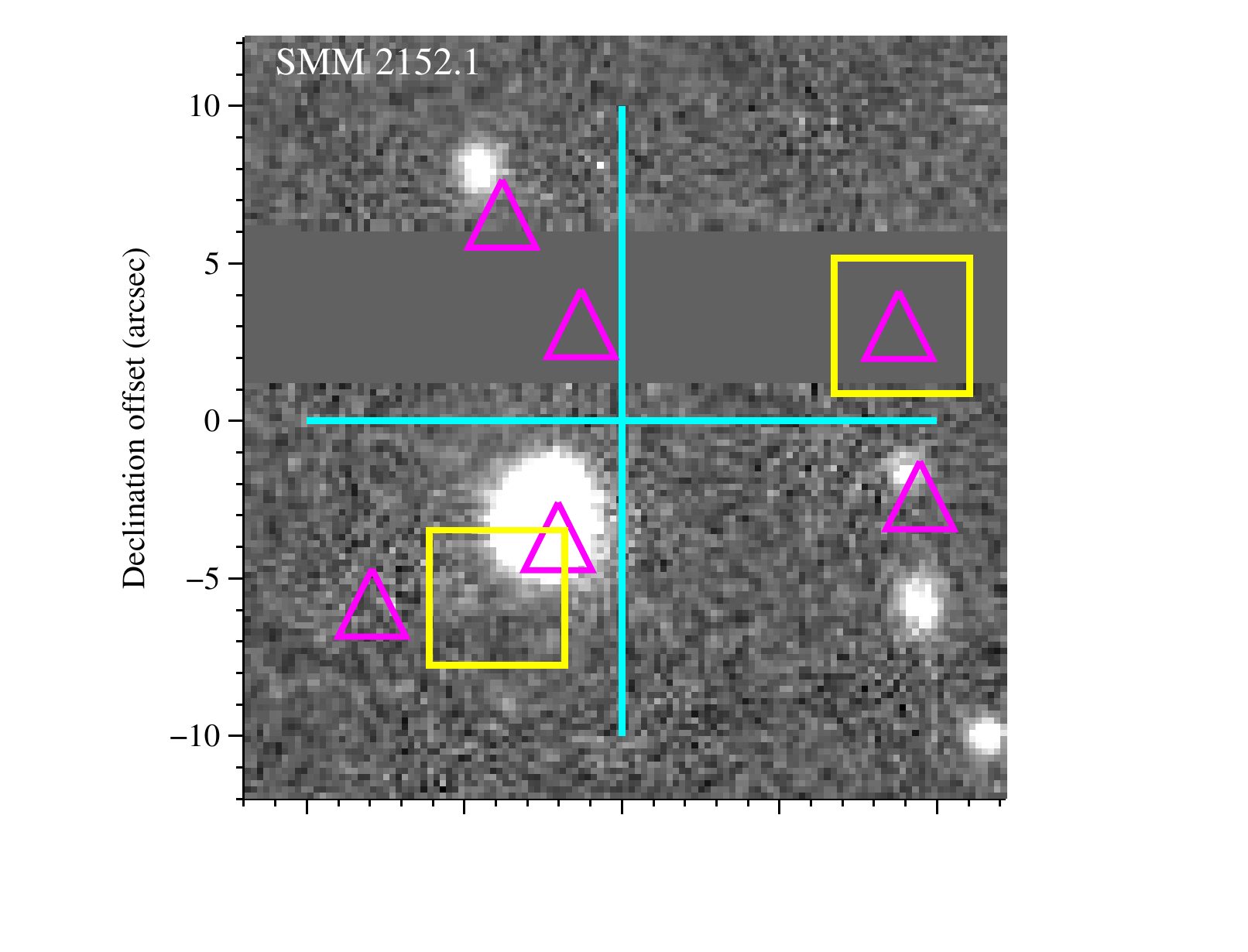}}
                     	\hspace{-27mm}
     \subfigure{\includegraphics[scale=0.42]{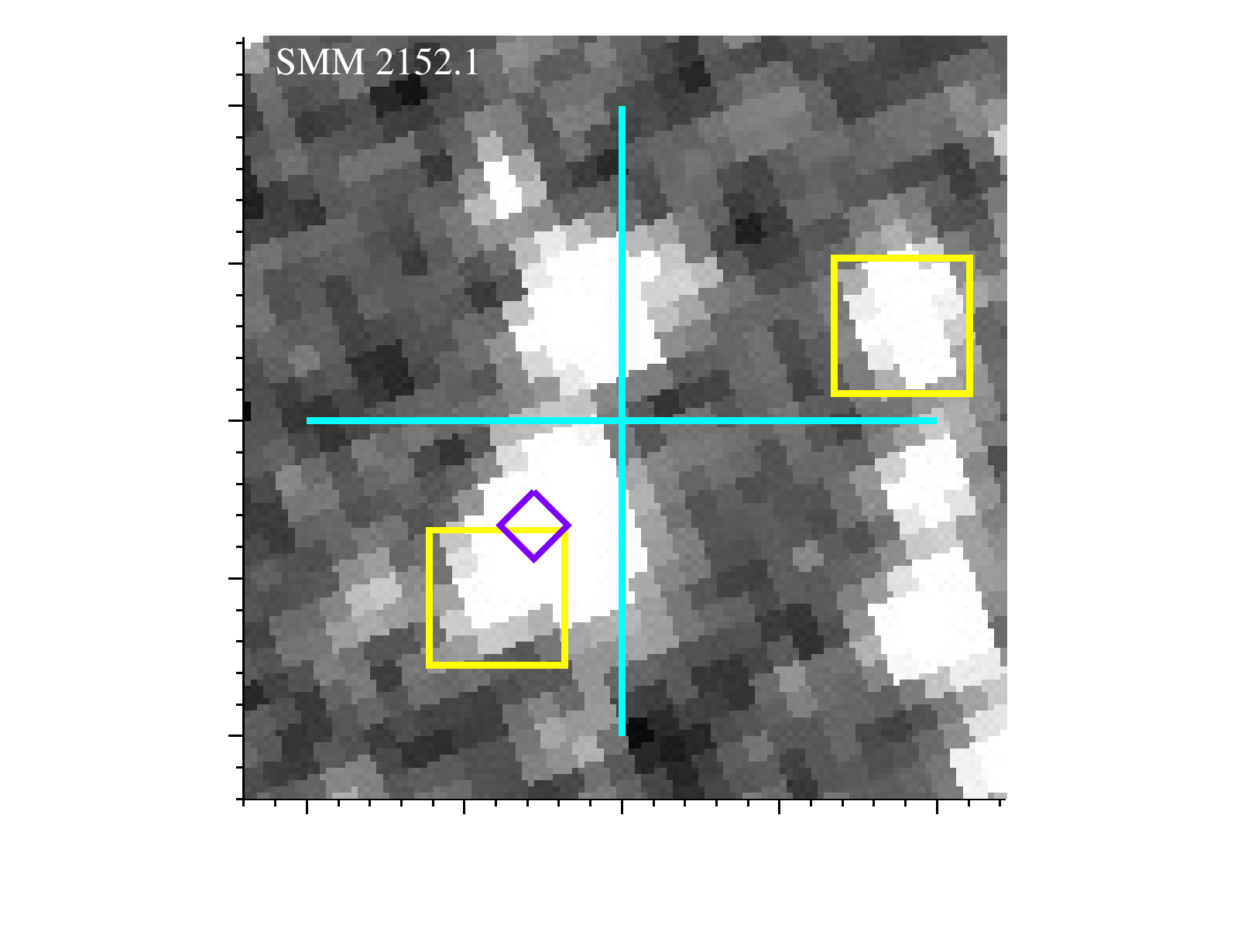}}
     		 \hspace{-27mm}
  	   	 \vspace{-12mm}
      \subfigure{\includegraphics[scale=0.42]{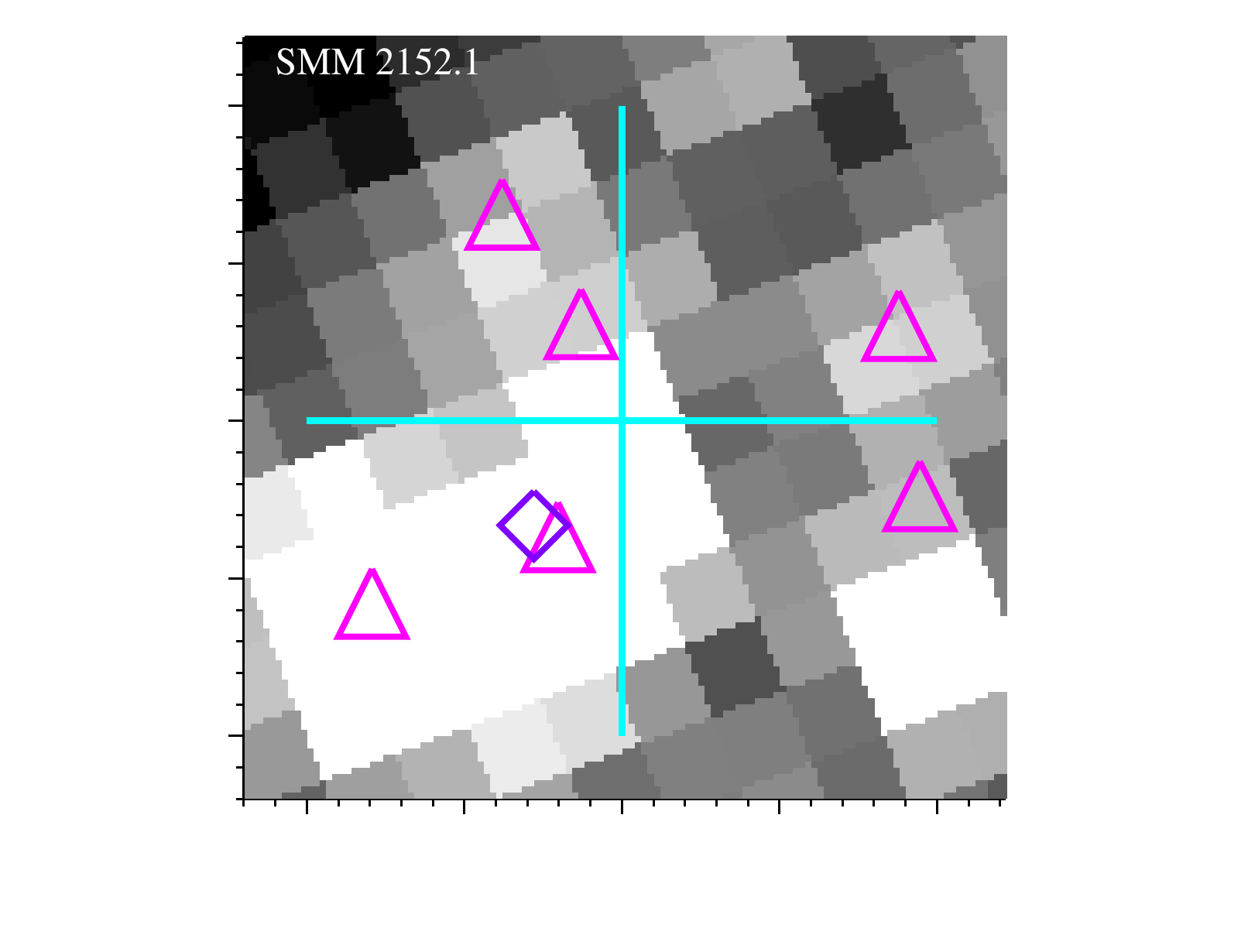}}
        \subfigure{\includegraphics[scale=0.42]{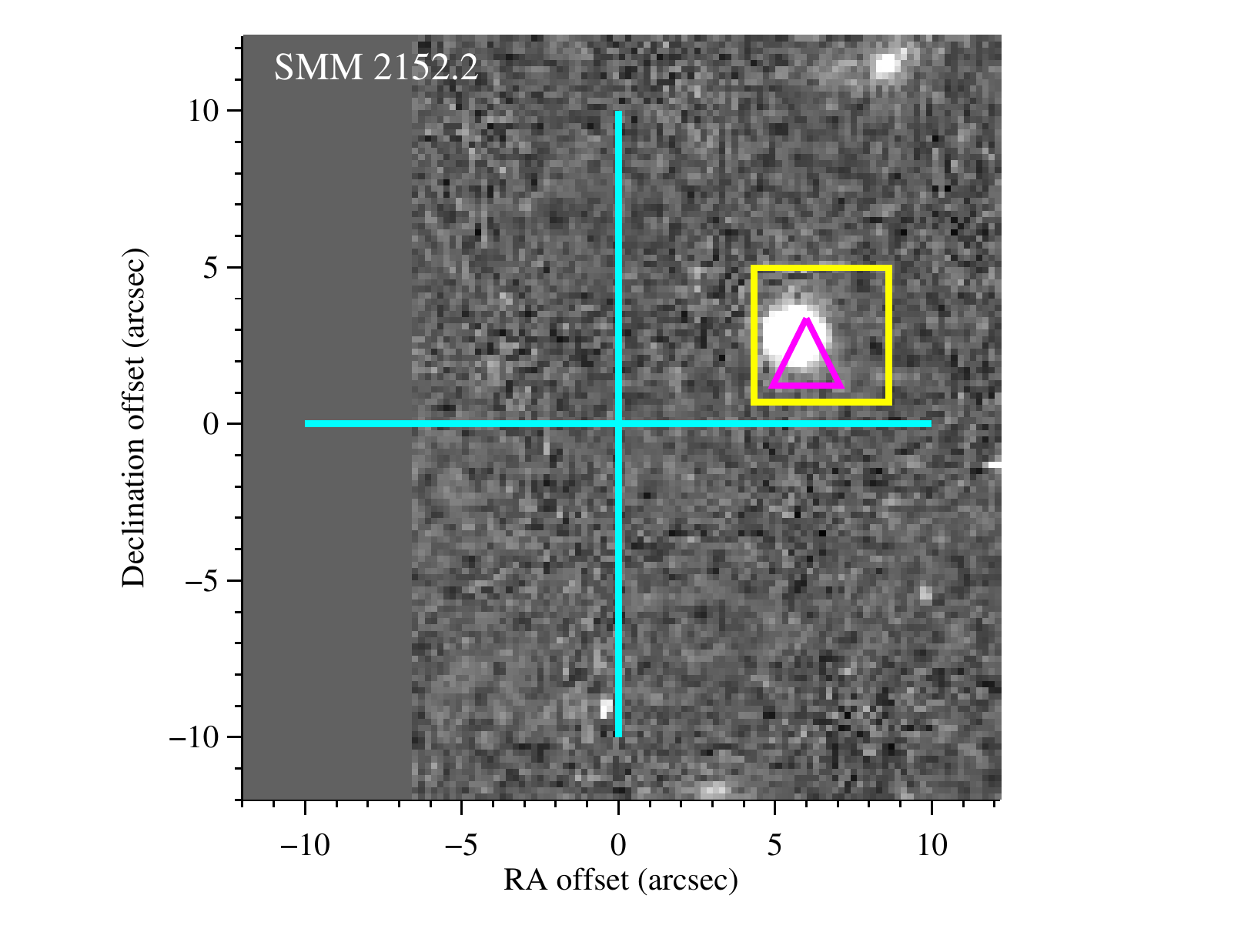}}
                      	\hspace{-27mm}
     \subfigure{\includegraphics[scale=0.42]{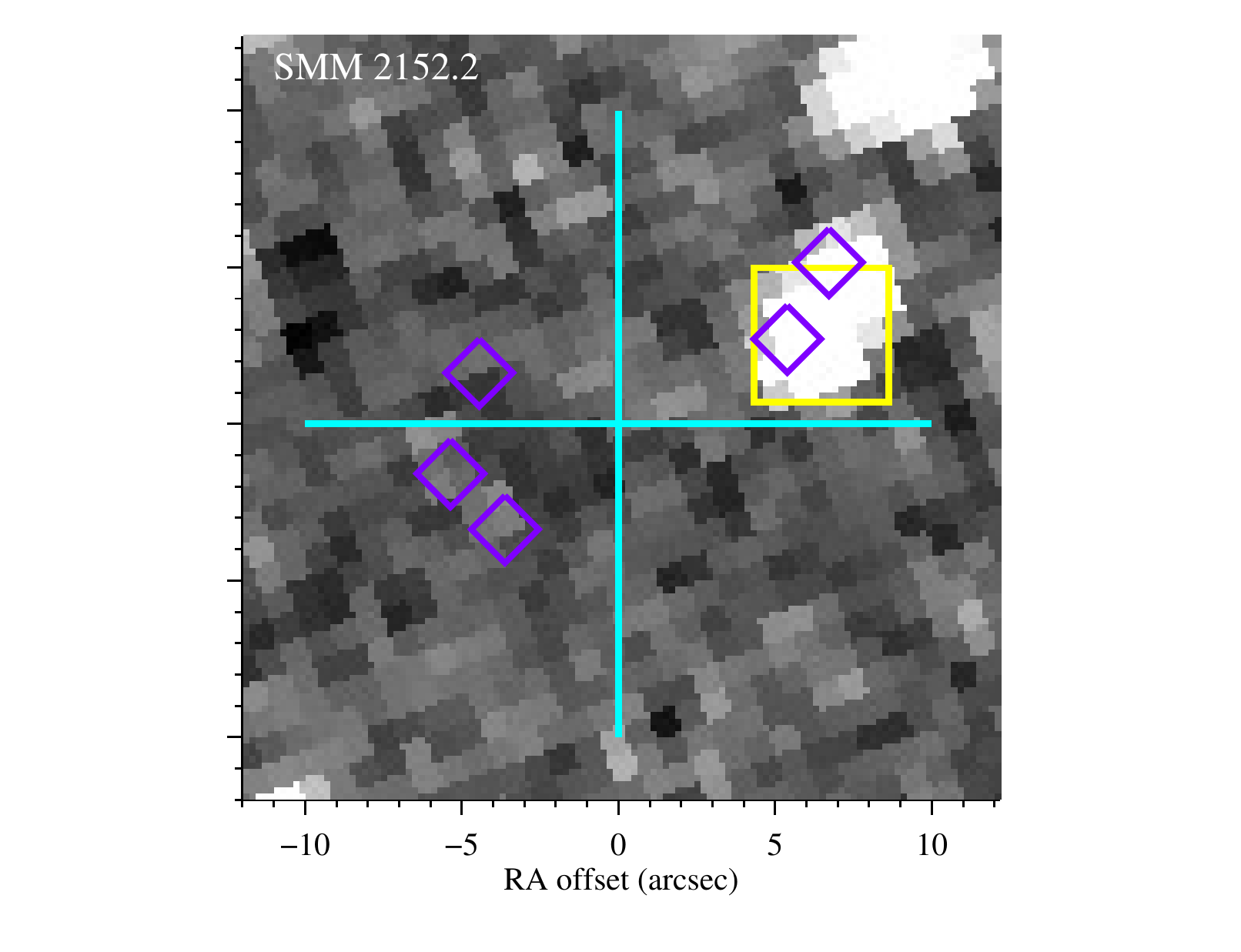}}
     		 \hspace{-27mm}
  	   	 \vspace{-4mm}
      \subfigure{\includegraphics[scale=0.42]{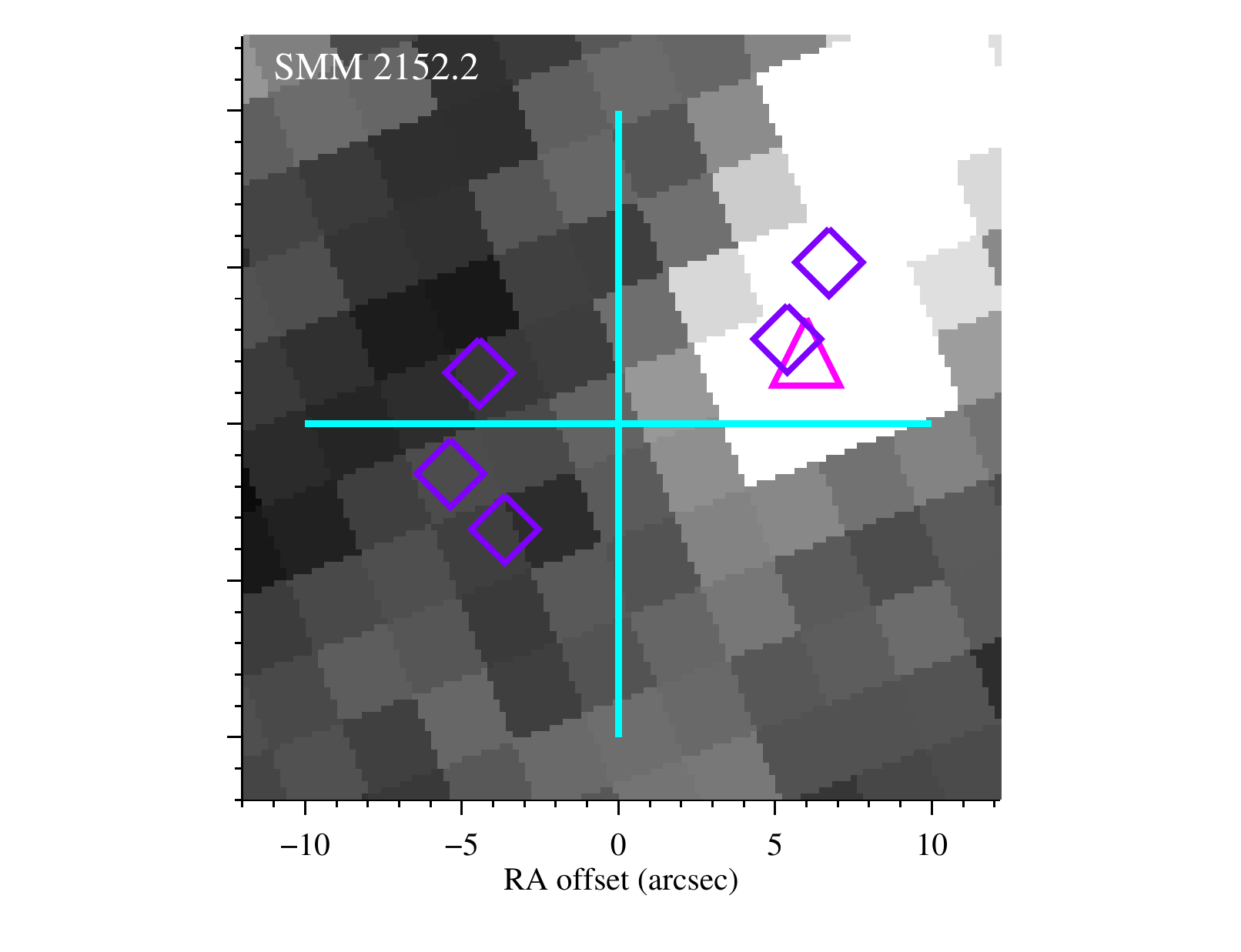}}
 
          \caption{24 $\times$ 24 arcsec$^2$ postage stamps for SMGs with ambiguous counterparts.  The grey-scale images and symbols are the same as those described in Fig.~\ref{fig:secure}.}
   \label{fig:tentative}
\end{figure*}

\begin{figure*}
   \centering
      \subfigure{\includegraphics[scale=0.38]{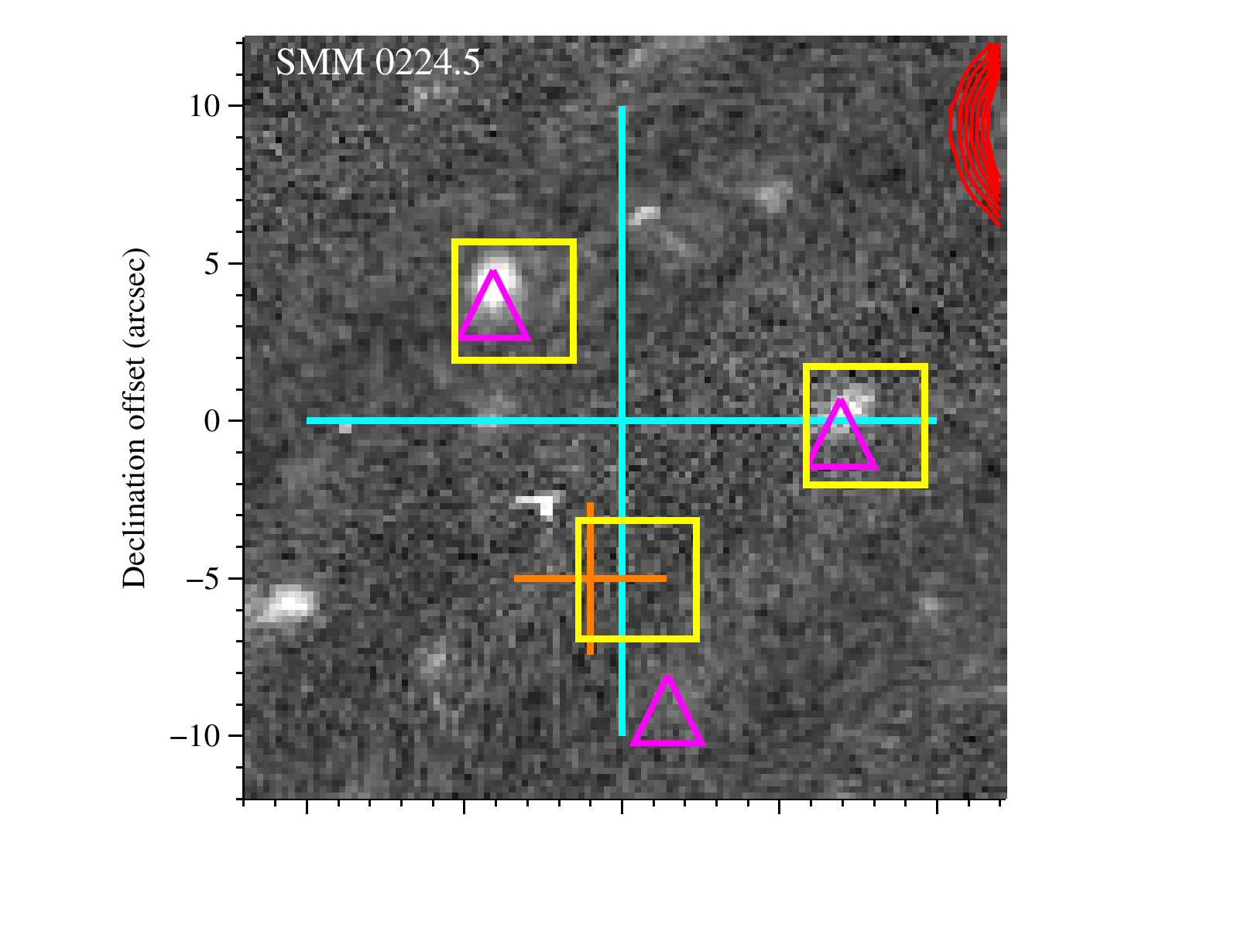}}
                   	\hspace{-24mm}
   \subfigure{\includegraphics[scale=0.38]{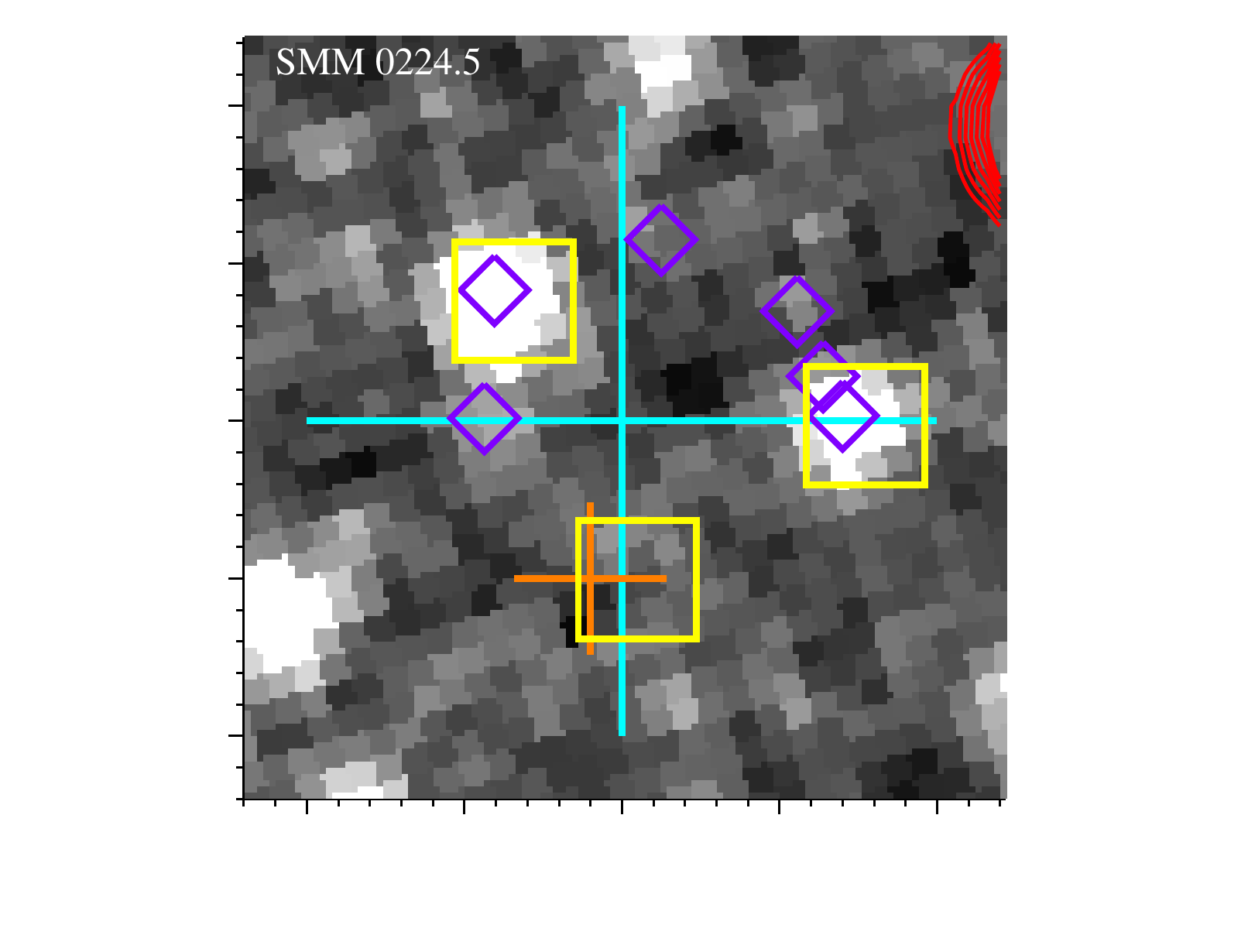}}
        		 \hspace{-24mm}
  	   	 \vspace{-12mm}
   \subfigure{\includegraphics[scale=0.38]{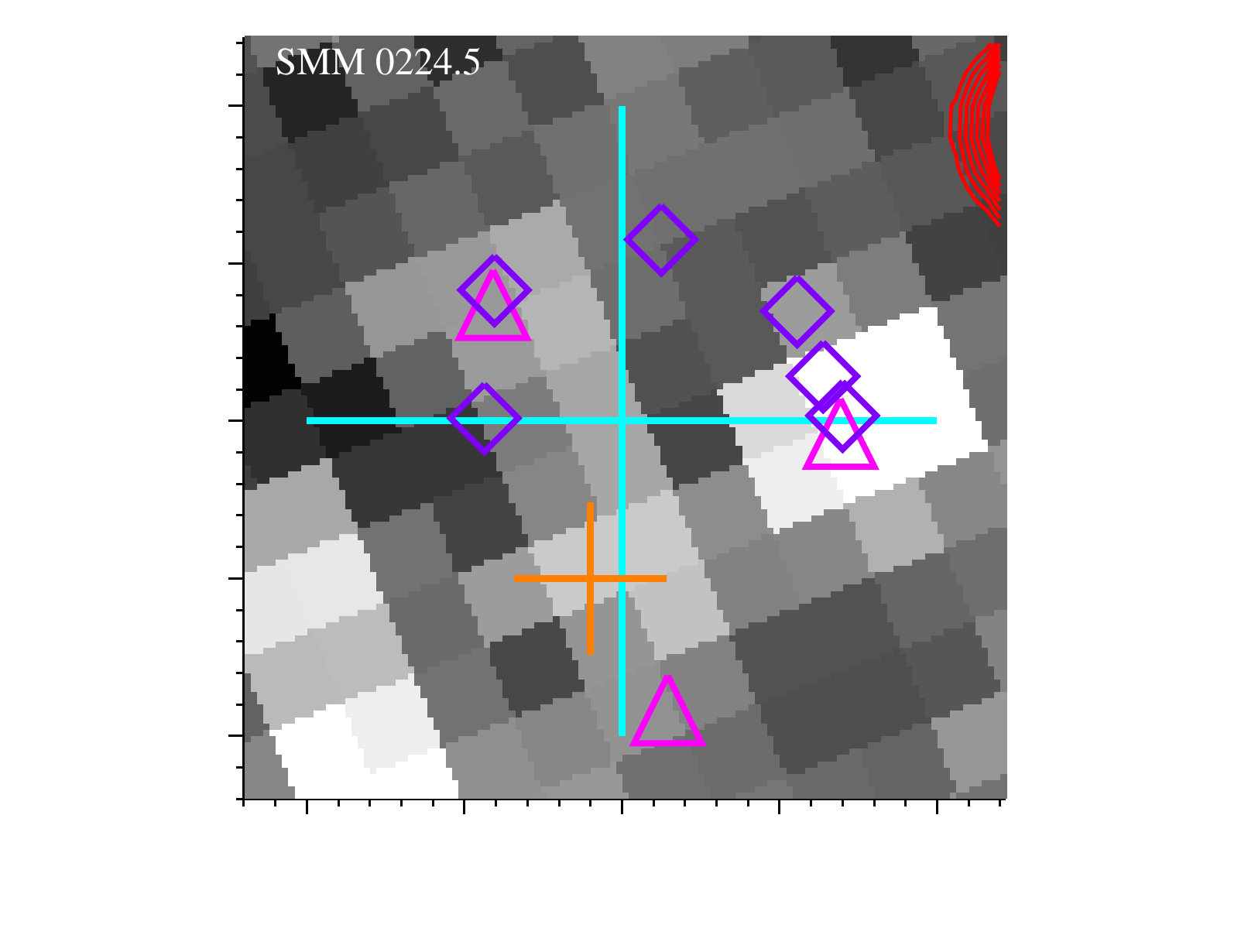}}
       \subfigure{\includegraphics[scale=0.38]{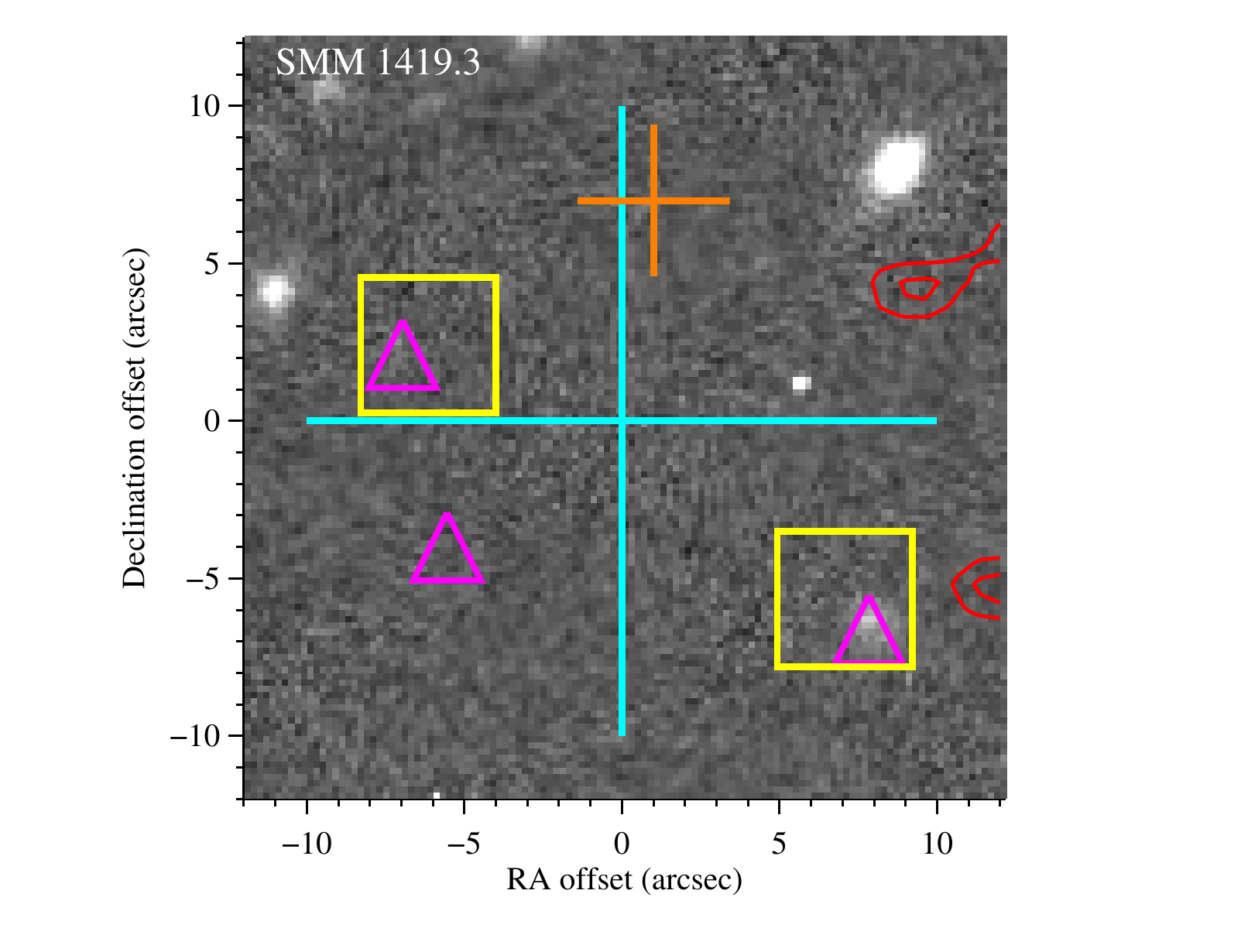}}
                          	\hspace{-24mm}
     \subfigure{\includegraphics[scale=0.38]{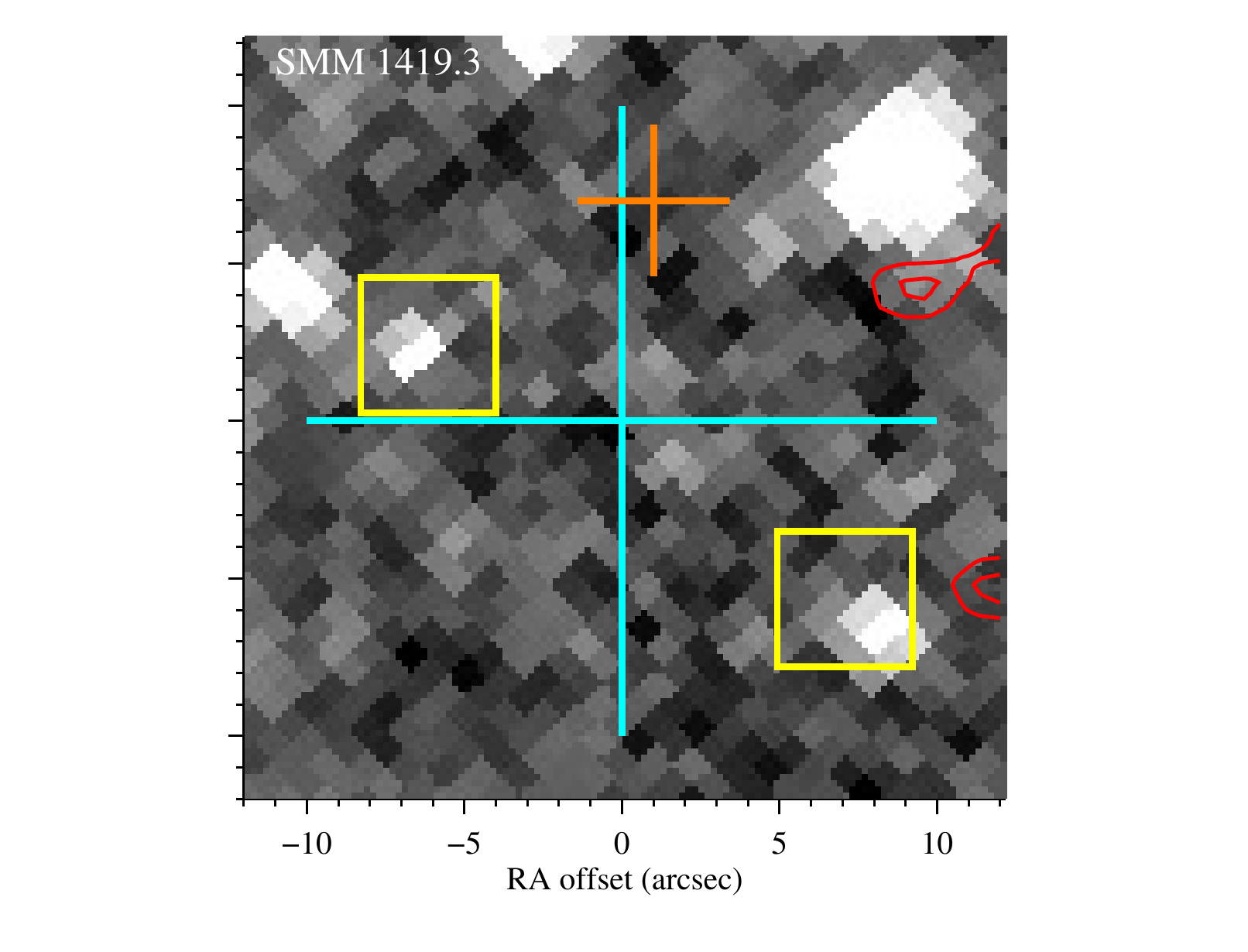}}
          		 \hspace{-24mm}
  	   	 \vspace{-4mm}
      \subfigure{\includegraphics[scale=0.38]{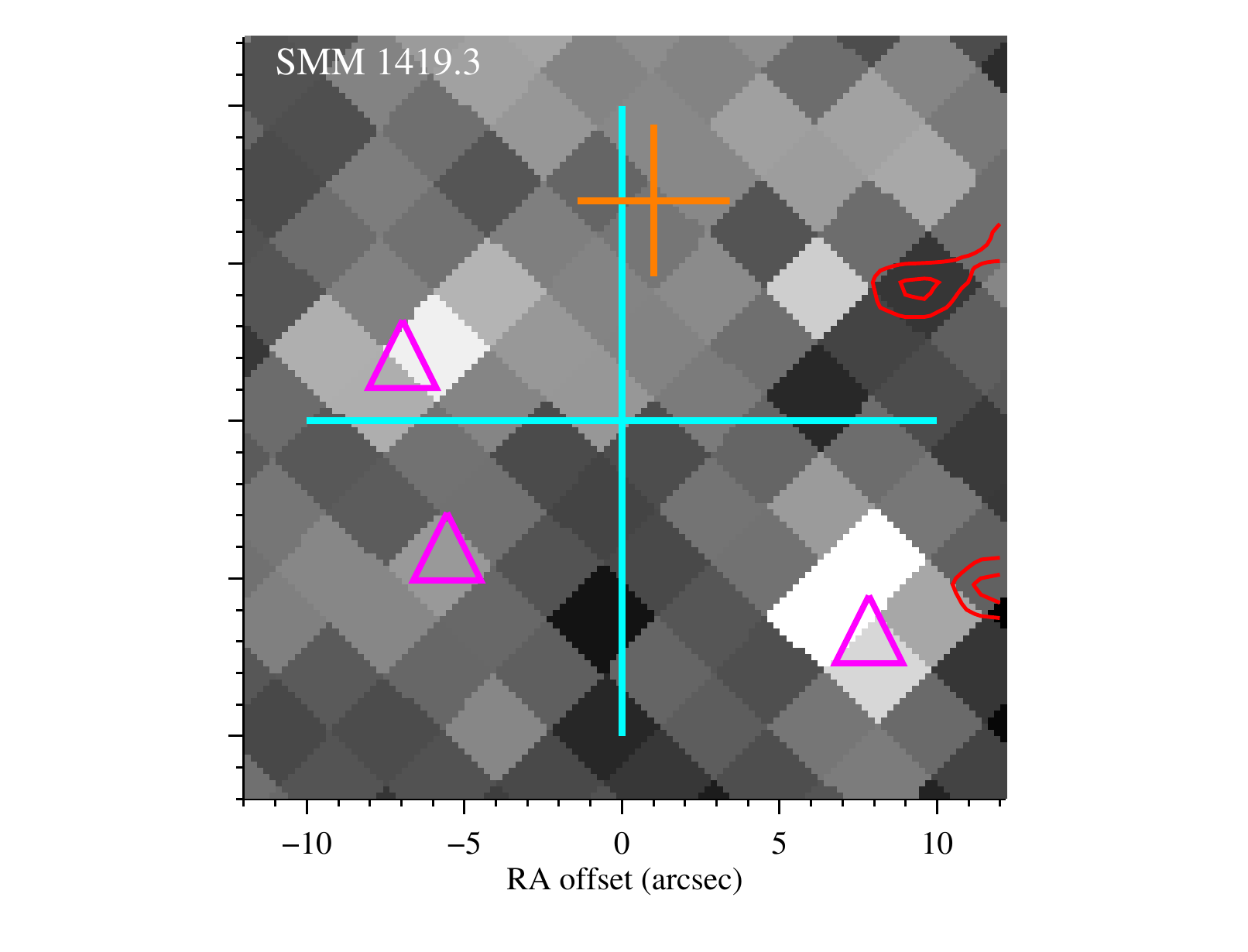}}
           \caption{24 $\times$ 24 arcsec$^2$ postage stamps for SMGs with tentative counterparts.  The grey-scale images and symbols are the same as those described in Fig.~\ref{fig:secure}.}
   \label{fig:tentative}
\end{figure*}
\nopagebreak
\begin{figure*}
   \centering
      \subfigure{\includegraphics[scale=0.38]{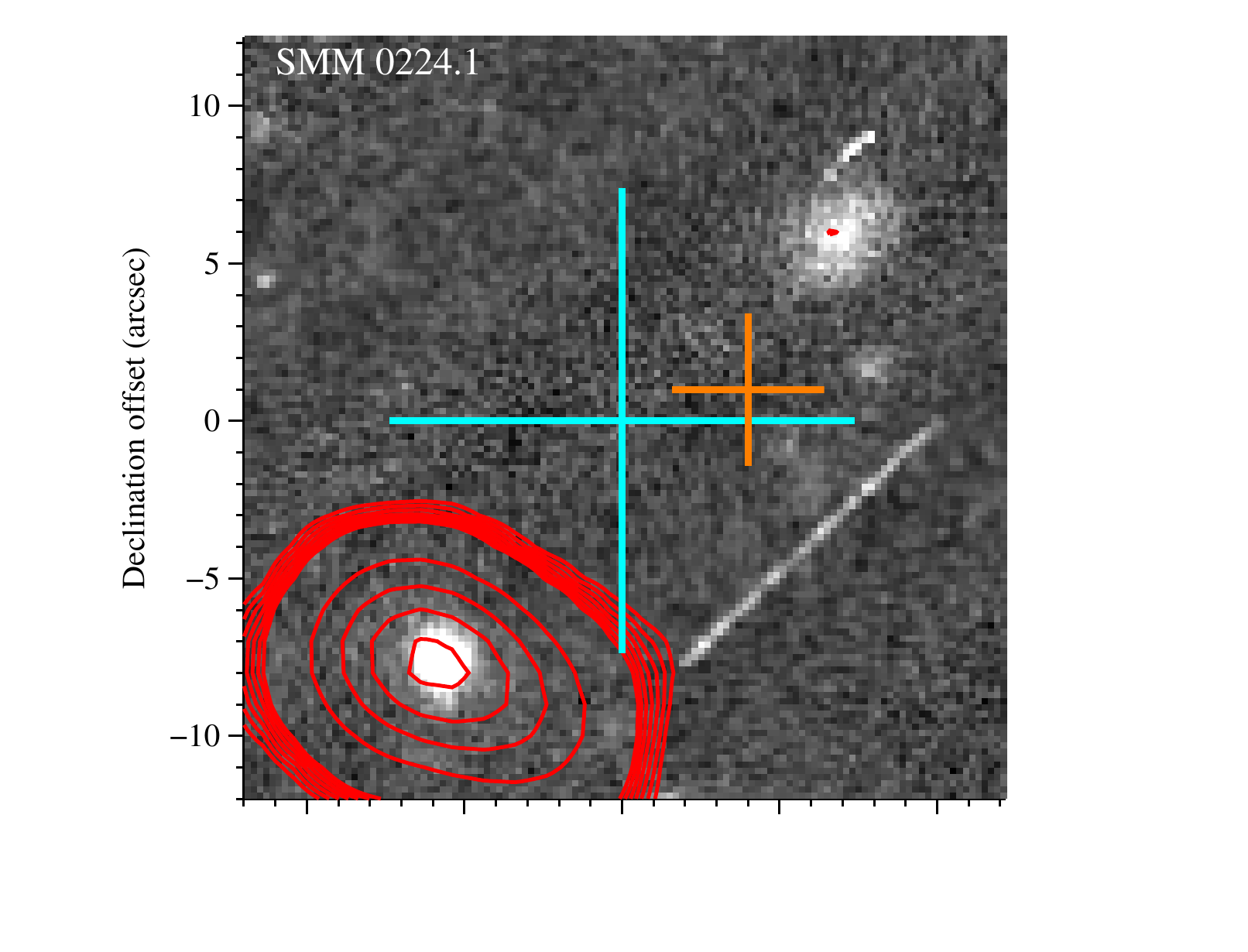}}
                         	\hspace{-24mm}
   \subfigure{\includegraphics[scale=0.38]{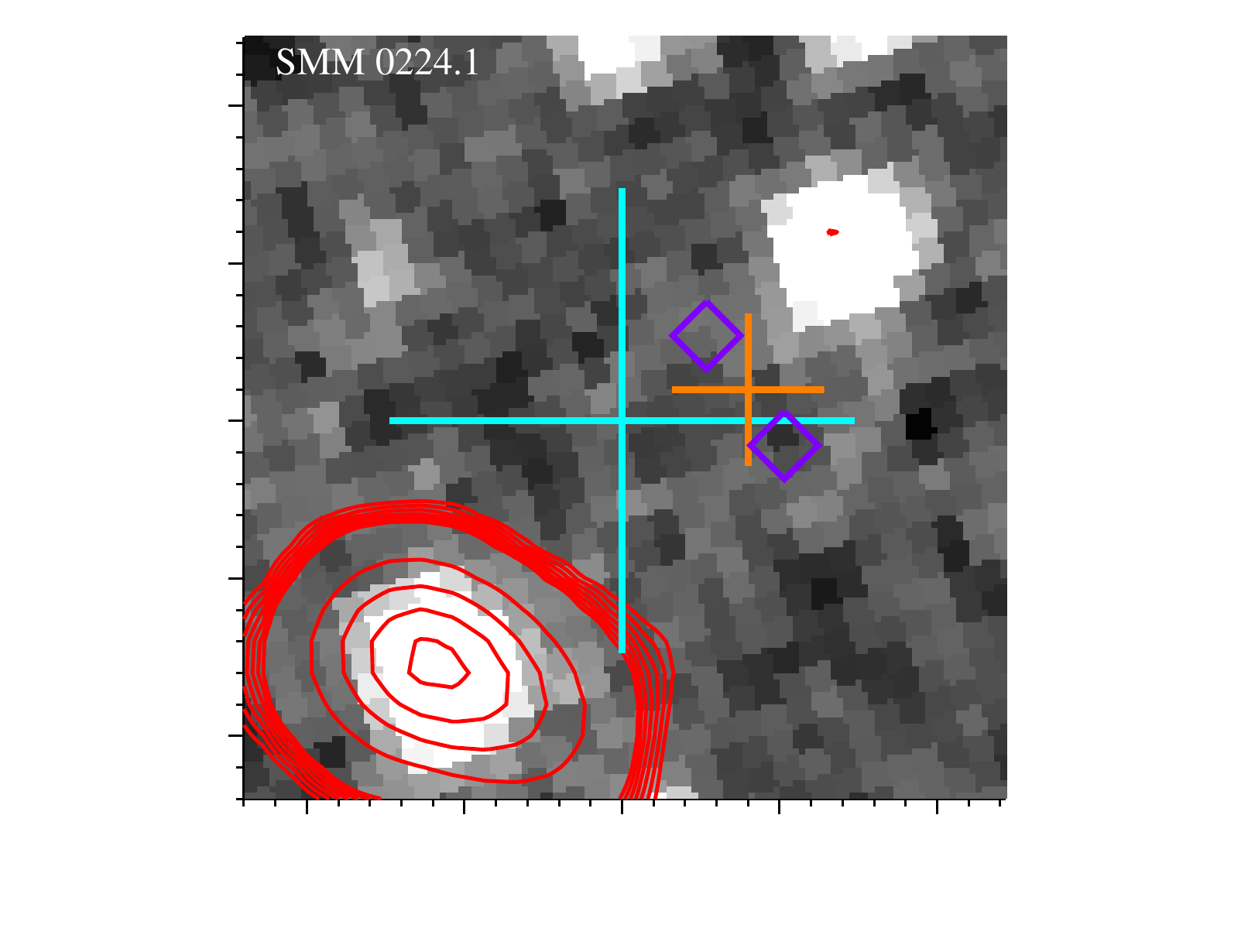}}
        		 \hspace{-24mm}
  	   	 \vspace{-12mm}
   \subfigure{\includegraphics[scale=0.38]{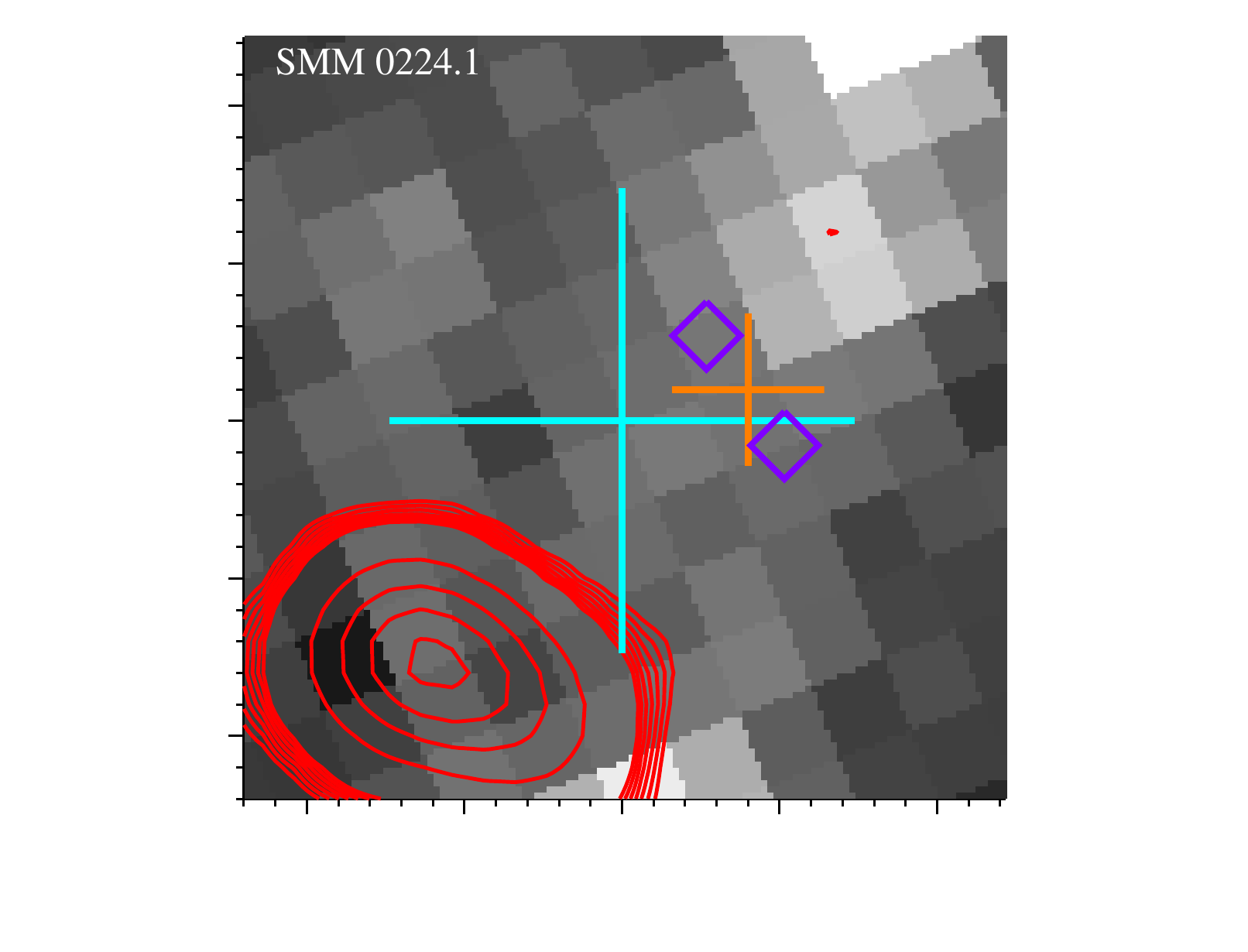}}
      \subfigure{\includegraphics[scale=0.38]{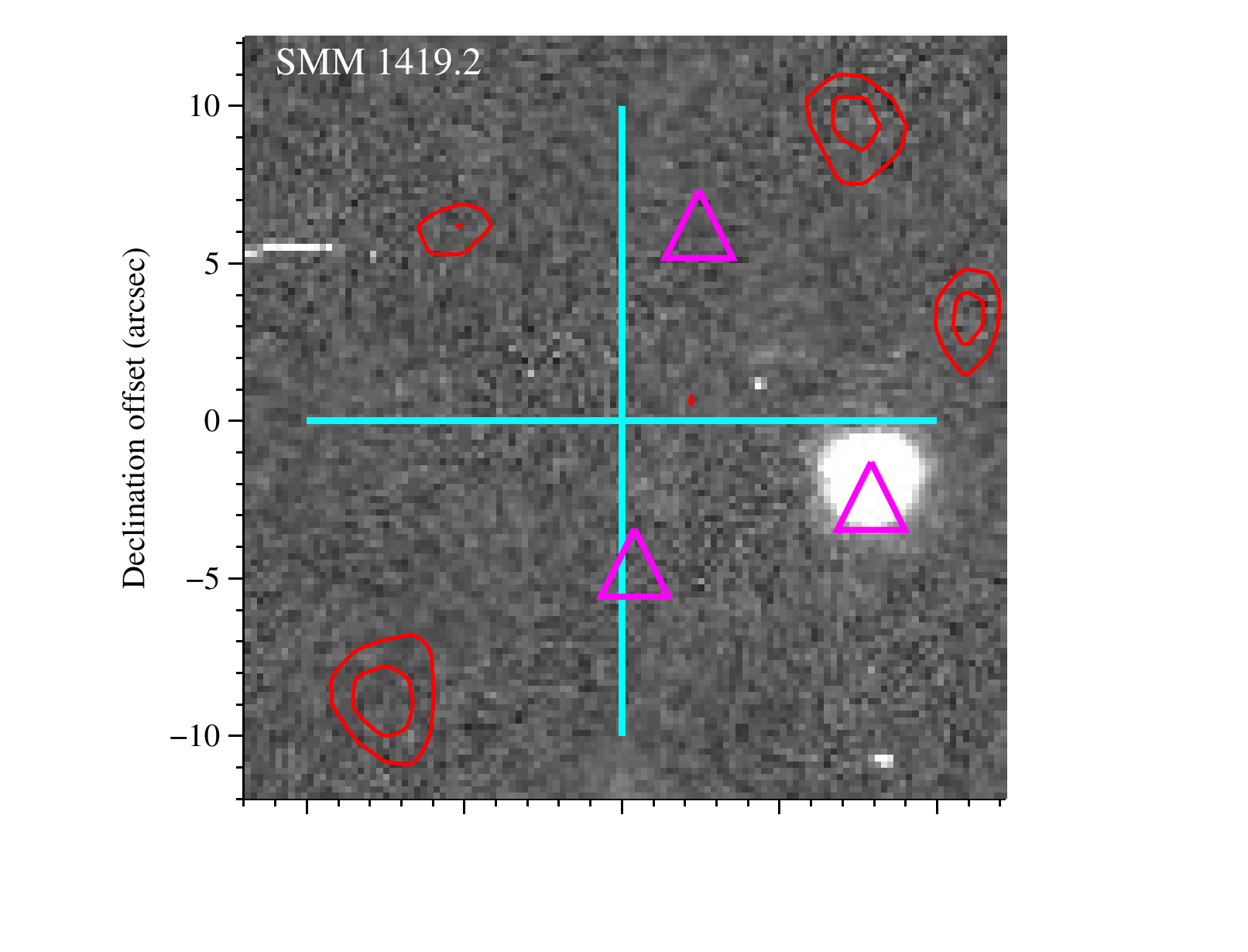}}
                         	\hspace{-24mm}
   \subfigure{\includegraphics[scale=0.38]{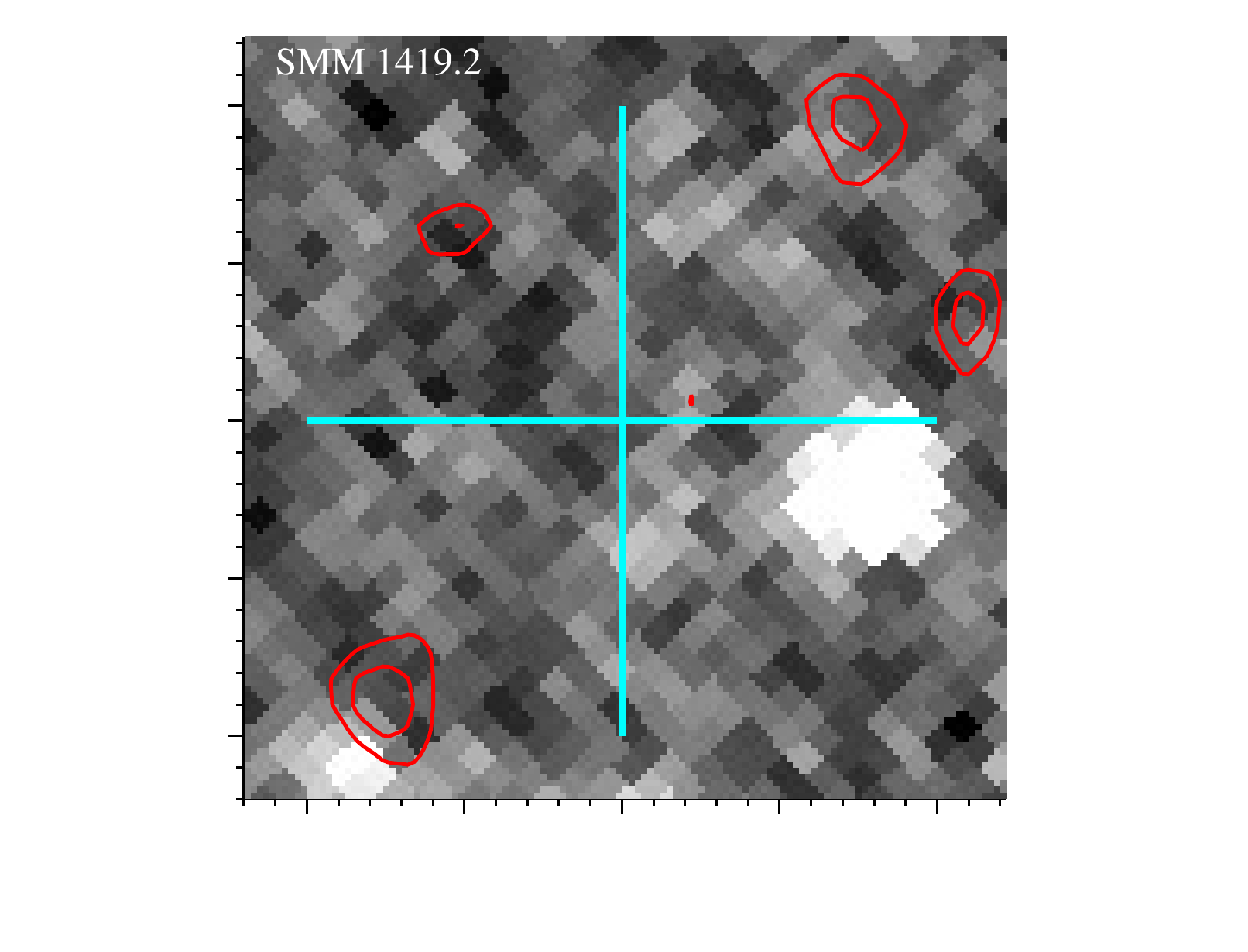}}
        		 \hspace{-24mm}
  	   	 \vspace{-12mm}
   \subfigure{\includegraphics[scale=0.38]{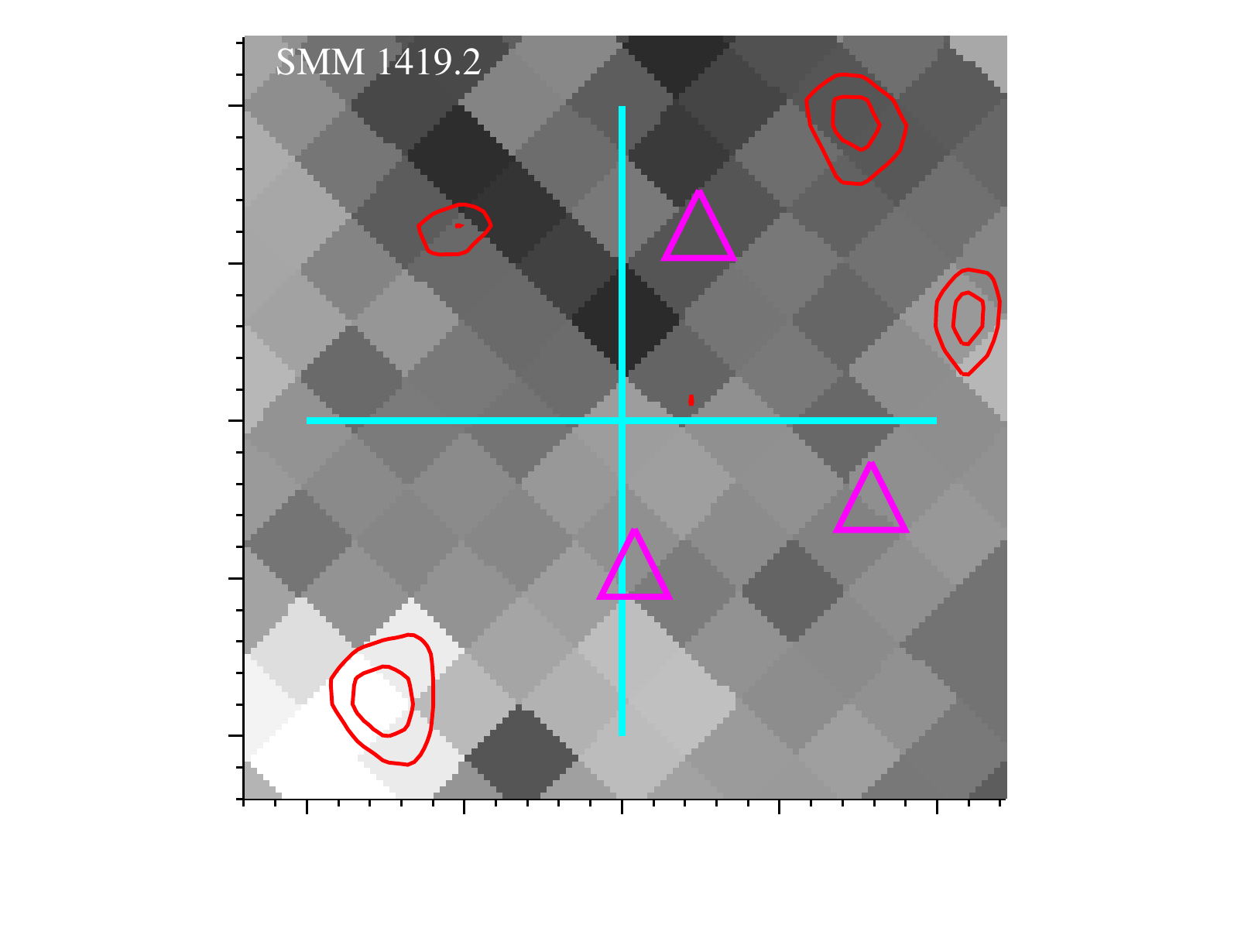}}
            \subfigure{\includegraphics[scale=0.38]{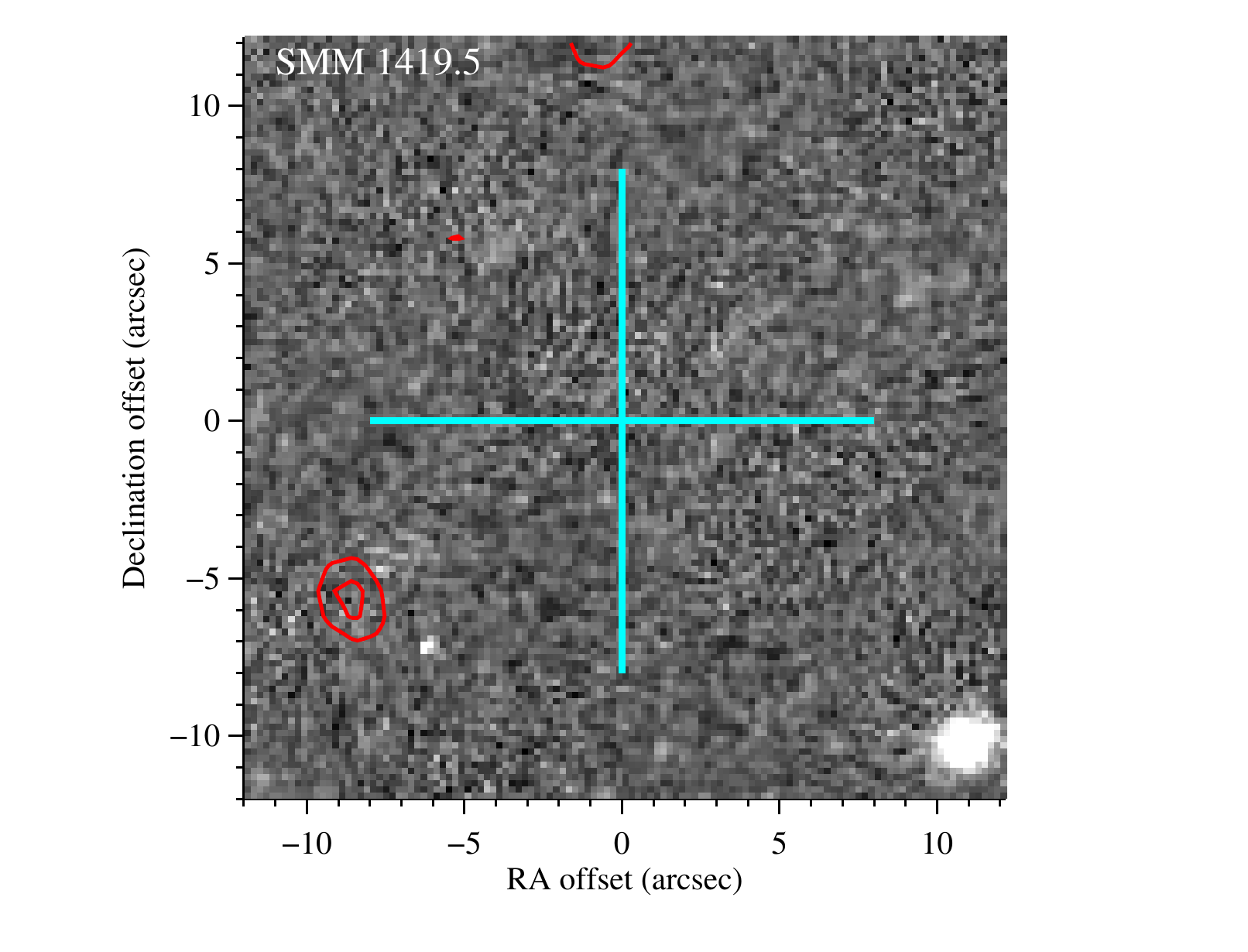}}
                               	\hspace{-24mm}
   \subfigure{\includegraphics[scale=0.38]{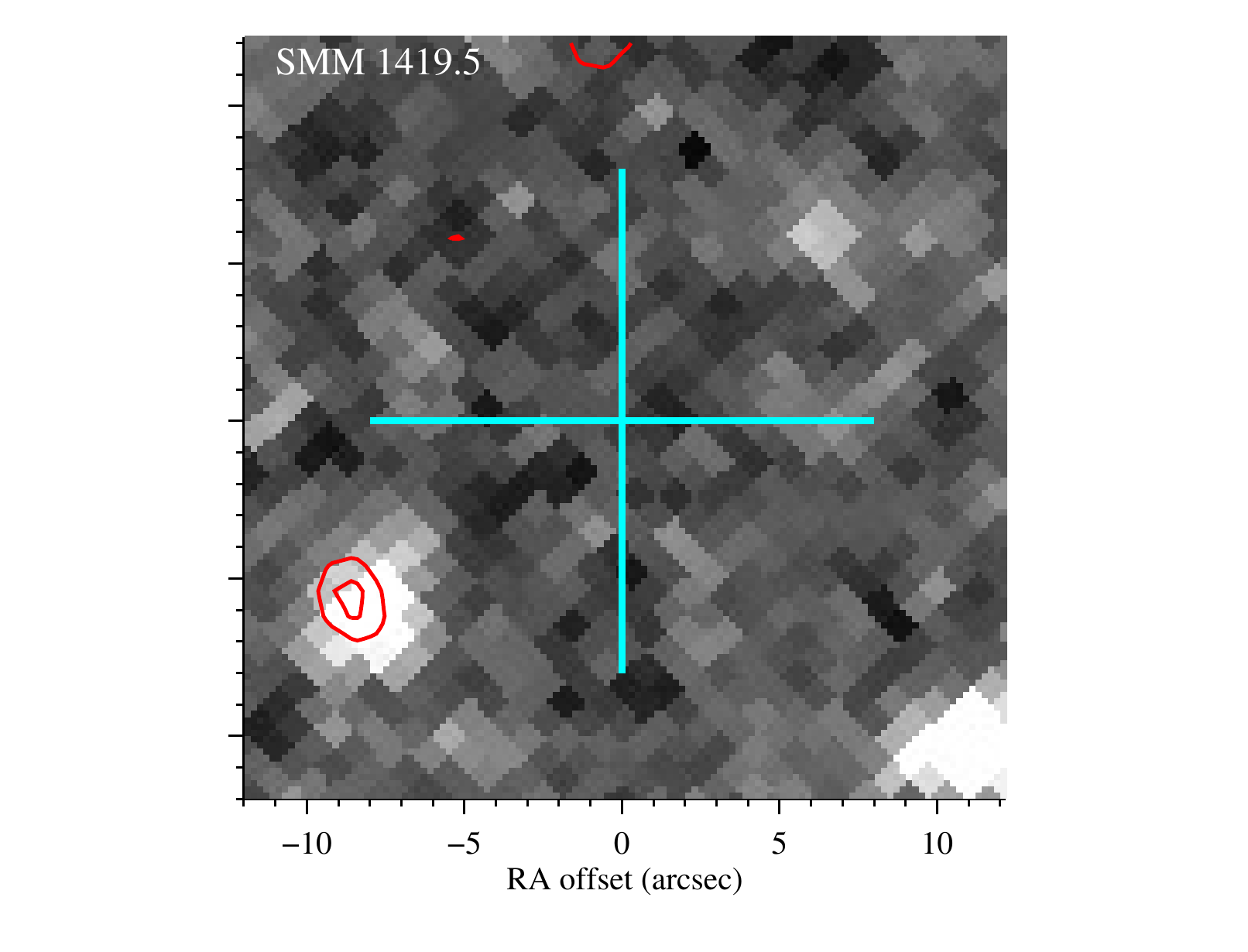}}
        		 \hspace{-24mm}
  	   	 \vspace{-4mm}
   \subfigure{\includegraphics[scale=0.38]{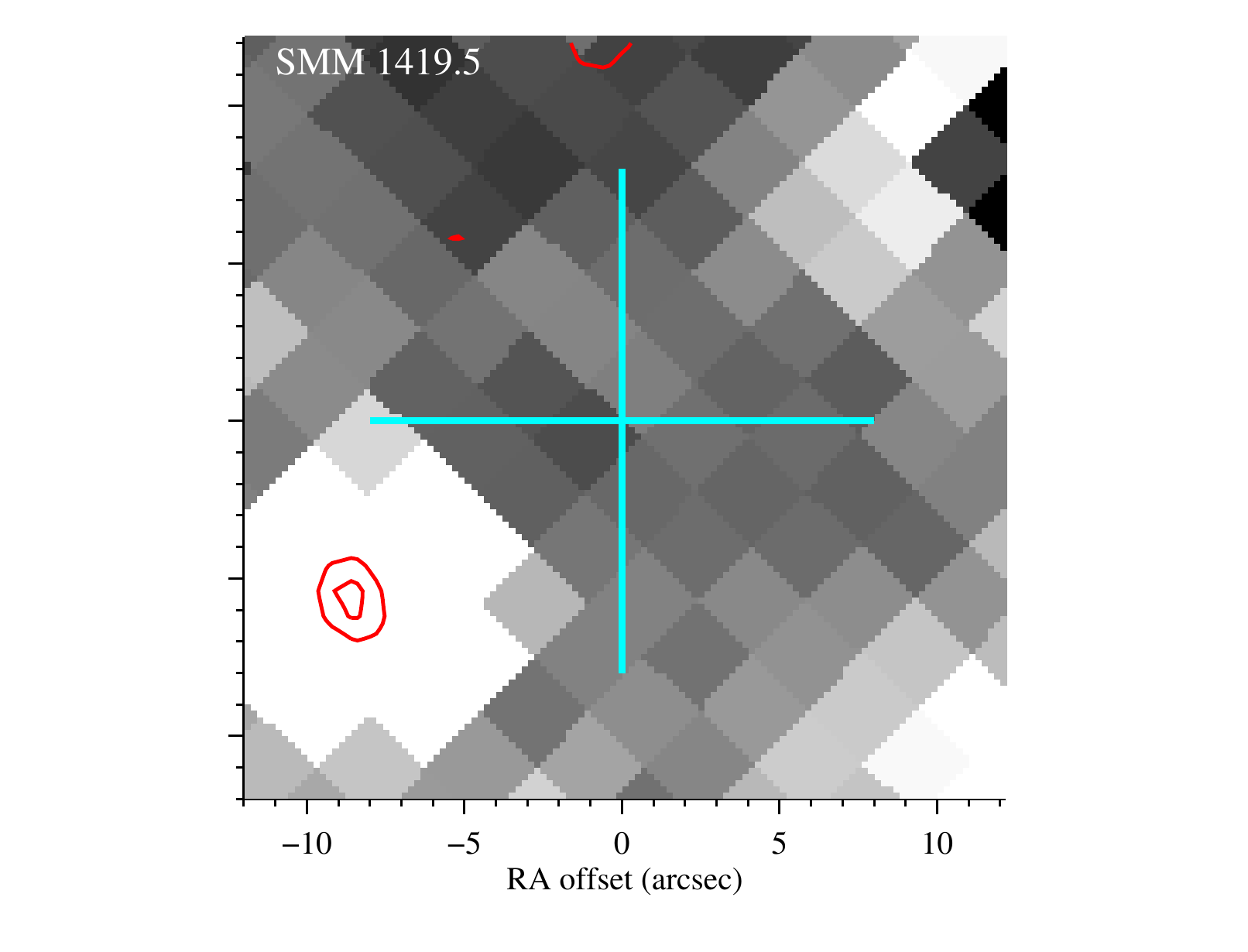}}
               \caption{24 $\times$ 24 arcsec$^2$ postage stamps for SMGs without any likely counterparts.  The grey-scale images and symbols are the same as those described in Fig.~\ref{fig:secure}.}
   \label{fig:none}
\end{figure*}

\begin{figure*}
\centering
\vspace{-4mm}
	\subfigure{\includegraphics[scale=0.36]{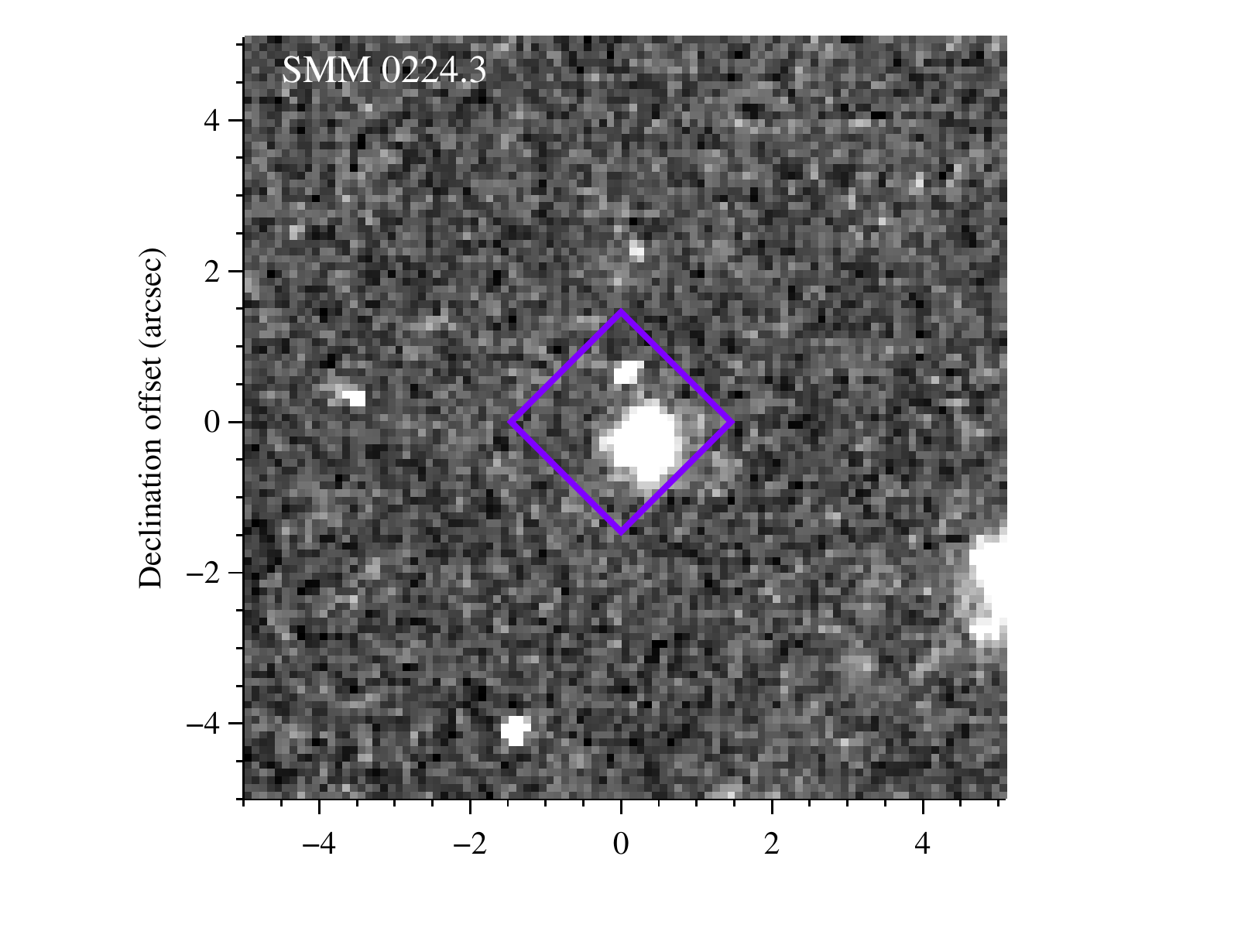}}
          	\hspace{-23mm}
  	   	 \vspace{-5mm}
	\subfigure{\includegraphics[scale=0.36]{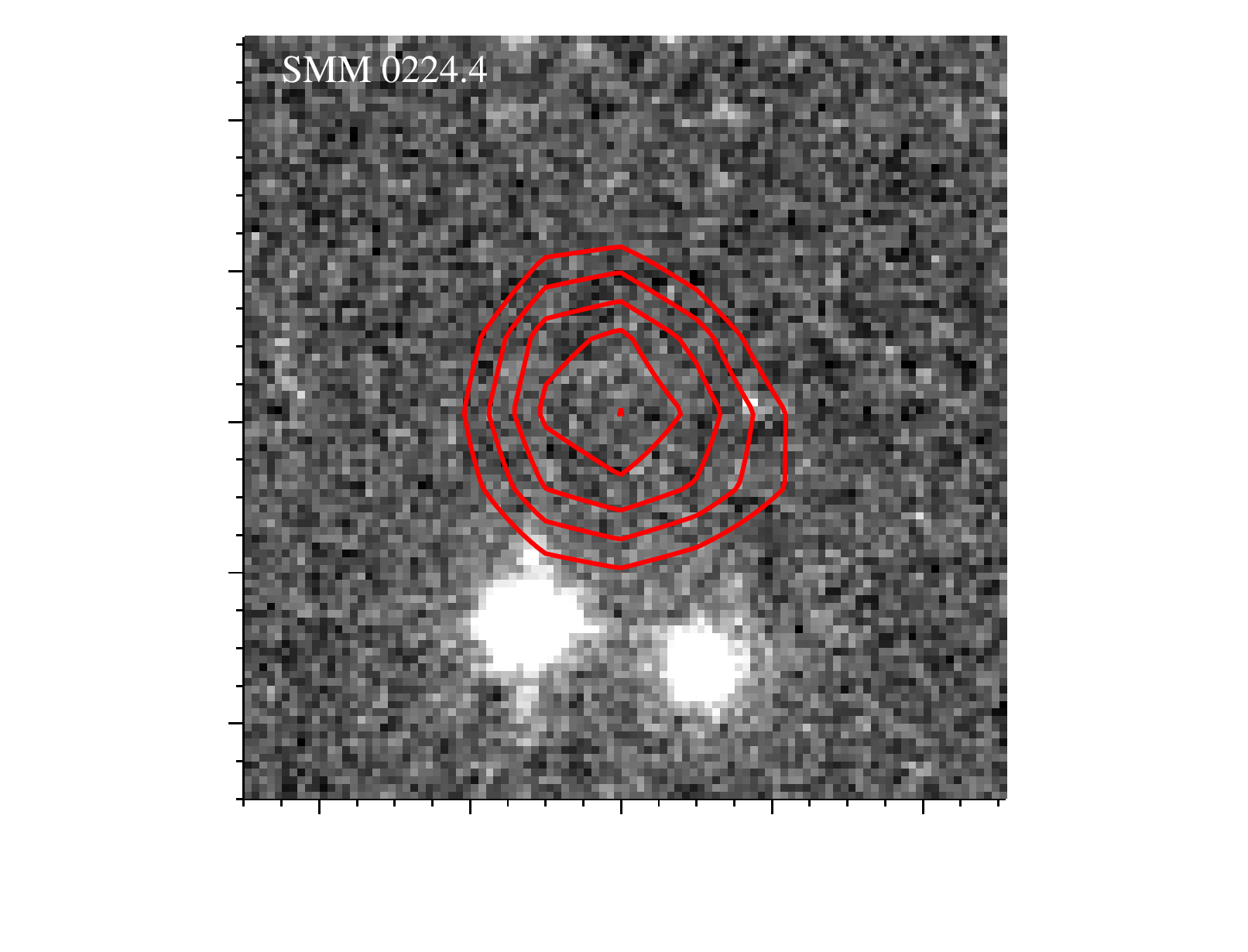}}
  	   	 \vspace{-6mm}
\\
	\subfigure{\includegraphics[scale=0.36]{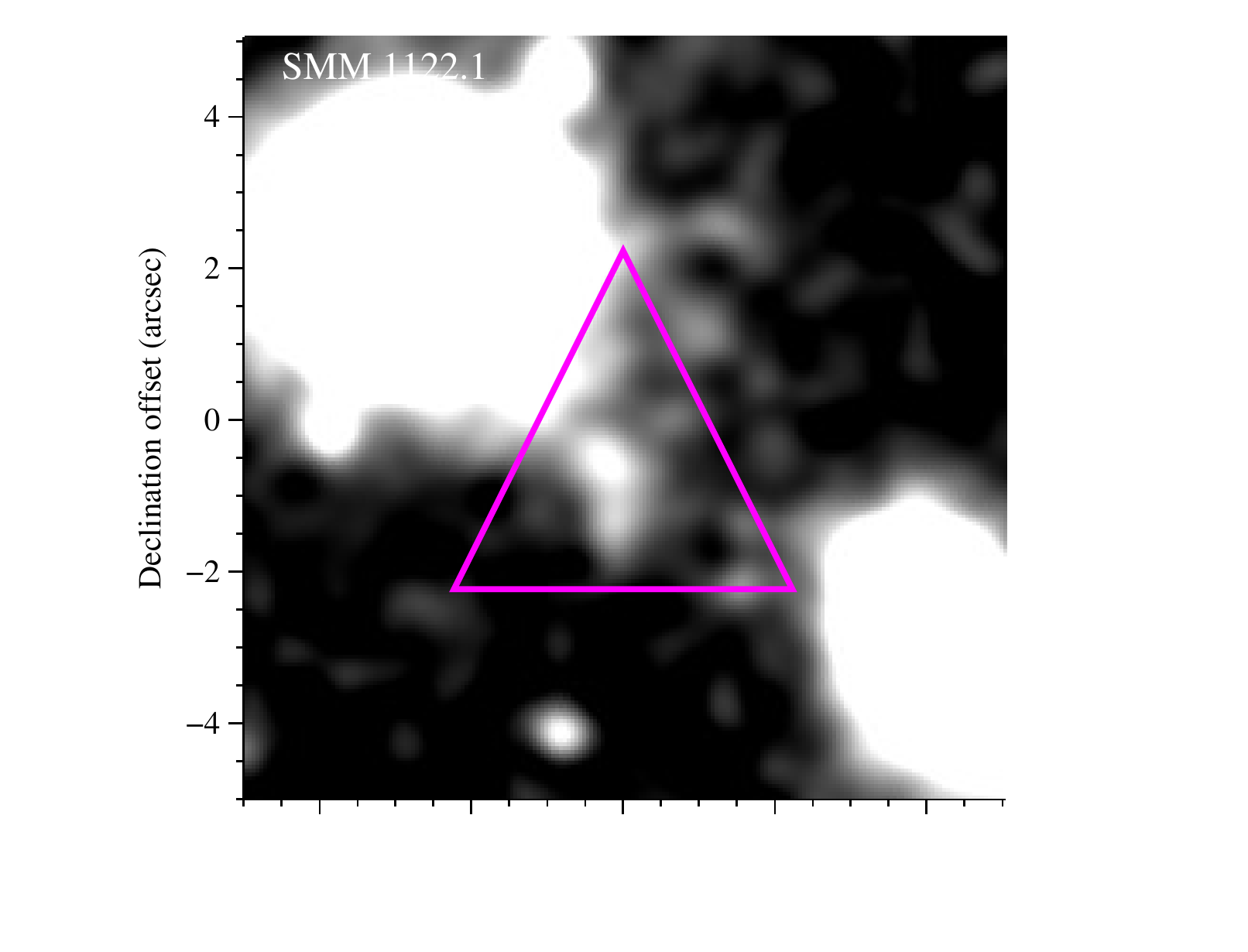}}
		 \hspace{-23mm}
  	   	 \vspace{-5mm}
	\subfigure{\includegraphics[scale=0.36]{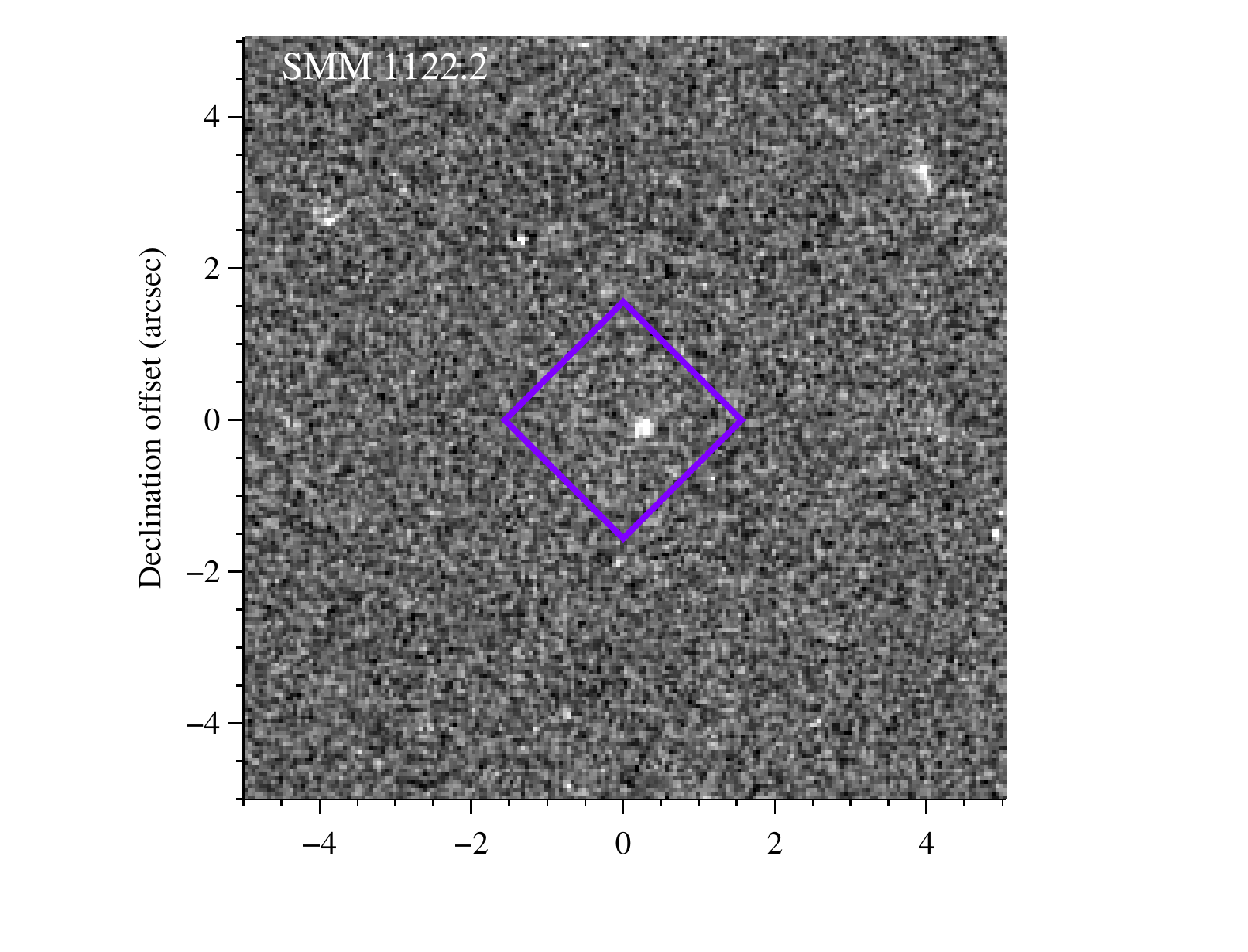}}
        		 \hspace{-23mm}
	\subfigure{\includegraphics[scale=0.36]{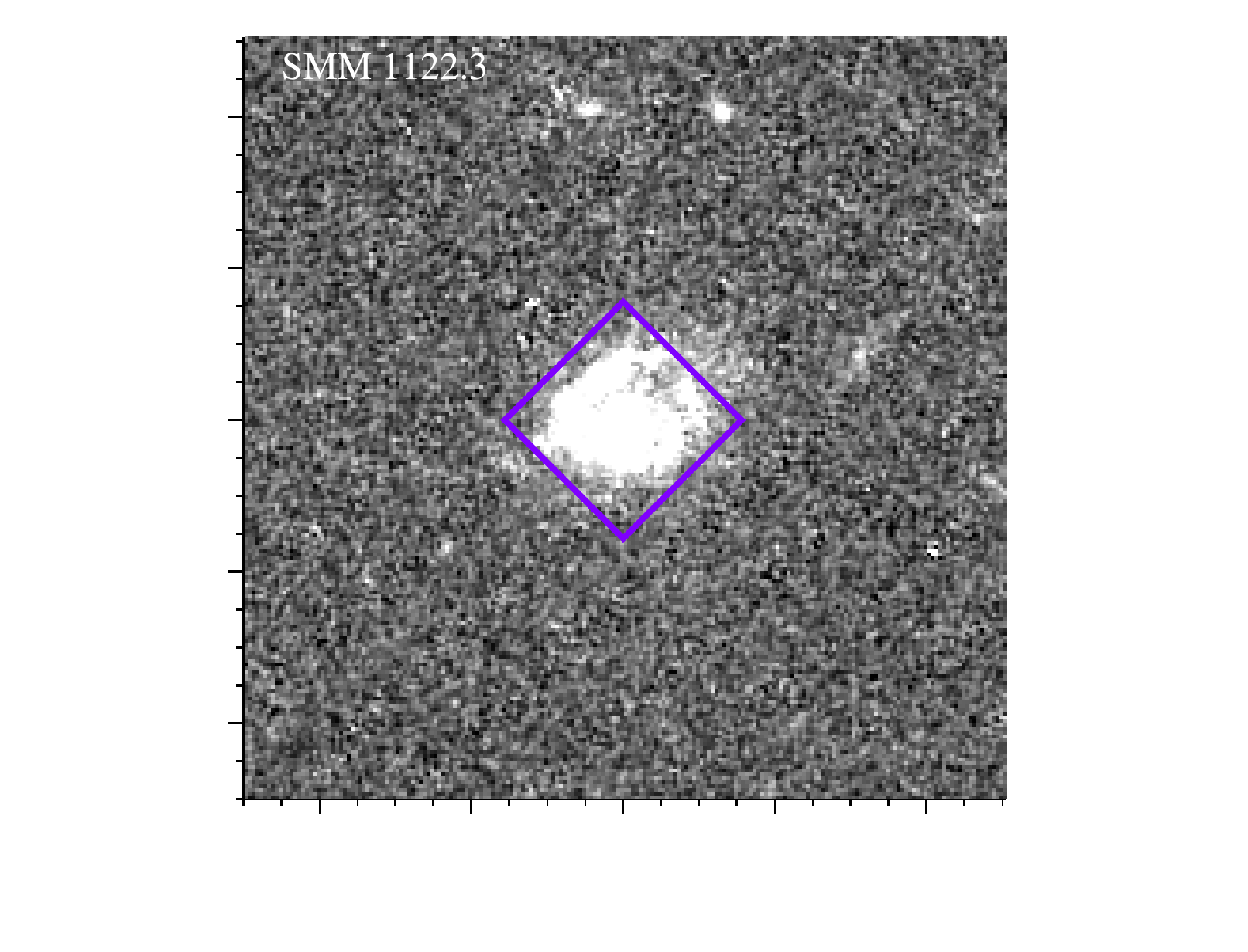}}
	   	 \vspace{-6mm}
\\
	\subfigure{\includegraphics[scale=0.36]{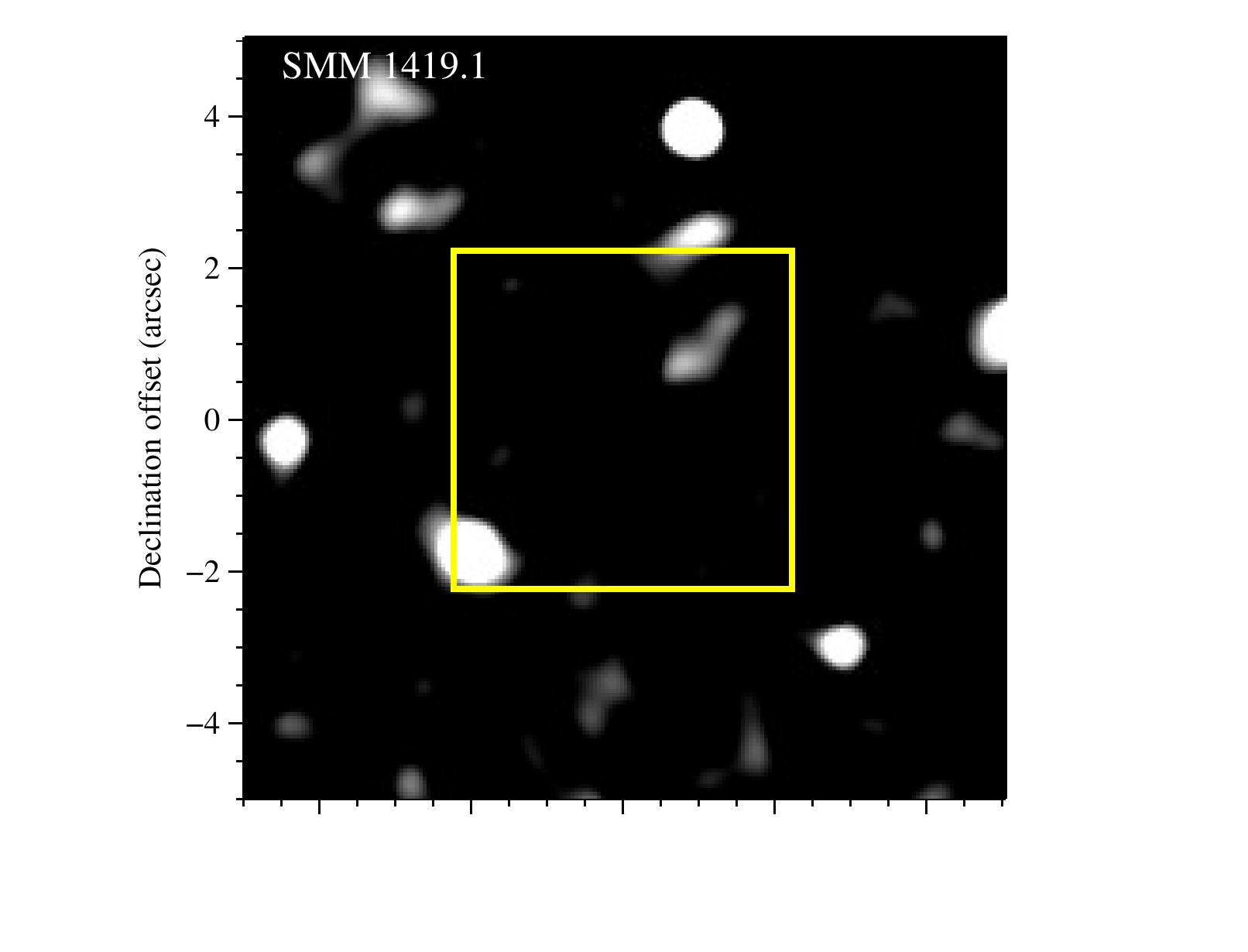}}
		 \hspace{-23mm}
  	   	 \vspace{-5mm}
	\subfigure{\includegraphics[scale=0.36]{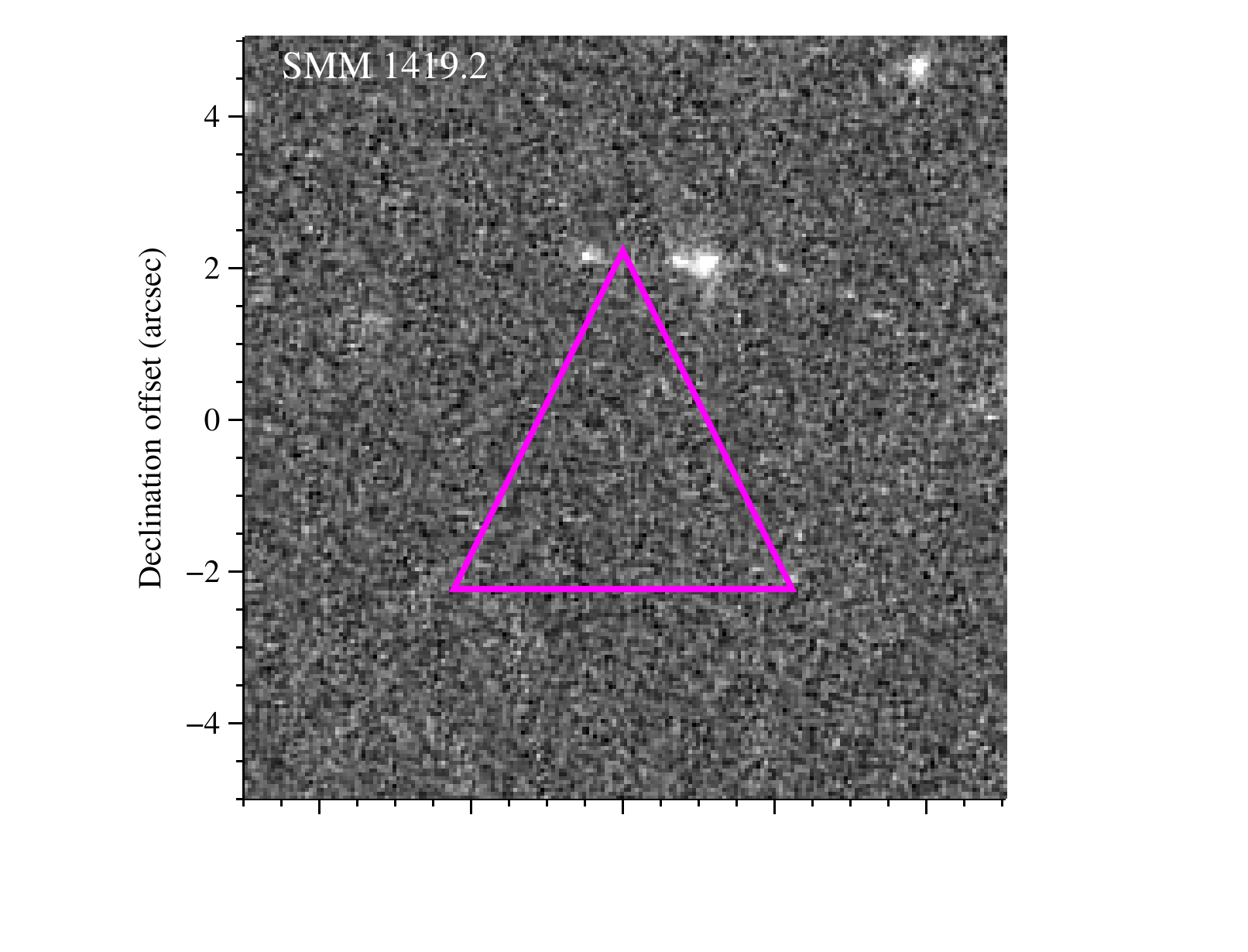}}
		 \hspace{-23mm}
	\subfigure{\includegraphics[scale=0.36]{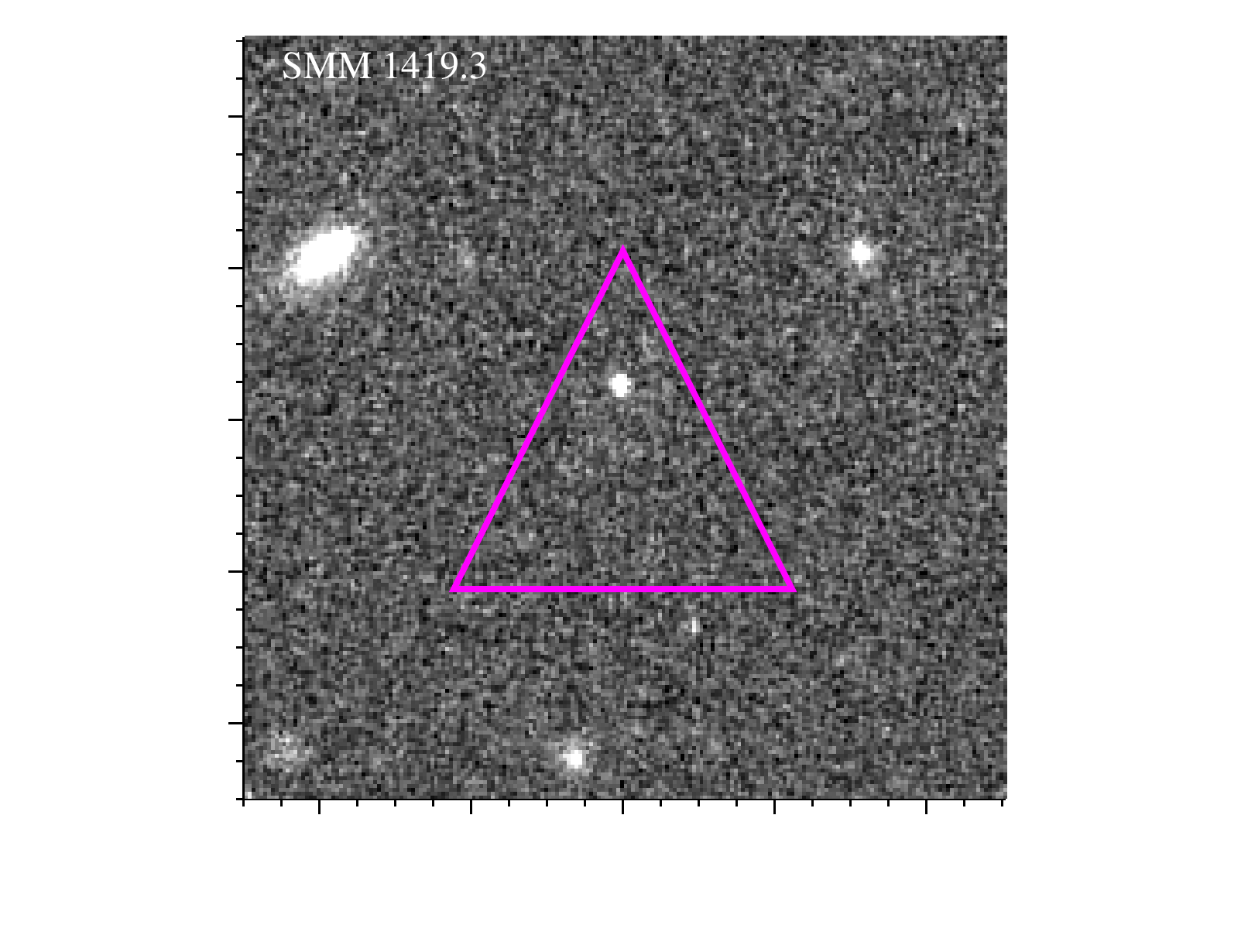}}
		\hspace{-23mm}
	\subfigure{\includegraphics[scale=0.36]{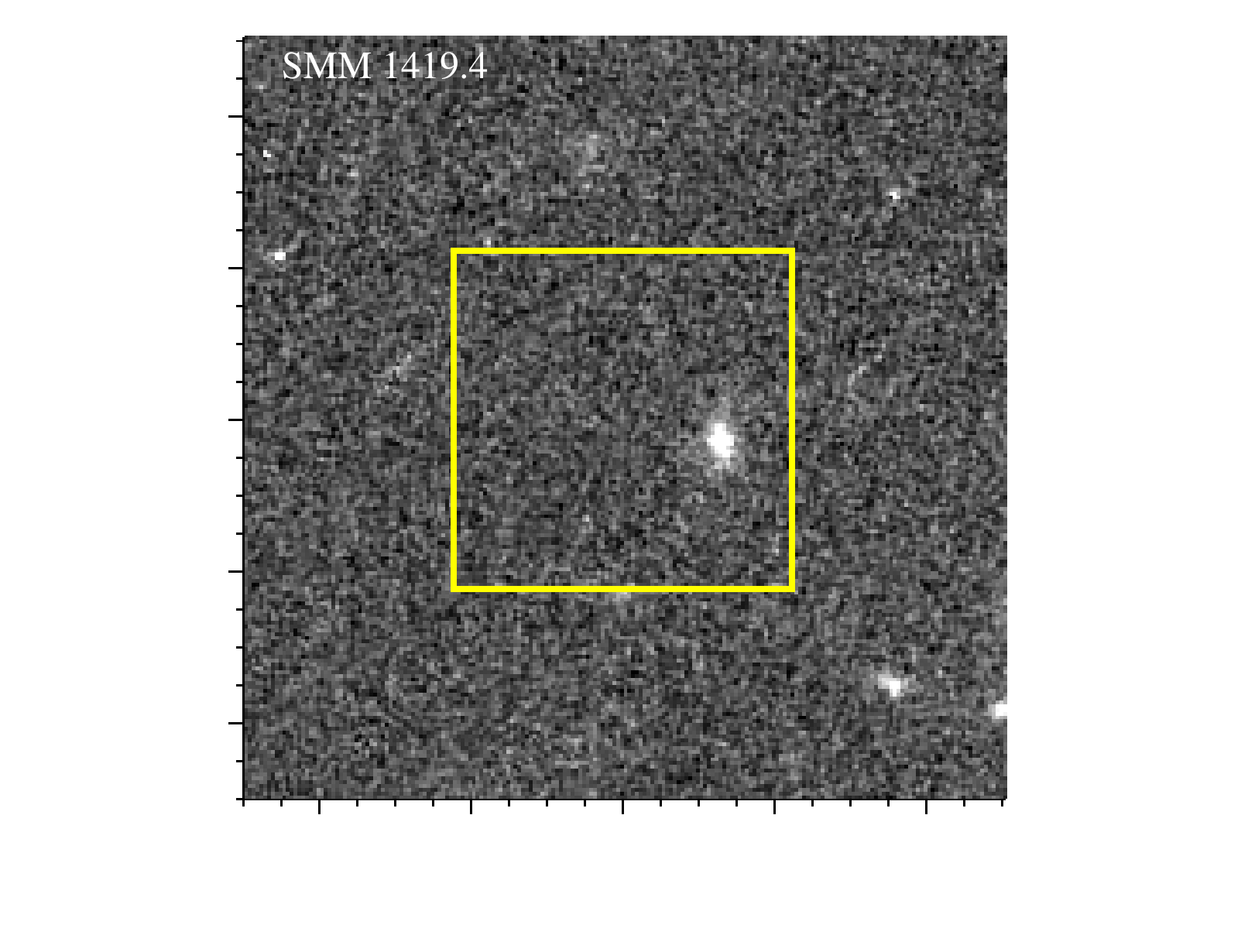}}
  	   	 \vspace{-6mm}
		 
	\subfigure{\includegraphics[scale=0.36]{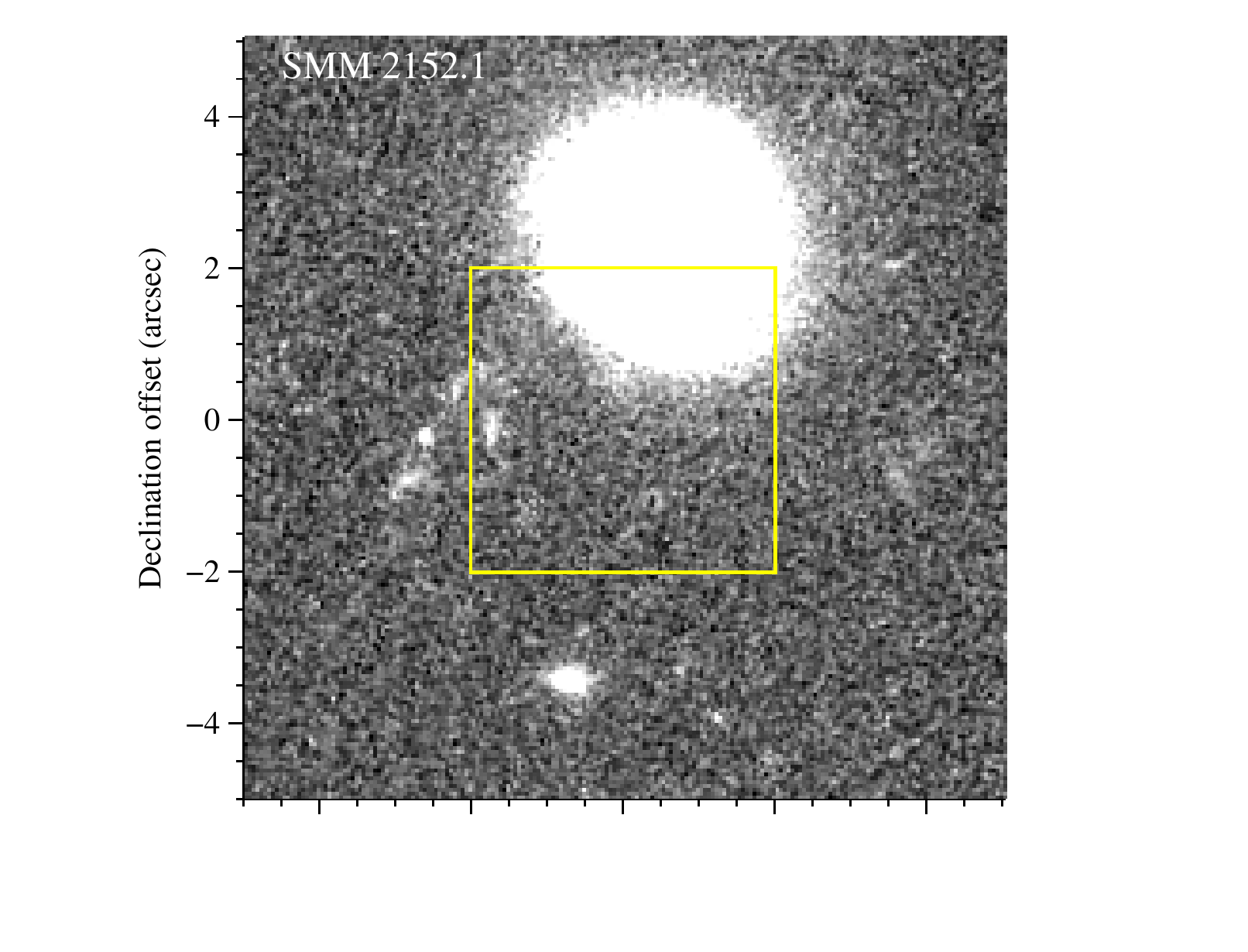}}
		 \hspace{-23mm}
  	   	 \vspace{-5mm}
	\subfigure{\includegraphics[scale=0.36]{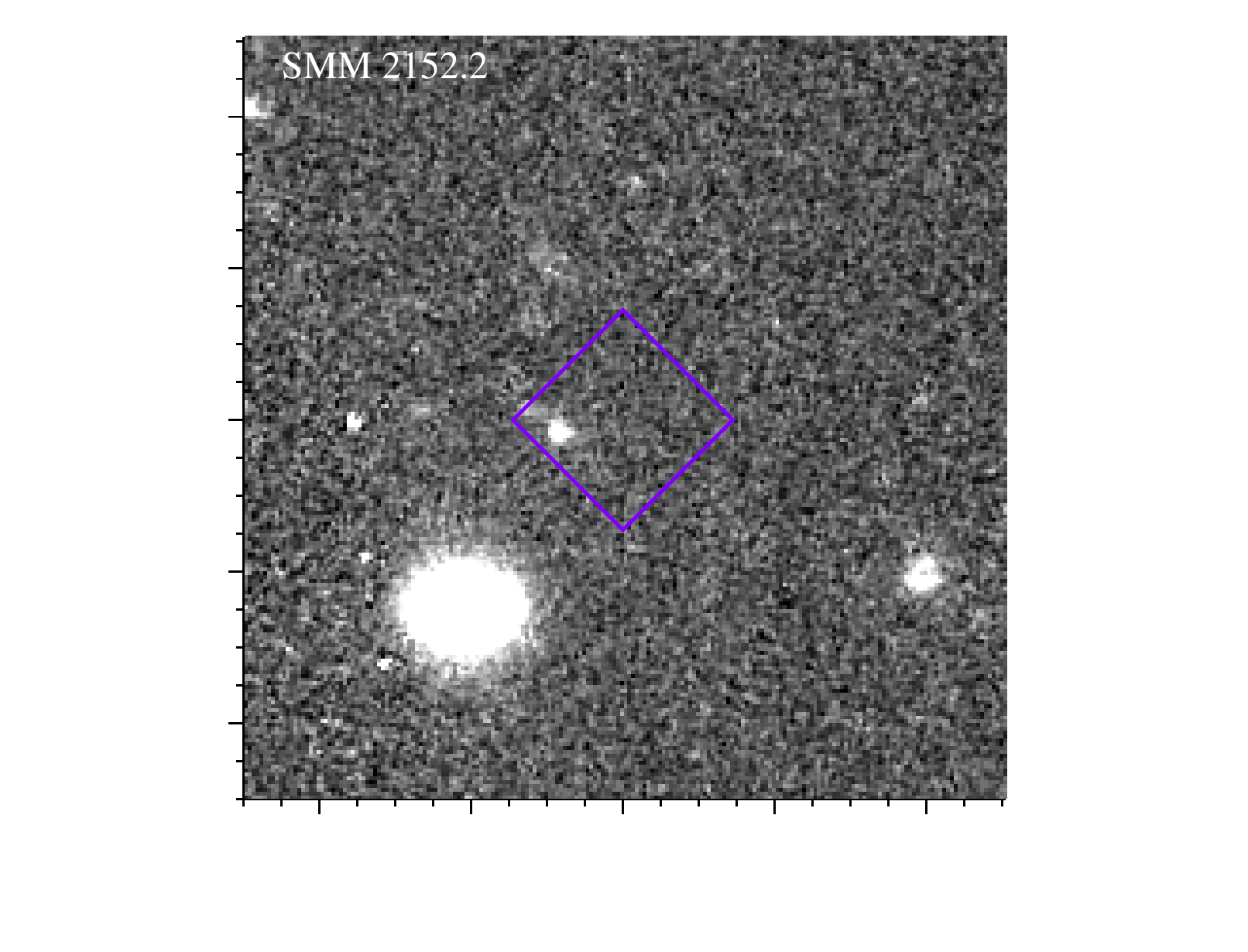}}
  	   	 \vspace{-6mm}

	\subfigure{\includegraphics[scale=0.36]{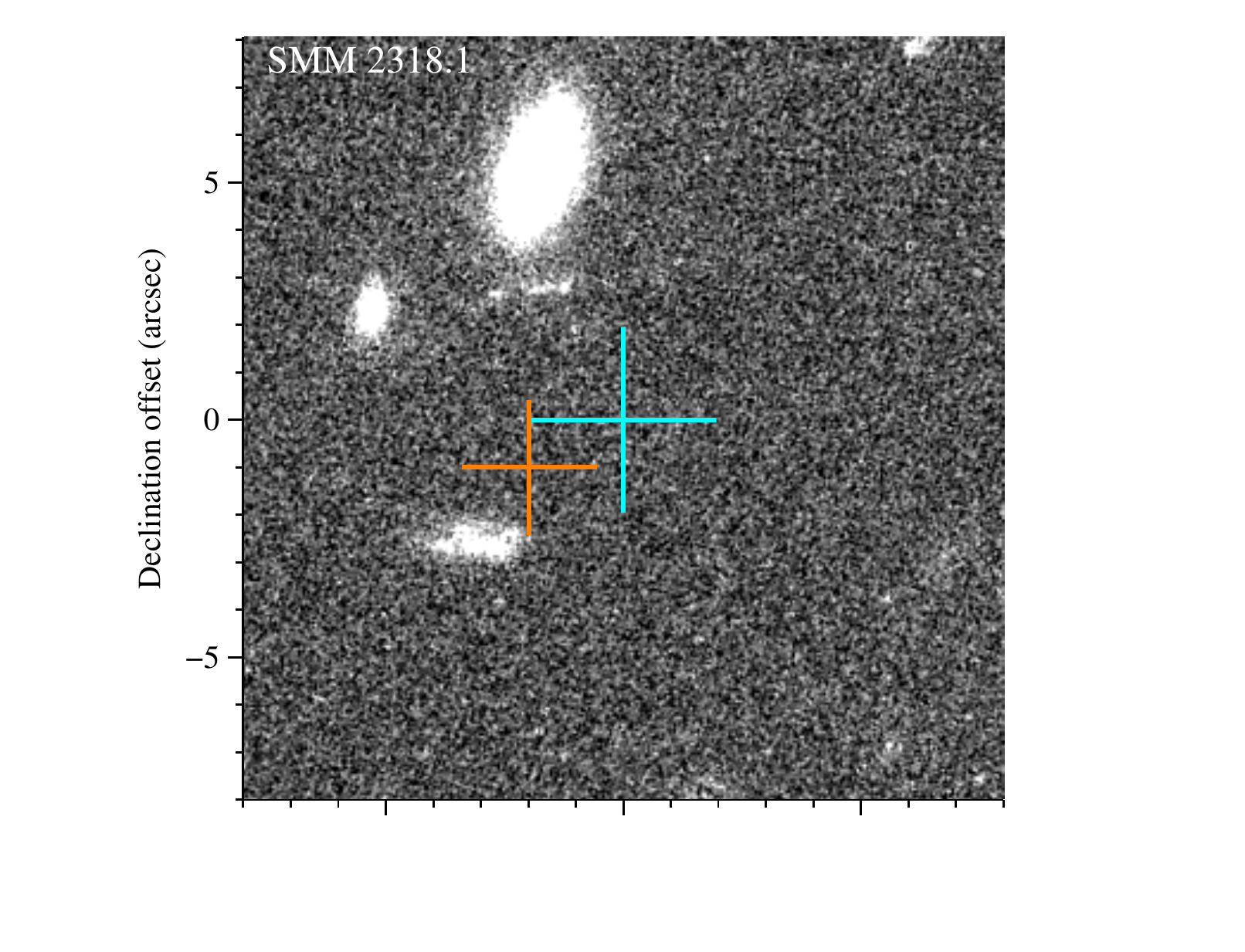}}
          	\hspace{-23mm}
  	   	 \vspace{-5mm}
	\subfigure{\includegraphics[scale=0.36]{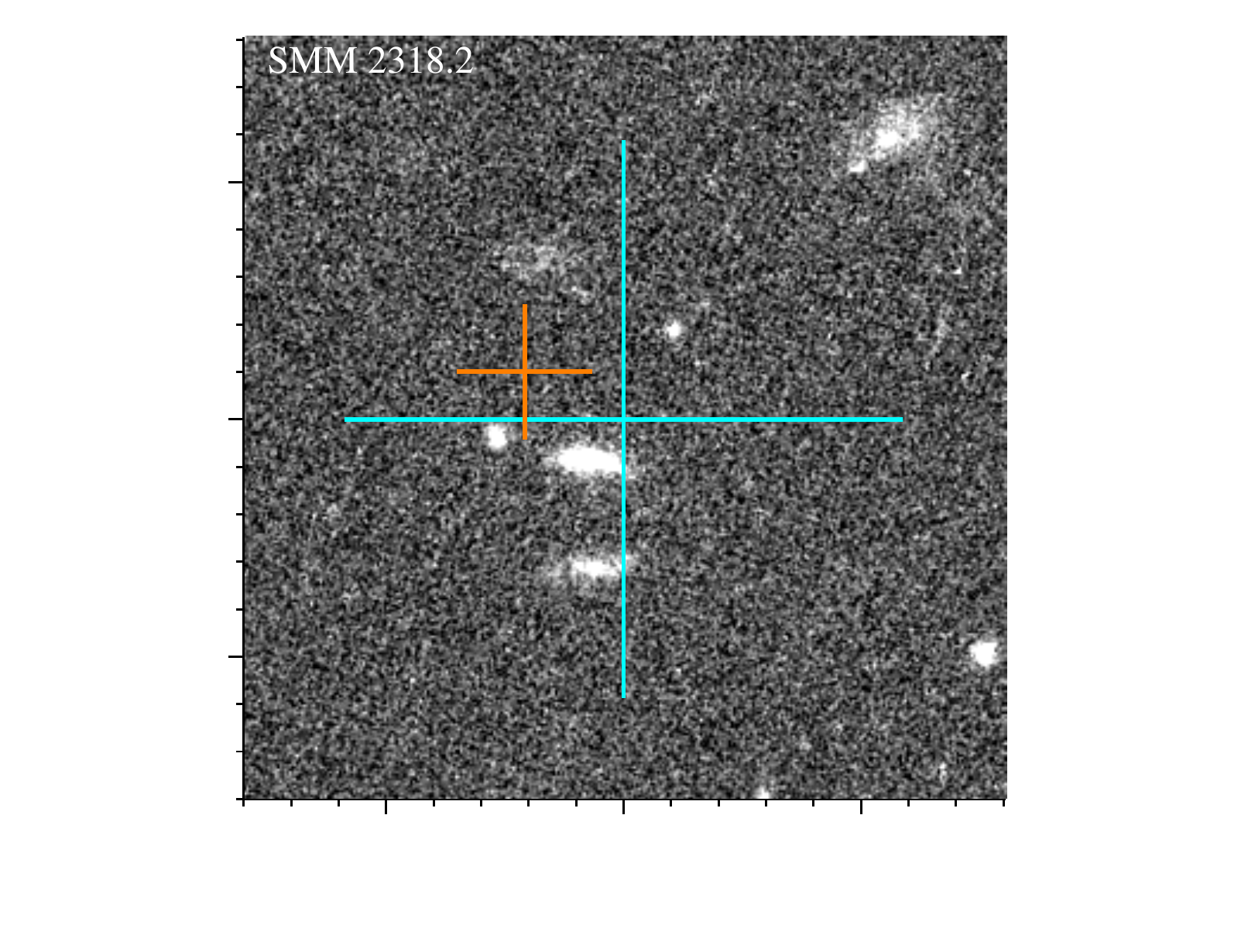}}
		 \hspace{-23mm}
  	   	 \vspace{-6mm}
	\subfigure{\includegraphics[scale=0.36]{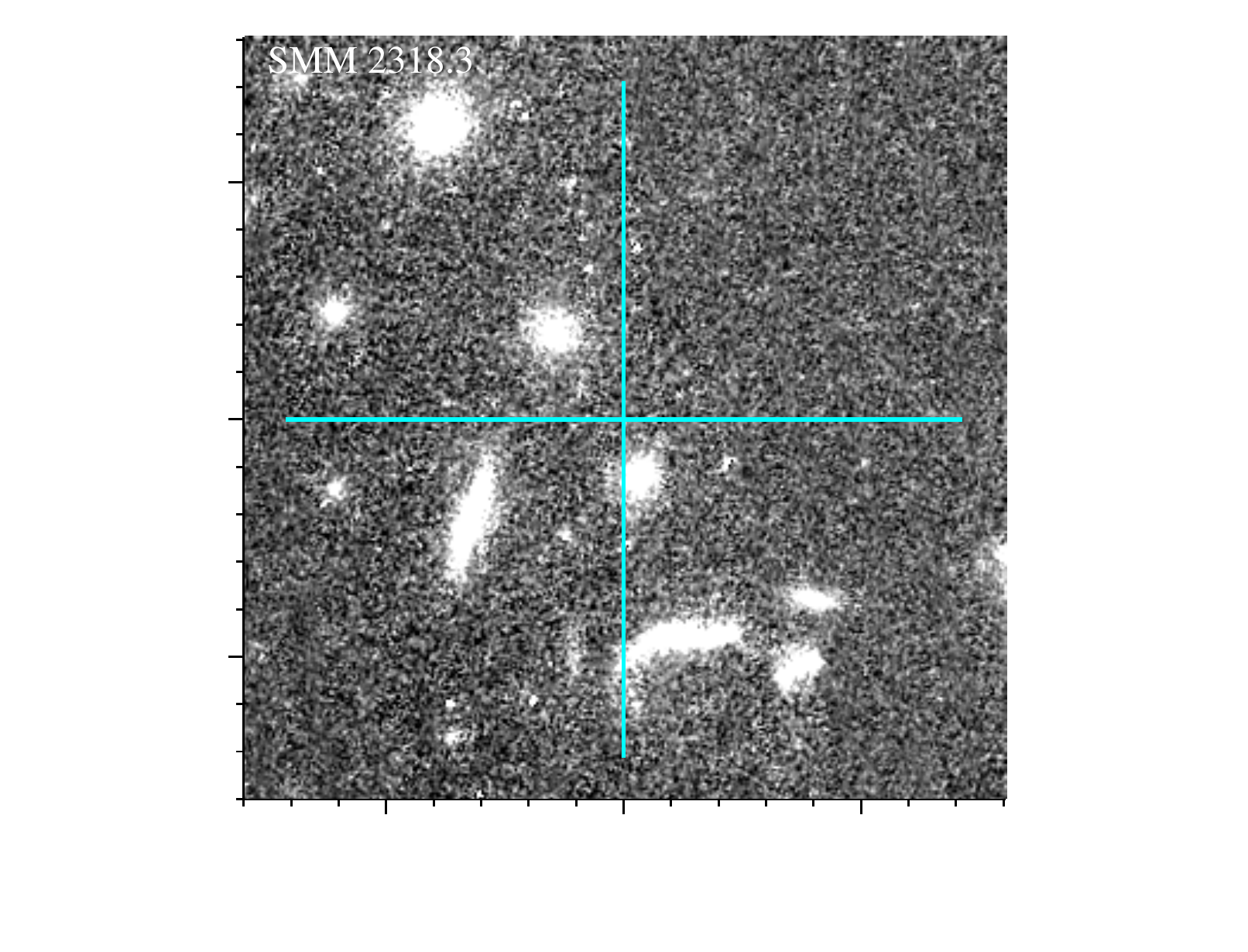}}
  	   	 \vspace{-0mm}	
		 
	\subfigure{\includegraphics[scale=0.36]{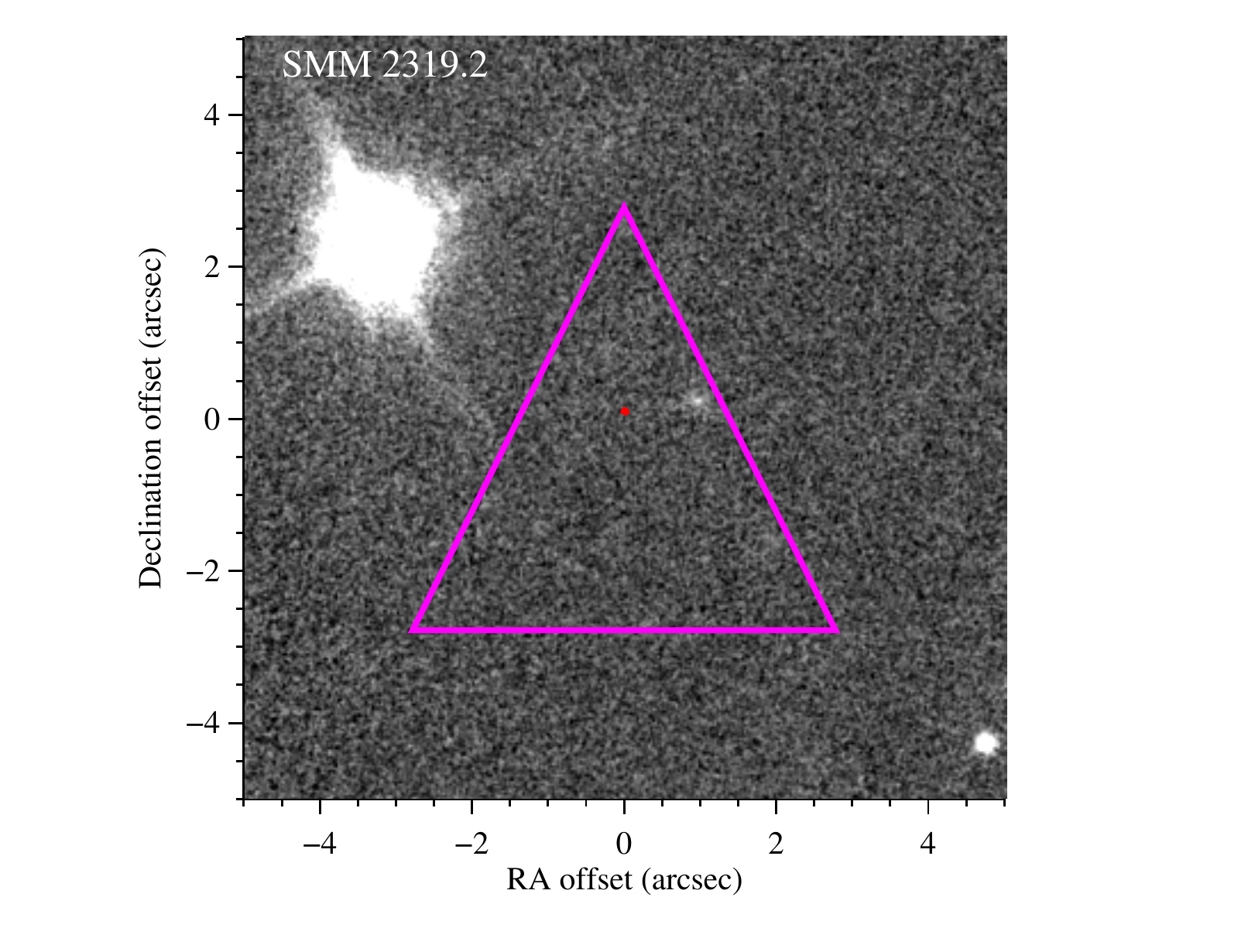}}
             	\hspace{-23mm}
 	   	 \vspace{-2mm}
	\subfigure{\includegraphics[scale=0.36]{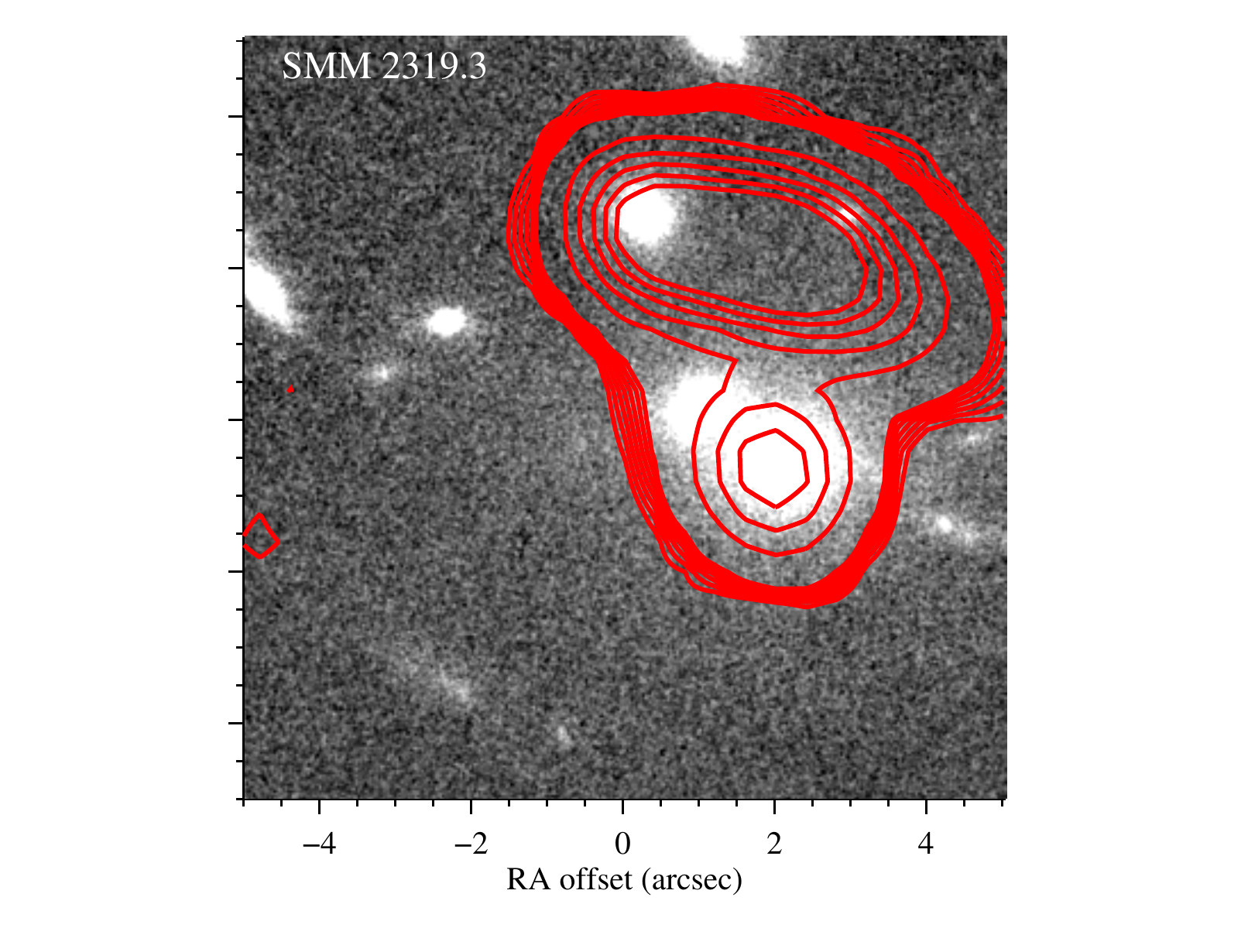}}
  	   	 \vspace{-2mm}
	 \caption{10 $\times$ 10 arcsec$^2$ \textit{HST} postage stamps for each SMG with \textit{HST} coverage; SMGs not listed lack archival \textit{HST} data.  The symbols are the same as listed in Fig.~\ref{fig:secure}.  Only one symbol is shown in each stamp to avoid crowding the images; each stamp is centred at the symbol shown.  SMMs 1122.1 and 1419.1 are smoothed with a 0.3\,arcsec Gaussian to better illustrate faint or diffuse emission.}
\label{fig:hstall}
\end{figure*}


\label{lastpage}

\end{document}